\documentclass[aip,jcp,preprint]{revtex4-2}
\usepackage{graphicx}
\usepackage{amssymb}
\usepackage{amsmath}
\usepackage{bm}
\usepackage{mathrsfs}
\usepackage{tcolorbox}
\usepackage{hyperref}
\usepackage{fancyhdr}
\newcommand{\bra}[1]{\langle #1 |}
\newcommand{\ket}[1]{| #1 \rangle}

\newcommand{\bk}[2]{\bra{#1} #2 \rangle}

\newcommand{\no}{\nonumber}

\newcommand{\ep}{\epsilon}

\newcommand{\rCT}{\mathrm{CT}}

\newcommand{\bmr}{\mathbf{r}}

\newcommand{\bF}{\mathbf{F}}

\newcommand{\mN}{\mathcal{N}}

\newcommand{\eqr}[1]{Eq.~\eqref{eq:#1}}

\newcommand{\eql}[1]{\label{eq:#1}}
\newcommand{\figr}[1]{Fig.~\ref{fig:#1}}
\newcommand{\figl}[1]{\label{fig:#1}}

\newcommand{\bse}{\begin{subequations}}
\newcommand{\ese}{\end{subequations}}
\newcommand{\mM}{\mathcal{M}}
\newcommand{\mcN}{\mathcal{N}}
\newcommand{\tabr}[1]{Table~\ref{tab:#1}}
\newcommand{\tabl}[1]{\label{tab:#1}}
\newcommand{\rLE}{\mathrm{LE}}

\begin{document}
\bibliographystyle{tim}

\title{%
Inverse molecular design from first principles: tailoring organic chromophore spectra for optoelectronic applications
}
\author{James D.\ Green}
\affiliation{Department of Chemistry, Christopher Ingold Building, University College London, WC1H 0AJ, UK}
\author{Eric G. Fuemmeler}
\affiliation{Baker Laboratory, 259 East Avenue, Cornell University, Ithaca, NY 14853, USA}
\author{Timothy J.\ H.\ Hele}
\email{t.hele@ucl.ac.uk}
\affiliation{Department of Chemistry, Christopher Ingold Building, University College London, WC1H 0AJ, UK}
\date{\today}

\begin{abstract}
The discovery of molecules with tailored optoelectronic properties such as specific frequency and intensity of absorption or emission is a major challenge in creating next-generation organic light-emitting diodes (OLEDs) and photovoltaics. This raises the question: how can we predict a potential chemical structure from these properties? Approaches that attempt to tackle this inverse design problem include virtual screening, active machine learning and genetic algorithms. However, these approaches rely on a molecular database or many electronic structure calculations, and significant computational savings could be achieved if there was prior knowledge of (i) whether the optoelectronic properties of a parent molecule could easily be improved and (ii) what morphing operations on a parent molecule could improve these properties. In this perspective we address both of these challenges from first principles. We firstly adapt the Thomas-Reiche-Kuhn sum rule to organic chromophores and show how this indicates how easily the absorption and emission of a molecule can be improved. We then show how by combining electronic structure theory and intensity borrowing perturbation theory we can predict whether or not the proposed morphing operations will achieve the desired spectral alteration, and thereby derive widely-applicable design rules. We go on to provide proof-of-concept illustrations of this approach to optimizing the visible absorption of acenes and the emission of radical OLEDs. We believe this approach can be integrated into genetic algorithms by biasing morphing operations in favour of those which are likely to be successful, leading to faster molecular discovery and greener chemistry. 
\end{abstract}

\maketitle

%
%

\tableofcontents

\section{Introduction}
\subsection{Designer molecules}
Conventional electronic structure theory\cite{sza89a,atk11a} has become highly successful at taking a molecular structure as an input and producing molecular properties as outputs [\figr{flow}(a)]. However, to our knowledge there is no widely-applicable general theory or algorithm for the reverse process, namely starting with a list of desired molecular properties and producing a molecule which satisfies them as an output --- the inverse design problem [\figr{flow}(b)]. This is a shame, since there is huge demand for molecules meeting a given set of criteria, especially in fields such as optoelectronics where there are many (often competing) requirements for the chromophore. These requirements can include a specific absorption/emission wavelength, high oscillator strength, slow internal conversion to the ground state, fast internal conversion to a triplet-triplet state (for singlet fission\cite{smi10a,smi13a,alv19a,zen14a,zen16a}), large singlet-triplet energy gap (for singlet fission), small singlet-triplet gap (for thermally-activated delayed fluorescence, TADF\cite{eva18a,fen20a,mor15a}) and so on. For light-emitting diodes (LEDs), further development of molecular materials will enable realisation of efficient and stable blue LEDs that meet the performance of established red and green pixels for displays,\cite{lee19a,xu20a} as well as in infrared devices for optical communications and bio-imaging\cite{vas21a}. Strong light absorption of molecules in the visible range is useful for indoor photovoltaics, where device efficiency can be optimised for lighting conditions by chemical design\cite{jah21a}. This emerging technology presents opportunities for the design of new molecules tailored for these applications and hence motivates a general theory for solving the inverse design problem. 

\begin{figure}[!h]
\includegraphics[width=0.8\textwidth]{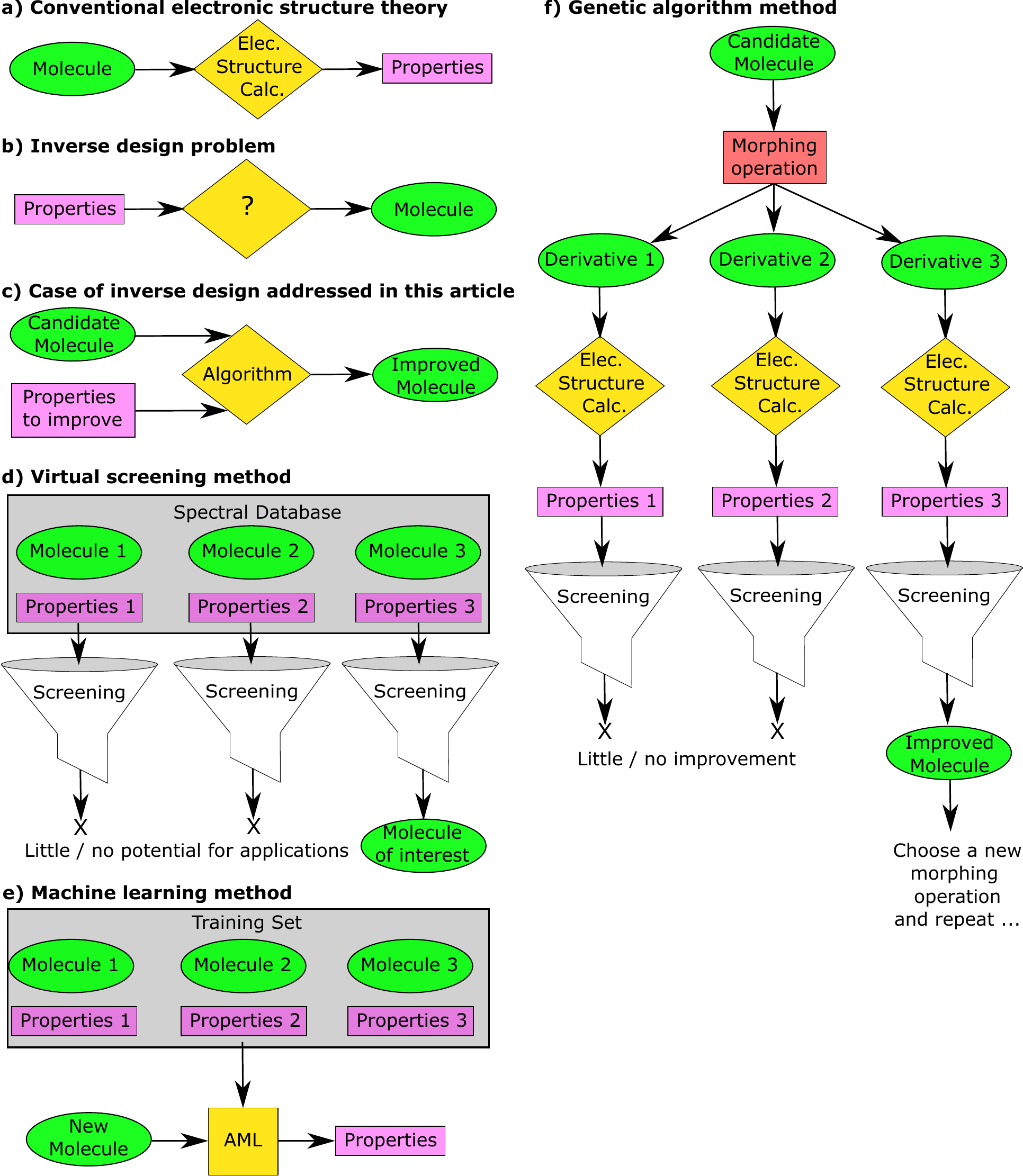}
 \caption{Flow diagrams illustrating the motivation of this article. (a) Standard electronic structure theory: computing properties from a known molecular structure. (b) The overall challenge: computing a molecular structure from a given set of properties. (c) The case of (b) considered in this article: taking a candidate (parent) molecule and properties to optimize, and finding an improved molecule. (d) A virtual screening approach using a known database, but which is constrained by the number of known and documented molecules. (e) A machine learning approach: training an algorithm to rapidly compute the properties of unseen molecules from known chemical data. (f) A genetic algorithm approach: morphing operations to a first generation molecule, computing the properties of the mutated molecules, screening and repeating for multiple generations.}
 \figl{flow}
\end{figure}

This article considers a more specific version of the inverse design problem in \figr{flow}(b), namely starting with (i) a parent molecule which has reasonable, but not ideal, properties, and (ii) knowledge of which of these properties require modification, and from these inputs producing (iii) an improved molecule which better satisfies the desired properties in (ii), as shown schematically in \figr{flow}(c). In this article we show how by combining electronic structure theory and quantum mechanical perturbation theory, we can provide a general theoretical and computational framework from first principles for achieving this \emph{without extensive computation or chemical synthesis}. Furthermore, by applying pre-existing theory to the inverse design problem we obtain three major principles of chromophore design, summarized in the box on page~\pageref{threeprin}. 

\begin{tcolorbox}
\begin{center}
 \textbf{Three Major Principles of Chromophore Design}
\end{center}

\begin{enumerate}\label{threeprin}
\item When considering different isomers of a chromophore, oscillator strength can neither be created nor destroyed but only moved from one region of the spectrum to another. \label{trk}
\item For substitution, excitations mix if and only if they differ by (at maximum) one orbital and the orbitals by which they differ both have amplitude on the substituted atom. \label{sub}
\item For addition and dimerization, using a zeroth-order Hamiltonian of separated monomers: \label{addi}
\begin{enumerate} 
\item Leads to zeroth-order states which have exclusively Local Excitation (LE) or Charge Transfer (CT) character, are strictly orthogonal and form a complete singly-excited basis, \label{zoh}
\item Leads to CT states which at zeroth-order are always dark and whose energies are given by Koopman's theorem, but at first order (and above) can borrow intensity from bright LE states, and whose first-order energy correction is dominated by Coulombic stabilisation. \label{ctbor}
\item LE states mix with CT states through one-electron terms (transfer integrals) which can be determined from examining monomer molecular orbitals, and from the relative geometry of monomer units in the dimer, \label{lect}
 \item LE states mix with other LE states, and CT states with other CT states, via electrostatic terms which can be estimated from the monomers' multipoles and their relative geometry. The LE/LE mixing can be approximated by the Kasha dipole-dipole interaction. \label{lelectct}
\end{enumerate}
\end{enumerate}
\end{tcolorbox}
In this article we derive these rules and provide examples of their applicability.

This perspective is far from the first article to address the inverse design of chromophores\cite{kun21a,ben20a,san18a} and current methods fall into three broad groups. The first is the virtual screening (VS) approach [\figr{flow}(d)] which comprises screening known molecules in a database against design criteria, using molecules' properties contained within the database or computed on the fly, as shown in \figr{flow}(d).\cite{baj02a} This approach is limited by the size of the database and the quality and quantity of data therein, and any database stored on a modern computer is likely to  include only a tiny fraction of chemical space. VS alone is limited in its scope as a prediction tool as it does not offer any new information about potential new molecules, even those which are similar to, or derivatives of, those which are in the database. 

The second approach is active machine learning (AML) [\figr{flow}(e)] where trends in chemical or physical properties are inferred by comparing the properties of a large set of molecules, called a training set, using statistical algorithms such as regression [e.g. Gaussian Processes Regression (GPR)], deep learning (DL), and artificial neural networks (ANN) among others. \cite{but18a}  The training set is taken from a database \cite{hac11a} or generated by successively applying morphing operations to a parent molecule. \cite{kun21a,pol21a} AML offers more predictive power than VS in the sense that it can `learn' from the data being given to it in order to make predictions about unseen data. \cite{car19a} However its predictive power is limited by the size and variation of data in the user-defined training set, as predictions about data outside the space of the training set (i.e.\ extrapolation) can be unreliable. Despite these limitations, AML algorithms have had much success in rapidly predicting chemical and physical properties of atoms and molecules, saving computational effort. \cite{hac11a,bar17a,dra21a}

Finally there is the genetic algorithm (GA) approach\cite{ven94a, lea01a} [\figr{flow}(f)] in which a parent molecule is altered using morphing operations, producing `second generation' molecules which have chemical properties computed for them. These second generation molecules are then screened against a design criteria and those which pass the screening process are then kept and the remainder discarded. The process is then repeated for many generations with many possible morphing operations until a suitable molecule is found. The screening is based on a fitness function, calculated from one or more properties of the molecule, which aims to quantitatively describe how well suited a molecule is for the applications. Those molecules which pass the screeening process must exceed a threshold value of the fitness function. The success of the GA approach is crucially dependent on: i) whether the properties of the parent molecule can be improved and ii) whether the chosen morphing operations can improve these properties. Hence there have been efforts in combining GA and AML for molecular design (see SI Fig. 1), where AML is used to predict which are the most productive morphing operations to be applied to each successive generation.\cite{ven96a,kun21a,pol21a,kwo21a} 
\subsection{Molecular design challenges}
Machine learning, genetic algorithms and combinations of the two are perhaps the best current approaches in the field of molecular design, but still have various challenges, namely that:
\begin{enumerate}
 \item It is not usually known at the outset whether or not, and to what extent, the properties of the first generation molecule can be improved before many calculations are run.

 \item It is not usually known whether the proposed morphing operations will, or can, lead to any improvement, and if they can, where on the molecule they should be made, or how many are required, without running many electronic structure calculations.
\end{enumerate}
Taken together these challenges mean that, until a full AML/GA calculation is run, it is difficult to know whether, and to what extent, it will be successful. Due to the large size of chemical space and the need for accurate property computation, the calculations can therefore be extremely expensive. It would therefore be very helpful to have an indication, in advance of a full calculation, of whether any improvement is possible and if so what morphing operations should be made. In this article we address both of the challenges given above for the case of electronic absorption and emission intensity (and energy) in specific regions of the spectrum, an area of huge interest for optoelectronic applications. 

We stress that the above challenges are not reasons to avoid machine learning and genetic algorithms --- quite the converse. We hope that the theoretical toolkit proposed in this article can be used to inform these algorithms and accelerate the discovery of new useful molecules for optoelectronic applications.

To address the first challenge, we show how the Thomas-Reiche-Kuhn sum rule can be used to provide much lower bounds on the total oscillator strengths of molecules than are usually quoted\cite{atk11a,zhe16a}, giving a more realistic maximum of the total low-energy absorption that can be expected. To address the second challenge, we combine electronic structure theory\cite{sza89a} with intensity borrowing perturbation theory\cite{rob67a} to construct a theoretical framework for describing how morphing operations to a parent molecule are likely to change the electronic structure of its offspring. Crucially, this requires only the knowledge of the electronic structure of the parent molecule and very basic knowledge of the alteration (for subtitution, which atoms will be changed and for addition and dimerization, the relative geometry of monomers). This does \emph{not} require a separate full electronic structure calculation for each possible morphing operation. Using this theoretical framework we show how it is possible to predict whether and which morphing operations are likely to increase the absorption or emission in a particular region of the spectrum (usually the visible), addressing the second challenge. It is also possible to determine where on the molecule the alterations should be made, and to give an approximate idea of the extent to which any given alteration will improve the molecule's properties. Combining these results can then lead to general rules for increasing chromophore absorption and emission. The general methodology is illustrated in \figr{flowc}. 

\begin{figure}[!h]
 \centering
\includegraphics[width=.7\textwidth]{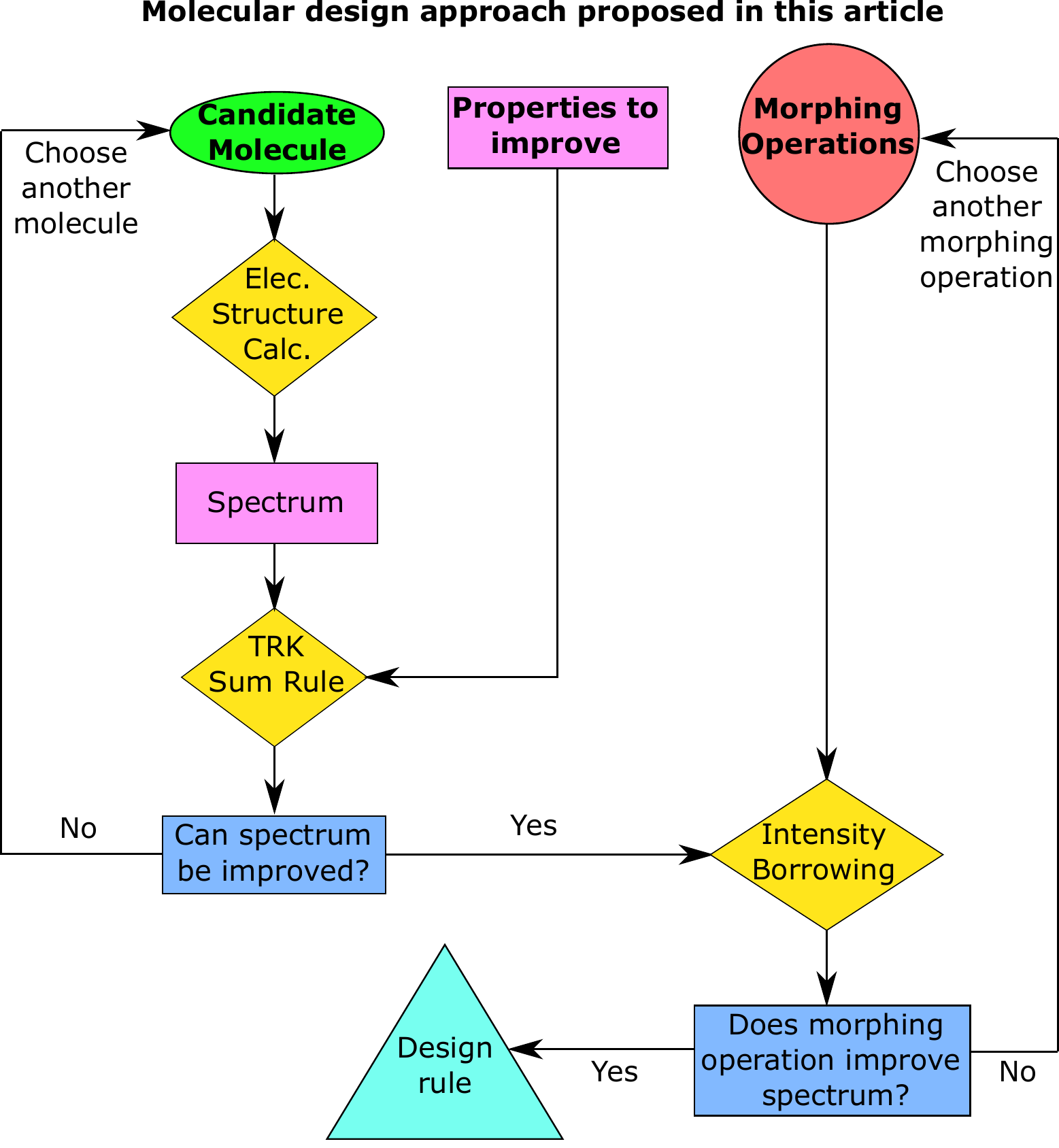}
 \caption{Flowchart illustrating the general methodology proposed in this article. Starting with a candidate (parent) molecule, properties to be improved and possible morphing operations, we show how by combining intensity borrowing perturbation theory and electronic structure theory, we can predict which morphing operations are likely to be successful, and thereby formulate design rules. }
 \figl{flowc}
\end{figure}


\subsection{What to optimize?}
For light-emitting optoelectronics, such as organic light-emitting diodes (OLEDs), emission is usually from the lowest bright excited state, usually S$_1$ for ground-state closed shell molecules and D$_1$ for radicals. The rate of emission is given by the Einstein transition coefficients\cite{atk11a} which in turn depend on the transition dipole moment of the S$_1$ $\to$ S$_0$ transiton (or D$_1$ $\to$ D$_0$ for monoradicals). A key requirement for high-efficiency OLEDs is for the rate of radiative decay, $k_{\rm r}$, to exceed the rate of non-radiative decay, $k_{\rm nr}$, such as reaching the ground state via an avoided crossing or conical intersection. Consequently, a goal for OLED design is to maximise the dipole moment associated with the S$_1$ $\to$ S$_0$ transition. An added complication for ground-state closed-shell species is the dark triplet state T$_1$, which is usually a loss pathway, though can be brightened by phosphorescence, or can emit indirectly through reverse intersystem crossing (RISC) in thermally-activated delayed fluorescence (TADF) devices\cite{eva18a,fen20a}.

Conversely, for light-absorbing optoelectronics such as photovoltaic cells, the priority is usually absorbing as much light as possible within the solar spectrum, which is principally concentrated in the visible (roughly 400--700nm in wavelength). The can be achieved by having a lowest-lying excited state with very broad absorption, or by having multiple absorptions in the visible. In the latter case, a molecule absorbing above the S$_1$ state usually undergoes rapid internal conversion to S$_1$, such that most photophysics of interest, such as charge generation, occurs from the S$_1$ state. Exceptions to this include intramolecular singlet fission\cite{smi10a,smi13a,san16b,san16a,san15a,tay17a,alv19a}, where the lowest bright state undergoes internal conversion to a (usually dark) triplet-triplet state. Many common organic molecules such as acenes have intense absorption in the UV (where there is less solar irradiance) but weak absorption in the visible, making them ideal candidate molecules for the spectral optimization discussed in this article.

Consequently, for OLEDs it is of particular interest to increase the transition dipole moment between the ground state and the lowest bright excited state, and for photovoltaics to increase absorption in the visible. These are therefore the main motivations of the theoretical framework in this article. Although we apply our methodology to the low-lying excited states of conjugated organic molecules (for which there is large interest for optoelectronics) the general theoretical principles can be applied to optimizing any property of a molecule which can be related to a quantum mechanical operator, such as spin-orbit coupling, permanent dipole, non-adiabatic coupling etc.

\subsection{The general idea}
Clearly there are an enormous range of chromophore alterations, and in this article we consider heteroatom substitution, addition and dimerization, where dimerization can be considered a form of addition where the adduct is identical to the parent chromophore. 

The general idea, as illustrated in \figr{flowc}, is to start with a molecule that has reasonable, but not ideal properties (such as pentacene for visible absorption), properties to improve (visible absorption) and morphing operations (substitution, addition, and dimerization). The electronic Hamiltonian of the parent molecule can be solved at an approximate level of theory, leading to the spectrum of the parent chromophore. In the language of perturbation theory, this defines our zeroth-order Hamiltonian and zeroth-order states. From this spectrum and the TRK sum rule we can deduce whether or not a molecule has significant potential for improvement (for pentacene, having large absorption in the UV which could be moved to the visible). For substitution, we then define the morphing operations as perturbations to the zeroth order states (for addition and dimerization the algebra is similar but more involved) and use intensity borrowing perturbation theory\cite{rob67a} to determine how morphing alterations will alter the spectrum. In the event that the morphing operation does not improve the spectrum, a different one can be tried, and if the spectrum is improved a design rule can usually be derived. 


By construction, a perturbative approach will not give an exact solution for the chromophore's electronic structure (even within an approximate Hamiltonian), but the qualitative results can then be used to guide which morphing operations are more likely to improve the desired property. Much research on organic chromophores has focused on low-energy HOMO$\to$LUMO transitions in organic molecules (whether local excitation or charge-transfer in nature)\cite{smi10a,smi13a,sch17a} and the theoretical results in this article are applicable to single excitations between any occupied and unoccupied orbitals. This can therefore include excitations in the visible which are higher-lying than the HOMO$\to$LUMO transition, and whose optimization is unlikely to substantially affect the pre-existing photophysics (such as ability to undergo singlet fission) since Kasha's rule\cite{kas50a} implies rapid internal conversion to the lowest excited singlet state.

In this article we restrict our attention to electronic intensity borrowing, with application to the (generally weaker) effects of vibronic (Herzberg-Teller) coupling\cite{atk11a,orl72a,ren04a} and spin-orbit interactions\cite{bel01a,mar12b,atk11a} left for future research. Although not the main focus of the article, a by-product of intensity borrowing perturbation theory are expressions for perturbed energies, and these can be used to guide how a given alteration can lead to spectral blue- or red-shifting. In addition, the spectra of organic molecules commonly contain vibrational stretching progressions which are of experimental and theoretical interest\cite{yam11a,hes15a}. These progressions spread the oscillator strength associated with a single transition over a number of vibronic peaks, but since the sum over Franck-Condon factors is equal to unity\cite{atk11a}, do not increase the total absorption associated with a single electronic transition, and are consequently not the focus of this article. We seek to provide general rules for suggesting which molecules are likely to have improved absorption/emission, rather than high-level calculation on a single molecule for which many methods already exist. Similarly, we do not consider zero-point-energy adjustments\cite{hel21a} which may be required to replicate accurately experimental spectra.

For construction of a suitable Hamiltonian, there are a wide range of electronic structure methods available of varying computational cost and accuracy, and here we consider the simplest model which describes the necessary physics of organic chromophores. While H\"uckel theory is arguably the simplest model for simulating arbitrary $\pi$ systems, it neglects two electron terms so (for example) does not account for the exchange energy of forming a singlet excitation or its Coulomb stabilization, leading to an inaccurate description of excited states\cite{par53a}. We therefore use Pariser-Parr-Pople theory\cite{par53a,par56a,pop53a,pop55a,kou67a}, which is similar to H\"uckel theory but also includes two-electron terms within the neglect of differential overlap (NDO) approximation. Early applications of PPP theory include simulating the spectra of acenes\cite{par56a} and their substituted derivatives\cite{suz73a,kou67c,suz76a} and it continues to be widely used\cite{woh01a,ary15a,ary15b,ary13a,alb98a,cha13a,cha99a,jai13a,hu15a,pac06a,lob99a,mon90a} for the simulation of large conjugated systems. In this article we use configuration interaction singles (CIS) to describe excited states, since the ground state of many organic chromophores can be reasonably well described by a single restricted-Hartree Fock (RHF) determinant as the dipole moment is a one-electron operator, meaning that most double and higher excitations are dark\cite{cor91a}. There are of course situations where double and higher excitations play an important role, for example in modelling the triplet-triplet state singlet fission, or the dark doubly-excited $2A_g^-$ state in $\beta$-carotene. However, in this article we concern ourselves with describing linear absorption spectra in which single excitations dominate.\cite{zen14a,sza89a} 
Nevertheless, PPP can be extended to double and higher excitations\cite{gro88a,son07a,zha11a}, as could intensity borrowing theory. As we shall see, PPP can accurately describe the spectral phenomena we consider in this article, though for weak interactions (such as Dexter) which rely on orbital overlap, different methods may be required.

There are, of course, many other electronic structure methods which could be used such as density functional theory (DFT) based approaches. Time-dependent DFT (TD-DFT) can sometimes predict inaccurate energies for CT states due to the incorrect long-range behaviour of local exchange-correlation functionals such as B3LYP, and therefore one has to employ a modified functional which is corrected for long-range behavior.\cite{dre03a,pea08a,ali20a,ste09a,won08a} This difficulty in predicting the energies of CT states is particularly problematic for acenes.\cite{won10a} Alternatively, complete active space methods such as CASSCF could in theory be used, but their high computational expense, the difficulty in selecting an active space,\cite{zen14a} and the need to incorporate perturbation theory in order to obtain accurate energies\cite{hel19a} makes them unsuited to simulating large numbers of excited states in many candidate molecules. 
 
Although the underlying theories in this article (TRK sum rule, intensity borrowing perturbation theory (IBPT), configuration interaction singles and PPP theory) have been around for decades, we believe this is the first time the TRK sum rule has been applied to the inverse molecular design problem, for which it could be used to formulate a fitness function. Combining PPP theory and IBPT to predict molecular spectra was, to our knowledge, first proposed in 2019\cite{hel19a} for the specific case of acene dimer absorption, after which it has been applied to the design of radical OLEDs\cite{abd20a}. However, to the best of our knowledge this is the first time this algebraic framework has been published generally and in full (i.e.\ not for a single specific application) and the first time it presented in the context of the inverse design problem or artifical intelligence approaches.

\subsection{Background chromophore theories}
There are clearly a large range of pre-existing chromophore theories which we briefly review here. 
Since the proposition of the chromophore theory of colour in 1876\cite{wit76a} there have been continued efforts to describe the effects of molecular structure upon optical absorption. Early developments, mainly concerned with excitations in crystals, include the eponymous research on excitons by Frenkel\cite{fre31a}, Wannier\cite{wan37a}, Mott\cite{mot38a} and Davydov\cite{dav49a}. For molecular absorption, Kasha's exciton model\cite{kas65a} describes chromophore interaction by a point-dipole approximation, and more theoretical approaches include the Thomas-Reiche-Kuhn (TRK) sum rule\cite{rei25a,tho25a,kuh25a,atk11a} and intensity borrowing perturbation theory\cite{mul39a,rob67a} which are used in this article. Textbook\cite{atk11a} approaches to chromophore properties include the particle-in-a-box model (predicting an intense HOMO$\to$LUMO transition which redshifts with increasing molecular size) and satisfying spin and point group symmetry.

Since then, there has been substantial interest in chromophore alteration and interaction in biological systems such as porphyrins\cite{gou59a,sey69a}, DNA\cite{wei71a} and green fluorescent protein\cite{moo13a,ton04a}, in photovoltaic applications\cite{qin08a,sch17a,mis09a} such as singlet fission\cite{smi13a,hav16a,bel13a,kum17a}, and in organic light-emitting diodes\cite{shu17a,bru01a} such as thermally activated delayed fluorescence\cite{eva18a,fen20a}, radical emitters\cite{ai18a,abd20a,hud21a} and carbene-metal-amides\cite{di17a,hel18a}. Examination of intermolecular interaction in crystals has led to explanation of crystallochromy\cite{kaz94a}, and investigation of single molecule junctions to the effects of conformation on intramolecular interaction\cite{ven06a,woi89a}. In addition, tuning the absorption frequency (colour) of a chromophore has been achieved by altering the HOMO-LUMO gap with various substitutents\cite{hag07a,leh16a,app10a,sch17a}, and design principles formulated by examining large groups of previously-synthesised or computed chromophores\cite{mis09a,sch17a}.

Despite this progress challenges remain. The absorption of organic molecules in the visible is often a small fraction of the maximum allowed by the TRK sum rule\cite{zhe16a} and it can be unclear how to optimize chromophore structure to increase absorption intensity (extinction coefficient)\cite{zol03a,chr15a}. Rational design principles would arguably have wider and clearer applicability than those obtained empirically\cite{mis09a}, and the large size of chemical space can render a trial-and-error approach inefficient. Nevertheless, the design of highly absorbent chromophores would be of considerable practical benefit such as enabling the construction of thinner solar cells without attenuating the absorption of light\cite{zhe16a}, thereby requiring smaller exciton diffusion and leading to greater photovoltaic efficiency\cite{for05a}. Similarly, designing OLED emitters with greater intensity of emission (transition dipole moment) would lead to faster radiative decay\cite{atk11a} and a greater likelihood of this outcompeting undesirable non-radiative processes such as internal conversion.\cite{hud21a}.

\subsection{Article structure}


In section~\ref{sec:chal1} we address the challenge of determining the extent to which a given molecule's spectrum can easily be improved. To do this we examine the Thomas-Reiche-Kuhn sum rule in section~\ref{ssec:TRK} and see how this leads to lower limits on UV-vis transitions than is sometimes quoted. We illustrate this by application to a variety of organic chromophores in section~\ref{ssec:typchromo}. We then address the second challenge of predicting where and how a molecule should be substituted, added to or dimerized in section~\ref{sec:chal2}. To do this we firstly recap pre-existing intensity borrowing theory (section~\ref{ssec:ibpt}). We then apply this theory, using the framework of configuration interaction singles and Pariser-Parr-Pople theory (see SI section III A 1 and 3), to substitution in section~\ref{ssec:sub}, addition in section~\ref{ssec:add} and dimerization in section~\ref{ssec:dimer}. Having addressed these challenges from a theoretical perspective, we then apply the methodology to real chromophores in section~\ref{sec:realmol}. We consider improving the visible absorption of acenes in section~\ref{ssec:acenes} followed by improving the emission of organic radicals in section~\ref{ssec:brightrad}. We conclude in section~\ref{sec:conclusion}.

\section{Challenge 1: How improvable is a molecule?}
\label{sec:chal1}
\subsection{Thomas-Reiche-Kuhn sum rule}
\label{ssec:TRK}
Here we consider the Thomas-Reiche-Kuhn sum rule\cite{tho25a,rei25a,kuh25a} (TRK) theoretically to determine approximate upper bounds to the absorbance of organic chromophores. For a system with $N_e$ electrons TRK is commonly quoted as\cite{zhe16a,atk11a}
\begin{align}
 \sum_{u>0} f_u = N_e \eql{trk}
\end{align}
where the oscillator strength $f_u$ of excitation to eigenstate $\ket{\Psi_u}$ is
\begin{align}
 f_u  = \frac{2m_{\rm e}}{3 \hbar^2} | \bra{\Psi_0} \hat \bmr \ket{\Psi_u}|^2 (E_u - E_0) \eql{osc}
\end{align}
where $m_{\rm e}$ is the mass of an electron and $\hat \bmr$ is a generalized co-ordinate vector. As presented in \eqr{trk}, the TRK refers to excitations from the ground state, though it also holds for any excited state. Qualitatively, \eqr{trk} means that oscillator strength is conserved, and that while oscillator strength can be moved from one area of the spectrum to another (as \eqr{trk} and \eqr{osc} give no limits on the energy of a transition), it can neither be created nor destroyed, i.e.\ structucal isomers will have the same total oscillator strength for all transitions from the ground state.

Computationally, it is found that low-energy (such as visible) excitations often constitute a tiny fraction of the total allowed oscillator strength\cite{zhe16a}, and since only the $N_{\rm v}$ valence electrons are expected to constribute to low-energy transitions \eqr{trk} is sometimes approximated as $N_{\rm v}$.\cite{atk11a}
This would suggest that for an organic chromophore with $N_{\pi}$ electrons in the $\pi$ system, the upper limit on oscillator strength would be $N_{\pi}$.
However, the derivation of TRK holds in each dimension $\{x,y,z\}$ separately\cite{atk11a} such that
\begin{align}
 \sum_{u>0} f_u^{(x)} = \frac{N_e}{3} \eql{trkx}
\end{align}
where
\begin{align}
 f_u^{(x)} = \frac{2m_{\rm e}}{3 \hbar^2}\sum_{u>0} | \bra{\Psi_0} \hat x \ket{\Psi_u}|^2 (E_u - E_0) \eql{oscx}
\end{align}
and likewise for $y$ and $z$. Clearly equations~\eqref{eq:trkx} and \eqref{eq:oscx} are consistent with, but more restrictive than, the TRK in \eqr{trk}, though in themselves do not suggest a new upper limit for chromophore absorption. 

Let us then consider a linear $\pi$-system, with atoms aligned along the $x$ axis. Using a basis of $p$ orbitals on each atom (as in H\"uckel and PPP theory), $\pi \to \pi^*$ transitions must be $x$-polarized. 
This means that the maximum oscillator strength for any linear organic chromophore, which carotenoids (polyenes) can be approximated to be\cite{zhe16a}, is $N_{\pi}/3$
where $f_u^{(x,\pi)}$ refers to $x$-polarized transitions involving $\pi$ electrons. 

Let us now consider a planar $\pi$-system such as pentacene or tetracene, with atoms in the $(x,y)$-plane. Using similar arguments to the above, transitions can only be $x$ or $y$-polarized, meaning that the maximum allowed oscillator strength is $2N_{\pi}/3$.
We see that both the linear and planar limits are smaller upper limits on absorption than $N_{\pi}$ for a general 3-dimensional chromophore, and are summarized in Table~\ref{tab:trk}, allowing us to appraise more accurately the extent to which the absorption of a chromophore in the visible (or any other region of the spectrum) is close to the maximum possible, or whether there may be scope to increase absorption further. 
\begin{table}[h]
 \begin{tabular}{c|c}
  Chromophore & $\sum_u f_u$ \\ \hline
  Linear $\pi$ system & $N_{\pi}/3$ \\
  Planar $\pi$ system & $2N_{\pi}/3$ \\
  General & $N_{\rm v}$
 \end{tabular}
 \caption{Approximate upper bounds on chromophore absorption. $N_{\pi}$ is the number of electrons in the $\pi$ system, $N_{\rm v}$ the number of valence electrons.}
\label{tab:trk}
\end{table}

These results allow us to define an `absorption efficiency' in a spectral region by comparing the integrated oscillator strength\cite{zhe16a} in that region with the approximate absorption upper bound to total absorption in Table~\ref{tab:trk}. For example, a planar chromophore absorbing in the visible would give
\begin{align}
\eta_{\rm abs} = & \frac{\sum_{\lambda=400nm}^{700nm}f(\lambda)}{\frac{2}{3}N_{\pi}} \\
= & \frac{2m_e c \epsilon_0}{N_{\mathrm{A}}N_{\pi} e^2} \int_{4.3\times 10^{14}\textrm{s}^{-1}}^{7.5\times 10^{14}\textrm{s}^{-1}} d\nu \epsilon(\nu) \eql{osceta}
\end{align}
where frequency is given in Hz. This can be calculated from both theory and experiment, and unlike molar extinction coefficient goes not grow simply by oligomerizing the chromophore.\cite{zol03a} Equation~\eqref{eq:osceta} gives a rough guide to whether much of the possible absorption is in the desired region of the spectrum and there is little scope for improving efficiency, or whether only a small fraction of absorption is in the visible, with substantial scope for improvement by `borrowing intensity' from high-energy transitions. In the context of machine learning, the absorption efficiency in \eqr{osceta} could be incorporated into a fitness function for optimizing organic chromophores.

\begin{figure}[h]
 \centering
 \includegraphics[width=.6\textwidth]{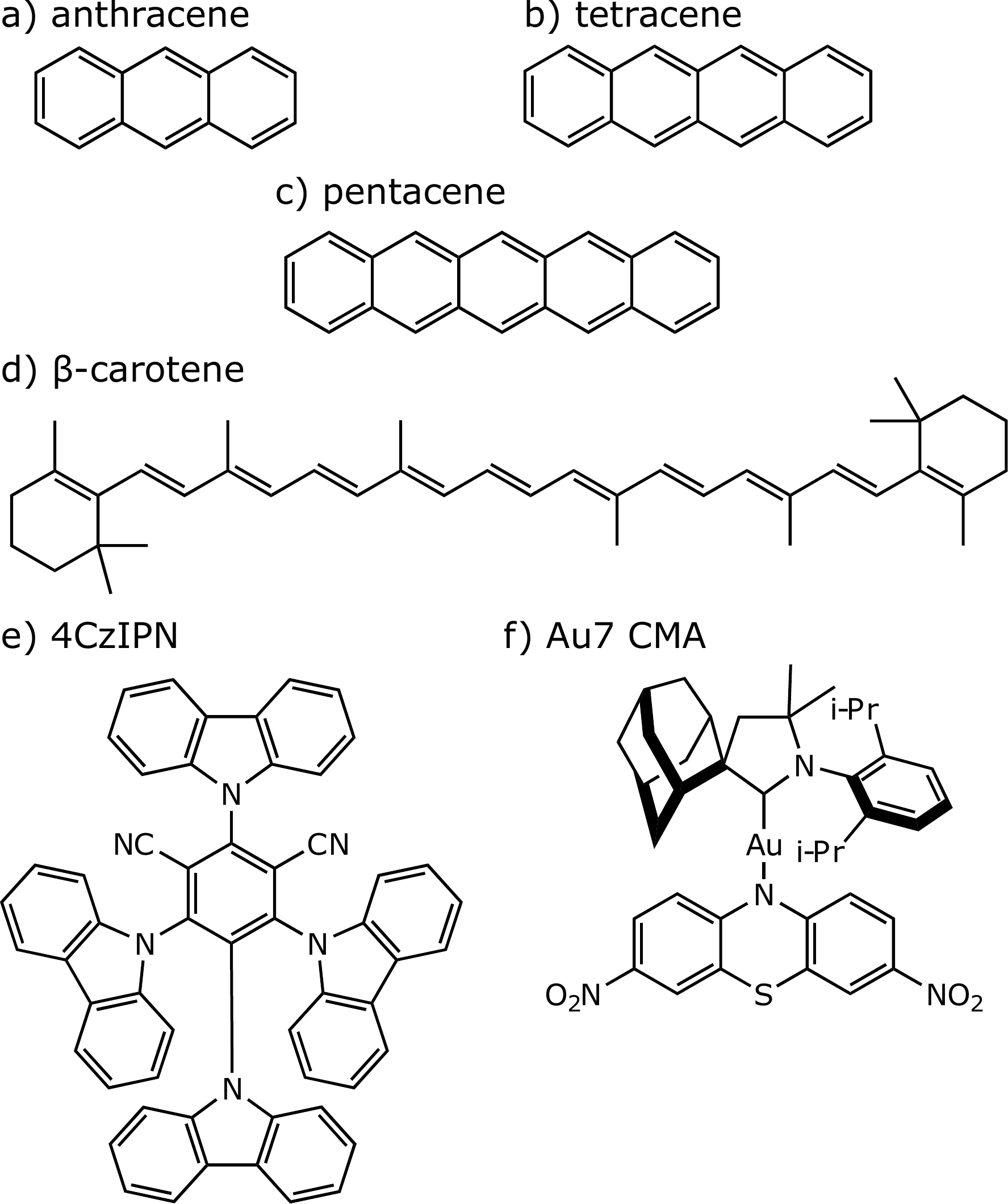}
 \caption{Structures of example chromophores whose absorption is considered in \tabr{typ_crm}. $\beta$-carotene is an example of a polyene, and acenes anthracene, tetracene and pentacene are presented. We also include the common TADF molecule 4CzIPN and a carbene-metal-amide (CMA) chromophore, Au7.}
 \figl{typ_chrphrs}
\end{figure}

\subsection{Typical Chromophores}
\label{ssec:typchromo}
\begin{table}[h]
\centering 
\vspace*{2mm} 
 \begin{tabular}{c|c|c|c|c|c}
  Chromophore & State & $ \lambda $ & $ f_{osc} $ & TRK Max. & \% of TRK Max. \\ \hline
  Ethene\cite{pla49a} & $1B_{3u}^+$ & 163 &  0.34 & 0.67 ($x$) & 51 \\
  Anthracene\cite{kle49a} & $1B_{2u}^+ $ & 379 & 0.1 &9.33 ($x,y$) & 1.1  \\  
                     & $1B_{3u}^+ $ & 256 &  2.28 & &  24 \\
  Tetracene\cite{kle49a}& $1B_{2u}^+ $ & 474 & 0.08 & 12 ($x,y$) & 0.7  \\ 
                     & $1B_{3u}^+ $ & 275 &  1.85 & & 15  \\
  Pentacene\cite{kle49a}& $1B_{2u}^+ $  & 585 & 0.08& 14.67 ($x,y$) & 0.5 \\ 
                  & $1B_{3u}^- $  & 417 & $\sim$0& & $\sim$0 \\
                  & $1B_{3u} ^+$ & 310 & 2.2 & &15  \\
   $\beta$-carotene \cite{ona99a,kan94a,kle09b} & $2A_{g}^-$ & $\sim$700 & 0 & 7.33 ($x$)& 0 \\
                    & $1B_{u}^+$ & 488 & 2.66 & 	 &  36\\
  4CzIPN (calc.)\cite{lin17a}  & $B$ & 474 & 0.05 & 68 ($x,y,z$) & 0.1 \\
  Au7 CMA (calc.)\cite{rom20a}&  $A_{1} $  &  434  & 0.26  &  36 ($x,y,z$) & 0.7 \\
 \end{tabular}
 \caption{Vertical excitation energies and oscillator strengths for a series of $\pi$-systems from experimental data and ab initio calculations (see SI Table I for more details).} 
\tabl{typ_crm}
\end{table}   

For example, it has been known since the 1930s that much of the oscillator strength of carotenoids is in the $S_0\to S_1$ transition \cite{mul39a} and higher-energy $\pi\to\pi^*$ transitions are relatively weak. Conversely, acenes have a relatively weak $S_0 \to S_1$ transition and a much more intense transition ($x$ polarized) at higher energies \cite{par56a}. However, in general the absorption of hydrocarbons in the visible falls well below the lower limits in Table~\ref{tab:trk}. In \figr{typ_chrphrs} we present a selection of common organic chromophores and in Table~\ref{tab:typ_crm} compare their absorption with the maxima given by the TRK sum rule. Tetracene, for example, has a transition around 474nm with oscillator strength (from early experiments \cite{kle49a}) of 0.08, corresponding to 0.7\% of the total possible oscillator strength in the visible region. In the near UV around 275nm, it has a transition with oscillator strength of 1.85, 15\% of the total allowed. Tetracene also has other transitions in the visible which are dark. This suggests that (for tetracene at least) there is substantial scope for improving the visible absorption, because (a) the total absorption in the visible is a tiny fraction of the total allowed, (b) there is substantial oscillator strength `nearby' in the spectrum that in theory could be borrowed and (c) there exist states in the visible which could in theory borrow intensity.  

From the results in Table~\ref{tab:typ_crm} and in the literature, we can approximately categorise many chromophores:
\begin{enumerate}
\item{Those whose lowest energy transitions are in the UV (such as ethene), suggesting that increasing their visible absorption is difficult.}\label{cat_noimpov}%
\item{Those which have intense visible absorption (such as $\beta$-carotene) but undesireable photochemical properties, such as a dark $S_1$ state} \label{cat_darkS1}
\item{Those which have weak absorption in the visible which could in theory be increased by borrowing intensity from transitions in the UV. Some of these molecules (such as acenes, 4czIPN, carbene-metal-amides)\cite{hir12a, di17a} also have favourable optoelectronic properties such as the ability to undergo singlet fission or TADF.} \label{cat_ib}
\item{Those whose lowest energy transition is in the near UV that could be redshifted into the visible (such as anthracene)\cite{jon47a}, therefore having potential for increased / tailored visible absorption.} \label{cat_redshift}
\item{Those whose lowest energy transition is in the near infra-red but which could be blue-shifted into the visible, such as TTM-1Cz (see section~\ref{ssec:radaza}).}\label{cat_blueshift}
\end{enumerate}

Therefore, we turn our focus away from molecules such as ethene and $\beta$-carotene in categories \ref{cat_noimpov} and \ref{cat_darkS1}, for which simple morphing operations are unlikely to significantly improve their optoelectronic properties, and we will leave the cases of red- or blue-shifting the lowest energy electronic transition in categories \ref{cat_redshift} and \ref{cat_blueshift} for future work. This leaves the molecules in category \ref{cat_ib} which are good candiates for improvement, and brings us to the motivation of this article --- how can we employ intensity borrowing perturbation theory to improve the weak visible absorption of such molecules?

\section{Challenge 2: How to improve a molecule?}
\label{sec:chal2}
In this section we consider what morphing operations are likely to work, where to make them upon a molecule and how many. To do this we firstly present the relevant equations from intensity borrowing perturbation theory\cite{rob67a} (IBPT). We then show how this theoretical framework can be applied to substitution, addition and dimerization of an arbitrary conjugated organic molecule, from which we formulate widely-applicable design rules. 

\subsection{Some background theory}
\label{ssec:ibpt}
Although the results presented here are applied to borrowing of linear absorption intensity, the perturbation expressions hold for any well-defined quantum mechanical operator, such as the spin-orbit coupling operator or nonadiabatic derivative coupling. In this article, as in the original IBPT article, \cite{rob67a} we consider perturbing states rather than orbitals.

As in standard perturbation theory\cite{atk11a} we define our system as
\begin{align}
 \hat H = \hat H_0 + \hat V \eql{tot_ham}
\end{align}
where $\hat H_0$ is our zeroth-order Hamiltonian and $\hat V$ the perturbation. We begin with a set of zeroth-order eigenstates $\ket{\Psi_u^{(0)}}$ where  $\hat H_0 \ket{\Psi_u^{(0)}} = E_u^{0} \ket{\Psi_u^{(0)}}$ and $\{u,v,w\}$ are indices for eigenstates, and we assume the eigenstates to be real, as in generally the case for stationary electronic structure calculations. For generality, and unlike Ref.~\onlinecite{rob67a}, we do not split $\hat H_0$ into a `molecule' and `perturber' part, nor factor $\ket{\Psi_u^{(0)}}$ into a product of molecule and perturber contributions.

As in the previous literature \cite{rob67a} we only consider transitions from the ground electronic state, though excitations between excited states can be treated similarly. From perturbation theory (see SI section III A 2) we find the leading first-order perturbation to the transition dipole moment of the state $\ket{\Psi_u^{(0)}}$ to be
\bse
\begin{align}
\bra{\Psi_0} \hat \mu \ket{\Psi_u^{(1)}} = & \sum_{v \neq u} \bra{\Phi_0} \hat \mu \ket{\Psi_v^{(0)}} \frac{ \bra{\Psi_v^{(0)}} \hat V \ket{\Psi_u^{(0)}}}{E_u^{0} - E_v^{0}}  \\
= & \sum_{v \neq u} \sum_{ij^\prime} S_{v,ij^\prime}^{(0)} \bra{\Phi_0} \hat \mu \ket{\Phi_i^{j^\prime}} \frac{ \sum_{kl^\prime} S_{v,kl^\prime}^{*(0)} \bra{\Phi_k^{l^\prime}} \hat V \sum_{rs^\prime} S_{u,rs^\prime}^{(0)} \ket{\Phi_r^{s^\prime}} }{E_u^{0} - E_v^{0} }, \eql{ptb_dip}
\end{align}
\ese
where we use the definition of the CIS expansion of the state
\begin{align}
\ket{\Psi_u^{(0)}} = \sum_{i,j}{\ket{\Phi_i^{j^\prime}}S_{u,ij^\prime}^{(0)}}. \eql{cis_def}
\end{align}
We must emphasize that throughout this article we will refer to the singlet spin-adapted configuration $\ket{\Phi_i^{j^\prime}}$ corresponding to the excitation of an electron from orbitals $i$ to $j^{\prime}$ as an \emph{excitation}, and a \emph{state} $\ket{\Psi_u}$ as a linear combination of \emph{excitations} [\eqr{cis_def}].
The first-order correction to the energy of state $\ket{\Psi_u^{(0)}}$ is 
\begin{align}
E_u^{(1)} = \bra{\Psi_u^{(0)}} \hat V \ket{\Psi_u^{(0)}}. \eql{ptb_e}
\end{align}
Unless stated otherwise, the molecular design rules in this perspective are all obtained using \eqr{ptb_dip}.

\subsection{Where to substitute?}
\label{ssec:sub}
Here we combine IBPT, CIS and PPP theory to inform where an organic chromophore should be substituted to achieve a desired spectral alteration. We firstly define the Zeroth order and perturbation Hamiltonians, followed by the zeroth-order eigenstates, and then see how they are mixed by substitution.

For cases of heteroatom substitution the zeroth-order Hamiltonian is simply that of the base molecule and, to a first approximation, for a set of atoms $\mathcal{G}$ which are substituted, the perturbation is\cite{kou67c}
\begin{align}
 \hat V = \sum_{\mu \in \mathcal{G}} \Delta \ep_{\mu} \hat n_{\mu} \eql{vsub}
\end{align}
where $\hat n_{\mu}$ is the number operator for the number of electrons on atom $\mu$,
\begin{align}
 \hat n_{\mu} = & \sum_{\sigma = \{ \uparrow,\downarrow\}} \hat n_{\mu,\sigma}; \\
 \hat n_{\mu,\sigma} = & \hat a^\dag_{\mu,\sigma} a_{\mu,\sigma},
\end{align}
where $\hat a^\dag_{\mu\sigma}$ and $\hat a_{\mu\sigma}$ are the creation and annihilation operators respectively for a spin orbital of spin $\sigma$ on atom $\mu$.
$\ep_\mu$ is the on-site energy, which for a purely hydrocarbon chromophore we can set to zero without affecting the energies of excited states. In quantitative applications of PPP theory for heteroatoms \cite{mcw57b}, heterosubstition will also change the $t_{\mu\nu}$ (hopping) and $\gamma_{\mu\nu}$ (repulsion) parameters. However, here we focus on the effect of changing the on-site energy (H\"uckel $\alpha$ parameter), which affects the diagonal elements of the Fock matrix only \cite{mcw57a}, as we find this to be sufficient to derive predictive design rules for substitution.

For substitution the zeroth-order orbitals $\ket{\phi_i}$ are those which diagonalize the zeroth-order Fock matrix obtained from $\hat H_0$, corresponding to the orbitals of the unsubstituted chromophore. The zeroth order eigenstates $\ket{\Psi_u}$ and expansion coefficients $S_{u,ij'}$ are obtained by diagonalising the CIS Hamiltonian for the unsubstituted chromophore (see SI section III A 1).

From \eqr{vsub}, the perturbation is a one-electron operator\cite{kou67c} so we can define the change in the Fock matrix as 
\begin{align}
 F_{ij}^{(1)} = &  \sum_{\mu \in \mathcal{G}} \Delta \ep_{\mu} C_{\mu i} C_{\mu j}. \eql{delfij}
\end{align}
where, as above, $\mathcal{G}$ is the set of substituted atoms. We find the perturbation to the Hamiltonian to be
\bse
\begin{align}
  \bra{\Phi_i^{j'}} \hat V - \Delta E_0\ket{\Phi_k^{l'}} = & \delta_{ik} F_{j'l'}^{(1)} - \delta_{j'l'} F_{ik}^{(1)},\eql{subijkl} \\
  \bra{\Phi_0} \hat V \ket{\Phi_i^{j'}} = & F_{ij'}^{(1)}, \\
  \Delta E_0 = & \sum_{i\ \mathrm{occ}} \sum_{\mu \in \mathcal{G}} \Delta \ep_{\mu\mu} C_{\mu i}^2.
\end{align}\eql{subpt}%
\ese
From this we immediately see the following:
\begin{tcolorbox}
\begin{center}
\textbf{Principles of chromophore design for substitution}
\end{center}
Substitution will only mix excitations:
\begin{enumerate}
 \item That differ by one orbital from each other.
 \item When the orbitals which differ both have amplitude on the atom which is being subtituted.
\end{enumerate}
\end{tcolorbox}
Both these conditions can be determined by inspection of the monomer spectrum and orbitals and do not require separate calculation for each possible derivative.

For states which are described by linear combinations of excitations, a similar analysis shows that they will only mix if there are excitations in each of the states which differ by at most one electron from each other. If the states consist of the same excitations, but with different expansion coefficients, substitution could alter the diagonal energy of each excitation [\eqr{subijkl}] and therefore the weighting of those states in the configuration interaction expansion, leading to alteration in absorption intensity. 

We note that the perturbations in \eqr{subpt} require minimal extra computation than a monomer calculation, and since the effect of substitution is (to first order) additive, contributions from different subsitutions can simply be added removing the need to simulate each possible derivative separately.

%
 
For the case of mixing two excitations which are of different symmetry in the parent chromophore, the for the mixing element to be nonzero the perturbation $\hat V$ must lower the symmetry of system such that in the new, lower, point group $\ket{\Phi_i^{j'}}$ and $\ket{\Phi_k^{l'}}$ transform as the same irreps. This is consistent with previous results examining the effects of aza-substitution on the polarization of transitions\cite{kou67c}. 

\subsection{Where to add?}
\label{ssec:add}
As we shall see, the algebra for addition and dimerization is more complex than for subsitution, but the general methodology is the same. We define a zeroth-order Hamiltonian, which in this case corresponds to the two monomers at infinite separation, and then a perturbation Hamiltonian of their interaction when bonded together. We find the zeroth-order orbitals and states, showing that they are exclusively Local Exciton (LE) or Charge-Transfer (CT) in character, and then see how addition mixes these. This leads naturally to many pre-existing chromophore theories. 

\subsubsection{Hamiltonian definitions}
\label{ssec:addhamdef}
For simplicity we consider an overall chromophore formed from adding one monomer $m$ with set of atoms $\mM$ to one other monomer $n$ with set of atoms $\mN$, though these ideas can be extended to oligomers.\cite{li17a}. We then imagine taking the two chromophores to infinite separation, such that they can be described as a sum of separate Hamiltonians,
\begin{align}
 \hat H_0 = \hat H_m + \hat H_n \eql{h0}
\end{align}
where
\begin{align}
 \hat H_{m} = & \sum_{\mu \in \mM } \ep_\mu \hat n_{\mu} +  U_{\mu\mu} \hat n_{\mu,\uparrow} \hat n_{\mu,\downarrow} - \sum_{\mu<\nu \in \mM} \sum_\sigma t_{\mu\nu} (\hat a^\dag_{\mu\sigma} \hat a_{\nu\sigma} + \hat a^\dag_{\nu\sigma} \hat a_{\mu\sigma}) \no\\
 & + \sum_{\mu<\nu, \in \mM} \gamma_{\mu\nu} (\hat n_\mu - Z_{\mu})(\hat n_{\nu} - Z_{\nu}) \eql{hmon}
\end{align}
and likewise for $H_n$, where the notation $\sum_{\mu<\nu, \in \mM}$ is taken to mean summation over all $\mu$ in $\mM$ and summation over all $\nu>\mu$ where $\nu$ is also in $\mM$. $U_{\mu\mu}$ is the on-site (Hubbard) repulsion and $\gamma_{\mu\nu}$ is the parameterized repulsion between an electron on atom $\mu$ and an electron on atom $\nu$, approximating the two-electron integral
\begin{align}
 \gamma_{\mu\nu} \simeq & (\mu\mu|\nu\nu) \no\\
 = & \int d\bmr_1 \int d\bmr_2\ \chi_{\mu}(\bmr_1) \chi_{\mu}(\bmr_1) \frac{1}{r_{12}} \chi_{\nu}(\bmr_2) \chi_{\nu}(\bmr_2)
\end{align}
where $\chi_\mu(\bmr)$ is the atomic spatial orbital on atom $\mu$, and we use the chemists' notation\cite{sza89a} for two-electron integrals. We then consider bringing the two chromophores back together from which to calculate the perturbation
\bse
\begin{align}
 \hat V & =:  \hat H - \hat H_0 \\
 & = -\sum_{\mu \in \mM, \nu \in \mcN} \sum_{\sigma} t_{\mu\nu} (\hat a^\dag_{\mu\sigma} \hat a_{\nu\sigma} + \hat a^\dag_{\nu\sigma} \hat a_{\mu\sigma}) + \sum_{\mu \in \mM, \nu\in\mcN} \gamma_{\mu\nu} (\hat n_\mu - Z_{\mu})(\hat n_{\nu} - Z_{\nu}). 
\end{align}\eql{vad}%
\ese%
The idea of describing multichromophore interaction as a sum of individual chromophore terms and their mutual interaction dates back to Kasha exciton theory\cite{kas65a}, though the form of the perturbation we obtain in \eqr{vad} is more complex than a point-dipole interaction and, as we shall see later, able to describe a wider range of phenomena. The Hamiltonian structure used in \eqr{hmon} and \eqr{vad} has been previously obtained by using PPP theory to describe interchromophore interactions\cite{ary13a,ary15a}, though here it is used to construct a zeroth-order Hamiltonian and perturbation which causes intensity borrowing, rather than used to generate an overall Hamiltonian for the numerical simulation of large systems. 


For addition, the Fock matrix associated with $H_m$, written in the atomic basis where $\mu, \nu \in \mM$, is
\bse
\begin{align}
F_{\mu\mu}^{(0)} = & \ep_{\mu} + \frac{1}{2}U_{\mu\mu} P_{\mu\mu} + \sum_{\lambda \neq \mu, \lambda \in \mM} (P_{\lambda\lambda}-Z_\lambda) \gamma_{\mu\lambda}, \eql{fmmM}\\
F_{\mu\nu}^{(0)} = & - t_{\mu\nu} - \frac{1}{2}P_{\mu\nu}\gamma_{\mu\nu} \eql{fmnM},
\end{align}\eql{f0}%
\ese
with similar expressions holding for $\mcN$. If the atoms are on different monomers ($\mu \in \mM$ and $\nu \in \mcN$), then $\gamma_{\mu\nu}=0$ and $t_{\mu\nu}=0$ at zeroth order, such that $F_{\mu\nu}^{(0)} = 0$ regardless of the orbital coefficients and the density matrix. Consequently the Fock matrix corresponding to $\hat H_0$ is block-diagonal,
\begin{align}
 \bF^{(0)} = \begin{pmatrix}
              \bF_{m}^{(0)} & 0 \\ 0 & \bF_{n}^{(0)}
             \end{pmatrix}, \eql{f0mat}
\end{align}
such that $\bF_{m}^{(0)}$ and $\bF_{n}^{(0)}$ can be solved separately for the orbitals on $\mM$ and $\mcN$. This is no surprise, since the electronic structure of two widely-separated monomers should simply be that of the isolated chromophores.

We therefore define the molecular orbitals $\{\phi_{ni}\}$ and $\{\phi_{mj}\}$ where $n$ and $m$ refer to the monomers upon which the orbitals are located and $i,j,k,l$ are arbitrary orbital indices. From this we can form our ground-state (unexcited) wavefunction as the Slater determinant
\begin{align}
 \ket{\Phi_0} = & |\phi_{n1}\alpha,\phi_{n1}\beta,\ldots,\phi_{ni}\alpha,\phi_{ni}\beta,\ldots,\phi_{n,K_n/2}\alpha,\phi_{n,K_n/2}\beta,\no\\
 & \ \phi_{m1}\alpha,\phi_{m1}\beta,\ldots,\phi_{mi}\alpha,\phi_{mi}\beta,\ldots,\phi_{mK_m/2}\alpha,\phi_{mK_m/2}\beta\rangle \eql{Phi0}
\end{align}
where $K_n$ is the number of electrons on monomer $n$ and likewise for $K_m$, and $\alpha$ and $\beta$ denote the spin component of the orbital. Note we assume here that the molecule is a ground-state singlet but this can be extended to systems such as radicals with non-zero ground-state spin.\cite{abd20a}

\subsubsection{Zeroth order excitations: Local Exciton or Charge Transfer}
\label{ssec:zob}
We write single excitations as $\ket{\Phi_{pi}^{qj'}}$, corresponding to a singlet spin-adapted\cite{sza89a} excitation from orbital $pi$ to $qj$ where $p,q \in \{n,m\}$ (see SI section III A 1). There are consequently two forms of excitation:
\begin{enumerate}
 \item $p=q$, such as $\ket{\Phi_{ni}^{nj'}}$. This corresponds to an intramonomer excitation, commonly called a local or Frenkel excitation\cite{fue16a,smi10a,li17a,bel13a}. We therefore define $\ket{\rLE_{ni}^{nj'}}=:\ket{\Phi_{ni}^{nj'}}$.
 \item $p\neq q$  such as $\ket{\Phi_{ni}^{mj'}}$, which is an intermonomer excitation, commonly called a charge-transfer (CT) excitation.\cite{smi10a,fue16a,bel13a,li17a} Note that these are sometimes referred to as charge-resonance excitations,\cite{pie07a,fen13a,gol09a} which we refrain from using to avoid confusion with the Coulson-Rushbrooke theorem. We therefore define $\ket{\rCT_{ni}^{mj'}}=\:\ket{\Phi_{ni}^{mj'}}$. 
\end{enumerate}

The idea of local and charge-transfer excitations has been extensively discussed in the literature\cite{mul39b,mul50a,fue16a,smi10a,li17a,bel13a} and the definitions used here are consistent with (for example) Ref.~\citenum{li17a}. They are sometimes referred to as `diabatic' states\cite{li17a,fue16a}, terminology we refrain from here since, strictly speaking, they do not necessarily diagonalize the nuclear kinetic energy operator.\cite{wor04a} Other literature defines CT excitations more generally\cite{mul39b} as any transition between a `neutral' ground and an `ionic' excited state. Here we show LE and CT excitations emerge naturally from our choice of zeroth-order Hamiltonian in \eqr{h0} and do not need to be specified \emph{a posteriori}. In addition, our definition of CT and LE states leads immediately to the following useful properties:
\begin{enumerate}
\item LE and CT excitations are strictly orthogonal as their constituent MOs are orthogonal:
\begin{align}
 \bk{\Phi_{pi}^{qj'}}{\Phi_{rk}^{sl'}} = \delta_{pr}\delta_{qs}\delta_{ik}\delta_{j'l'}, \eql{orthog1}
\end{align}
and therefore (for example)
\begin{align}
 \bk{\rLE_{mi}^{mj'}}{\rCT_{mk}^{nl'}} = 0 \eql{orthog2}
\end{align}
for all possible $\{i,j,k,l\}$. This property does not necessarily hold if charge transfer excitations are defined in terms of \emph{dimer} molecular orbitals,\cite{pie07a,mul39b} which when rotated to the \emph{monomer} basis may have local exciton character.  


\item LE and CT exctations form a complete basis of singly-excited states. 
This result, while useful, should be interpreted with caution since bases of singly excited states constructed from different MOs do not necessarily span the same part of Hilbert space. We also note that, upon extension to double and higher excitations, excitations can have both local and charge-transfer character such as $\ket{\Phi_{ninj}^{nk'ml'}}$.

\item The definition of LE and CT excitations is unique for two given monomers and independent of their relative orientation, since there are no intermonomer interaction terms in $\hat H_0$ [\eqr{hmon}] or $\bF^{(0)}$ [Eqs~\eqref{eq:f0} and \eqref{eq:f0mat}]. This property does not usually hold if LE and CT states are determined from a RHF (or DFT) calculation on the dimer or oligomer, and it simplifies comparison between different isomeric dimers since they will have the same zeroth-order bases. More practically, the independence of the basis to intermonomer geometry can be useful when the relative orientation of monomeric units is variable or unknown, such as in amorphous films.

\item The zeroth-order basis functions and their zeroth-order energies (including the CT excitations) can be determined from calculations on the respective, independent, monomer units and does not require a separate calculation for each orientation or isomer of the dimer.

\item The transition dipole moment for an LE excitation is
\begin{align}
 \bra{\Phi_0} \hat \mu \ket{\rLE_{ni}^{nj'}} = \sqrt{2}\bra{\phi_{ni}} \hat \mu \ket{\phi_{nj'}}
\end{align}
which is the dipole moment of the monomer transition, as to be expected, where the factor of $\sqrt{2}$ arises from spin-adaptation of the singly excited wavefunction.\cite{sza89a} This means that for a LE to be bright, the excitation must be allowed in the point group of the monomer, which is generally a more restrictive requirement than for a lower-symmetry dimer. 

\item CT excitations are always dark at zeroth order:
\bse
\begin{align}
 \bra{\Phi_0} \hat \mu \ket{\rCT_{ni}^{mj'}} = & \sqrt{2}\bra{\phi_{ni}} \hat \mu \ket{\phi_{mj'}} \\
 = & -\sqrt{2}\sum_{\mu \in \mM} C_{\mu, ni} C_{\mu,mj} \bmr_{\mu} - \sqrt{2}\sum_{\mu \in \mcN}C_{\mu, ni} C_{\mu,mj} \bmr_{\mu} \eql{splitsum} \\
 = & 0. \eql{zdm}
\end{align}\eql{ctdm}%
\ese 
In the first line we have used the standard result for a one-electron operator\cite{sza89a}, and in the second line the PPP expression for the dipole moment\cite{par56a}, choosing to sum over each monomer separately and the minus sign arises from the negative charge of the electron\cite{sza89a}. However, for the first term in \eqr{splitsum} $C_{\mu,ni}=0$ for $\mu \in \mM$ (since a monomer orbital entirely on $n$ has no amplitude on any atom on $m$), and in the second term $C_{\mu,mj}=0$ for $\mu \in \mcN$ (since a monomer orbital entirely on $m$ has no amplitude on any atom on $n$). This result leads to significant simplifications in the following algebra, since the absence of oscillator strength at zeroth order means that CT excitations must borrow intensity from a LE in order to appear bright, and that intensity cannot be `borrowed' from CT states. This result does not usually hold if LE and CT excitations are determined from dimer MOs where the dimer orbitals may not be spatially separated, and the CT excitations computed in the dimer orbitals (which may be bright) correspond to states which are mixtures of CT and LE excitations in the monomer orbitals. This definition also means that if the zeroth-order orbitals are used to compute the dimer excited states at zeroth order, they are exclusively LE or CT, and when computed at first order, it is straightforward to unambigously compute the extent of LE or CT character. It is clear that the zeroth-order states alone are generally not sufficient to describe the excited states of the dimer, for example it is known that CT states (strictly speaking, states with CT character) are generally not dark\cite{mul39b,mul50a,lei20a,bha93a,kul06a,tan13a} and we will go on to show how perturbation theory accounts for this.
\end{enumerate}
These results are true for monomers at infinite separation with and without NDO, and true for monomers brought together within PPP theory. For a general calculation (at finite separation and without NDO), \eqr{ctdm}, \eqr{orthog1} and \eqr{orthog2} are expected to hold approximately, though may be non-zero due to the basis functions of one monomer being non-zero in the same region of space as the basis functions of the other monomer.

\subsubsection{Zeroth order eigenstates: Exclusively LE or CT}
The results in section~\ref{ssec:zob} are derived for the \emph{basis} of LE and CT excitations, and we now determine the zeroth-order \emph{eigenstates}, obtained by diagonalising the CIS Hamiltonian.

To simplify the zeroth-order CIS matrix, we firstly note that at zeroth order $\gamma_{\mu\nu}=0$ unless $\mu$ and $\nu$ are on the same monomer, such that two electron integrals are zero unless of the form $(ninj|nknl)$ or $(mimj|mkml)$ for all $\{i,j,k,l\}$. Similarly, the density matrix elements $P_{\mu\nu}$ are zero if $\mu$ and $\nu$ are on different monomers, since for any monomer orbital $pi$ either $C_{\mu pi}$ or $C_{\nu pi}$ will be zero.

Since the zeroth order orbitals (by construction) diagonalize the zeroth-order Fock matrix, the zeroth-order Hamiltonian matrix elements are:
\bse
 \begin{align}
  \bra{\rLE_{ni}^{nj'}} \hat H_0 - E_0 \ket{\rLE_{nk}^{nl'}} = & \delta_{ik}\delta_{j'l'}(F_{nj'nj'} - F_{nini}) + 2(nj'ni|nknl') - (nj'nl'|nkni), \eql{zon} \\
  \bra{\rLE_{ni}^{nj'}} \hat H_0 \ket{\rCT_{nk}^{ml'}} = & 0, \eql{zo1} \\
  \bra{\rLE_{ni}^{nj'}} \hat H_0 \ket{\rCT_{mk}^{nl'}} = & 0, \eql{zo2} \\
  \bra{\rLE_{ni}^{nj'}} \hat H_0 \ket{\rLE_{mk}^{ml'}} = & 0, \eql{zo3} \\ 
  \bra{\rCT_{ni}^{mj'}} \hat H_0 - E_0 \ket{\rCT_{nk}^{ml'}} = & \delta_{ik}\delta_{j'l'}(F_{mj'mj'} - F_{nini}), \eql{zof} \\
  \bra{\rCT_{ni}^{mj'}} \hat H_0 \ket{\rCT_{mk}^{nl'}} = & 0, \eql{zos} \\
  \bra{\Phi_0} \hat H_0 \ket{\rLE_{nk}^{nl'}} = & 0, \eql{zo7} \\
  \bra{\Phi_0} \hat H_0 \ket{\rLE_{nk}^{ml'}} = & 0, \eql{zo8}
 \end{align}\eql{zom}%
\ese
and these are general, holding without the approximations of PPP theory for monomers at infinite separation. At finite separation they hold only within NDO (and therefore PPP), but are likely to be approximately true for a general basis set. 

Equation~\eqref{eq:zon} shows that the interaction between two excitations on the \emph{same} monomer is simply the CIS matrix element if that monomer were considered in isolation, as to be expected for an intramolecular Frenkel excitation. Equations~\eqref{eq:zo1} and \eqref{eq:zo2} show that there is no mixing between Frenkel and CT excitations, and \eqr{zo3} that there is no interaction between Frenkel excitations on different monomers at zeroth order. Equations~\eqref{eq:zof} and \eqref{eq:zos} shows that CT excitations interact with no other state except themselves, and that their diagonal energy is simply the ionisation energy of orbital $i$ ($-F_{ii}$) minus the electron affinity of orbital $j'$ ($-F_{j'j'}$), as would be expected upon forming a separated cation and anion using Koopmans' theorem\cite{sza89a}.

The results in \eqr{zom} mean that the zeroth order eigenstates are simply the monomer CIS eigenstates and CT excitations from each monomer orbital to each orbital on the other monomer, with an energy given by the orbital energy difference. Since a CIS calculation on the monomers automatically calculates orbital energies, the complete set of zeroth-order eigenstates and their energies can be obtained at no greater computational cost than calculations on the separate monomers. Since CT and LE \emph{excitations} do not mix with each other at zeroth order, the zeroth-order eigenstates will be purely LE or CT in character, such that the results obtained in section~\ref{ssec:zob} still apply. 
Often electronic states of interest are dominated by a single excitation, though the results can easily be extended to states which are linear combinations of excitations using \eqr{cis_def}.
These results easily extend to oligomers (more than two monomers) using a methodology similar to Ref.~\citenum{li17a}. 

These results are summarized in \tabr{comp} where we compare the CT and LE states obtained here from diagonalizing the zeroth-order Hamiltonian to running a dimer calculation and inferring from it which transitions may have LE or CT character. 

\begin{table}[h]
 \begin{tabular}{l|c|c}
  Are the LE and CT states$\ldots$ & From Monomer orbitals & From Dimer orbitals \\ \hline
  Orthogonal? & Yes & Not necessarily \\
  Eigenstates of \emph{monomer} Fock matrix? & Yes & No \\
  Eigenstates of \emph{dimer} Fock matrix? & No & Yes \\
  A complete singly excited basis? & Yes & Sometimes \\
  Uniquely defined, independent of orientation? & Yes & No \\
  Determined from monomer calculation? & Yes & No \\
  Always dark if CT? & Yes & Not necessarily
 \end{tabular}
\caption{Properties of the LE and CT states obtained from the zeroth-order Hamiltonian, which correspond to excitations between monomer orbitals (used in this article) to those found from excitations between dimer orbitals.}
\tabl{comp}
\end{table}

\subsubsection{How does addition perturb LE and CT states?}
\label{ssec:teint}
We now consider bringing our two monomers together and how this causes the zeroth-order LE and CT states to mix. To do this we first briefly consider how addition alters two-electron integrals and the dimer Fock matrix.

Using NDO\cite{par56a} and the spatial separation of monomer orbitals we find 
\begin{align}
 (pi qj | rk sl) = \delta_{pq} \delta_{rs} (pipj| rkrl) 
\end{align}
where $\{p,q,r,s\}$ to refer to arbitrary monomers from $\{n,m\}$. 
Although it might appear that integrals such as $(ninj|mkml)$ have to be obtained from a dimer electronic structure calculation, they can be estimated from monomer calculation and relative intermonomer geometry, as shown in the SI section III B 2 b.



From \eqr{f0}, the perturbation to the Fock matrix in the atomic orbital basis will be, for $\mu \in \mM$, 
\bse
\begin{align}
F_{\mu\mu}^{(1)} = & \sum_{\lambda \in \mcN} (P_{\lambda\lambda}-Z_\lambda) V_{\mu\lambda}, \eql{fmmM1}\\
F_{\mu\nu}^{(1)} = & \left\{ 
\begin{array}{cc}
 0 & \nu \in \mM, \\
 H_{\mu\nu} & \nu \notin \mM.
\end{array}
\right. 
\end{align}\eql{f1}%
\ese
There is no $ - \frac{1}{2}P_{\mu\nu}V_{\mu\nu}$ term in \eqr{f1} since $P_{\mu\nu}=0$ if $\mu$ and $\nu$ are on different monomers.

For the Fock matrix in the basis of molecular orbitals, where the MOs are on the same monomer $n$, we use \eqr{f1} to find 
\begin{align}
F_{ni,nj}^{(1)} = & \sum_{\mu \in \mcN} C_{\mu, ni}C_{\mu, nj} \sum_{\lambda \in \mM} (P_{\lambda\lambda} - Z_{\lambda}) \gamma_{\mu \lambda}\eql{delfij1}
\end{align}
We show in SI section III B 2 c that this can be written as the electrostatic interaction between two charge distributions, and can be estimated from a multipole expansion\cite{buc67a,sto13a}, and therefore approximated without requiring a separate calculation on the dimer. In particular, if there is zero net charge density on monomer $m$ (as in an alternant hydrocarbon\cite{pop53a}) then $F_{ni,nj}^{(1)}=0$ for all $ni$ and $nj$.

%

If the MOs are on different monomers, \eqr{f1} gives 
\bse
\begin{align}
F_{ni,mj}^{(1)} = & -\sum_{\mu \in N} \sum_{\nu \in M} C_{\mu, ni}C_{\nu, mj} t_{\mu\nu} \eql{fabdif} \\
= & -t_{\mu^* \nu^*} C_{\mu^*, ni}C_{\nu^*, mj}, \eql{fnm}
\end{align}\eql{fnimj}%
\ese%
where $\mu^*$ and $\nu^*$ are the atoms through which the two monomers are joined (we assume that the monomers are only joined through one atom, but this can clearly be generalized to more complex bonding geometries). The results in \eqr{delfij1} and \eqr{fnm} mean that the perturbed Fock matrix is off-diagonal. This means that Brillouin's theorem no longer holds and that the ground state $\ket{\Phi_0}$ can mix with singly excited states (for alternant hydrocarbon dimers only with CT states), but as noted before\cite{cor91a} the effect of this is likely to be small due to the large energy gap (see SI section III A 1).

Using these results and the NDO assumptions of PPP theory, the perturbation to the CIS Hamiltonian matrix is therefore 
\begin{subequations}
 \begin{align}
  \bra{\rLE_{ni}^{nj'}} \hat V - \Delta E_0 \ket{\rLE_{nk}^{nl'}} = & \delta_{ik}F_{nj',nl'}^{(1)} - \delta_{j'l'}F_{ni,nk}^{(1)}, \eql{foo} \\
  \bra{\rLE_{ni}^{nj'}} \hat V \ket{\rCT_{nk}^{ml'}} = & F_{nj',ml'}^{(1)} \delta_{ik}, \eql{fot} \\
  \bra{\rLE_{ni}^{nj'}} \hat V \ket{\rCT_{mk}^{nl'}} = & -F_{ni,mk}^{(1)} \delta_{j'l'}, \eql{for} \\
  \bra{\rLE_{ni}^{nj'}} \hat V \ket{\rLE_{mk}^{ml'}} = & 2(nj'ni|mkml'), \eql{fou} \\ 
  \bra{\rCT_{ni}^{mj'}} \hat V -\Delta E_0 \ket{\rCT_{nk}^{ml'}} = & \delta_{ik}F_{mj',ml'}^{(1)} - \delta_{j'l'} F_{ni,nk}^{(1)} -(mj'ml'|nkni), \eql{fof} \\
  \bra{\rCT_{ni}^{mj'}} \hat V \ket{\rCT_{mk}^{nl'}} = & 0, \eql{fos} \\
  \bra{\Phi_0} \hat V \ket{\rLE_{ni}^{nj'}} = & F_{ni,nj'}^{(1)}, \eql{fov} \\
  \bra{\Phi_0} \hat V \ket{\rCT_{ni}^{mj'}} = & F_{ni,mj'}^{(1)}, \eql{foe} \\
  \bra{\Phi_0} \hat V \ket{\Phi_0} = & \sum_{ni \ \rm{occ}} F_{ni,ni}^{(1)} +  \sum_{mj \ \rm{occ}} F_{mj,mj}^{(1)} =: \Delta E_0. \eql{fon}
 \end{align}\eql{pcis}%
\end{subequations}

From this a number of results immediately follow. Firstly, Frenkel excitations within the same monomer are perturbed by an uneven charge distribution on the other monomer\cite{li17a}. CT states mix with Frenkel excitations and from \eqr{fnm} this is only through the H\"uckel-like $t_{\mu^*\nu^*}$ terms and not through two-electron terms, which are exactly zero within NDO but can be approximated to be zero at other levels of theory\cite{li17a,bel13a}. Although these results are derived in the context of PPP theory, similar results can be obtained at higher levels of theory\cite{li17a}.

For the interaction of two Frenkel excitations on different monomers, we show in SI section III B 2 that (up to first order in the charge distribution on each monomer)
\begin{align}
 2(nj'ni|mkml') \simeq \frac{\bm{\mu}_1 \cdot \bm{\mu}_2}{r^3} - 3\frac{(\bmr\cdot\bm{\mu}_1)(\bmr\cdot\bm{\mu}_2)}{r^5}, \eql{kasderiv}
\end{align}
where $\bm{\mu}_1 =: \bra{\Phi_0} \hat \mu \ket{\rLE_{ni}^{nj'}}$, $\bm{\mu}_2 =: \bra{\Phi_0} \hat \mu \ket{\rLE_{mk}^{ml'}}$, $\bmr$ is the (vector) displacement between the centres of the two chromopores and $r = |\bmr|$. The RHS of \eqr{kasderiv} is precisely the energy of interaction in Kasha's point dipole model\cite{kas65a}. Note that \eqr{kasderiv} is very approximate and \eqr{fou} can include dipole-quadrupole and higher terms. However, we will see that these higher-order terms can in practice be ignored (for the case of acenes) while still yielding qualitatively predictive results. 

From \eqr{fof}, the diagonal energy of a CT excitation $\ket{\rCT_{ni}^{mj'}}$ will change by $F_{mj'mj'}^{(1)} - F_{nini}^{(1)} - J_{mj',ni}$, where $J_{mj',ni}$ is the Coulomb stabilization of an electron in orbital $mj'$ and a hole in $ni$. In the SI we show that, for uncharged initial ground-state chromophores, the leading order contribution from the $F^{(1)}$ terms in \eqr{fof} is at most $\mathcal{O}(r^{-2})$, but the Coulomb term is approximately $1/r$, consistent with the stabilization of charge-transfer states given by Mott in 1938\cite{mot38a}, and still sometimes used as a point charge approximation\cite{bre04a}. Because Coulomb integrals are always positive (or zero for orbitals at infinite separation)\cite{roo51a}, the energy of CT states is therefore usually lowered at first order. It appears that CT excitations involving different from/to monomers do not mix, but they can be mixed indirectly via Frenkel states.

We note that Kasha's exciton theory approximates \eqr{fou} by a point-dipole interaction\cite{kas65a}, but does not consider CT states and therefore omits interactions such as \eqr{fot}, \eqr{for}, \eqr{fov} and \eqr{foe}. Frenkel-CT interactions similar to those in \eqr{pcis} have been incorporated in more advanced forms of exciton theory and to describe singlet fission phenomena\cite{bel13a,smi10a,li17a,yam11a}. These cases usually only consider states composed of HOMO/LUMO excitations [i.e. $i=1$ and $j=1$ in \eqr{pcis}] and here we provide a more general formulation. In addition, the formulation here shows that certain terms [such as \eqr{fos}] vanish rigorously within NDO, justifying their neglect at other levels of theory\cite{bel13a,li17a}, and that the mixing of Frenkel and CT states can be evaluated solely by considered the coefficients of the relevant orbitals on the atoms at which the monomers touch.

Lastly, all terms in \eqr{pcis} can be estimated from monomer calculation and the relative geometry of the monomers using the results of this section and the electrostatics given in the SI section III B 2, with \eqr{fot}, \eqr{for} and \eqr{foe} known exactly. This means that when considering the results of dimerization at different geometries, the approximate effects of chromophore alteration can be determined without separate calculation for each dimer geometry. The interactions in \eqr{pcis} are illustrated in \figr{ptzf}.

\begin{figure}[tb]
 \includegraphics[width=.7\textwidth]{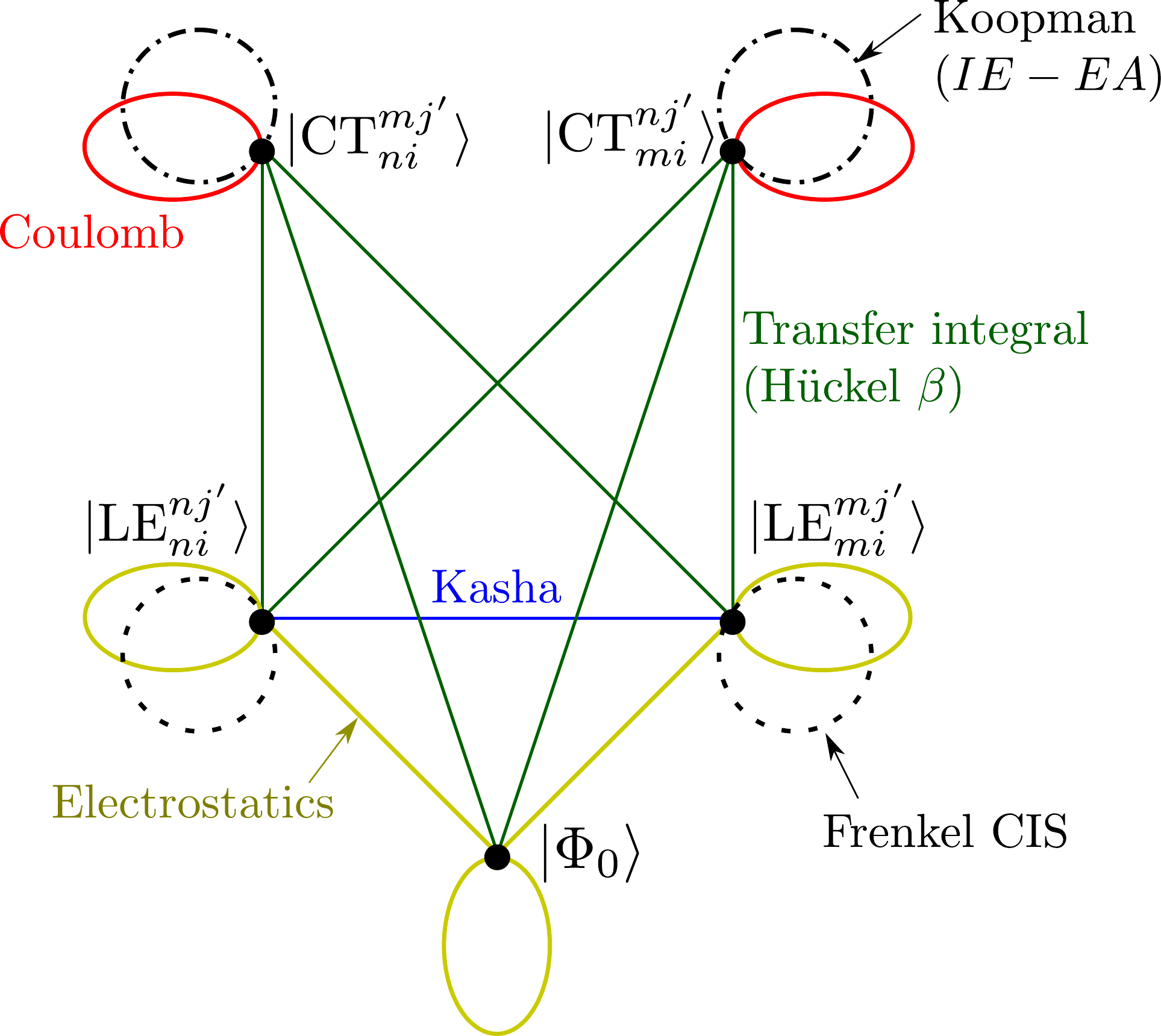}
 \caption{Perturbation theory diagram depicting zeroth and first order interactions [\eqr{pcis} and \eqr{zom}], illustrating how previously-known interactions emerge naturally using the Hamiltonian definitions in this article. Zeroth-order interactions [\eqr{zom}] are shown as dashed lines corresponding to monomer configuration interaction singles (CIS)\cite{sza89a} for Frenkel excitations and Koopman's theorem (ionization energy -- electron affinity)\cite{sza89a} for CT excitations. At first order, interactions correspond to electrostatic forces between monomers\cite{sto13a,buc67a,vol04a}, Kasha's dipole-dipole interaction\cite{kas65a}, Coulombic stabilization of CT excitations\cite{mot38a,bre04a} and the transfer integral (hopping term)\cite{bel13a}. For this dimer model, there is no direct interaction between CT states going from/to different monomers at either zeroth or first order.}
 \figl{ptzf}
\end{figure}

\subsection{Where to dimerize?}
\label{ssec:dimer}
Here we apply the general methodology to dimerization, which can be considered a special case of addition where the two monomers are identical, such that $\hat H_{\mM} = \hat H_{\mcN}$ and each molecular orbital on one monomer will be degenerate with a molecular orbital on another monomer.

One could consequently immediately define dimer orbitals as linear combinations of monomer orbitals, similar to the ``Dimer molecular orbital linear combination of fragment molecular orbital'' (DMO-LCFMO) method\cite{pie07a,fen13a}, but this is not performed here as it complicates subsequent assignment of intermolecular and intramolecular transitions. This means that the monomer orbitals are not symmetry `pure' (they do not necessarily transform as an irrep of the dimer's point group) but it is possible and advantageous to symmetry-adapt the resulting excitations. Bringing together two monomers to form a dimer and analysis of the resultant spectrum has been considered before in (for example) Kasha's exciton theory \cite{kas65a}. However, this only considered dipole-dipole interactions and, as we shall see, the perturbation considered here [\eqr{vad}] allows for description of a richer variety of interactions, including between two excitations where one (or both) has no transition dipole moment from the ground state.

For the case of homodimers, we immediately see that all zeroth-order eigenstates will be (at least) doubly degenerate. We therefore have to find the `good' eigenfunctions. This could be achieved by calculating all the mixing elements between the degenerate states, but as discussed in SI section III A 2, group theory can often be used. For the case of a $C_2$ rotation interconverting the two monomers the eigenstates will transform as $A$ and $B$, letting us define
\bse
\begin{align}
 \ket{\rLE_{k}^{l',\substack{A\\B}}} = & \frac{1}{\sqrt{2}}(\ket{\rLE_{nk}^{nl'}} \pm \ket{\rLE_{mk}^{ml'}}), \\
 \ket{\rCT_{k}^{l',\substack{A\\B}}} = & \frac{1}{\sqrt{2}}(\ket{\rCT_{mk}^{nl'}} \pm \ket{\rCT_{nk}^{ml'}}),
\end{align}\eql{hdim}%
\ese
where LE refers to Frenkel excitation and CT to charge transfer. The notation is taken to mean that on the LHS either $A$ or $B$ symmetry are chosen and the upper or lower sign respectively is taken on the RHS.

If there are further degeneracies between zeroth-order states then linear combinations of these excitations can be taken, but only within a particular irrep. Going forward, we assume that the dimer has $C_2$ symmetry although this methodology can easily be applied to other point groups.

To determine the effect of the perturbation between the `good' degenerate eigenstates in \eqr{hdim}, we apply \eqr{pcis} to find
\bse
\begin{align}
 \bra{\rLE_{i}^{j',\substack{A\\B}}} \hat V \ket{\rLE_{k}^{l',\substack{A\\B}}} = & \delta_{ik}F_{nj',nl'}^{(1)} - \delta_{j'l'}F_{ni,nk}^{(1)} \pm 2(nj'ni|mkml') \\
 \bra{\rLE_{i}^{j',\substack{A\\B}}} \hat V \ket{\rCT_{k}^{l',\substack{A\\B}}} = & F_{nj',ml'}^{(1)} \delta_{ik} \mp F_{ni,mk}^{(1)} \delta_{j'l'} \\
 \bra{\rCT_{i}^{j',\substack{A\\B}}} \hat V \ket{\rCT_{k}^{l',\substack{A\\B}}} = & \delta_{ik}F_{mj',ml'}^{(1)} - \delta_{j'l'} F_{ni,nk}^{(1)} -(mj'ml'|nkni)\\
 \bra{\Phi_0^A} \hat V \ket{\rLE_{i}^{j',A}} = & \sqrt{2}F_{ni,nj'}^{(1)} \\
 \bra{\Phi_0^A} \hat V \ket{\rCT_{i}^{j',A}} = & \sqrt{2}F_{ni,mj'}^{(1)}.
\end{align}\eql{hdimpt}%
\ese
The notation on the first three lines of \eqr{hdimpt} is taken to mean that on the LHS either both bra and ket are $A$ symmetry or both are $B$ symmetry (since there is no mixing between excitations of different irreps) and the upper or lower sign respectively is taken on the RHS.
$\ket{\Phi_0}$ is $A$ symmetry and only interacts with $A$ states, such that $\Delta E_0$ is the same as for addition. The CT$\leftrightarrow$CT mixing between $A$ and $B$ irreps is identical since there is no direct coupling in \eqr{fos}. 

Somewhat surprisingly, we find that the symmetry-adaptation of excitations leads to arguably simpler results and that the rules obtained earlier for Frenkel-CT mixing still hold for homodimers.

We note that the energy difference between two $A/B$ Frenkel excitations is $4(nj'ni|mimj')$ which is the Davydov splitting\cite{dav49a} that is approximated by a point-dipole model in Kasha's exciton theory\cite{kas65a}.


To give an example of intensity borrowing, consider a HOMO$\to$LUMO CT excitation borrowing intensity from a HOMO$\to$LUMO LE. Assuming the two monomers to be joined only through one bond we find
\begin{align}
 \bra{\rLE_{1}^{1',\substack{A\\B}}} \hat V \ket{\rCT_{1}^{1',\substack{A\\B}}} = & -t_{\mu^*\nu^*} (C_{\mu^*,n1'}C_{\nu^*,m1'} \mp C_{\mu^*,n1}C_{\nu^*,m1})
\end{align}
and the extent of borrowing will depend on the interference between the product of orbital amplitudes at the joining carbon. If the monomer orbitals are defined such that they all have the same sign at the joining atom, and the HOMO and LUMO coefficients are approximately equal on the joining atoms (which is the case for an alternant hydrocarbon\cite{cou40a,pop53a}) then
\bse
\begin{align}
 \bra{\rLE_{1}^{1',\substack{A}}} \hat V \ket{\rCT_{1}^{1',\substack{A}}} \simeq & 0,  \\
 \bra{\rLE_{1}^{1',\substack{B}}} \hat V \ket{\rCT_{1}^{1',\substack{B}}} = & -2 t_{\mu^*\nu^*} C_{\mu^*,n1}C_{\nu^*,m1}.
\end{align}\eql{hlhl}
\ese
A similar result to \eqr{hlhl} has previously been obtained in the context of the Davydov splitting in crystalline pentacene\cite{bel13a}, and here we show is a special case of \eqr{hdimpt}. 


The results for addition and dimerization can be summarized as:
\begin{tcolorbox}
\begin{center}
\textbf{Principles of chromophore design for addition and dimerization}
\end{center}
\begin{enumerate}
 \item CT states are always dark at zeroth-order.
 \item In order for CT states to appear in the spectrum, they must (at first order) borrow from state containing a Frenkel excitation which differs by only one orbital from the CT excitation.
 \item This Frenkel excitation must possess a dipole moment at zeroth order ($\bra{\Phi_0} \hat \mu \ket{\rLE_{mk}^{ml'}} \neq 0$, cannot be forbidden by spin symmetry or group theory using the \emph{monomer} point group).
 \item The \emph{monomer} orbitals by which the CT and LE excitations differ must have orbital amplitude at the two atoms through which the monomers are bonded ($C_{\mu^*,mk}C_{\nu^*,ni} \neq 0$).
 \item There must be a nonzero H\"uckel resonance term between the two joining atoms ($t_{\mu^*\nu^*} \neq 0$). Given that the overlap of two $p$ orbitals approximately scales as $\cos(\theta)$ where $\theta$ is the dihedral angle\cite{woi89a,ven06a}, this means that chromophores should be as planar as possible.
\end{enumerate}
\end{tcolorbox}

\section{Trying it out on real molecules}
\label{sec:realmol}
In the preceding sections we have determined algebraic expressions for the mixing of states, and from this guidelines for altering molecules such that one transition (usually a dark state at a favourable energy) borrows intensity from another transition (which must be bright, and is usually at an unfavourable energy). The same perturbation theory framework can also be used to perturb the \emph{energies} of states as well as their intensity, leading to a qualitative picture of how molecular alteration can alter the absorption and emission spectra of a molecule. Clearly there are a huge range of possible molecular structures which this methodology can be applied to, and here we focus on the absorption of acenes and the emission of radicals, both areas of substantial current interest from the perspective of singlet fission\cite{smi10a,smi13a,alv19a,kum17a} and organic light-emitting diodes\cite{ai18a,abd20a,hel21b,cui20a,hud21a} respectively. For both classes of molecules we show how aza-substitution and addition can be used to improve their optoelectronic properties and how the methodology in this article can guide where on these molecules substitution and addition should occur.

\subsection{Visibly improving acenes}
\label{ssec:acenes}
\subsubsection{Background}
The electronic structure of acenes has been studied since at least the 1940s\cite{par56a,mcw52a,cou48a} and is still the subject of considerable interest\cite{hel19a,son07a,kum17a,zen14a} due to the ability of acenes to undergo singlet fission\cite{smi10a,smi13a}. These molecules generally contain a low-energy, $y$-polarized transition (labelled $1B_{2u}^+$ in Ref.~\citenum{par56a}) and an intense, high-energy $x$-polarized transition ($1B_{3u}^+$). They also possess a very weak $x$-polarized transition ($1B_{3u}^-$) between these two absorptions which is predicted to be dark by PPP theory.

The weak visible absorption and intense UV absorption of these molecules, as shown in Table~\ref{tab:typ_crm}, motivates considering whether some of the absorption intensity could be `moved' from the UV to the visible. Increasing the absorption (extinction coefficient) of these molecules in the visible would require less material in a photovoltaic device, which would in turn mean a thinner device such that excitons did not need to diffuse so far to be harnessed and would therefore increase photovoltaic efficiency\cite{hel19a}.

Using the foregoing design principles, in order for acenes to increase their visible absorption, the molecules must be perturbed such that transitions in the visible can `borrow intensity' from the UV absorption. Note that borrowing intensity from the pre-existing visible aborption ($1B_{2u}^+$), by, for example, a Kasha exciton splitting, would simply move absorption between different regions in the visible rather than increase overall visible absorption.

One idea could be to perturb the molecule such that the $1B_{2u}^+$ transition borrows intensity from the $1B_{3u}^+$. However, there is a large energy separation between these excitons, which from perturbation theory is likely to reduce their mixing, and they are orthogonal by point group symmetry, such that the alteration would have to substantially reduce the molecule's symmetry in order for there to be any possibility of mixing.

Alternatively, the molecule could be altered such that the dark $1B_{3u}^-$ state could borrow intensity from the bright $1B_{3u}^+$ state. As we will show, this can be achieved by aza-substitution, which causes these excitations to mix. However, addition or dimerization of another alternant hydrocarbon (such as another acene) will \emph{not} cause these excitations to mix, as such an alteration preserves pseudoparity meaning that the $+$ and $-$ states remain orthogonal\cite{par56a}. 

While addition cannot cause $1B_{3u}^-$ to mix with $1B_{3u}^+$, this does not mean that addition cannot improve the spectrum: instead, addition leads to the formation of CT states, the lowest-energy of which can be in the visible, and which can (subject to correct bonding geometry) borrow intensity from $1B_{3u}^+$. We show how the theory in this article produces design rules to determine the correct geometry, and compare this with experimental results. 

\subsubsection{Computational details}
We present simulated UV/Visible spectra of acenes using four methods based on the algebra in this article, and compare these results to experimental data. These four methods are, in order of increasing computational cost:
\begin{tcolorbox}
\begin{center}
\textbf{Methods used for the simulation of spectra}
\end{center}
\begin{enumerate}
\item{Zeroth-order calculation on isolated monomers} \label{mthd_zo}
\item{Algebraic first-order perturbation,} \label{mthd_alg}
\item{First-order Configuration Interaction Singles (CIS),}\label{mthd_1o}
\item{Full dimer Self-Consistent Field (SCF).}\label{mthd_ful}
\end{enumerate}
\end{tcolorbox}

Method \ref{mthd_zo} is formally the solution to the zeroth-order Hamiltonian, which for aza-substitution involves a PPP SCF calculation on the unsubstituted molecule and for addition/dimerization involves PPP SCF calculations on the isolated monomers, and subsequent CIS calculations to find the excited states. This is not expected to show any intensity borrowing but can provide a starting point for applying the TRK sum rule (see above) and indicating which states intensity can be borrowed from and to. In method \ref{mthd_alg}, the dipole moments of the zeroth-order excited states are perturbed by \eqr{ptb_dip} using the zeroth-order orbitals. This method is only slightly more computationally expensive than \ref{mthd_zo}. The energies of the states are also perturbed to first order according to \eqr{ptb_e}. In method \ref{mthd_1o}, the first-order Fock matrix (\eqr{delfij} for susbtitution, \eqr{delfij1} and \eqr{fnimj} for addition/dimerization) and first-order CIS Hamiltonian (\eqr{subijkl} for susbtitution, \eqr{pcis} and \eqr{hdimpt} for addition/dimerization) are calculated, using the zeroth-order orbitals. The first-order correction is added to the zeroth-order CIS Hamiltonian and the resulting Hamiltonian is diagonalised to give the first-order excited states. This method is more computationally expensive than \ref{mthd_alg}, but significantly less expensive than recalculating the two-electron integrals and the orbitals in method \ref{mthd_ful}. For addition/dimerization, the two-electron integrals $(ninj|mkml)$ between monomers in \eqr{fou} and \eqr{fof} are calculated approximately using charge distributions and the relative orientation of the monomers (as per SI section III B 2) which is shown to be a suitable approximation in order to predict the change to the spectrum. Method \ref{mthd_ful} is a full calculation of the perturbed Hamiltonian, which is a full SCF and CIS PPP calculation run on the substituted molecule (for substitution) or dimer (for addition/dimerization). We note that for the case of aza-substitution, the $\epsilon_{\mathrm{N}}$ (H\"uckel $\alpha$) parameter used for method \ref{mthd_ful} is for the 3-parameter model for N (where $\epsilon_{\mu}$ and the repulsion parameters $U$ and $r_0$ are changed) according to Ref.~\citenum{mat57a}, whereas for perturbation theory calculations in methods \ref{mthd_alg} and \ref{mthd_1o} (where only $\epsilon_{\mu}$ is changed) the value of $\epsilon_{\mathrm{N}}$ is taken from Ref.~\citenum{hin71a} (see SI section IV B).


\subsubsection{Aza-substitution}
\label{ssec:aceneaza}
For the case of substitution, we therefore consider whether the molecule can be perturbed such that the dark $1B_{3u}^-$ transition borrows intensity from the bright $1B_{3u}^+$ transition. These transitions are both $x$-polarized, and have different PPP pseudoparity such that they cannot mix if the molecule remains an alternant hydrocarbon\cite{par56a}. However, aza-substitution breaks Coulson-Rushbrooke symmetry such that these excitations can mix. 

For the case of pentacene the $1B_{3u}^-$ and $1B_{3u}^+$ transitions are $\ket{\Phi_{14}^-}$ and $\ket{\Phi_{14}^+}$ respectively,\cite{hel19a} we find
\begin{align}
 \bra{\Phi_{14}^+}\hat V \ket{\Phi_{14}^-} = \sum_{\nu\ \textrm{sub}} \Delta\epsilon_{\nu} (C_{\nu 4}^2 - C_{\nu 1}^2), \eql{Koutecky}
\end{align}
where the summation is only over the substituted atoms.

\begin{figure}[!h]
\centering
 \includegraphics[width=.7\textwidth]{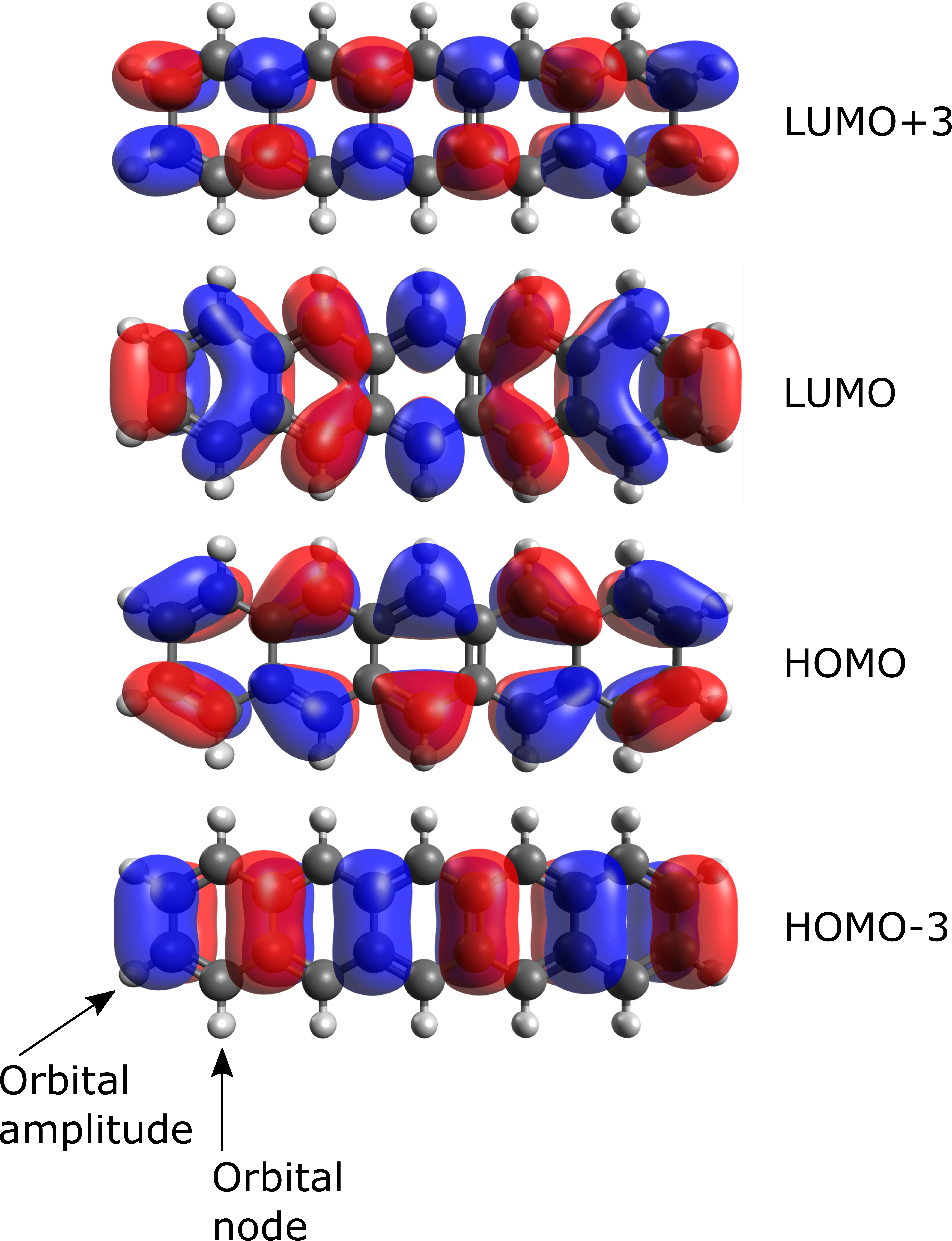}
 \caption{The four orbitals of significance to intensity borrowing in pentance: HOMO--3 (bottom), HOMO, (second from bottom), LUMO (second from top) and LUMO+3 (top) obtained from a DFT calculation. These four orbitals and their symmetry properties were described by Coulson\cite{cou48a} and Pariser\cite{par56a} among others, and are found in \emph{all} acenes. Note however that the HOMO--3 and LUMO+3 in pentacene are \emph{not} always the HOMO--3 and LUMO+3 in other acenes.}
\figl{hd_mc_orbs}
\end{figure}


This means that in order for the excitations to mix, the acene should be substituted on the atoms which have substantial difference in the density of orbitals 4 and 1, i.e.\ the HOMO--3 and the HOMO. Inspection of the relevant orbitals\cite{hel19a} shows that the HOMO--3 has a node on all long-axis (peri) carbons, whereas the HOMO has amplitude over all carbon atoms. We recall that in general there exists a unique pair of orbitals $j$ (occupied) and $j^\prime$ (vacant) in acenes, which have a node at the long-axis positions and a constant amplitude, alternating in sign, at the short-axis positions (see \figr{hd_mc_orbs}).\cite{par56a} For the case of pentacene these orbitals correspond to orbitals 4 and 4$^\prime$ (HOMO--3 and LUMO+3 respectively). Also, we show in SI III B 4 that the state 
\begin{align}
\ket{\Phi_{1}^{j^\prime, +}} = \frac{1}{\sqrt{2}}(\ket{\Phi_{1}^{j^\prime}} + \ket{\Phi_{j}^{1^\prime}})
\end{align}
where $j$ and $j^\prime$ are the unique orbitals described above, will always be $x$-polarised and have a significant dipole-moment and therefore suggest that this always corresponds to the $1B_{3u}^+$ transition. Therefore we expect this logic to hold for all acenes from which we suggest a general design rule: 

\begin{tcolorbox}
\begin{center}
\textbf{Design rule for aza-substitution of acenes}\\
For increased, low-energy absorption 
when aza-substituting acenes, place the nitrogen atoms in long-axis (peri) positions.
\end{center}
\end{tcolorbox}

Aza-substitution of acenes has been considered since at least the 1960s\cite{kum67a} where it was noted that as the number of nitrogens increased, so did the relative height of the (originally dark) $1B_{3u}^-$ transition. This study\cite{kum67a} only considered substitution at the peri positions. Around the same time research by Koutecky\cite{kou66a,kou67a,kou67b,kou67c} presented \eqr{Koutecky} for heteroatom substitution of otherwise alternant hydrocarbons\cite{kou67a}. Since then there has been a large literature on aza-substituting acenes amongst other modifications\cite{gey15a,kum67a,leh17a,lia11b,lia11a,lix15a,lix14a,liu10a,mia08a,mia09a,you12a,zis16a}.

\begin{figure}[!h]
\includegraphics[width=.8\textwidth]{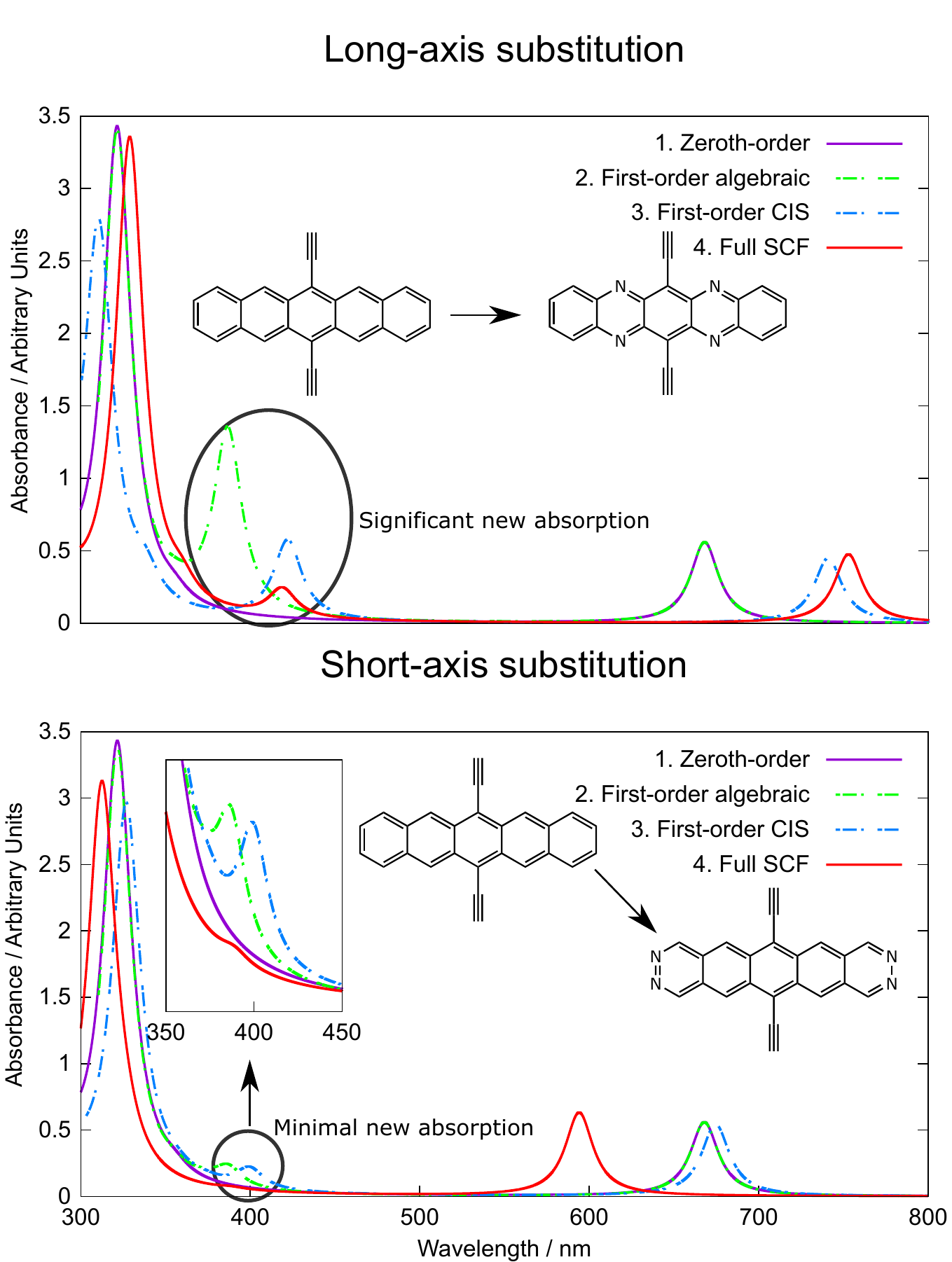}
\caption{Intensity borrowing upon tetra-aza-substitution of TIPS-Pentacene at long-axis positions (top) and short-axis positions (bottom). The unsubstituted TIPS-Pentacene spectrum is shown as `Zeroth Order' along with predicted spectra of i) TIPS-5,7,12,14-tetraazapentacene (TIPS-TAP) (top) and ii) TIPS-2,3,9,10-tetraazapentacene (bottom) at varying levels of theory. A new absorption can be seen around 430 nm after aza- substitution at the 5, 7, 12 and 14 positions as clearly shown by calculations at all levels of theory (top). Only a small new absorption can be seen after aza-substitution at the 2, 3, 9 and 10 positions (bottom). }
\label{fig:ttp}
\end{figure}

We compute the spectrum of TIPS-5,7,12,14-tetraazapentacene (TIPS-TAP, substitution on peri positions) at all four levels of theory where we can clearly see a new absorption emerge around 420 in broad agreement with experimental results\cite{gey15a}. We then consider the case of substitution at the short-axis (cata) positions where we observe, as predicted by theory and experiment\cite{lix14a,lix15a}, that no significant new low-energy absorption appears in the spectrum. These results are presented in \figr{ttp}. Although the algebraic perturbation and first order methods have quantitative differences in spectra from the full SCF calculation, namely that the perturbation theory methods exaggerate the intensity of the new absorption, in all three cases they capture the essential photophysics: a new absorption upon substitution at long-axis positions. A very small new absorption is seen upon substitution at short-axis positions which is replicated by perturbation theory but this does not significantly change the spectrum. 




\subsubsection{Addition and dimerization}
\label{ssec:acenedimer}
Similar to the case of aza-substitution, here we consider how addition and dimerization of acenes can increase their low-energy absorption intensity. Provided that the molecule being added to the starting monomer is also alternant, then the dimer/oligomer will also be alternant, and PPP psuedoparity will be preserved. This means that, for this type of modification, it will not be possible to mix `plus' and `minus' states in the same was as discussed earlier for aza-substitution. 

\begin{figure}[!h]
\includegraphics[width=.8\textwidth]{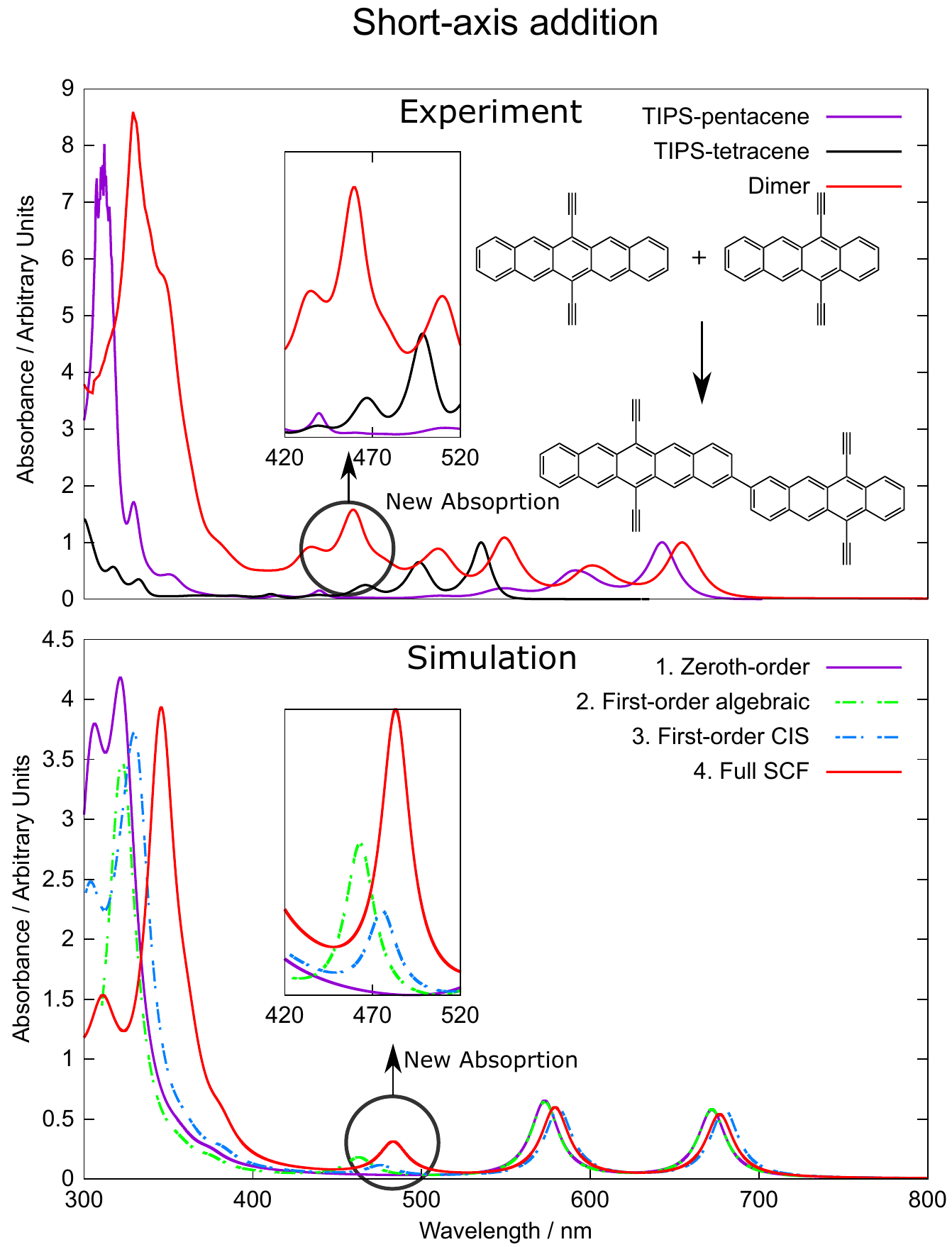}
\caption{Simulated UV/Visible spectrum of TIPS-2,2$^{\prime}$-pentacene-tetracene (bottom) from PPP-SCF / CIS and intensity-borrowing perturbation theory calculations along with the experimental spectra of the dimer the monomers (top). A new absorption is seen around 460nm in the experimental spectrum, predicted by our calculations to be between 460 and 485nm and seen at all levels of theory above zeroth order.}
\label{fig:22pt}
\end{figure}

\begin{figure}[!h]
\includegraphics[width=.8\textwidth]{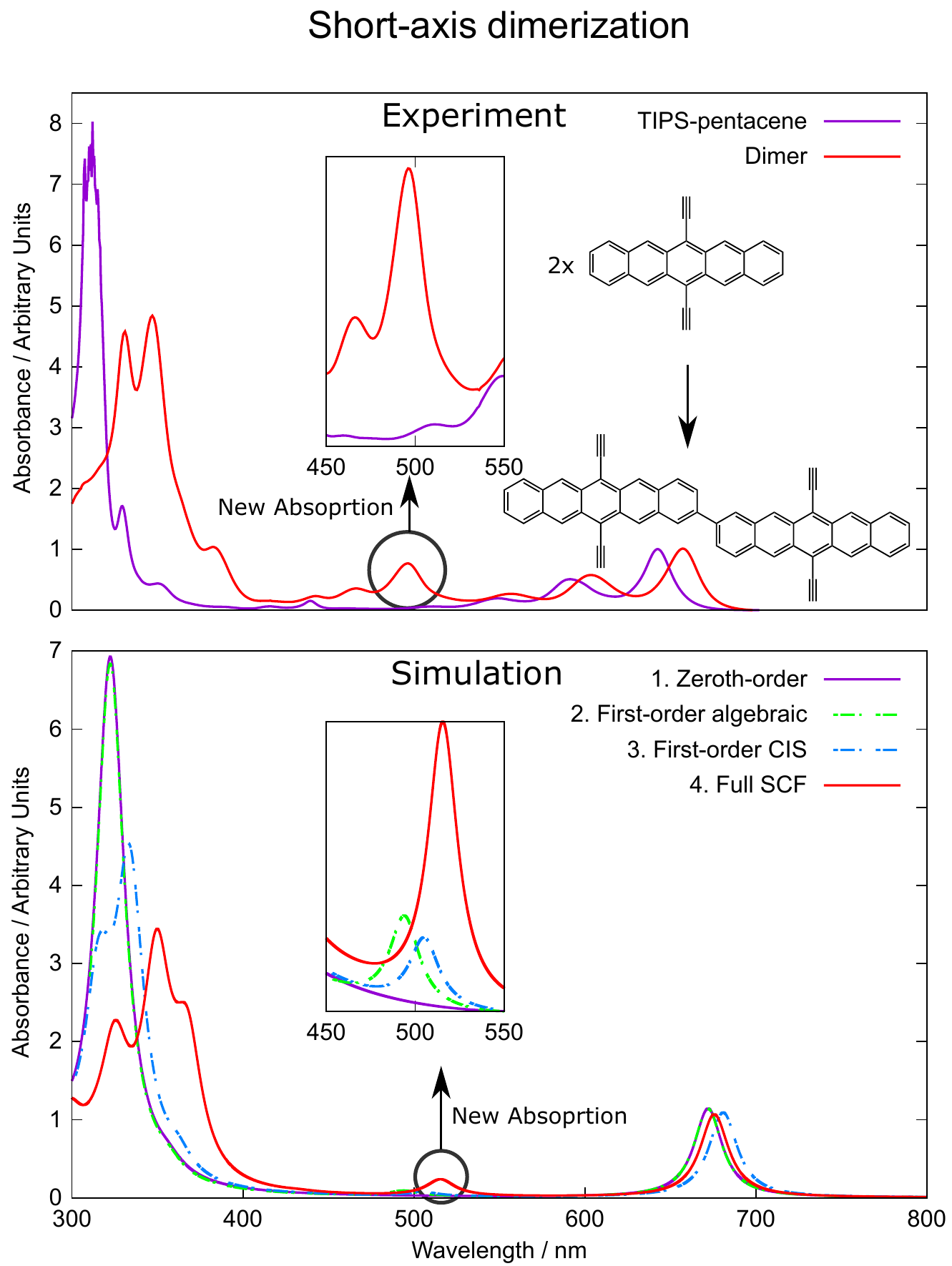}
\caption{Simulated UV/Visible spectrum of TIPS-2,2$^{\prime}$-bipentacene (bottom) from PPP-SCF / CIS and intensity-borrowing perturbation theory calculations along with the experimental spectra of the dimer and monomer (top). A new absorption is seen around 500nm in the experimental spectrum, predicted by our calculations to be between 490 and 520nm and seen at all levels of theory above zeroth order.}
\label{fig:22bp}
\end{figure}

Nevertheless, upon dimerizing it may be possible for the HOMO to LUMO CT excitation (suitably symmetry adapted) to borrow intensity from the intense UV peak. As we have shown previously\cite{hel19a} for the case of pentacene dimers, the mixing of the HOMO-LUMO CT excitation and the high-energy UV absorption is
\begin{align}
 \bra{\Psi_0} \hat \mu\ket{\textrm{CT}_{11}^{B,(1)}}
 =
 -\bra{\Psi_0} \hat \mu\ket{\textrm{LE}_{14}^{+,B}} 
 \frac{\sqrt{2}t_{\nu^*\sigma^*}C_{\nu^*,n1}C_{\sigma^*,m4}}{E(\textrm{CT}_{11}^{B}) - E(\textrm{LE}_{14}^{+,B})} \eql{fobor}
\end{align}
where $\nu^*$ and $\sigma^*$ are the atoms through which the dimer is bonded, $t_{\nu^*\sigma^*}$ is the relevant hopping term, $C_{\nu^*,n1}$ is the HOMO coefficient of atom $\nu^*$ on monomer $n$ and $C_{\sigma^*,m4}$ the HOMO--3 coefficient on monomer $m$.
As described above in section \ref{ssec:aceneaza} and in more detail in SI III B 4, the same reasoning can be applied to all acenes\cite{hel19a,par56a}, which leads to a simple design rule for increased low-energy absorption:\cite{hel19a}
\begin{tcolorbox}
\begin{center}
\textbf{Design rule for addition/dimerization of acenes}\\
 For a new, intense low-energy absorption in acene dimers, join via a short-axis (cata) carbon atom.
\end{center}
\end{tcolorbox}
This can also be extended using symmetry arguments\cite{hel19a} that there should be no long-axis symmetry plane through adjacent monomers. We illustrate this design rule computationally with application to two pentacene 2,2$^{\prime}$- dimers: pentacene-tetracene 2,2$^{\prime}$-dimer and pentacene 2,2$^{\prime}$-homodimer whose spectra are presented in Figs. \ref{fig:22pt} and \ref{fig:22bp}. 

In both cases a new, low-energy absorption is seen in accordance with experimental findings\cite{san16b,hel19a}. The energy of this new absorption in the experimental data is well reproduced by full SCF calculation (method \ref{mthd_ful}) but also by all approximate methods above zeroth-order, however it is difficult to compare the intensities of the experimental spectra with those of the simulated spectra. When comparing the approximate models to the full calculation in method \ref{mthd_ful}, the intensity of the new absorption is predicted most accurately for method \ref{mthd_alg} which predicts the highest intensity for all four dimers, but method \ref{mthd_1o} reproduces the energies of the new absorpion most accurately when compared to method \ref{mthd_ful}. In the SI we present the corresponding spectra for the 1,1$'$-analogues, none of which show a new low-energy absorption in accordance with our theory, and whose simulated spectra at all levels of theory are almost identical. We also present the spectra of pentacene-anthracene and pentacene-hexacene 2,2$^{\prime}$ dimers in the SI. Although the algebraic first-order perturbation \ref{mthd_alg} and first-order Hamiltonian diagonalisation method \ref{mthd_1o} have slight numerical differences in the energies and intensities when compared to the full calculation, they capture the qualitative change in the spectra upon dimerization: a new absorption upon dimerization at the 2,2$^{\prime}$- positions but no new absorption upon dimerization at the 1,1$^{\prime}$- positions. This confirms the use of the methodology presented in this perspective for spectral prediction and molecular design. 



\clearpage 

\subsection{Brightening dark radicals}
\label{ssec:brightrad}
Here we show how the design methodology given in this article can and has been applied to design highly efficient and emissive radical-based organic light-emitting diodes\cite{ai18a,abd20a}. This area has already been extensively reviewed from an organic chemistry\cite{cui20a} and applied physics perspective\cite{abd20a} and been the subject of numerous experimental\cite{ai18a,abd20a,guo19a}, computational\cite{he19a,cho20a,cho21a,guo19a,abe21a} and theoretical\cite{hel21b} studies and here we consider the problem from an inverse design perspective.

\subsubsection{To emit or not to emit}
\label{ssec:rademit}
Stable organic radicals have been known since 1900 with the discovery of the triphenylmethyl (TPM) radical\cite{cui20a}. Various other stable organic radicals have also been known for some time, such as the phenalenyl radical and TEMPO\cite{cui20a}. However, these radicals are not emissive. Attempts were made to stabilise and alter organic radicals, such as by chlorination of the TPM radical, leading to TTM and PTM radicals\cite{cui20a}, which were also not emissive. Further phenyl groups were bonded to the TPM radical\cite{li54a}, leading to derivatives with very weak D$_1$ (lowest excited doublet) absorption. Since the rate of spontaneous emission (fluoresence) is proportional to the absorption for that transition\cite{atk11a}, these investigations led to the belief that no stable radicals were fluorescent\cite{hir87a,cui20a}.

It was then noticed, somewhat surprisingly, that carbazole-, triarylamine- and other aza-derivatives of the TTM radical could in fact be emissive in the red and near-infrared regions of the spectrum. \cite{vel07a,hec09a,hat14a} In a further development in 2015, Peng at al. demonstrated an emissive radical in a functioning OLED based on TTM-1Cz [(4-N-carbazolyl-2,6-dichlorophenyl)bis(2,4,6-trichlorophenyl)methyl radical].\cite{pen15b} Even more surprisingly, in 2018 Ai \emph{et al.}\ announced a radical OLED which was 27\% efficient at 710nm, believed to be the highest efficiency of any known LED in that region of the spectrum, and which was based on the TTM-3NCz [tris-(2,4,6-trichlorophenyl)methyl 3-substituted-9-(naphthalen-2-yl)-9H-carbazole] molecule.\cite{ai18a}

These discoveries seemed particularly surprising given that the emissive radicals were derivatives of the non-emissive triphenylmethyl radical, and that previous attempts to functionalise the triphenylmethyl radical had resulted in non-emissive molecules. Since 2018 numerous other emissive radicals have also been reported.\cite{hud21a,cui20a}

In 2020 it was shown (be rediscovering algebraic theory from the 1950s\cite{lon50a,dew54a,lon55b} and applying it to OLED emission) that if the radical is an alternant hydrocarbon (i.e. it only has even membered rings and no heteroatoms within the conjugated structure) then it will have a vanishingly small transition dipole moment for the D$_1$ $\to$ D$_0$ transition, meaning that the radiative rate will be slow and usually outcompeted by non-radiative decay\cite{abd20a}. This therefore led to a simple design rule for emissive radical OLEDs: 
\begin{tcolorbox}
\begin{center}
\textbf{Major design rule for light-emitting radicals}\\
The emitter must not be an alternant hydrocarbon
\end{center}
\end{tcolorbox}
Comparison of emissive and non-emissive triphenylmethyl radical derivatives\cite{hud21a} corroborated these theoretical results in finding that all emissive radicals were non-alternant, and that the non-emissive derivatives of the triphenylmethyl radical were all alternant. 

In the language of machine learning, what had happened (most likely unintentionally) was that the morphing operations applied to the TPM radical kept it within the space of alternant hydrocarbons, molecular structures which could be shown on theoretical grounds would be extremely unlikely to emit. This led to the erroneous belief that there were no stable emissive organic radicals,\cite{hir87a} and arguably delayed progress in the field. We believe that this highlights the importance of determining whether or not the proposed morphing operations can ever lead to the properties that are desired from the molecule (in this case, what alterations to the TPM radical could cause it to have a significant D$_1$ $\to$ D$_0$ transition dipole moment and therefore be emissive).

The rediscovery of algebraic expressions for radical absorption\cite{abd20a} explained that organic radicals must not be alternant in order to emit, but did not give a clear prescription as how this should best be achieved. Alternant hydrocarbon radicals usually have a dark D$_1$ state which is\cite{lon55b,hel21b} 
\begin{align}
 \ket{D_1} \simeq \ket{\Phi_{10}^{-}} = \frac{1}{\sqrt{2}} (\ket{\Phi_{\bar 1}^{\bar 0}} - \ket{\Phi_{0}^{1'}})
\end{align}
that is, an out-of-phase combination of HOMO to SOMO and SOMO to LUMO excitations, which has a vanishing dipole moment. Conversely, the plus combination
\begin{align}
 \ket{\Phi_{10}^{+}} = \frac{1}{\sqrt{2}} (\ket{\Phi_{\bar 1}^{\bar 0}} + \ket{\Phi_{0}^{1'}})
\end{align}
usually has a substantial transition dipole moment\cite{lon55b}.

In practice there are two ways of forming a bright D$_1$ state: aza-substituting the radical which causes $\ket{\Phi_{10}^{-}}$ to mix with $\ket{\Phi_{10}^{+}}$ and borrow intensity from it, or addition of a non-alternant donor group which leads to a low-energy donor-HOMO to radical-SOMO excitation (beneath $\ket{\Phi_{10}^{-}}$ in energy) which can then borrow intensity from $\ket{\Phi_{10}^{+}}$. Note that adding a donor group which is alternant to the radical does does not increase emission, but simply leads to a lower-lying dark state.\cite{hud21a}

\subsubsection{Aza-substitution}
\label{ssec:radaza}
Aza substitution breaks PPP pseudoparity and cases the originally dark $\ket{\Phi_{10}^{-}}$ D$_1$ state to borrow intensity from the bright $\ket{\Phi_{10}^{+}}$ state,
\bse
\begin{align}
 \bra{\Phi_{10}^{+}} \hat V \ket{\Phi_{10}^{-}} = & \sum_{\nu\ \textrm{sub}} \Delta \epsilon (|C_{0 \nu}|^2 - |C_{1 \nu}|^2) \\
 \therefore \bra{\Phi_0} \hat \mu \ket{D_1^{(1)}} \simeq & \bra{\Phi_0} \hat \mu \ket{\Phi_{10}^{+}} \frac{\sum_{\nu\ \textrm{sub}} \Delta \epsilon (|C_{0 \nu}|^2 - |C_{1 \nu}|^2)}{E(\Phi_{10}^{-}) - E(\Phi_{10}^{+})} 
\end{align}\eql{azarad}
\ese
This suggests that the radical should be substituted on atoms whose HOMO and SOMO coefficients different significantly in magnitude. To our knowledge, TTM has been aza-substituted at the para position to make PyBTM\cite{hud21a}, but not at the meta position where there is vanishing SOMO amplitude\cite{ai18a}. We suggest that this may lead to brighter radicals.

Aza-substitution can also be used to blue-shift emission of a donor-acceptor molecule.\cite{abd20a} While forming the donor-acceptor species can be described using the framework in this article for addition (see next section), the effect of aza-substitution on the donor-acceptor species can be described using the substitution algebra. We denote the acceptor (TTM) as monomer $n$ and the donor (usually Carbazole or a derivative thereof) to be $m$. Approximating the D$_1$ state as $\ket{\textrm{CT}_{m1}^{n0,(1)}}$, aza-substituting the donor perturbs the donor HOMO energy by
\begin{align}
 \Delta E = \Delta \epsilon \sum_{\nu \textrm{sub}} C_{\nu m1}^2
\end{align}
as $\Delta \epsilon < 0 $ for nitrogen (due to its higher electronegativity than carbon), this lowers the energy of the donor HOMO and thereby increases the energy of the $\ket{\textrm{CT}_{m1}^{n0,(1)}}$ transition. This has been applied successfully to blue-shift emission of TTM-1Cz from 687nm to 612--643nm depending on the location of the substitution.\cite{abd20a} Furthermore, the extent of blue-shifting can be predicted approximately by inspection of the donor HOMO orbitals: the larger the HOMO coefficient on a given atom the greater the likely blueshift. In practice\cite{abd20a} this is only a qualitative guide, since the shape of the donor orbitals changes upon aza-substitution, and a more reliable predictor of blue-shifting can be obtained from the energies of the HOMOs of the aza-substituted donor (which is still substantially computationally cheaper than an excited-state calculation on the full radical).

\subsubsection{Addition}
\label{ssec:radadd}
For addition, to break the alternacy symmetry a non-alternant moiety (usually an electron-rich donor group) is bonded to the TTM group (knows as the acceptor)\cite{hud21a}. This leads to a charge-transfer excitation $\ket{\textrm{CT}_{m1}^{n0}}$ which is (or rather, can be) beneath the energy of the dark $\ket{\textrm{LE}_{n1n0}^{-}}$ state of TTM. However, as discussed earlier, to a first approximation CT states are themselves dark, such that \emph{prima facie} this process simply leads to a different dark D$_1$ state. However, it is possible for this state to borrow intensity from the intense $\ket{\Phi_{n1n0}^{+}}$ state of TTM\cite{abd20a,hel21b},
\begin{align}
 \bra{\Phi_0} \hat \mu \ket{\textrm{CT}_{m1}^{n0,(1)}} \simeq -\frac{1}{\sqrt{2}} \bra{\Phi_0} \hat \mu \ket{\textrm{LE}_{n1n0}^+} \frac{
  t_{\nu^*\rho^*} C_{n1,\nu^*} C_{m1,\rho^*} \cos(\theta)
 }{E(\textrm{CT}_{m1}^{n0,(0)}) - E(\textrm{LE}_{n1n0}^+)} \eql{radbor}
\end{align}
where the monomers are joined through atoms $\nu^*$ and $\rho^*$ with dihedral angle $\theta$. This can be summarized as a design rule:\cite{abd20a}
\begin{tcolorbox}
\begin{center}
\textbf{Design rule for light-emitting radicals by addition}\\
There must be substantial orbital amplitudes on the radical HOMO and donor HOMO on the atoms by which they are joined.
\end{center}
\end{tcolorbox}
This is found to hold experimentally in emissive donor-acceptor radicals\cite{abd20a}.

\subsection{Different molecules, similar rules}
\label{ssec:diffsim}
This section on radicals has considered two very different motivations for tailoring spectra: increasing absorption in the visible (not necessarily of the S$_1$ absorption) for photovoltaics and singlet fission, and increasing emission from the D$_1$ state of radical organic light-emitting diodes. Although seemingly very different goals, the algebraic expressions for aza-substition, \eqr{Koutecky} and \eqr{azarad}, and addition, \eqr{fobor} and  \eqr{radbor}, in both cases, are algebraically similar. This illustrates how the theoretical formalism of this article can be applied without substantial modification to a wide variety of problems. 

The numerical examples given in this article clearly show that the theoretical methodology can make the correct qualitative predictions for spectral changes upon molecular subsitution, addition and dimerization. However, there can be occasions where this could break down. One is for accidental degeneracies where the perturbation to the zeroth-order states is not small and perturbation theory itself breaks down. In this case quasidegenerate perturbation theory can be used. Another case could be if the orbitals of the altered chromophore are substantially different to the orbitals of the unaltered chromophore, such that perturbing excitations only (not orbitals) may be insufficient. One possible example of this is 6,6$'$-bipentacene which is explored in SI section IV B 3.

\section{To conclude}
\label{sec:conclusion}
We can summarize the results of this article into the three major principles of chromophore design given in page \pageref{threeprin} of the introduction.

Results similar to these have previously been obtained in very different contexts, and here we show they can be obtained within a consistent theoretical framework. The conservation of oscillator strength, principle~\ref{trk}, is an adaptation of the textbook Thomas-Reiche-Kuhn sum rule\cite{atk11a}. Excitations mixing upon substitution if and only if they differ my one orbital, and the orbitals by which the excitations differ having amplitude on the substituted atom, principle \ref{sub}, has been obtained algebraically in the context of transition polarizations in aza-substituted acenes\cite{kou67c}. 


The form of the zeroth-order Hamiltonian in principle \ref{addi} has been used in (for example) simulating polymer-fullerene blends\cite{ary10a}, and similar LE and CT definitions in principle~\ref{zoh} appear in a DFT exciton model\cite{li17a}. CT states acquiring intensity by mixing between excitations in principle~\ref{ctbor} has already been shown for acene dimers\cite{hel19a} and is consistent with (for example) Mulliken's theory of charge-transfer spectra\cite{mul39b,mul50a} (see SI section III B 3), and Coulombic stabilization of a CT excitation dates from the 1930s\cite{mot38a}. Here we also show it is approximately a charge-dipole term if the two excitations differ by one orbital and a dipole-dipole term if they differ by two orbitals. LE/CT mixing through one-electron terms in principle~\ref{lect} is commonly used for discussing HOMO/LUMO interactions in singlet fission\cite{ber13a,bel13a}, and here we show how it can extend to general excitations and be determined from examining orbital coefficients on the joining atoms. Mixing between LE states in principle~\ref{lelectct} builds upon Kasha exciton theory\cite{kas65a}, though the LE/LE interaction derived here operates between different (bright) LEs and not solely between the same LE on each monomer as in the original formulation\cite{kas65a}.


The results given here are independent of the molecule considered, provided its chromophore is satisfactorily described by a $\pi$ system and therefore by the PPP model. They also do not require presumption of a specific parameterization scheme for PPP, or for the hydrocarbon to be alternant\cite{cou40a}.

In future work we propose extending this methodology to dimerization via a linker or `wire', which has been the subject of recent experimental and theoretical studies\cite{fue16a,kum17a}. By consideration of the ground state mixing of CT excitations through M\"ulliken's theory\cite{mul39b,mul50a} and of double and higher excitations we could extend our methodology to a wider variety of charge-transfer dimers and complexes. The dimerization methodology can be extended further to oligomers and polymers, and a combination of different alterations considered to observe their combined effect. From a theoretical viewpoint, there is the possibility of extending this analysis to other observables, using an operator other than the dipole moment in the perturbation expression [\eqr{ptb_dip}]. 

As already discussed, some of the design rules in this manuscript, such as for acene dimer absorption\cite{hel19a} and efficient radical emission\cite{abd20a}, have already been published in conjunction with experimental results. It is our hope that the methodology in this article can guide synthetic efforts in the future.

We also hope these findings may be implemented in artifical intelligence calculations such as machine learning or genetic algorithms, in order to bias morphing operations in favour of those with a greater likelihood of achieving the desired spectral alteration, increasing the rate of molecular discovery and enabling inverse molecular design.


\section*{Supplementary Material}
The supplementary material comprises: supplementary flow diagrams for molecular design methodology, supplemntary data for typical chromophores, further background theory, implementation details, supplementary figures, computational details and parameterization, and atomic co-ordinates of the simulated chromophores. It is available via $<$insert link$>$.

\section*{Acknowledgements}
TJHH acknowledges a Royal Society University Research Fellowship number URF\textbackslash R1\textbackslash 201502 and a startup grant from University College London. 
TJHH would like to that Roald Hoffmann for helpful discussions. 
The authors would also like to thank Nandini Ananth for advice and helpful comments on the manuscript, and Emrys Evans for useful discussions. We are very grateful to KR Parenti and LM Campos for providing their experimental UV/Vis absorption spectra for the pentacene dimers presented in \figr{22pt}, \figr{22bp} and in the SI.

\section*{Data Availability Statement}
The data that supports the findings of this study are available within the article and its supplementary material.

\bibliography{refbig_persp2_short}
\end{document}


\bibliographystyle{tim}

\title{Supporting Information for: `Inverse molecular design from first principles: tailoring organic chromophore spectra for optoelectronic applications'}
\author{James D.\ Green}
\affiliation{Department of Chemistry, Christopher Ingold Building, University College London, WC1H 0AJ, UK}
\author{Eric G. Fuemmeler}
\affiliation{Baker Laboratory, 259 East Avenue, Cornell University, Ithaca, NY 14853, USA}
\author{Timothy J.\ H.\ Hele}
\email{t.hele@ucl.ac.uk}
\affiliation{Department of Chemistry, Christopher Ingold Building, University College London, WC1H 0AJ, UK}
\date{\today}
\maketitle

\tableofcontents


\section{Supplementary figures for Introduction}
\subsection{Designer Molecules}
\begin{figure}[!h]
 \centering
\includegraphics[width=.8\textwidth]{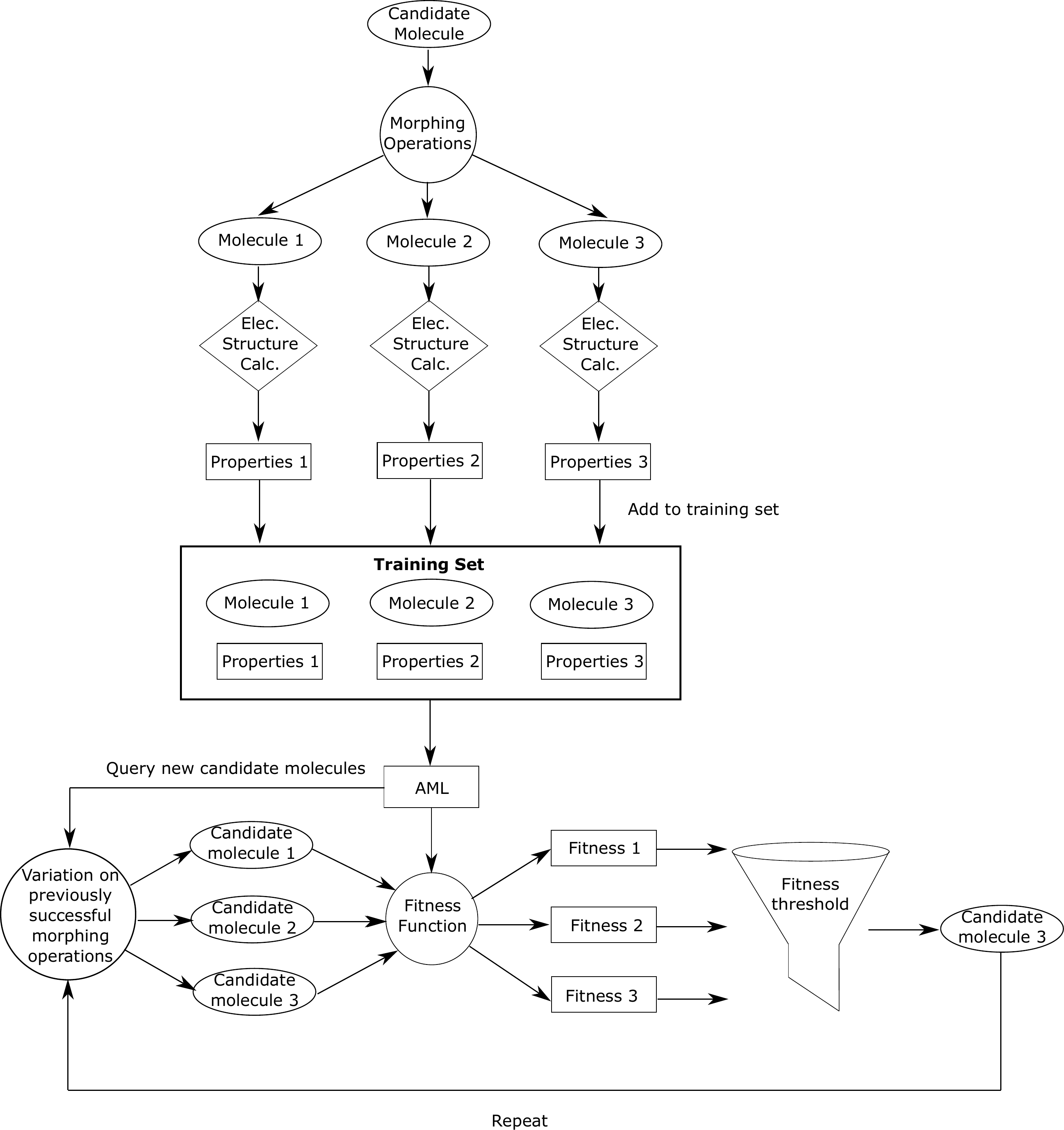}
 \caption{Flowchart illustrating one of the current and most successful approaches in the field of inverse molecular design: the machine learning - genetic algorithm approach. \cite{ven96a,kun21a,pol21a,kwo21a} Training data comprises of molecules generated by morphing operations on an initial candidate molecule, and properties computed using explicit electronic structure calculations. This training data is fed into the AML model which estimates the fitness function for new promising candidate molecules (predicted by AML) from within the chemical space of the training set, and the genetic algorithm baises the most successful molecules which exceed the threshold of the fitness function.}
\end{figure}
\begin{figure}[!h]
 \centering
\includegraphics[width=.8\textwidth]{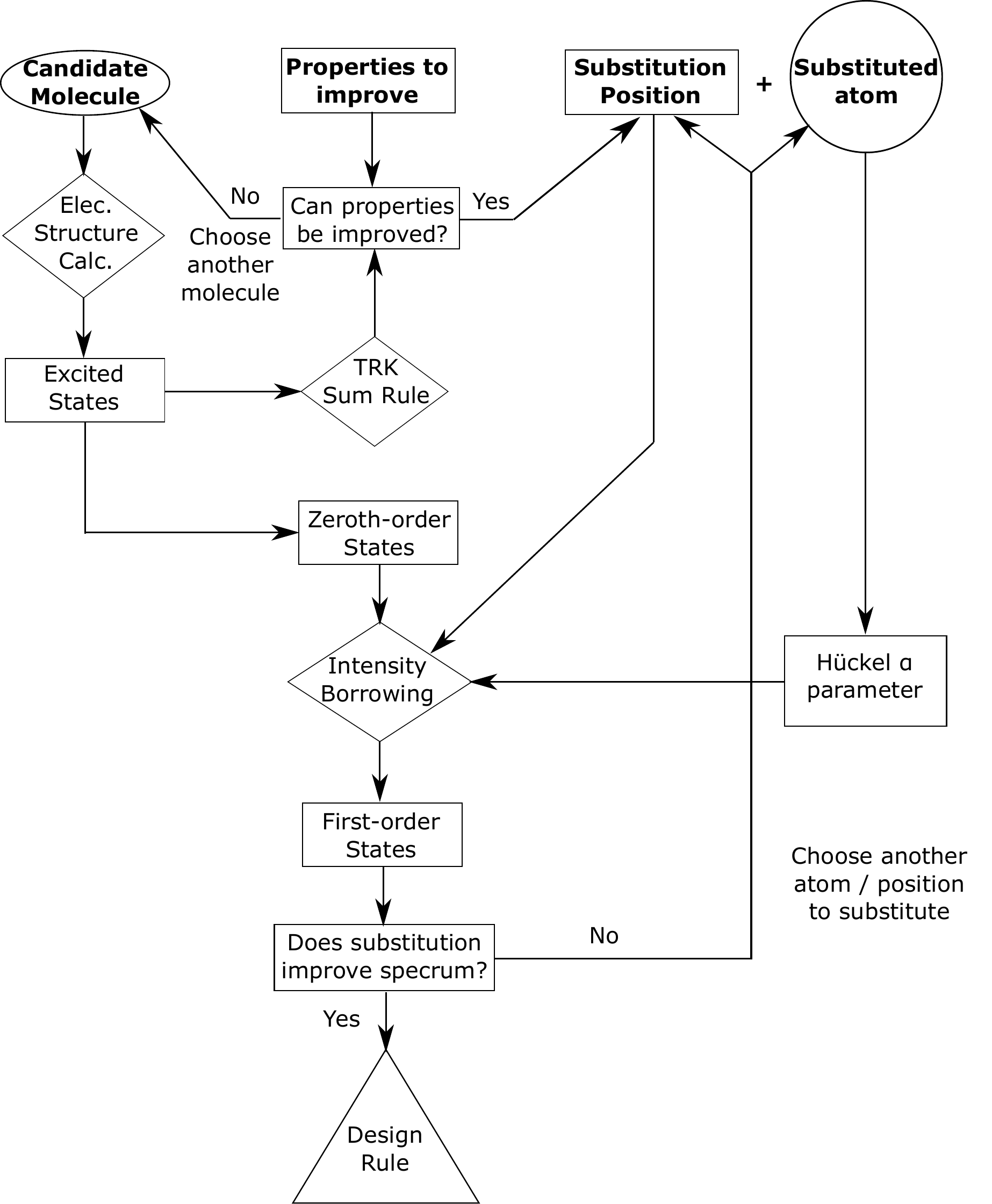}
 \caption{A diagram showing the methodology for tailoring the optoelectronic properties of a candidate (parent) molecule by substitution, as proposed in this article (see Fig. 1. c) in main text). }
\end{figure}

\clearpage
\begin{figure}[!h]
 \centering
\includegraphics[width=.8\textwidth]{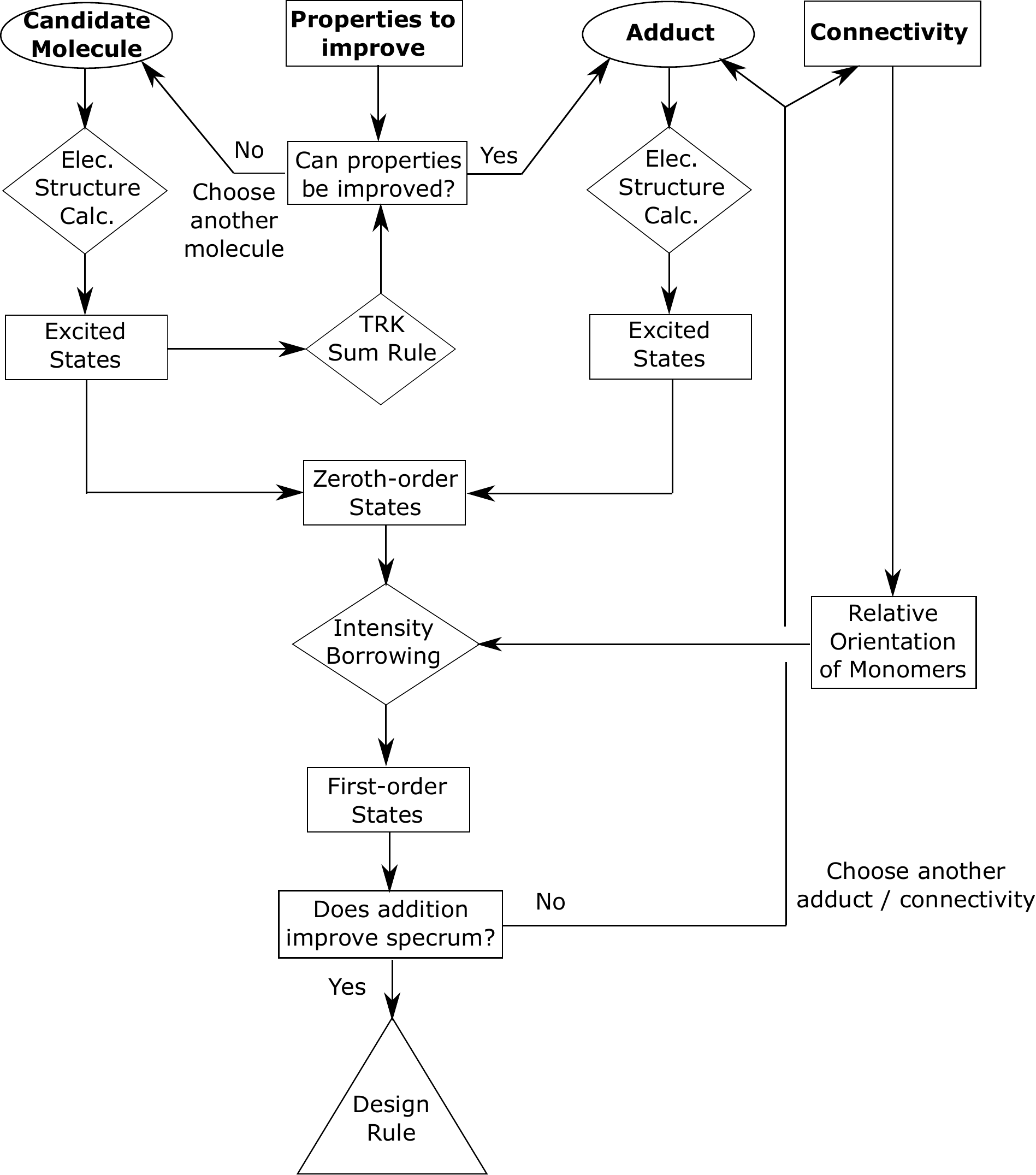}
 \caption{A diagram showing the methodology for tailoring the optoelectronic properties of a candidate (parent) molecule by addition, as proposed in this article (see Fig. 1. c) in main text). }
\end{figure}
\begin{figure}[!h]
 \centering
\includegraphics[width=.8\textwidth]{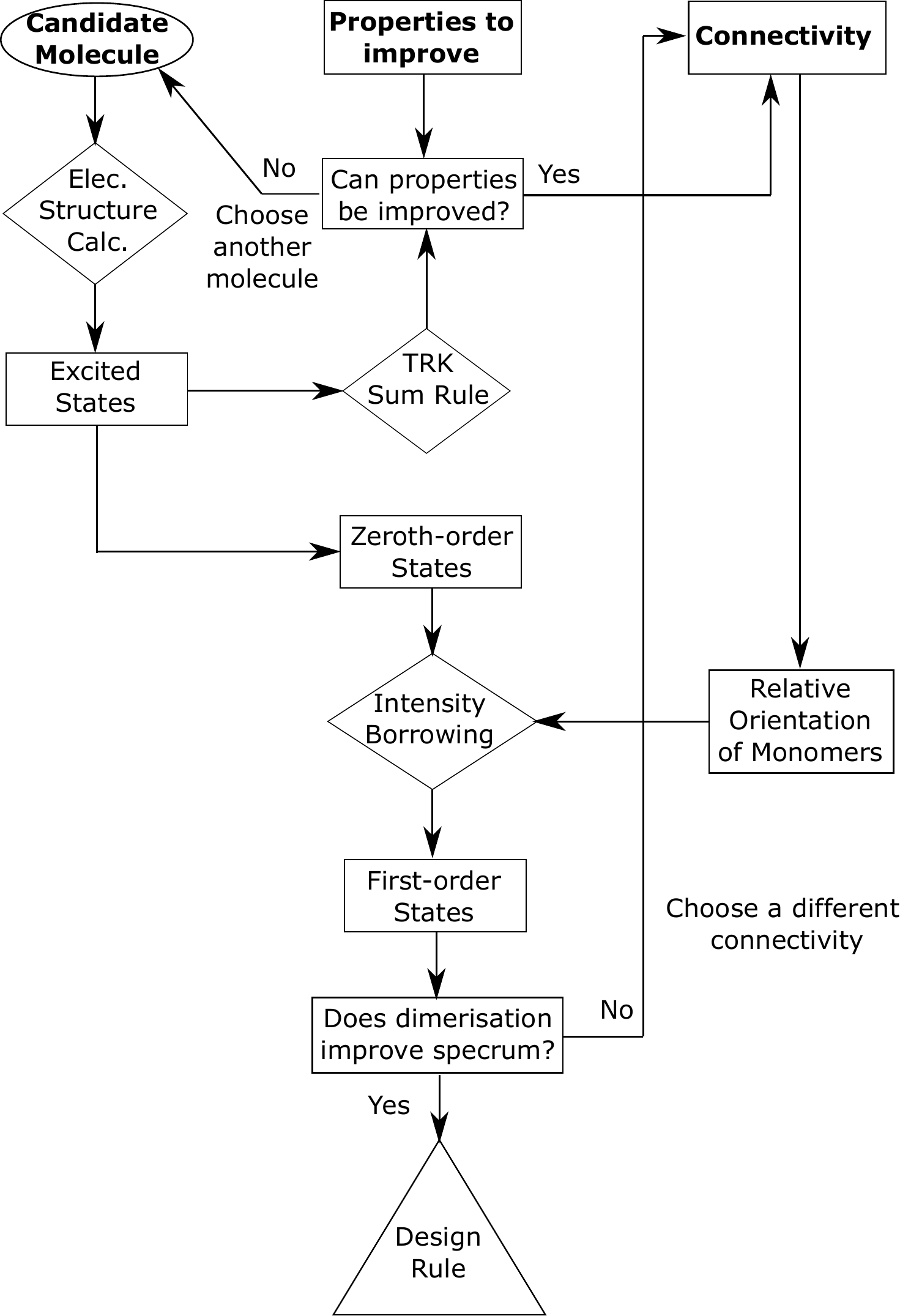}
 \caption{A diagram showing the methodology for tailoring the optoelectronic properties of a candidate (parent) molecule by dimerisation, as proposed in this article (see Fig. 1. c) in main text). }
\end{figure}
\clearpage
\section{Supplementary data for `Challenge 1: How improvable is a molecule?'}
\subsection{Typical Chromophores}
\begin{table}[h]
\begin{center} 
\caption{Vertical excitation energies and oscillator strengths for a series of $\pi$-systems, calculated by various ab initio quantum chemistry methods at varying levels of theory in comparison to the results of PPP calculations and experimental values. \protect \footnotemark[1]} 
\label{tab:si}
\vspace*{2mm}
\begin{tabular}{p{2.5cm}|p{1.2cm}|p{1.2cm}|p{1.1cm}|p{2.0cm}|p{2.9cm}|p{2.4cm}} 
  Chromophore & State & $ \lambda $ & $ f_{osc} $ & TRK Max. & \% of TRK Max. & Method \\ \hline
  Ethene\footnotemark[2]  & \boldsymbol{$1B_{3u}^+$} & \textbf{163} &  \textbf{0.34} & 0.67 (x)& \textbf{51} &  \textbf{Exp.}\cite{pla49a}\\
   & $1B_{3u}^+$ & 167 & 0.24&  & 36  & TD-DFT \cite{mat01a} \\ 
   & $1B_{3u}^+$ & 160 & 0.51  & &77 & CIS \cite{wib92a}\\
   &  $1B_{3u}^+$ & 156 & 0.16 &  & 24 & CASPT2 \cite{ser93a} \\
   & $1B_{3u}^+$ & 188 & 0.51 &  & 77 & PPP \\
  Naphthalene\footnotemark[3] & $1B_{3u}^-$ & 313 & 0.002 & 6.67 ($x,y$) &  $\sim0$ & Exp.\cite{kle49a} \\ 
                     &$1B_{2u}^+$ &289& 0.18 & &  2.7&\\
                     &$1B_{3u}^+ $  &220 &1.7  & &  25 &\\ 
& $1B_{2u} ^+$ & 285 & 0.06& & 0.9  &TD-DFT \cite{kad06a}\\
& $1B_{3u}^- $ & 279 & 0& &0 &\\
                     & $1B_{3u}^+ $ & 212 & 1.25 &    & 19 & \\
& $1B_{3u}^- $ & 308& 0&  & 0  & CASPT2 \cite{rub94a} \\
 & $1B_{2u} ^+$ & 272& 0.05&  & 0.7   & \\
                     & $1B_{3u} ^+$ & 224 & 1.34  & & 20& \\
 & $1B_{3u} ^-$ & 302 & 0&  & 0 & PPP \\
& $2B_{2u} ^+$ & 302 & 0.28&  & 4.2 & \\
                     & $1B_{3u}^+ $ & 234 & 1.96 & & 29 & \\
  Anthracene\footnotemark[3]& \boldsymbol{$1B_{2u}^+$}  & \textbf{379} &  \textbf{0.1} & 9.33 ($x,y$)  &  \textbf{1.1} &\textbf{Exp.}\cite{kle49a} \\ 
                     & \boldsymbol{$1B_{3u}^+$}  &  \textbf{256} &  \textbf{ 2.28} & &  \textbf{ 24} &\\ 	
  & $1B_{2u}^+ $ & 386  & 0.06 & & 0.6 & TD-DFT\cite{kad06a}\\
  & $1B_{3u}^- $ & 323 & 0 &         & 0 & \\
                     & $1B_{3u}^+ $ & 241 & 1.99 &         & 21 & \\
  & $1B_{2u}^+ $ & 399 & 0.32&  & 3.4& PPP\\
& $1B_{3u}^- $ & 338 & 0&         & 0& \\
                     & $1B_{3u}^+ $ & 267 & 2.68&         & 29& \\
\end{tabular}
\vspace*{2mm}
\begin{tabular}{c}
TABLE \ref{tab:si} continued on next page ...\\
\end{tabular}
\end{center}
\end{table}
\clearpage
\begin{table}[h]
\begin{center}
\begin{tabular}{c}
TABLE \ref{tab:si} continued from previous page ... \\
\end{tabular}
\vspace*{2mm}
\begin{tabular}{p{2.5cm}|p{1.2cm}|p{1.2cm}|p{1.1cm}|p{2.0cm}|p{2.9cm}|p{2.4cm}} 
 Chromophore & State & $ \lambda $ & $ f_{osc} $ & TRK Max. & \% of TRK Max. & Method \\ \hline 
  Tetracene\footnotemark[3]  & \boldsymbol{$1B_{2u}^+$ }& \textbf{474} & \textbf{0.08 }& 12 ($x,y$)   & \textbf{0.7}  &\textbf{Exp.}\cite{kle49a}\\
                     & \boldsymbol{$1B_{3u}^+ $} & \textbf{275} &  \textbf{1.85 }& & \textbf{15} & \\
  & $1B_{2u}^+$ & 508  & 0.05  &   & 0.4& TD-DFT\cite{kad06a}\\
                     & $1B_{3u}^- $ & 358 & 0 & &  0& \\
                     & $1B_{3u}^+ $ & 268  & 2.69 & & 22 & \\
  & $1B_{2u}^+  $ & 496& 0.33&  & 2.7& PPP \\
& $1B_{3u} ^-$ &363 &0& & 0& \\
                     & $1B_{3u}^+ $ & 289 & 3.36& & 28& \\
Pentacene\footnotemark[3]& \boldsymbol{$1B_{2u}^+ $}  & \textbf{585} & \textbf{0.08}& 14.67 ($x,y$) & \textbf{0.5} & \textbf{Exp.}\cite{kle49a}\\ 
                  & \boldsymbol{$1B_{3u}^- $}  & \textbf{417} & \boldsymbol{$\sim$}\textbf{0}& & \boldsymbol{$\sim$}\textbf{0}& \\
                  & \boldsymbol{$1B_{3u}^+ $} & \textbf{310} & \textbf{2.2} & &\textbf{15}&  \\
 & $1B_{2u}^+  $ & 653  & 0.04 &   &0.3  &TD-DFT\cite{kad06a} \\
& $1B_{3u}^-	$ & 422  & 0  &  & 0 & \\
                    & $1B_{3u}^+	$ & 292  & 3.35  &  & 23 & \\
 & $1B_{2u}^+ $ & 537 & 0.03 &  &	0.2& CASSCF \cite{zen14a}\\ 
 & $1B_{2u}  ^+$ & 589 & 0.32&   &2.2& PPP\\
& $1B_{3u}^-	$ & 380 & 0&  & 0& \\
                    & $1B_{3u}^+	$ & 313 & 3.94&  & 27& \\
  $\beta$-carotene\footnotemark[4] & \boldsymbol{ $2A_{g}^-$} &  \boldsymbol{$\sim$}\textbf{700} \footnotemark[5] & \textbf{0} & 7.33 ($x$)& \textbf{0}&\textbf{Exp.}
\cite{ona99a,kan94a,kle09b} \\
                    &  \boldsymbol{$1B_{u}^+$} & \textbf{488} & \textbf{2.66} & 	 &  \textbf{36}& \\
&$2A_{g}^-$ & 849  & 0 &   & 0 & DFT-MRCI\cite{kle09a}\\
                    & $1B_{u}^+$ &512 &3.66  & 	 & 50  & \\
  & $2A_{g}^-$ & 747& 4.72&   & 64 & PPP\\
                    & $1B_{u}^+$ & 503& 0.00 & 	 & 0 & \\
\end{tabular}
\vspace*{2mm}
\begin{tabular}{c}
TABLE \ref{tab:si} continued on next page ...\\
\end{tabular}
\end{center}
\end{table}
\clearpage
\begin{table}[h]
\begin{center}
\begin{tabular}{c}
TABLE \ref{tab:si} continued from previous page ... \\
\end{tabular}
\vspace*{2mm}
\begin{tabular}{p{2.5cm}|p{1.2cm}|p{1.2cm}|p{1.1cm}|p{2.0cm}|p{2.9cm}|p{2.4cm}} 
 Chromophore & State & $ \lambda $ & $ f_{osc} $ & TRK Max. & \% of TRK Max. & Method \\ \hline 
  4CzIPN\footnotemark[6] & \boldsymbol{$B$} & \textbf{474}  & \textbf{0.05}  &68 ($x,y,z$)  & \textbf{0.1} & \textbf{TD-DFT}\cite{lin17a} \\
  Au7 CMA\footnotemark[7] &  \boldsymbol{$A_{1}} $  & \textbf{434}  & \textbf{0.26}  &  36 ($x,y,z$)& \textbf{0.7} & \textbf{TD-DFT} \cite{rom20a}\\
 \end{tabular}

\footnotetext[1]{The rows in bold are the data presented in the main text in Table II. Where sufficient experimental data could be found these data were presented in the main text in Table II. For the molecules where sufficient experimental data was not found, the simulated data was presented in the main text in Table II.}
\footnotetext[2]{$D_{2h}$ point group. $x$ is the bond axis. } 
\footnotetext[3]{$D_{2h}$ point group. Molecule oriented in $x,y$ plane, with long axis $x$ and short axis $y$.} 
\footnotetext[4]{Approximately $C_{2h}$ point group. Molecule is approximately linear and oriented along $x$. PPP calculation on planar, all-trans undecaene skeleton of $\beta$-carotene.}
\footnotetext[5]{Estimated from fluorescence data.\cite{ona99a}}
\footnotetext[6]{Total oscillator strength is based on 68 $\pi$-electrons including nitrile groups. $C_{2}$ point group. See Figure 3 of the main text for the structure.}
\footnotetext[7]{Total oscillator strength is based on 36 $\pi$-electrons including all three benzene rings, N and S atoms and four electrons in $5d_{xz}$ and $5d_{yz}$ on Au according to \citenum{hel18a}. Approximately $C_{2v}$ point group. See Figure 3 of the main text for the structure.}
\end{center}
\end{table}
\clearpage

\section{Supplementary information for `Challenge 2: How to improve a molecule?'}

\subsection{Background Theories}

\subsubsection{Configuration interaction singles}
\label{ssec:cis}
Here we briefly review configuration interaction singles (CIS) and define our notation; the notation easily extends to higher excitations if required\cite{sza89a}. In accordance with convention\cite{par56a,hel19a,hel21b}, we number the occupied (bonding) orbitals 1,2,3,$\ldots$ from the HOMO downwards in energy and the unoccupied (antibonding) orbitals $1',2',3',\ldots$ upwards in energy, defining $\ket{\Phi_i^{j'}}$ to be a singlet spin-adapted excitation from orbital $i$ to orbital $j'$.\cite{sza89a}\footnote{For a monoradical we define the SOMO to be orbital 0}. Our orthogonal CIS basis is then formally $\{\ket{\Phi_i^{j'}}\}$ for all $i$ occupied and $j$ unoccupied. A given (singlet) wavefunction $\ket{\Psi_u}$ can therefore be written as\cite{par56a}
\begin{align}
 \ket{\Psi_u} = \sum_{i \ \mathrm{occ}} \sum_{j' \ \mathrm{unocc}} S_{u,ij'} \ket{\Phi_i^{j'}} \eql{cisexp}
\end{align}
where $S_{u,ij'}$ is an expansion coefficient. The CIS eigenstates will then be the eigenstates of the CIS matrix $\mathbf{H}^{\rm cis}$ whose elements are\cite{sza89a}
\bse
\begin{align}
 \mathbf{H}^{\rm cis}_{ij',kl'} =: & \bra{\Phi_i^{j'}} \hat H - E_0\ket{\Phi_k^{l'}} \no \\
 = & \delta_{ik}F_{j'l'} - \delta_{j'l'}F_{ik} + 2(j'i|kl') - (j'l'|ki), \\
 \mathbf{H}^{\rm cis}_{0,ij'} =: &  \bra{\Phi_0} \hat H - E_0\ket{\Phi_i^{j'}} \no\\
 = & F_{ij'}
\end{align}\eql{cish}%
\ese
where the energy of the ground-state (unexcited) determinant $\ket{\Phi_0}$ is
\begin{align}
 E_0 = & \bra{\Phi_0} \hat H \ket{\Phi_0} \no \\
 = & \sum_{i \ \mathrm{occ}} h_{ii} + F_{ii}.\eql{e0def}
\end{align}
In \eqr{e0def} $h_{ii}$ is the one-electron component of the Born-Oppenheimer Hamiltonian, Fock matrix elements are given by
\begin{align}
 F_{ij} = h_{ij} + \sum_{k \mathrm{occ}} 2(ij|kk) - (ik|kj), \eql{fdef}
\end{align}
and we use the chemists' notation for two-electron integrals\cite{sza89a}
\begin{align}
 (ij|kl) = \int d\bmr_1 \int d\bmr_2\ \phi_{i}(\bmr_1) \phi_{j}(\bmr_1) \frac{1}{r_{12}} \phi_{k}(\bmr_2) \phi_{l}(\bmr_2), \eql{teintdef}
\end{align}
where we assume the orbitals are real, and in \eqr{teintdef} can be either occupied or unoccupied in the ground state.

As we are considering ground-state closed-shell systems containing light atoms and therefore without significant spin-orbit coupling, in this article we restrict ourselves to spin-singlet excitations and do not include the triplet manifold. However, triplet states can easily be included as an extension of the basis (and can be described using PPP\cite{par56a}). In addition, this methodology can (and has been\cite{abd20a,hel21b}) applied to other spin systems such as radical (doublet) molecules. 

These results are general to configuration interaction singles (do not require presumption of PPP theory) and do not require the Fock matrix to be diagonal (which will become useful later). When the Fock matrix is diagonal (i.e.\ the same Fock matrix is used to construct CIS states as the one used to generate the orbitals from a converged SCF calculation), \eqr{cish} reduces to\cite{sza89a} 
\bse
\begin{align}
\bra{\Phi_i^{j'}} \hat H - E_0\ket{\Phi_k^{l'}} = & \delta_{ik}\delta_{j'l'}(F_{j'j'} - F_{ii}) + 2(j'i|kl') - (j'l'|ki),\eql{cishsimp1} \\
\bra{\Phi_0} \hat H - E_0\ket{\Phi_i^{j'}} = & 0.
\end{align}\eql{cishsimp}%
\ese
The physical meaning of the terms is evident if we consider the diagonal energy $\bra{\Phi_i^{j'}} \hat H - E_0\ket{\Phi_i^{j'}} = F_{j'j'} - F_{ii} + 2K_{ij'} - J_{ij'}$, where $K_{ij'} = (ij'|ij')$ is the exchange integral and $J_{ij'} = (ii|j'j')$ the Coulomb integral\cite{sza89a,roo51a}. Using Koopman's theorem\cite{sza89a}, the Fock matrix elements correspond to removing an electron from orbital $i$ and placing it in orbital $j'$. In addition there is a Coulombic stabilization $J_{ij'}$ of the proximity of the electron and hole and an exchange `penalty' $2K_{ij'}$ for the electron spins being paired, which is not present if the singly-excited wavefunctions are triplets \cite{par56a}. The second line of \eqr{cishsimp} (Brillouin's theorem) means that at zeroth-order, the ground state does not mix with any singly excited states. Like \eqr{cish}, \eqr{cishsimp} is general and not specific to PPP theory.

The low-lying zeroth-order eigenstates of small organic chromophores are often dominated by one excitation (or one PPP plus/minus state\cite{par56a,son07a}) and in what follows we usually assume that this is the case. More complicated scenarios can clearly be handled using expansions such as \eqr{cisexp}.

\subsubsection{Perturbation theory}
Here we use perturbation theory to derive the first-order correction to the transition dipole moment given in Eq. (8) in the main text.
\paragraph{Non-degenerate states}
As in standard perturbation theory\cite{atk11a} we define our system as
\begin{align}
 \hat H = \hat H_0 + \hat V
\end{align}
where $\hat H_0$ is our zeroth-order Hamiltonian and $\hat V$ the perturbation. We begin with a set of zeroth-order eigenstates $\ket{\Psi_u^{(0)}}$ where  $\hat H_0 \ket{\Psi_u^{(0)}} = E_u \ket{\Psi_u^{(0)}}$ and $\{u,v,w\}$ are indices for eigenstates, and we assume the eigenstates to be real, as in generally the case for stationary electronic structure calculations. For generality, and unlike Ref.~\onlinecite{rob67a}, we do not split $\hat H_0$ into a `molecule' and `perturber' part, nor factor $\ket{\Psi_u^{(0)}}$ into a product of molecule and perturber contributions. For brevity we define
\bse
\begin{align}
 \mu_{uv} = & \bra{\Psi_u^{(0)}} \hat \mu \ket{\Psi_v^{(0)}}, \eql{mudef} \\
 V_{uv} = & \bra{\Psi_u^{(0)}} \hat V \ket{\Psi_v^{(0)}}, \eql{vdef} \\
 E_{uv} = & E_u^{(0)} - E_v^{(0)}. \eql{edef}
\end{align}
\ese
The standard first and second-order corrections to the wavefunction (neglecting degeneracies which are dealt with below) are\cite{atk11a,rob67a}$^,$\footnote{We note that Wilse Robinson does not appear to include the first term of the second-order perturbation theory expression, though this term, which can be considered a form of normalization, does not significantly affect the eventual results.}:
\bse
\begin{align}
 \ket{\Psi_u^{(1)}} = & \sum_{v\neq u} \ket{\Psi_v^{(0)}} \frac{V_{vu}}{E_{uv}} \eql{fopt} \\
 \ket{\Psi_u^{(2)}} = & \sum_{v\neq u} \ket{\Psi_v^{(0)}} \left[ -\frac{V_{uu}V_{vu}}{E_{uv}^2} + \sum_{w\neq u} \frac{V_{vw}V_{wu}}{E_{uv}E_{uw}} \right] - \frac{1}{2} \sum_{v\neq u} \ket{\Psi_u^{(0)}} \frac{|V_{uv}|^2}{E_{uv}^2} \eql{sopt} 
\end{align}\eql{pteqs}%
\ese
These equations are equally applicable to the ground state ($u=0$) as to excited states.

\paragraph{Degenerate states}
\label{ssec:degstat}
For a system with a subset $\mathcal{D}$ of degenerate states, we can define the `good' eigenstates as\cite{atk11a}
\begin{align}
 \ket{\ti\Psi_v^{(0)}} = \sum_{u \in \mathcal{D}} c_{uv} \ket{\Psi_{u}^{(0)}}
\end{align}
from which we obtain the secular equations\cite{atk11a}
\begin{align}
  \sum_{u \in \mathcal{D}} (\delta_{wu} E_v^{(1)} - V_{wu}) c_{uv} = 0 \eql{degstat}
\end{align}
where $w \in \mathcal{D}$, meaning that we must find diagonalize the perturbation matrix $\mathbf{V}$ (where $(\mathbf{V})_{wu} = V_{wu}$) in the subspace of degenerate states $\mathcal{D}$, from which the eigenvalues are $\{E_v^{(1)}\}$ and eigenvectors $\{c_{uv}\}$. For systems with many degenerate subspaces, as we shall encounter below, this procedure can be performed in each subspace separately.

If there is no direct coupling between the degenerate states, i.e.\ $V_{wu} = 0 \ \forall w,u \in \mathcal{D}$, then  $\{E_v^{(1)}\} = 0$ and there is no first-order correction to the energy. To determine the good eigenstates we then go to the second order energy correction, finding\cite{lan77a}
\begin{align}
  \sum_{u \in \mathcal{D}} \left( \sum_{w \notin \mathcal{D}} \frac{V_{u'w}V_{wu}}{E_{uw}} - E_v^{(2)}\delta_{vu} \right) c_{uv} = & 0 
\end{align}
so we now need to diagonalize the matrix with elements
\begin{align}
 (\mathbf{\ti V})_{u'u} = & \sum_{w \notin \mathcal{D}} \frac{V_{u'w}V_{wu}}{E_{uw}}. \eql{vmat}
\end{align}
Once the `good' eigenstates $\{ \ket{\ti \Psi_v^{(0)}}\}$ have been found they can be perturbed as in \eqr{pteqs}, though the sums over $v$ and $w$ now exclude states in the degenerate subspace $\mathcal{D}$.


Frequently symmetry arguments can be used to determine the good eigenstates, since $\hat V$ usually transforms as the totally symmetric representation in the point group of the perturbed chromophore. This means that $\bra{\Psi_u^{(0)}} \hat V \ket{\Psi_v^{(0)}} = 0$ unless $\Psi_u^{(0)}$ and $\Psi_v^{(0)}$ are of the same irrep, such that finding the irreducible representations\cite{atk11a} of the zeroth-order eigenstates can sometimes be sufficient to diagonalize $V_{wu}$ in \eqr{degstat} or $\mathbf{\ti V}$ in \eqr{vmat}.


We note that previous uses of intensity borrowing, such as Kasha's point-dipole model, have used special cases of the above perturbation theory, such as for two identical monomers where only dipole-dipole interactions are considered \cite{kas65a}. For strong perturbations where \eqr{fopt} breaks down we can diagonalize in the quasidegenerate basis\cite{gou59a,low51a} which is a similar procedure to \eqr{degstat}.

In general we are not interested in determining the exact value of the perturbation for a specific alteration (otherwise a non-perturbative electronic structure calculation should be run) but instead to consider whether a perturbation is nonzero or not, and what chemical factors are likely to influence its size (such as molecule geometry or planarity).

\paragraph{Dipole moment}
As in the previous literature \cite{rob67a} we only consider transitions from the ground electronic state, though excitations between excited states can be treated similarly. We now expand the dipole moment for the $0\to u$ transition to second order in the perturbation parameter $\lambda$, 
\bse
\begin{align}
 \bra{\Psi_0} \hat \mu \ket{\Psi_u} = & \bra{\Psi_0^{(0)}} \hat \mu \ket{\Psi_u^{(0)}} \\
 & + \lambda(\bra{\Psi_0^{(0)}} \hat \mu \ket{\Psi_u^{(1)}} + \bra{\Psi_0^{(1)}} \hat \mu \ket{\Psi_u^{(0)}}) \\
 & + \lambda^2(\bra{\Psi_0^{(0)}} \hat \mu \ket{\Psi_u^{(2)}} + \bra{\Psi_0^{(1)}} \hat \mu \ket{\Psi_u^{(1)}} + \bra{\Psi_0^{(2)}} \hat \mu \ket{\Psi_u^{(0)}}) + \mathcal{O}(\lambda^3).
\end{align}\eql{petdm}
\ese
We can now insert \eqr{pteqs} and set $\lambda = 1$ 
giving
\begin{align}
  \bra{\Psi_0} \hat \mu \ket{\Psi_u} \simeq & \mu_{0u} + \sum_{v\neq u} \mu_{0v} \frac{V_{vu}}{E_{uv}} + \sum_{v \neq 0} \frac{V_{0v}}{E_{v0}} \mu_{vu} \no\\ 
  &  + \sum_{v\neq u}\sum_{w\neq u} \frac{V_{vw}V_{wu}}{E_{uv}E_{uw}}\mu_{0v} + \sum_{v \neq 0} \sum_{w \neq u} \frac{V_{0v}V_{wu}}{E_{0v}E_{uw}}\mu_{jk} + \sum_{v \neq 0} \sum_{w \neq 0} \frac{V_{0k}V_{vw}}{E_{0v}E_{0k}} \mu_{wu} \no\\
  &  - \frac{1}{2} \sum_{v\neq u} \frac{V_{uv}V_{vu}}{E_{uv}^2}\mu_{0u} - \frac{1}{2} \sum_{v \neq 0} \frac{V_{0v}V_{v0}}{E_{0v}^2} \mu_{0u}  -\sum_{v \neq u} \frac{V_{uu}V_{vu}}{E_{uv}^2} \mu_{0v} -\sum_{v \neq 0} \frac{V_{0v}V_{00}}{E_{0v}^2}\mu_{vu} \eql{dipmompt}
\end{align}
To give a very rough guide concerning the importance of the perturbation terms, they have been ordered firstly by the order of the perturbation. For the same order of perturbation, they have been ordered by whether they are `normalization' terms or not \cite{rob67a}, i.e.\ whether there exists a loop in the perturbation theory diagram (see \figr{ptwr}) and if one does exist, whether removing it would give a direct or perturbed transition. We then sort terms by how many perturbation interactions are from (or to) the ground state. This is because for many organic chromophores the ground state energy is substantially beneath the first excited state compared to the energy gaps between excited states (cf. Kasha's rule\cite{kas50a}) and perturbations to/from the ground state are therefore likely  to be smaller than perturbations within excited states. One can then evaluate terms in \eqr{dipmompt} in order and it is likely (though of course not guaranteed) that the first non-zero term will be the major contributor to the spectrum. 
\begin{figure}
 \includegraphics[width=.7\textwidth]{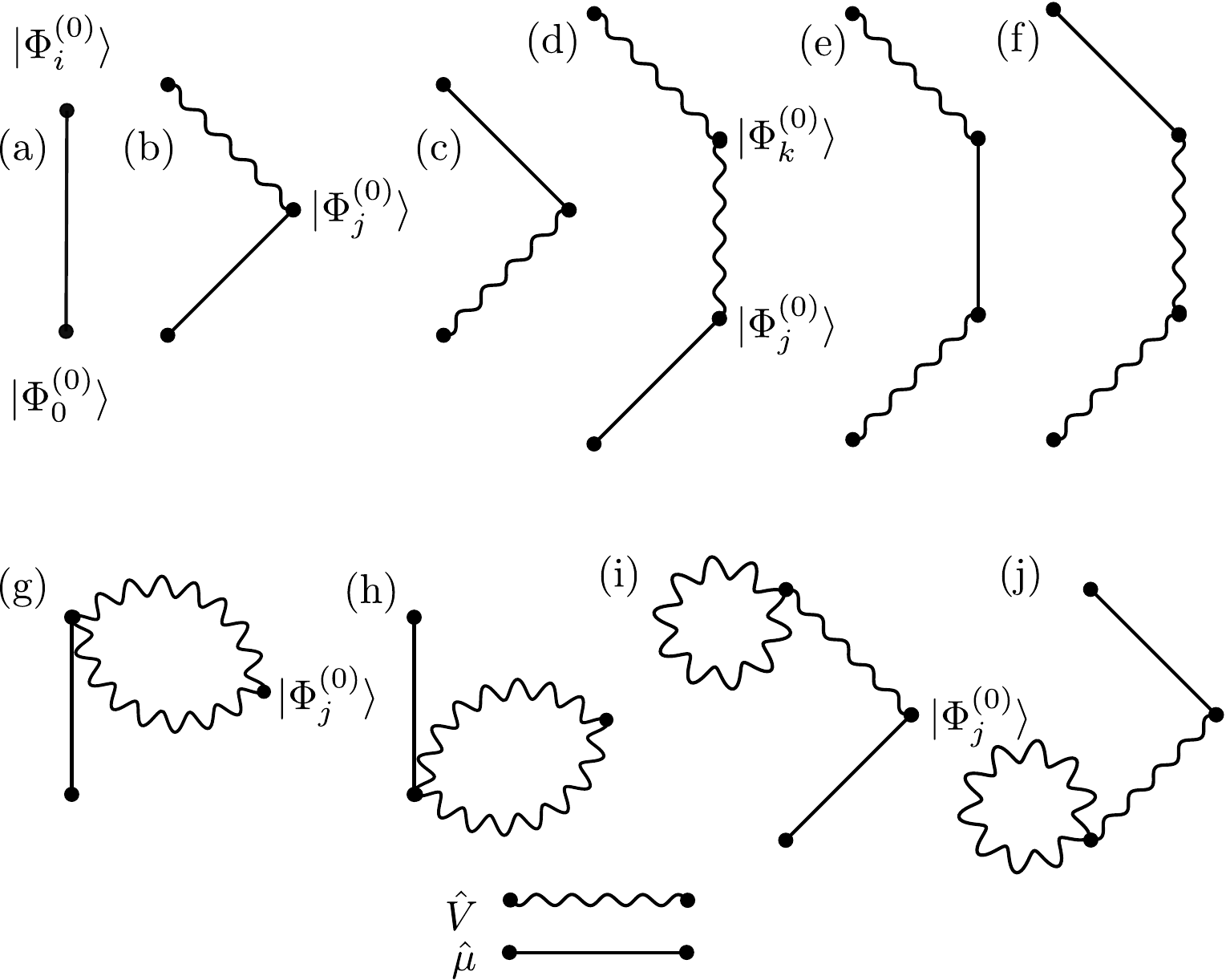}
 \caption{Intensity borrowing perturbation theory diagrams, given in the same order as in \eqr{dipmompt}. Consistent with Ref.~\citenum{rob67a}, each state is represented by a circle, the dipole moment operator $\hat \mu$ is represented by a straight line and the perturbation $\hat V$ by a wavy line. In each diagram the ground state is at the bottom of the diagram and the state borrowing intensity at the top. States which mediate the intensity borrow are to the right of these two states.}
 \figl{ptwr} 
\end{figure}

In \figr{ptwr} we draw perturbation theory diagrams corresponding to \eqr{dipmompt}. For comparison with Ref.~\onlinecite{rob67a}, the ground state $\ket{\Psi_0^{(0)}}$ is at the bottom of each diagram and the final state $\ket{\Psi_u^{(0)}}$ at the top, and these are not summed over in the perturbation expansion. If there is only one intermediate state it is $\ket{\Psi_v^{(0)}}$, and if there are two the `lowest' (only one connection to $\ket{\Psi_0^{(0)}}$) is $\ket{\Psi_v^{(0)}}$ and the `highest' (only one connection to $\ket{\Psi_u^{(0)}}$) is labelled $\ket{\Psi_w^{(0)}}$. This does not necessarily mean that $E_v < E_w$. Intermediate states are draw to the right of the initial/final state and are summed over. Note that these diagrams are different to original perturbation picture in Ref.~\onlinecite{rob67a} since we do not discretize the system into molecule and perturber, and Ref.~\onlinecite{rob67a} appears to omit terms corresponding to (i) and (j).

Analysing the perturbation terms in \figr{ptwr}, (a) is the original (zeroth-order) transition, and is likely to be zero or very small in order for intensity borrowing to be of interest. (b) corresponds to state $i$ mixing with state $j$, which has a dipole moment to the ground state, and this is likely to be the dominant contributor to intensity borrowing. (c) requires the perturbation to mix the ground state with an excited state which can happen but is likely to be small due to a large energy denominator. If there is no direct mixing giving rise to intensity, (d) is likely to be the dominant term where the intensity borrowing is `mediated' by state $k$, with (e) and (f) providing other second order possibilities, though acquiring one and two energy denominators corresponding to ground state excitation respectively and therefore likely to be small. (g) and (h) are normalization terms \cite{rob67a} and will be zero if (a) is also zero. (i) will be zero if (b) is also zero for borrowing via state $j$ and 
mutatis mutandis for (j) and (c). In this article we focus on intensity borrowing via the mixing in (b) and find this is sufficient to explain the spectral alterations we consider. We re-write the second term in \eqr{dipmompt} corresponding to (b) in \figr{ptwr}, and insert \eqr{cisexp} giving (Eq. (8) in the main text)
\begin{align}
\bra{\Psi_0} \hat \mu \ket{\Psi_u^{(1)}} = & \sum_{v \neq u} \bra{\Phi_0} \hat \mu \ket{\Psi_v^{(0)}} \frac{ \bra{\Psi_v^{(0)}} \hat V \ket{\Psi_u^{(0)}}}{E_u - E_v } \nonumber \\
= & \sum_{v \neq u} \sum_{ij^\prime} S_{v,ij^\prime}^{(0)} \bra{\Phi_0} \hat \mu \ket{\Phi_i^{j^\prime}} \frac{ \sum_{kl^\prime} S_{v,kl^\prime}^{*(0)} \bra{\Phi_k^{l^\prime}} \hat V \sum_{rs^\prime} S_{u,rs^\prime}^{(0)} \ket{\Phi_r^{s^\prime}} }{E_u - E_v }. \eql{alg_perturb}
\end{align}


These perturbation diagrams can also be applied to degenerate states providing that the `good' eigenstates are used and the perturbations do not include other good eigenstates within the same degenerate subspace. While this article mainly focuses on absorption or emission \emph{intensity} there are textbook\cite{atk11a} expressions for perturbed energies which we can use to guide the extent to which a proposed alteration will alter the absorption or emission \emph{frequency} of a molecule.

\subsubsection{Pariser-Parr-Pople theory}
\label{ssec:ppp}
The Pople-Parr-Pariser Hamiltonian for an arbitrary $\pi$ system, written in second quantization notation, is\cite{pop53a,par56a,ary10a} 
\begin{align}
 \hat H = & \sum_{\mu} \ep_{\mu} \hat n_{\mu} +  U_{\mu\mu} \hat n_{\mu,\uparrow} \hat n_{\mu,\downarrow} - \sum_{\mu<\nu} \sum_\sigma t_{\mu\nu} (\hat a^\dag_{\mu\sigma} \hat a_{\nu\sigma} + \hat a^\dag_{\nu\sigma} \hat a_{\mu\sigma}) \no\\
 & + \sum_{\mu<\nu} \gamma_{\mu\nu} (\hat n_\mu - Z_{\mu})(\hat n_{\nu} - Z_{\nu}), \eql{horig}
\end{align}
where $\hat n_{\mu}$ is the number operator for the number of electrons on atom $\mu$,
\begin{align}
 \hat n_{\mu} = & \sum_{\sigma = \{ \uparrow,\downarrow\}} \hat n_{\mu,\sigma}; \\
 \hat n_{\mu,\sigma} = & \hat a^\dag_{\mu,\sigma} a_{\mu,\sigma}
\end{align}
and $\hat a^\dag_{\mu\sigma},\hat a_{\mu\sigma}$ are the creation and annihilation operators respectively for a spin orbital of spin $\sigma$ on atom $\mu$. $\ep_\mu$ is the on-site energy, which for a purely hydrocarbon chromophore we can set to zero without affecting the energies of excited states, and $U_{\mu\mu}$ is the on-site (Hubbard) repulsion. $\gamma_{\mu\nu}$ is the parameterized repulsion between an electron on atom $\mu$ and an electron on atom $\nu$, approximating the two-electron integral
\begin{align}
 \gamma_{\mu\nu} \simeq & (\mu\mu|\nu\nu) \no\\
 = & \int d\bmr_1 \int d\bmr_2\ \chi_{\mu}(\bmr_1) \chi_{\mu}(\bmr_1) \frac{1}{r_{12}} \phi_{\nu}(\bmr_2) \phi_{\nu}(\bmr_2)
\end{align}
where $\chi_\mu(\bmr)$ is the atomic spatial orbital on atom $\mu$, and we use the chemists' notation\cite{sza89a} for two-electron integrals as in \eqr{teintdef}. $\gamma_{\mu\nu}$ is sometimes written as $V_{\mu\nu}$,\cite{ary10a} terminology we refrain from here to avoid confusion with $V_{uv}$ in \eqr{vdef}. 

Consistent with Ref.~\citenum{pop53a}, we use Greek letters to denote atomic (spatial) orbitals $\{\chi_\mu\}$ and roman letters to denote molecular (spatial) orbitals $\{\phi_i\}$, such that molecular orbitals are given by\cite{par56a}
\begin{align}
 \phi_i = \sum_{\mu} C_{\mu i} \chi_\mu.
\end{align}

The corresponding Fock matrix elements in the atomic orbital basis are\cite{pop53a}
\bse
\begin{align}
F_{\mu\mu} = & \ep_{\mu} + \frac{1}{2}P_{\mu\mu}U_{\mu\mu} + \sum_{\lambda \neq \mu} (P_{\lambda\lambda}-Z_\lambda) \gamma_{\mu\lambda}, \eql{fmm2}\\
F_{\mu\nu} = & -t_{\mu\nu} - \frac{1}{2}P_{\mu\nu}\gamma_{\mu\nu}, \eql{fmn}
\end{align}\eql{fmppp}%
\ese
where the density matrix elements are defined as\cite{pop53a,sza89a}
\begin{align}
 P_{\mu\nu} = 2\sum_{i \ \rm{occ}} C_{\mu i}C_{\nu i} \eql{denmat}
\end{align}
and the sum in \eqr{denmat} is over all occupied spatial orbitals. 

In this article we do not focus on a specific parameterization of PPP theory in order to show that the results hold generally. However, we make the physically reasonable assumptions that $t_{\mu\nu}$ is positive (an electron on one atom is attracted to the nuclear charge on the adjacent atom), and do not require it to be only between nearest neighbours (except in examples where we apply the Coulson-Rushbrooke theorem\cite{cou40a}). We also assume that the two-electron repulsion term $\gamma_{\mu\nu}$ behaves Coulombically at long distances
\begin{align}
 \lim_{|\bmr_{\nu} - \bmr_{\mu}|\to \infty} \gamma_{\mu\nu} = \frac{1}{\kappa |\bmr_{\nu} - \bmr_{\mu}|}
\end{align}
where $\kappa$ can be loosely interpreted as a dielectric constant (c.f.~Refs~\citenum{ary13a,alv19a}).

\subsection{Implementing the theories}
\subsubsection{Implementation of first-order algebraic perturbation expression}
Calculation of the transition dipole moment of the ground state $\ket{\Psi_0}$ to the state $\ket{\Psi_u^{(1)}}$ to first-order given by \eqr{alg_perturb}, at first glance would seem to require five nested loops for indices $u$, $v$, $ij^\prime$, $kl^\prime$ and $rs^\prime$. However, \eqr{alg_perturb} can be split into four parts where we define
\begin{align}
M_{0v} = \sum_{ij^\prime} S_{v,ij^\prime}^{(0)} \bra{\Phi_0} \hat \mu \ket{\Phi_i^{j^\prime}}, \eql{m0v}
\end{align}
\begin{align}
U_{kl^\prime,u} = \sum_{rs^\prime}S_{u,rs^\prime}^{(0)}\bra{\Phi_k^{l^\prime}} \hat V \ket{\Phi_r^{s^\prime}} \eql{ukl}
\end{align}
and
\begin{align}
V_{vu} = &\sum_{kl^\prime} S_{v,kl^\prime}^{*(0)}U_{kl^\prime,u} \eql{vvu}
\end{align}
so \eqr{alg_perturb} becomes
\begin{align}
\bra{\Psi_0} \hat \mu \ket{\Psi_u^{(1)}} = \sum_{v \neq u} \epsilon_{uv} M_{0v} V_{vu} \eql{alg_ptb_2}
\end{align}
where 
\begin{align*}
\epsilon_{uv} = \frac{1}{E_u - E_v }.
\end{align*}
\eqr{m0v}, \eqr{ukl}, \eqr{vvu} and \eqr{alg_ptb_2} require two, three, three and two nested loops respectively.

\subsubsection{Simplification of two-electron integrals}
\label{ap:el}
Here we show how the two-electron integrals arising in the perturbation expressions can be simplified into charge and dipole interactions, which are used in the main text to relate the results to pre-existing models of chromophore interaction. These are similar to standard results for the computation of electrostatic contributions to intermolecular forces between ground-state molecules\cite{buc67a,sto13a,vol04a}. We firstly give results in terms of general molecular orbitals before sketching how they also apply in PPP theory.
\paragraph{General procedure}
\label{ap:genproc}
Prima facie, computation of the two-electron integrals in Eq. 32 and Eq. 35 in the main text require an electronic structure computation on the chromophores combined. The two electron integrals in the perturbation mixing elements are all of the form 
\begin{align}
 (ninj|mkml) = \int d\bmr_1 \int d\bmr_2\ \phi_{ni}(\bmr_1) \phi_{nj}(\bmr_1) \frac{1}{r_{12}} \phi_{mk}(\bmr_2) \phi_{ml}(\bmr_2) \eql{teint}
\end{align}
where the two orbitals on the left-hand side are on one monomer and the two on the right-hand side on the other, though the orbitals can be either bonding, antibonding, or combinations of the two. We consequently define the charge distributions $\rho_n(\bmr_1) = \phi_{ni}(\bmr_1) \phi_{nj}(\bmr_1)$ and $\rho_m(\bmr_2) = \phi_{mk}(\bmr_2) \phi_{ml}(\bmr_2)$, which are non-overlapping. One charge distribution we arbitrarily set to be centered at the origin, the other at a distance $\bmr$ away. Without approximation, we can rewrite \eqr{teint} in terms of integrating over the charge distribution of $\mathcal{N}$ and integrating over the charge distribution of $\mathcal{M}$, such that $\bmr_1 = \bx_1$ and $\bmr_2 = \bmr + \bx_2$,  
\begin{align}
 (ninj|mkml) = \int d\bx_1 \int d\bx_2 \rho_n(\bx_1) \frac{1}{|\bmr - \bx_1 + \bx_2|} \rho_m(\bx_2) \eql{estat}
\end{align}
which has converted the two-electron integal into an electrostatics problem. We then approximate \eqr{estat} by expanding $1/|\bmr - \bx_1 + \bx_2|$ around $\bmr - \bx_1$, assuming $|\bx_2| \ll |\bmr - \bx_1|$,
\begin{align}
 \frac{1}{|\bmr - \bx_1 + \bx_2|} \simeq \frac{1}{|\bmr - \bx_1|} - \frac{\bx_2\cdot(\bmr-\bx_1)}{|\bmr - \bx_1|^3} + \ldots \eql{rexp}
\end{align}
which we place into \eqr{estat} to give
\begin{align}
 (ninj|mkml) = \int d\bx_1 \rho_n(\bx_1)   \left[\frac{1}{|\bmr - \bx_1|} \int \rho_m(\bx_2) d\bx_2 - \frac{(\bmr-\bx_1)}{|\bmr - \bx_1|^3}\cdot \int \bx_2 \rho_m(\bx_2) d\bx_2 + \ldots \right].
\end{align}
We then notice that
\begin{align}
 \int \rho_m(\bx_2) d\bx_2 = & \bk{mk}{ml} = \delta_{kl}, \\
 \int \bx_2 \rho_m(\bx_2) d\bx_2 = & -\bra{mk}\hat \mu \ket{ml} = -\frac{1}{\sqrt{2}}\bra{\Phi_0} \hat \mu \ket{\rLE_{mk}^{ml}} \eql{mu2}
\end{align}
where the minus sign in \eqr{mu2} arises from the negative charge of the electron. 
For notational simplicity we define $\bra{\Phi_0} \hat \mu \ket{\rLE_{mk}^{ml}}=: \bm{\mu}_2$ such that
\begin{align}
 (ninj|mkml) = \int d\bx_1 \rho_n(\bx_1)   \left[\frac{\delta_{kl}}{|\bmr - \bx_1|}  + \frac{\bm{\mu}_2 \cdot (\bmr-\bx_1)}{\sqrt{2}|\bmr - \bx_1|^3} + \ldots \right]
\end{align}
We now expand $1/|\bmr - \bx_1|$ and $(\bmr-\bx_1)/|\bmr - \bx_1|^3$ around $\bmr$, and applying the foregoing reasoning obtain
\begin{align}
 (ninj|mkml) \simeq \frac{\delta_{ij}\delta_{kl}}{|\bmr|} + \frac{1}{\sqrt{2}|\bmr|^3}(\delta_{ij} \bmr\cdot \bm{\mu}_2 - \delta_{kl} \bmr\cdot\bm{\mu_1}) + \frac{1}{2} \left[\frac{\bm{\mu}_1 \cdot \bm{\mu}_2}{|\bmr|^3} - 3\frac{(\bmr\cdot\bm{\mu}_1)(\bmr\cdot\bm{\mu}_2)}{|\bmr|^5}\right] + \ldots \eql{bigthree}
\end{align}
where $\bm{\mu}_1 =: \bra{\Phi_0} \hat \mu \ket{\rLE_{ni}^{nj}}$. The first term in \eqr{bigthree} is a charge-charge interaction, followed by charge-dipole and dipole-charge terms, followed by a dipole-dipole term.
\begin{figure}
\centering
 \includegraphics[width=.6\textwidth]{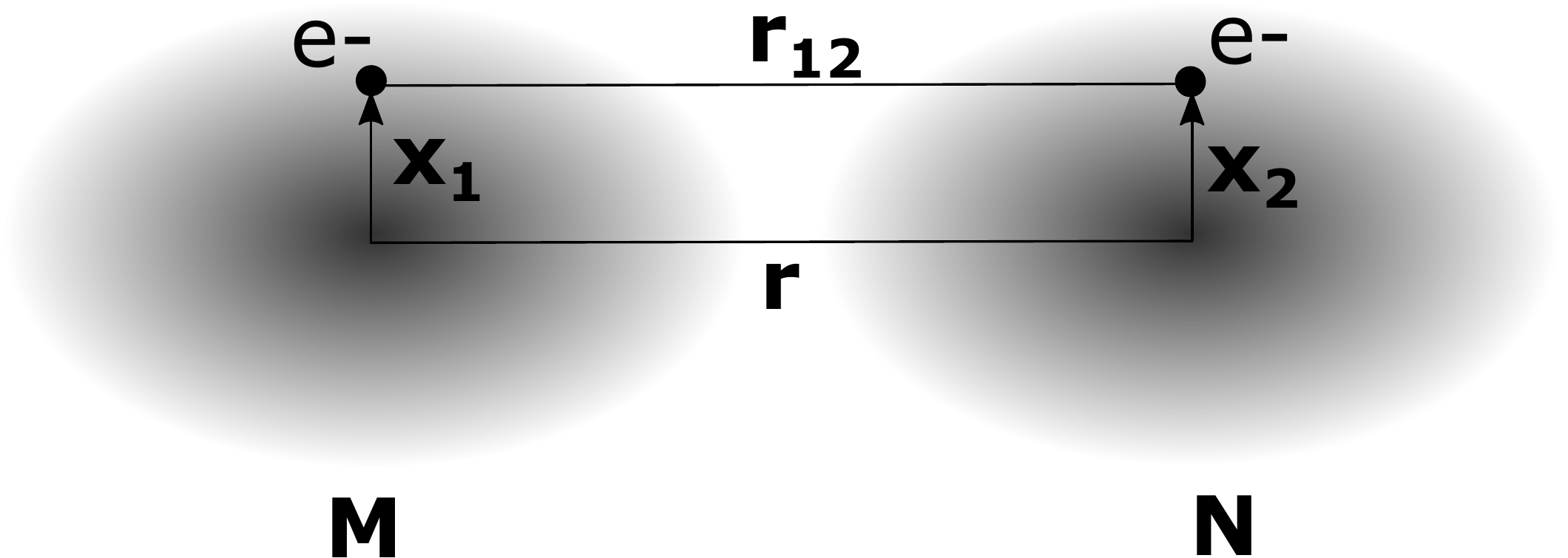}
 \caption{Schematic of the charge distributions of the two acene monomers $\mathcal{M}$ and $\mathcal{N}$, where $\bmr_{12}$ is the vector separating the two electrons, $\bx_1$ and $\bx_2$ are the vectors of electrons on $\mathcal{M}$ and $\mathcal{N}$ from the centres of their respective charge distributions and $\bmr$ is the vector separating these centres.}
 \figl{charge_dist}
\end{figure}
\clearpage

\paragraph{Two-electron integrals upon addition}
\label{ap:teiua}
To evaluate  $(ninj|mkml)$ within PPP, we can identify the charge distribution of two overlapping orbitals $ni$ and $nj$ as 
\begin{align}
 \rho_{ni,nj;\mu} = C_{\mu, ni}C_{\mu, nj} \eql{rhoninj}
\end{align}
and likewise for $\rho_{mk,ml;\nu}$, therefore
\begin{align}
 (ninj|mkml) = \sum_{\mu \in \mcN} \sum_{\nu \in \mM} \ \rho_{ni,nj;\mu} \ \gamma_{\mu,\nu} \ \rho_{mk,ml;\nu} \eql{cdist}
\end{align}
which is the electrostatic interaction between two charge distributions\cite{sto13a}.

\paragraph{Fock matrix perturbation}
\label{ap:fep}
If we now consider the first-order perturbation to a Frenkel excitation from Eqn. 58a) written out explicitly\cite{sza89a}
\begin{align}
 F_{ninj}^{(1)} = -\int d\bmr_1 \phi_{ni}(\bmr_1) \sum_{A \in \mM_a} \frac{Z_A}{r_{1A}} \phi_{nj}(\bmr_1)  + 2 \sum_{k\ {\rm occ}} \int d\bmr_1 \int d\bmr_2 \phi_{ni}(\bmr_1)\phi_{nj}(\bmr_1) \frac{1}{r_{12}} \phi_{mk}(\bmr_2)^2 \eql{delf1}
\end{align}
where $\mM_a$ is the set of atoms on monomer $m$, and there is no exchange term since orbitals on different monomers are non-overlapping. 

To evaluate \eqr{delf1}, we note that the net charge on monomer $m$ is
\begin{align}
 Q_m = \sum_{A \in \mM_a} Z_A - 2 N_{\mathrm{occ},m}
\end{align}
where $N_{\mathrm{occ},m}$ is the number of occupied spatial orbitals on monomer $m$. We also find the (permanent, ground state) dipole moment\cite{sza89a} of monomer $m$ to be
\begin{align}
 \bm{\mu}_{m,0} = - 2 \sum_{k\ {\rm occ}} (mk|\hat \bmr|mk) + \sum_{A \in \mM_a} Z_A \bmr_A
\end{align}
Considering the charge distributions of the nuclei and electrons on monomer $m$ similarly to evaluating the two-electron integral above gives
\begin{align}
 F_{ninj}^{(1)} \simeq -\frac{\delta_{ij}Q_m}{|\bmr|} + \frac{Q_m}{|\bmr|^3}\frac{\bmr\cdot\bm{\mu}_1}{\sqrt{2}} + \frac{\delta_{ij} \bmr\cdot \bm{\mu}_{m,0}}{|\bmr|^3} + \frac{1}{\sqrt{2}}\left[\frac{\bmu_{m,0}\cdot\bmu_1}{|\bmr|^3} - 3\frac{(\bmr\cdot\bmu_{m,0})(\bmr\cdot\bmu_1)}{|\bmr|^5}\right] + \ldots
\eql{eq:1o_fock_approx}
\end{align}

\paragraph{Application to PPP theory}
In the Mataga-Nishimoto (MN) parameterisation of PPP theory,\cite{mat57a} the Coulombic repulsion between two electrons is given by 
\begin{align}
\gamma = \frac{U}{1 + \frac{|\bmr|}{r_0}}.
\end{align}
We now present the approximate two-electron integrals $(ninj|mkml)$ and Fock matrix elements $F_{ninj}^{(1)}$ derived above in this parameterisation of PPP theory. Expanding out the terms of the Taylor expansion of 
\begin{align*}
\frac{1}{1 + \frac{|\bmr+\bm{x_1}+\bm{x_2}|}{r_0}}
\end{align*}
to first order in powers of $\bx_2$, simplifying the integral in \eqr{teint}, and then expanding to first order in terms of $\bx_1$ (cf. Eq. \ref{eq:estat} - Eq. \ref{eq:bigthree}) gives the expression for the two-electron integrals
\begin{align} 
(ninj|mkml) \simeq \delta_{ij}\delta_{kl}\frac{U}{a} + \frac{U}{\sqrt{2}r_0a^2|\bmr|}(\delta_{ij} \bmr \cdot \bm{\mu}_2 -
\delta_{kl}  \bmr \cdot \bm{\mu}_1) + \frac{U}{2r_0a^2|\bmr|}(\bm{\mu}_1 \cdot \bm{\mu}_2)  \nonumber \\
-\frac{U}{r_0a^2|\bmr|^2} \left[ \frac{1}{r_0a} + \frac{1}{2|\bmr|} \right]  ( \bmr \cdot \bm{\mu}_1)( \bmr \cdot \bm{\mu}_2) + \ldots,
\label{eqn:approx_2el}
\end{align}
where we define the dimensionless quantity
\begin{align*}
a = 1 + \frac{| \bmr|}{r_0}.
\end{align*}
In the MN parameterisation\cite{mat57a} the Fock matrix elements in \eqr{eq:1o_fock_approx} become
\begin{align}
 F_{ninj}^{(1)} \simeq -\frac{\delta_{ij}UQ_m}{a} + \frac{UQ_m}{\sqrt{2}r_0a^2|\bmr|}\bmr\cdot\bm{\mu}_1 + \frac{\delta_{ij}U}{r_0a^2|\bmr|}\bmr\cdot \bm{\mu}_{m,0} + \frac{U}{\sqrt{2}r_0a^2|\bmr|}(\bm{\mu}_1 \cdot \bm{\mu}_{m,0}) \nonumber \\
- \frac{\sqrt{2}U}{r_0a^2|\bmr|^2} \left[ \frac{1}{r_0a} + \frac{1}{2|\bmr|} \right]  ( \bmr \cdot \bm{\mu}_1)( \bmr \cdot \bm{\mu}_{m,0}) + \ldots
\end{align}
Within PPP, the two-electron integral in \eqr{teint} can be written as\cite{par56a}
\begin{align}
 (ninj|mkml) = \sum_{\nu\in\mcN} \sum_{\mu \in \mM} C_{\nu,ni}C_{\nu,nj} \gamma_{\mu\nu} C_{\mu,mj} C_{\mu,mk}.
\end{align}
We note
\begin{align}
 \sum_{\nu\in\mcN} C_{\nu,ni}C_{\nu,nj} = & \delta_{ni,nj}, \\
  \sum_{\nu\in\mcN} C_{\nu,ni}C_{\nu,nj} \bmr_{\nu} = & -\frac{1}{\sqrt{2}} \bra{\Phi_0} \hat \mu \ket{\Phi_{ni}^{nj'}}.
\end{align}
Provided that the electron repulsion parameterization obeys
\begin{align}
 \lim_{|\bmr_{\nu} - \bmr_{\mu}|\to \infty} = \frac{1}{\kappa |\bmr_{\nu} - \bmr_{\mu}|} 
\end{align}
where $\kappa$ can be interpreted as a dielectric constant\cite{ary13a}, the derivations in sections~\ref{ap:genproc} and \ref{ap:fep} hold, subject to scaling by $1/\kappa$.
To see how $F_{ni,nj}^{(1)}$ in PPP theory can be written as an electrostatics problem, we start with 
\begin{align}
F_{ni,nj}^{(1)} = & \sum_{\mu \in \mcN} C_{\mu, ni}C_{\mu, nj} \sum_{\lambda \in \mM} (P_{\lambda\lambda} - Z_{\lambda}) \gamma_{\mu \lambda}\eql{delfij1} \\ \nonumber
=& \Delta F_{ni,nj}
\end{align}
We now identify the charge density of atom $\lambda$ on monomer $m$ as
\begin{align}
 \varrho_{m;\lambda} = P_{\lambda\lambda} - Z_{\lambda}
\end{align}
and using \eqr{rhoninj}, \eqr{delfij1} becomes
\begin{align}
 F_{ni,nj}^{(1)} = \sum_{\mu \in \mcN} \sum_{\lambda \in \mM} \ \rho_{ni,nj;\mu}\ \gamma_{\mu \lambda} \ \varrho_{m;\lambda}
\end{align}
which, like \eqr{cdist}, is the electrostatic interaction between two charge distributions, and can be estimated from a multipole expansion\cite{buc67a,sto13a}, and therefore approximated without requiring a separate calculation on the dimer. In particular, if there is zero net charge density on monomer $m$ (as in an alternant hydrocarbon\cite{pop53a}) then $\varrho_{m;\lambda}=0$ for all $\lambda$ and $F_{ni,nj}^{(1)}=0$ for all $ni$ and $nj$.

\subsubsection{Charge-Transfer Definitions}
Intensity borrowing perturbation theory can be applied to M\"ulliken's theory of charge-transfer states. Charge-transfer states may be bright due to charge-transfer excitations mixing with bright local excitations as we have discussed in the main manuscript, but may also appear bright due to mixing with the ground state.\cite{mul39b,mul50a} 

\paragraph{M\"ulliken charge-transfer in our notation}
We first consider a system of two non-interacting molecules $m$ and $n$ at infinite separation, whose orbitals are disjoint (meaning that each orbital either has its entire amplitude on $n$ and zero amplitude on $m$, or vice versa). We consider two states of the system, the (neutral) HF ground state $\ket{\Psi_0^{(0)}} = \ket{\Phi_0}$ and an excited ionic configuration $\ket{\Psi_{\rCT}^{(0)}} = \ket{\rCT}$ ($= \ket{\Phi_{m1}^{n1^\prime}}$). We have assumed the the CT excitation corresponds to transfer of an electron from the HOMO of $m$ to the LUMO of $n$ but it could be any arbitrary excitation resulting from the the transfer of an electron from an occupied orbital on one molecule to a vacant orbital on the other. The transition dipole moment to the CT state $\ket{\Psi_{\rCT}^{(0)}}$ at zeroth-order is
\begin{align}
\mu_{0\rCT} & = \bra{\Psi_0^{(0)}} \mu \ket{\Psi_{\rCT}^{(0)}} \\ \nonumber
& = \bra{\Phi_0} \hat \mu \ket{\rCT} \\ \nonumber
&= 0
\end{align}
as CT excitations are dark at zeroth-order (see main text). When $m$ and $n$ are brought together so that they interact, $\ket{\Psi_0^{(0)}}$ and $\ket{\Psi_{\rCT}^{(0)}}$ mix such that there is a small proportion of $\ket{\rCT}$ in the ground state, and an equally small proportion of $\ket{\Phi_0}$ in the excited state:
\bse
\begin{align}
\ket{\Psi_0^{(1)}} &= c_{00}\ket{\Phi_0} + c_{\rCT0}\ket{\rCT} \\
\ket{\Psi_{\rCT}^{(1)}} &=c_{0\rCT}\ket{\Phi_0} + c_{\rCT\rCT}\ket{\rCT}.
\end{align}
\ese
The transition dipole moment becomes 
\begin{align}
\mu_{0\rCT}^{(1)} =& \bra{\Psi_0^{(1)}} \hat \mu \ket{\Psi_{\rCT}^{(1)}}  \\ \nonumber
=&  (c_{00}\bra{\Phi_0} + c_{\rCT0}\bra{\rCT})\hat \mu (c_{0\rCT}\ket{\Phi_0} + c_{\rCT\rCT}\ket{\rCT}) \\ \nonumber
=&c_{00}c_{0\rCT}\bra{\Phi_0}\hat \mu \ket{\Phi_0} + c_{00}c_{\rCT\rCT}\bra{\Phi_0} \hat \mu \ket{\rCT} \\ \nonumber
&+  c_{\rCT0}c_{0\rCT}\bra{\rCT} \hat \mu \ket{\Phi_0} + c_{\rCT0}c_{\rCT\rCT}\bra{\rCT}\hat \mu \ket{\rCT} \\ \nonumber
=&c_{00}c_{0\rCT}\mu_{00} + c_{\rCT0}c_{\rCT\rCT}\mu_{\rCT\rCT},
\end{align}
where $\mu_{\rCT\rCT}$ is the excited state permanent dipole moment. We will assume that the permanent dipole moment of the the neutral ground state $\mu_{00} = 0$ as did M\"ulliken. This gives
\begin{align}
\mu_{0\rCT}^{(1)}& = \bra{\Psi_0^{(1)}} \hat \mu \ket{\Psi_{\rCT}^{(1)}} \\ \nonumber
& = c_{\rCT0}c_{\rCT\rCT}\mu_{\rCT\rCT}\\ \nonumber
\end{align}
We will assume that in the excited state, the two molecules have equal and opposite charges of $+e$ and $-e$ and that they are connected by a vector $\bmr$. Considering the two molecules as point charges, the excited state dipole moment becomes
\begin{align}
\mu_{\rCT\rCT} &= e\bmr,
\end{align}
which simply equals $ \bmr$ in atomic units. Therefore, the transition dipole moment resulting from mixing of the ground state is simply
\begin{align}
\mu_{0\rCT}^{(1)} = c_{\rCT0}c_{\rCT\rCT}\bmr.
\end{align}
The mixing is assumed to be small so $c_{\rCT\rCT} \simeq 1$, which leaves
\begin{align}
\mu_{0\rCT}^{(1)} \simeq c_{\rCT0}\bmr. \eql{ctmix0}
\end{align}
M\"ulliken finds that
\begin{align}
\mu_{0\rCT}^{(1)} \simeq S\bmr \eql{mul_ct}
\end{align}
for large $r$, where $S$ is the overlap integral between the molecular orbitals of $m$ and $n$ which are involved in the transition. \cite{mul39b,mul50a} 
\paragraph{Intensity borrowing perturbation theory approach}
One arrives at a similar result from intensity borrowing theory. We consider the first three terms in the expansion in \eqr{dipmompt}:
\begin{align}
\mu_{0\rCT}^{(1)} & = \bra{\Psi_0} \hat \mu \ket{\Psi_{\rCT}} \\ \nonumber
& \simeq  \mu_{0\rCT} + \sum_{v\neq \rCT} \mu_{0v} \frac{V_{v\rCT}}{E_{\rCT v}} + \sum_{v \neq 0} \frac{V_{0v}}{E_{v0}} \mu_{v\rCT}.
\end{align}
Firstly $\mu_{0\rCT} = 0$ as CT excitations are dark at zeroth-order. We now let $v=0$ in the second term and $v=\rCT$ in the third, and remove the summations as we are only considering these two states. This gives
\begin{align}
\mu_{0\rCT}^{(1)} \simeq \mu_{00} \frac{V_{0\rCT}}{E_{\rCT 0}} + \frac{V_{0\rCT}}{E_{\rCT0}} \mu_{\rCT\rCT}.
\end{align}
The leading term vanishes as $\mu_{00} = 0$, giving
\begin{align}
\mu_{0\rCT}^{(1)} \simeq \frac{V_{0\rCT}}{E_{\rCT0}} \mu_{\rCT\rCT}.
\end{align}
$\mu_{\rCT\rCT} = \bmr$ so
\begin{align}
\mu_{0\rCT}^{(1)} \simeq \frac{V_{0\rCT}}{E_{\rCT0}}\bmr. \eql{ptb_ctmix0}
\end{align}
Here the CT state borrows intensity as shown in \figr{ptwr} c). We notice that both expressions for the transition dipole moment $\mu_{0\rCT}^{(1)}$ of the CT state, given by M\"ulliken's theory in \eqr{mul_ct} and by intensity borrowing perturbation theory in \eqr{ptb_ctmix0} are of the form
\begin{align}
\mu_{0\rCT}^{(1)} = C\bmr 
\end{align}
where C is some factor determining the extent of mixing of the ground state $\ket{\Psi_0^{(0)}}$ and the CT state $\ket{\Psi_{\rCT}^{(0)}}$. This similarity between the results of M\"ulliken  and IBPT suggests that IBPT is compatible for molecules which exhibit M\"ulliken-type charge-transfer.\cite{mul39b,mul50a}

\subsubsection{The intense UV transition in acenes}

According to \citenum{par56a,cou48a}, all acenes exhibit an intense $x$-polarised transition in the UV which is assigned to the symmetry $1B_{3u}^+$. This transition corresponds to the excited state
\begin{align}
\ket{\Phi_{i}^{j^\prime, +}} = \frac{1}{\sqrt{2}}(\ket{\Phi_{i}^{j^\prime}} + \ket{\Phi_{j}^{i^\prime}}) \eql{b3u+}
\end{align}
where in our notation orbital $i$ is the HOMO, $i^\prime$ is the LUMO and $j$ and $j^\prime$ (which are occupied and vacant respectively) are a special type of orbital described by \citenum{par56a,cou48a} which have a node on the even-numbered atoms and a constant amplitude, alternating in sign, on the odd-numbered atoms. The form of these orbitals is depicted in Fig. 5 in the main text. 
\paragraph{Orbital numbering}
In Pariser's notation, the atoms in the positive $y$-axis (with the molecule lying in the $xy$ plane with its centre at the origin, the long axis along $x$ and the short axis along $y$) are labelled by unprimed indices and the atoms in the negative $y$-axis with primed indices (see \figr{par_num}). Pairs of carbon p-orbitals on atoms $\mu,\mu^\prime$ of the acene are sorted into two sets of symmetry orbitals, one set which are symmetric under reflection in the $xz$ plane and designated `$+$' and the other which are antisymmetric and designated `$-$'.
\bse
\begin{align}
\sigma_\mu^{+} &= \chi_\mu + \chi_{\mu^\prime} \\ 
\sigma_\mu^{-} &= \chi_\mu - \chi_{\mu^\prime}
\eql{symorb}
\end{align}
\ese
The molecular orbitals are defined in terms of these symmetric and antisymmetric symmetry orbitals and are sorted into two sets, `$+$' and `$-$', according to their symmetry under reflection in $xz$, and each set is numbered by a positive integer $\lambda$ running from 1 to $2n+1$ where $n$ is the number of rings in the acene (\eqr{par_orbs}). 
\begin{align}
\phi_\lambda^\pm = \sum_{\mu=1}^{2n+1}C_{\mu,\lambda}^\pm \sigma_\mu^\pm \eql{par_orbs}
\end{align}
\begin{figure}[h!]
\centering
 \includegraphics[width=.6\textwidth]{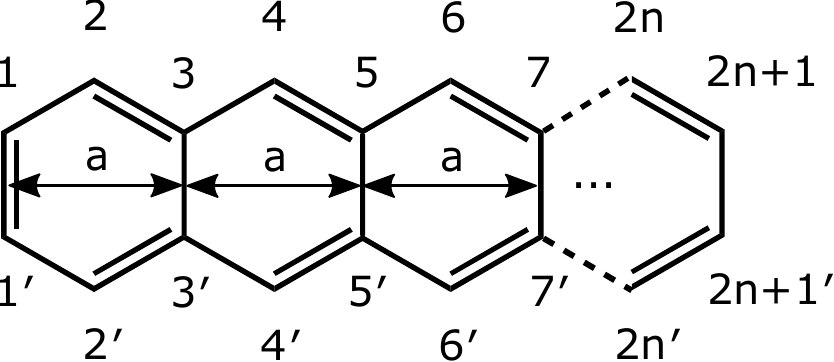}
 \caption{Atom numbering in an acene of $n$ rings according to \citenum{par56a}. Also shown is the distance $a$ between odd-numbered atoms.}
 \figl{par_num}
\end{figure}
\clearpage
The orbital coefficients $C_{\mu,\lambda}^{\pm}$ are given by 
\begin{align}
C_{\mu,\lambda}^\pm = \left[\frac{k_\lambda \mp 1}{(2k_\lambda \mp 1)(n+1)}\right]^{\frac{1}{2}}\sin{\frac{\mu\lambda\pi}{2(n+1)}}\left[\cos^2{\frac{\mu\pi}{2}}+\left(\frac{k_\lambda}{k_\lambda \mp 1}\right)^{\frac{1}{2}}\sin^2{\frac{\mu\pi}{2}}\right], \eql{par_coeff}
\end{align}
where $k_\lambda^\pm$ is
\begin{align}
k_\lambda^\pm = \pm \frac{1}{2} + \frac{1}{2}\cos{\frac{\lambda\pi}{2(n+1)}}\left[16+\sec^2{\frac{\lambda\pi}{2(n+1)}}\right]^{\frac{1}{2}} \eql{k_lambda}
\end{align}
and relates to the orbital energy, according to the relation
\begin{align}
E = \alpha+ k\beta, \eql{Ek}
\end{align}
where $\alpha$ and $\beta$ are the H\"uckel parameters. The special orbitals $j$ (occupied) and $j^\prime$ (vacant) correspond to the $(n+1)$th `+' and `-' orbitals respectively, and their coefficients are given by \eqr{par_coeff} where $\lambda = n+1$ which simplifies to
\begin{align}
C_{\mu,n+1}^\pm = \frac{1}{\sqrt{2(n+1)}}\sin{\frac{\mu\pi}{2}}.
\end{align}
We now must find the two values of $\lambda$ which correspond to the HOMO and LUMO. We first re-write \eqr{k_lambda} in a more suggestive way
\begin{align}
k_\lambda^\pm = \pm \frac{1}{2} + \frac{1}{2}\textrm{sgn} \left( \cos{\frac{\lambda\pi}{2(n+1)}} \right) \left[16\cos^2{\frac{\lambda\pi}{2(n+1)}} + 1 \right]^{\frac{1}{2}}
\end{align}
and we notice the following:
\bse
\begin{align}
k_\lambda^+ &< \frac{1}{2},\: \lambda > n+1 \\
k_\lambda^+ &> \frac{1}{2},\: \lambda < n+1 \\
k_\lambda^- &< -\frac{1}{2},\: \lambda > n+1 \\
k_\lambda^- &> -\frac{1}{2},\: \lambda < n+1 \\
\end{align}
\ese
and with reference to \eqr{Ek} we notice that:
\begin{enumerate}
\item The 1st `$+$' orbital $\phi_{1}^{+}$ is the lowest energy orbital.
\item The $2n+1$th `$-$' orbital $\phi_{2n+1}^{-}$ is the highest energy orbital. 
\item The HOMO is the orbital with the lowest positive value of $k_\lambda^\pm$.
\item The LUMO is the orbital with the highest negative value of $k_\lambda^\pm$. 
\end{enumerate}
Therefore the HOMO and LUMO must correspond to a $k_\lambda^\pm$ value of either $k_n^-$ or $k_{n+2}^+$. To determine which is which we find the signs of $k_n^-$ and $k_{n+2}^+$. Lets suppose that $k_n^- < 0$. 
\begin{align}
 -\frac{1}{2} + \frac{1}{2}\textrm{sgn}\left( \cos{\frac{n\pi}{2(n+1)}} \right)\left[16\cos^2{\frac{n\pi}{2(n+1)}} + 1 \right]^{\frac{1}{2}} &< 0 \\ \nonumber
 -\frac{1}{2} + \frac{1}{2}\left[16\cos^2{\frac{n\pi}{2(n+1)}} + 1 \right]^{\frac{1}{2}} &< 0 \\ \nonumber
\frac{1}{2}\left[16\cos^2{\frac{n\pi}{2(n+1)}} + 1 \right]^{\frac{1}{2}} &<  \frac{1}{2} \\ \nonumber
\left[16\cos^2{\frac{n\pi}{2(n+1)}} + 1 \right]^{\frac{1}{2}} &<  1 \\ \nonumber
16\cos^2{\frac{n\pi}{2(n+1)}} + 1 &<  1 \\ \nonumber
16\cos^2{\frac{n\pi}{2(n+1)}}  &<  0\\ \nonumber
\end{align}
There is a contradiction as we know that 
\begin{align}
16\cos^2{\frac{n\pi}{2(n+1)}}  \geq  0
\end{align}
so $k_{n}^- > 0$. Therefore the HOMO is the $n$th `$-$' orbital $\phi_{n}^{-}$. Now lets suppose that $k_{n+2}^+ > 0$ so we can again prove that the opposite is true by contradiction. 
\begin{align}
 \frac{1}{2} + \frac{1}{2}\textrm{sgn}\left( \cos{\frac{(n+2)\pi}{2(n+1)}} \right)\left[16\cos^2{\frac{(n+2)\pi}{2(n+1)}} + 1 \right]^{\frac{1}{2}} &> 0 \\ \nonumber
 \frac{1}{2} - \frac{1}{2}\left[16\cos^2{\frac{(n+2)\pi}{2(n+1)}} + 1 \right]^{\frac{1}{2}} &> 0  \\ \nonumber
  - \frac{1}{2}\left[16\cos^2{\frac{(n+2)\pi}{2(n+1)}} + 1 \right]^{\frac{1}{2}} &>  -\frac{1}{2} \\ \nonumber
-\left[16\cos^2{\frac{(n+2)\pi}{2(n+1)}} + 1 \right]^{\frac{1}{2}} &>  -1 \\ \nonumber
\left[16\cos^2{\frac{(n+2)\pi}{2(n+1)}} + 1 \right]^{\frac{1}{2}} &<  1 \\ \nonumber
16\cos^2{\frac{(n+2)\pi}{2(n+1)}} + 1 &<  1 \\ \nonumber
16\cos^2{\frac{n\pi}{2(n+1)}}  &<  0
\end{align}
Again we find the same contradiction as we know that 
\begin{align}
16\cos^2{\frac{n\pi}{2(n+1)}}  \geq  0
\end{align} 
so $k_{n+2}^+< 0$.

Therefore the LUMO is the $(n+2)$th `$+$' orbital $\phi_{n+2}^{+}$. The relationship between $k_\lambda^\pm$ and $\lambda$ is illustrated graphically for acenes of 2, 3 and 4 rings in \figr{k_lam}.

\begin{figure}[h!]
\centering
 \includegraphics[width=.6\textwidth]{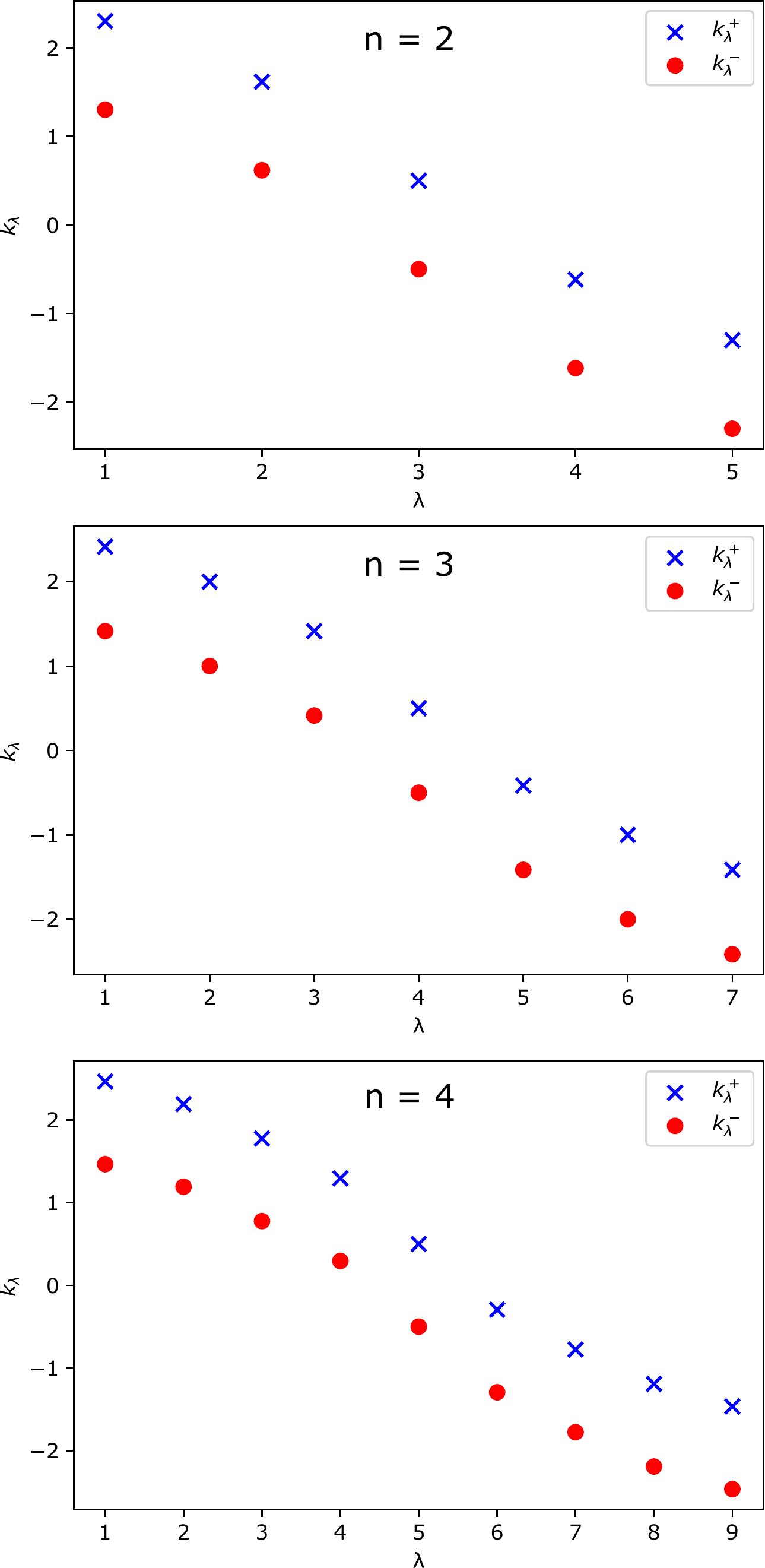}
 \caption{Plot of $k_\lambda$ agaist $\lambda$ for acenes of 2, 3, and 4 rings.}
 \figl{k_lam}
\end{figure}
\paragraph{Group theory}
To show that the transition from the ground state $\ket{\Phi_0}$ to $\ket{\Phi_{i}^{j^\prime, +}}$ is in fact the bright $1B_{3u}^+$ transition, we must first find the irreducible representation of this state from group theory. We recall that in order for a transition to be symmetry allowed, i.e.
\begin{align}
\bra{\Phi_0} \hat\mu \ket{\Phi_{i}^{j^\prime}} &= \\ \nonumber
\bra{i} \hat\mu \ket{j^\prime} &\neq 0,
\end{align}
the product of the irreduclible representations of the orbital $i$, the dipole moment operator and the orbital $j^\prime$ must equal the totally symmetric representation of the point group.
\begin{align}
\Gamma_i  \times \Gamma_{ \hat\mu} \times \Gamma_{j^\prime} = \Gamma_{\textrm{Totally Symmetric}}
\end{align}
Therefore for the transition to be allowed
\begin{align}
\Gamma_i  \times \Gamma_{j^\prime} = \Gamma_{x, y, z}.
\end{align}
The dipole moment of the state $\ket{\Phi_{i}^{j^\prime, +}}$  [\eqr{b3u+}] from the ground state $\ket{\Phi_0}$ is
\begin{align}
\bra{\Phi_0} \hat\mu \ket{\Phi_{i}^{j^\prime, +}} &= \frac{1}{\sqrt{2}}\bra{\Phi_0} \hat\mu (\ket{\Phi_{i}^{j^\prime}} + \ket{\Phi_{j}^{i^\prime}}) \\ \nonumber
& = \frac{1}{\sqrt{2}}(\bra{\Phi_0} \hat\mu \ket{\Phi_{i}^{j^\prime}} + \bra{\Phi_0}  \hat\mu \ket{\Phi_{j}^{i^\prime}}). \eql{dip_b3u+}
\end{align}
The following rules can be used to asign the irreducible representation of each state according to \citenum{par56a}:
For $\phi^+$:
\begin{enumerate}
\item If $\lambda$ is odd, orbital belongs to $B_{1u}$,
\item If $\lambda$ is even, orbital belongs to $B_{2g}$.
\end{enumerate}
For $\phi^-$:
\begin{enumerate}
\item If $\lambda$ is odd, orbital belongs to $B_{3g}$,
\item If $\lambda$ is even, orbital belongs to $A_{u}$.
\end{enumerate}
Therefore if $n$ is odd:
\begin{enumerate}
\item $\phi^-_{n}$ belongs to $B_{3g}$.
\item $\phi^-_{n+1}$ belongs to $A_{u}$.
\item $\phi^+_{n+1}$ belongs to $B_{2g}$.
\item $\phi^+_{n+2}$ belongs to $B_{1u}$.
\end{enumerate}

And if $n$ is even:
\begin{enumerate}
\item $\phi^-_{n}$ belongs to $A_{u}$.
\item $\phi^-_{n+1}$ belongs to $B_{3g}$.
\item $\phi^+_{n+1}$ belongs to $B_{1u}$.
\item $\phi^+_{n+2}$ belongs to $B_{2g}$.
\end{enumerate}

We start by finding the polarisation of the first term in \eqr{dip_b3u+}
\begin{align}
 \bra{\Phi_0} \hat\mu  \ket{\Phi_{i}^{j^\prime}} &= \sqrt{2} \bra{i} \hat\mu  \ket{{j^\prime}} \\ \nonumber
&= \sqrt{2} \bra{\phi_{n}^{-}} \hat\mu \ket{\phi_{n+1}^{-}}.
\end{align}
For all $n$
\begin{align}
\Gamma_{\phi_{n}^{-}} \times \Gamma_{\phi_{n+1}^{-}} & = B_{3g} \times A_{u} \\ \nonumber
& = B_{3u}.
\end{align}
Therefore the first term is symmetry allowed and $x$-polarised. We then do the same for the second dipole moment term
\begin{align}
 \bra{\Phi_0} \hat\mu  \ket{\Phi_{j}^{i^\prime}} &= \sqrt{2} \bra{j} \hat\mu  \ket{{i^\prime}} \\ \nonumber
&= \sqrt{2} \bra{\phi_{n+1}^{+}} \hat\mu \ket{\phi_{n+2}^{+}}.
\end{align}
For all $n$
\begin{align}
\Gamma_{\phi_{n+1}^{+}} \times \Gamma_{\phi_{n+2}^{+}} & = B_{2g} \times B_{1u} \\ \nonumber
& = B_{3u}.
\end{align}
Therefore the second term is also symmetry allowed and $x$-polarised. This is a neccesary but not sufficient criteria for the state $\ket{\Phi_{i}^{j^\prime, +}}$ to be bright. We must also prove that the two dipole moments do not cancel out and are not vanishingly small.

\paragraph{Dipole moment}
We now derive the dipole moment by first re-writing \eqr{dip_b3u+}
\begin{align}
\bra{\Phi_0} \hat\mu \ket{\Phi_{i}^{j^\prime, +}} &= \frac{1}{\sqrt{2}}\bra{\Phi_0} \hat\mu (\ket{\Phi_{i}^{j^\prime}} + \ket{\Phi_{j}^{i^\prime}}) \\ \nonumber
& = \frac{1}{\sqrt{2}}(\bra{\Phi_0} \hat\mu \ket{\Phi_{i}^{j^\prime}} + \bra{\Phi_0}  \hat\mu \ket{\Phi_{j}^{i^\prime}}) \\ \nonumber
&= -\frac{1}{\sqrt{2}}(\bra{\Phi_0} \bmr \ket{\Phi_{i}^{j^\prime}} + \bra{\Phi_0}  \bmr \ket{\Phi_{j}^{i^\prime}}) \\ \nonumber
&= -\frac{1}{\sqrt{2}}(\sqrt{2}\bra{i} \bmr  \ket{{j^\prime}} + \sqrt{2}\bra{j} \bmr  \ket{{i^\prime}}) \\ \nonumber
& = - \bra{i} \bmr  \ket{{j^\prime}} - \bra{j} \bmr  \ket{{i^\prime}} \\ \nonumber
&= -\bra{\phi_{n}^{-}} \bmr \ket{\phi_{n+1}^{-}} - \bra{\phi_{n+1}^{+}} \bmr \ket{\phi_{n+2}^{+}}.
\end{align}
As the transition is $x$-polarised we find the $x$-component
\begin{align}
\bra{\Phi_0} \hat\mu_x \ket{\Phi_{i}^{j^\prime, +}} &= -\bra{\phi_{n}^{-}} x \ket{\phi_{n+1}^{-}} - \bra{\phi_{n+1}^{+}} x \ket{\phi_{n+2}^{+}} \\ \nonumber
& = - \sum_{\mu=1}^{N}C_{\mu,n}^{-}C_{\mu,n+1}^{-}x_\mu - \sum_{\mu=1}^{N}C_{\mu,n+1}^{+}C_{\mu,n+2}^{+}x_\mu \\ \nonumber
&= -\left[\sum_{\mu=1}^{2n+1}C_{\mu,n}^{-}C_{\mu,n+1}^{-}x_\mu +\sum_{\mu^\prime=1}^{2n+1}C_{\mu^\prime,n}^{-}C_{\mu^\prime,n+1}^{-}x_{\mu^\prime} \right]\\ \nonumber
&- \left[ \sum_{\mu=1}^{2n+1}C_{\mu,n+1}^{+}C_{\mu,n+2}^{+}x_\mu +\sum_{\mu^\prime=1}^{2n+1}C_{\mu^\prime,n+1}^{+}C_{\mu^\prime,n+2}^{+}x_{\mu^\prime} \right]
\end{align}
Due to symmetry in the $xz$ mirror plane,
\begin{align}
x_\mu = x_{\mu^\prime}
\end{align}
and by definition
\begin{align}
C_{\mu,\lambda}^{\pm} = C_{\mu^\prime,\lambda}^{\pm}
\end{align}
therefore
\begin{align}
\bra{\Phi_0} \hat\mu_x \ket{\Phi_{i}^{j^\prime, +}} = -2\left[\sum_{\mu=1}^{2n+1}C_{\mu,n}^{-}C_{\mu,n+1}^{-}x_\mu + \sum_{\mu=1}^{2n+1}C_{\mu,n+1}^{+}C_{\mu,n+2}^{+}x_\mu \right]
\end{align}
We now note that, due to being an alternant hydrocarbon
\begin{align}
C_{\mu,\lambda}^{+} = C_{\mu,\gamma}^{-}
\end{align}
where $\lambda$ and $\gamma$ correspond to bonding and antibonding orbitals $j$ and $j^\prime$ of equal and opposite energies relative to the energy of free p-orbital. Therefore,
\begin{align}
C_{\mu,n+2}^{+} &= C_{\mu,n}^{-}  \\ \nonumber
C_{\mu,n+1}^{+} &= C_{\mu,n+1}^{-}
\end{align}
and
\begin{align}
\bra{\Phi_0} \hat\mu_x \ket{\Phi_{i}^{j^\prime, +}} = -4 \sum_{\mu=1}^{2n+1}C_{\mu,n}^{-}C_{\mu,n+1}^{-}x_\mu.
\end{align}
We use the fact that $C_{\mu,n+1}^{-} = 0$ for even $\mu$ (see Fig. 5 in main text), so we only sum over odd numbered atoms
\begin{align}
\bra{\Phi_0} \hat\mu_x \ket{\Phi_{i}^{j^\prime, +}} = -4\sum_{\mu=1, odd}^{2n+1}C_{\mu,n}^{-}C_{\mu,n+1}^{-}x_\mu,
\end{align}
and we use the fact that successive odd numbered atoms are separated by a constant distance $a$, such that 
\begin{align}
x_\mu = \frac{\mu-1}{2}a
\end{align}
to give
\begin{align}
\bra{\Phi_0} \hat\mu_x \ket{\Phi_{i}^{j^\prime, +}}&= -4\sum_{\mu=1, odd}^{2n+1}C_{\mu,n}^{-}C_{\mu,n+1}^{-}\frac{\mu-1}{2}a \\ \nonumber
&= -2a\sum_{\mu=1, odd}^{2n+1}C_{\mu,n}^{-}C_{\mu,n+1}^{-}(\mu-1).
\end{align}
We notice that for odd $\mu$, $\cos^2{\frac{\mu\pi}{2}} = 0$ and $\sin^2{\frac{\mu\pi}{2}} = 1$, so $C_{\mu,n}^{-}$ becomes
\begin{align}
C_{\mu,n}^{-} &= \left[\frac{k_n^{-} + 1}{(2k_n^{-} + 1)(n+1)}\right]^{\frac{1}{2}}\sin{\frac{\mu n\pi}{2(n+1)}}\left[\cos^2{\frac{\mu\pi}{2}}+\left(\frac{k_n^{-}}{k_n^{-} + 1}\right)^{\frac{1}{2}}\sin^2{\frac{\mu\pi}{2}}\right] \\ \nonumber
& = \left[\frac{k_n^{-} + 1}{(2k_n^{-} + 1)(n+1)}\right]^{\frac{1}{2}}\sin{\frac{\mu n\pi}{2(n+1)}}\left(\frac{k_n^{-}}{k_n^{-} + 1}\right)^{\frac{1}{2}} \\ \nonumber
&= \left[\frac{k_n^{-}}{(2k_n^{-} + 1)(n+1)}\right]^{\frac{1}{2}}\sin{\frac{\mu n\pi}{2(n+1)}}.
\end{align}
For simplicity we define the normalisation constants
\bse
\begin{align}
N_{n+1}^{\pm} &= \frac{1}{\sqrt{2(n+1)}}, \\
N_n^{-} &= \left[\frac{k_n^{-}}{(2k_n^{-} + 1)(n+1)}\right]^{\frac{1}{2}}.
\end{align}
\ese
We can now write the $x$-component of the dipole moment as
\begin{align}
 \bra{\Phi_0} \hat\mu_x \ket{\Phi_{i}^{j^\prime, +}} = -2aN_{n+1}^{\pm}N_n^{-}\sum_{\mu=1, odd}^{2n+1}(\mu-1)\sin{\frac{\mu\pi}{2}}\sin{\frac{\mu n\pi}{2(n+1)}}.
\end{align}
We should now introduce a new index $\nu$ where $\mu=2\nu-1$ and runs over only the odd numbered atoms, so we arrive at
\begin{align}
 \bra{\Phi_0} \hat\mu_x \ket{\Phi_{i}^{j^\prime, +}} = -2aN_{n+1}^{\pm}N_n^{-}\sum_{\nu=1}^{n+1}(2\nu-2)\sin{\frac{(2\nu-1)\pi}{2}}\sin{\frac{(2\nu-1)n\pi}{2(n+1)}}. \eql{dip_sum}
\end{align}
We evaluate \eqr{dip_sum} for linear acenes of $n=2$ to $100$ rings assuming all C-C bonds to have a length of 1.39{\AA}\cite{par56a} and present these results in \figr{dip_sum}. Clearly, the model proposed by \citenum{par56a} and \citenum{cou48a} predicts the $x$-component of the dipole moment of the $1B_{3u}^+$ transition to be significantly large ($>9D$), and also that its magnitude increases with increasing number of rings and plateaus off around 20 D. 
\begin{figure}[h!]
\centering
 \includegraphics[width=.8\textwidth]{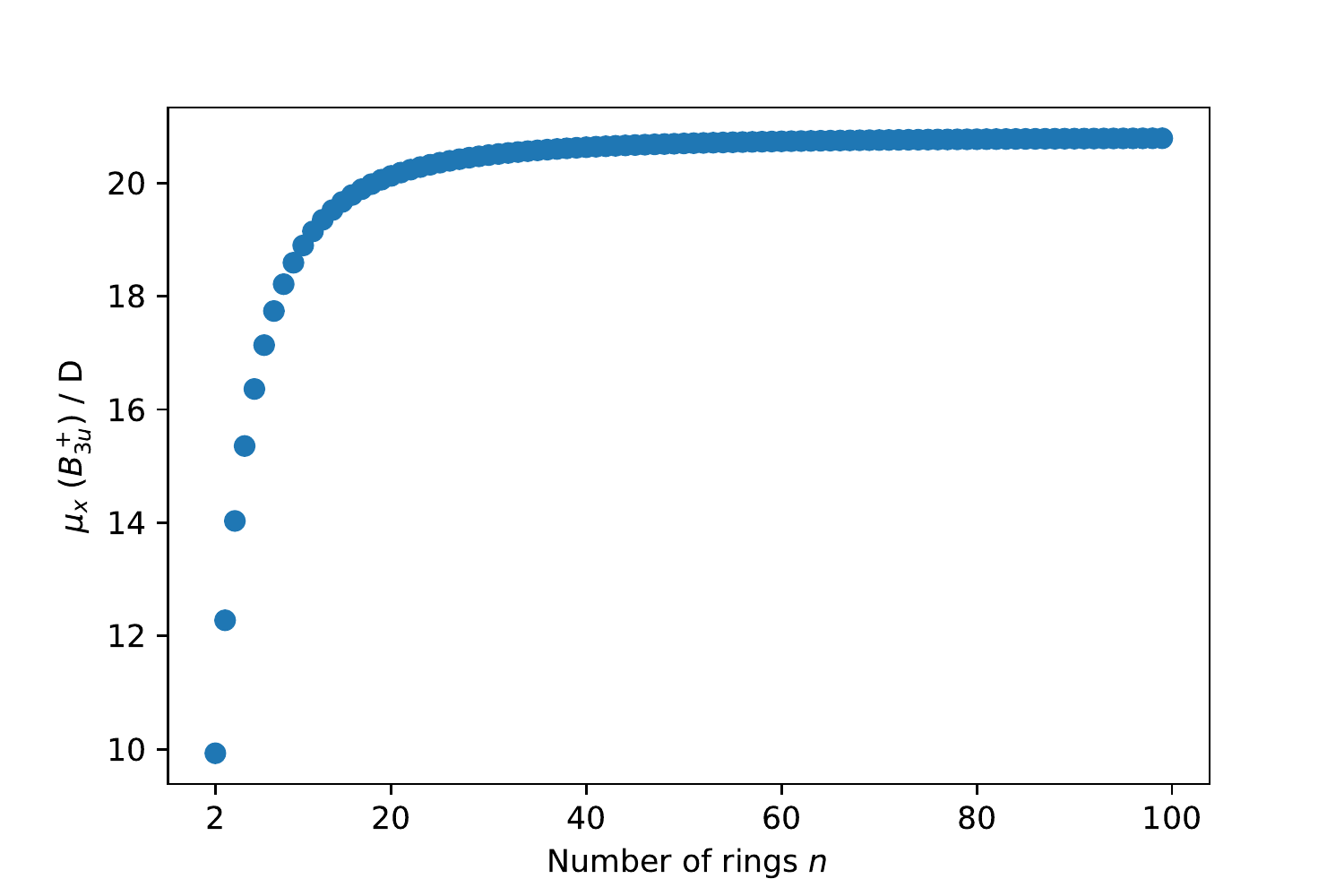}
 \caption{Dipole moment of the $1B_{3u}^+$ transition in units of Debye as predicted by \eqr{dip_sum} for acenes of 2-100 rings.\cite{par56a}}
 \figl{dip_sum}
\end{figure}
\clearpage
\subsection{Supplementary figures for `Where to substitute?'}
\begin{figure}[h!]
  \includegraphics[width=0.3\textwidth]{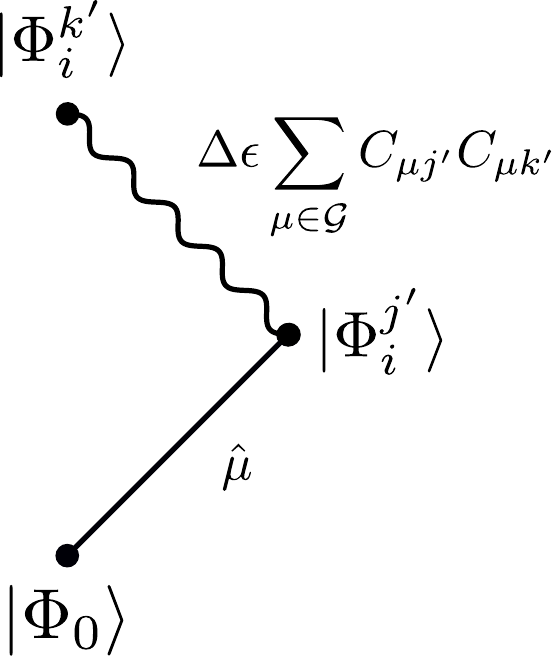}
  \caption{Example intensity borrowing interaction upon substitution. Consistent with Ref.~\citenum{rob67a}, each state is represented by a circle, the dipole moment operator $\hat \mu$ is represented by a straight line and the perturbation $\hat V$ by a wavy line.}
  \figl{sub_intbor}
 \end{figure} 

\subsection{Supplementary figures for `Where to add?'}
\begin{figure}[h!]
 \includegraphics[width=.6\textwidth]{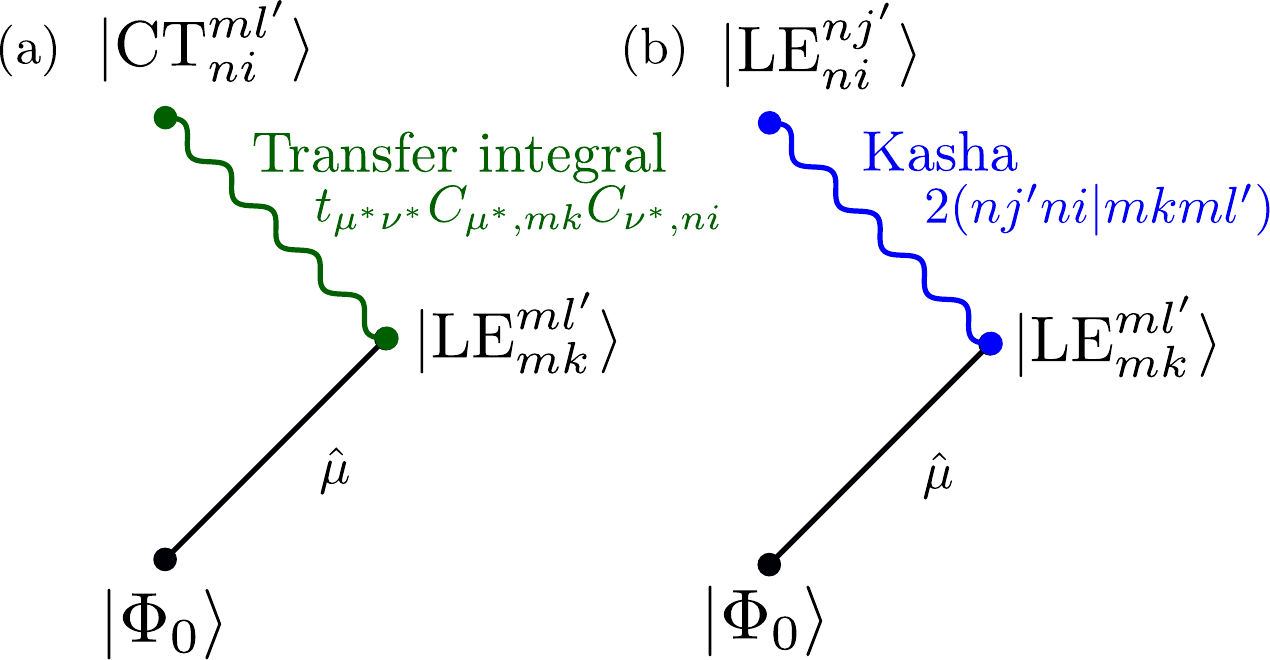}
 \caption{Example perturbation theory diagrams for addition of two chromophores, showing (a) a charge-transfer excitation $\ket{\rCT_{ni}^{mj'}}$ and (b) a local excitation $\ket{\rLE_{ni}^{nj'}}$ borrowing intensity from the same local excitation $\ket{\rLE_{mk}^{ml'}}$. The colouration is the same as in Fig. 4 in the main text.}
 \figl{add_intbor}
\end{figure}
\subsection{Supplementary figures for `Where to dimerize?'}
\begin{figure}[h!]
 \includegraphics[width=.6\textwidth]{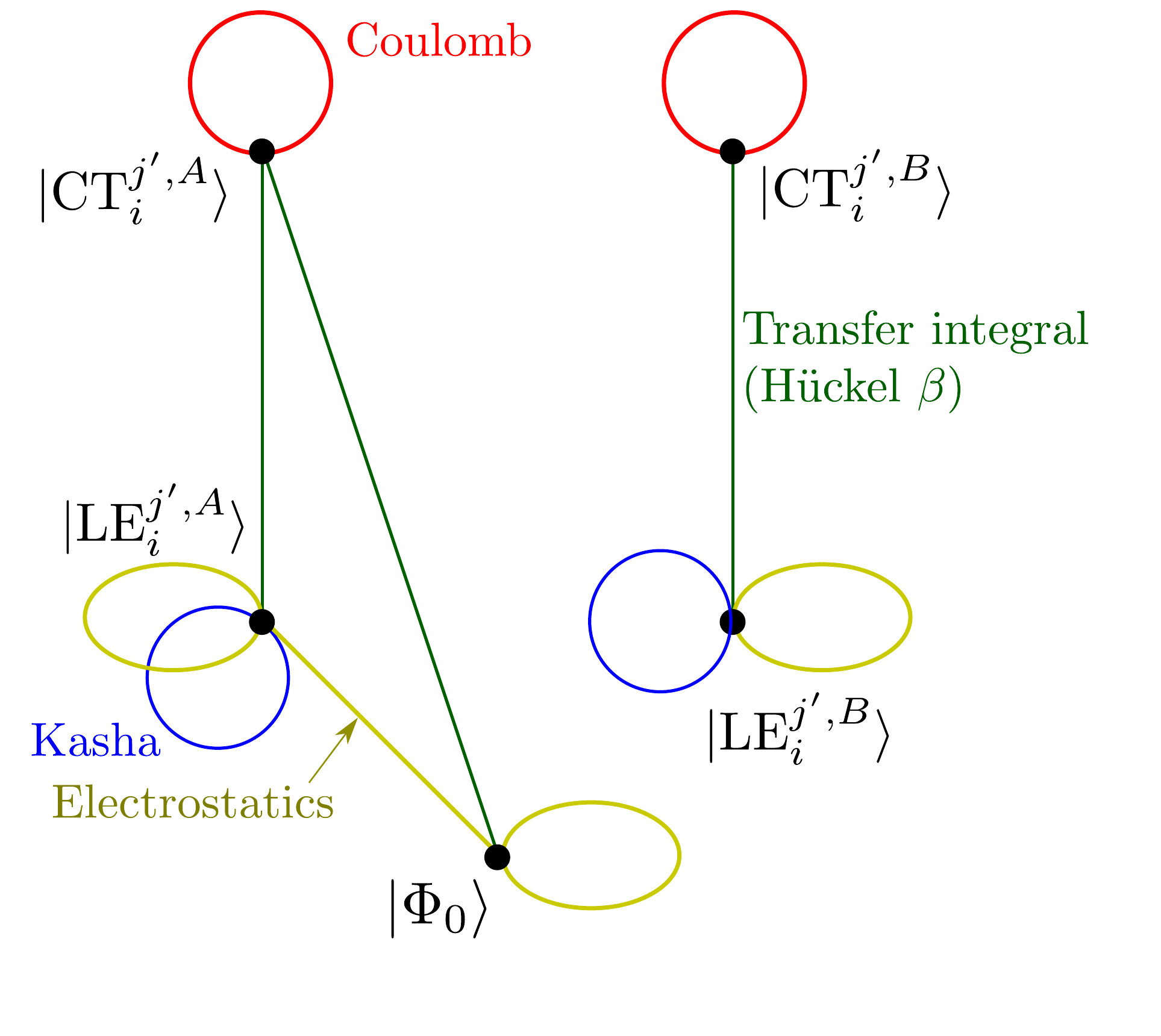}
 \caption{Perturbation diagram for symmetrized `good' eigenstates in Eq. 35 in the main text. Electrostatic interactions in $F^{(1)}$ affect $A$ and $B$ Frenkel excitations similarly, but the Kasha dipole-dipole interaction [Eq. 32 d) in the main text] splits them in energy. Since $\hat V$ is symmetric there is no interaction between different irreps. The zeroth-order interactions are the same as in Fig. 4 in the main text and omitted for clarity.}
 \figl{ptab}
\end{figure}
\begin{figure}[h!]
 \includegraphics[width=.7\textwidth]{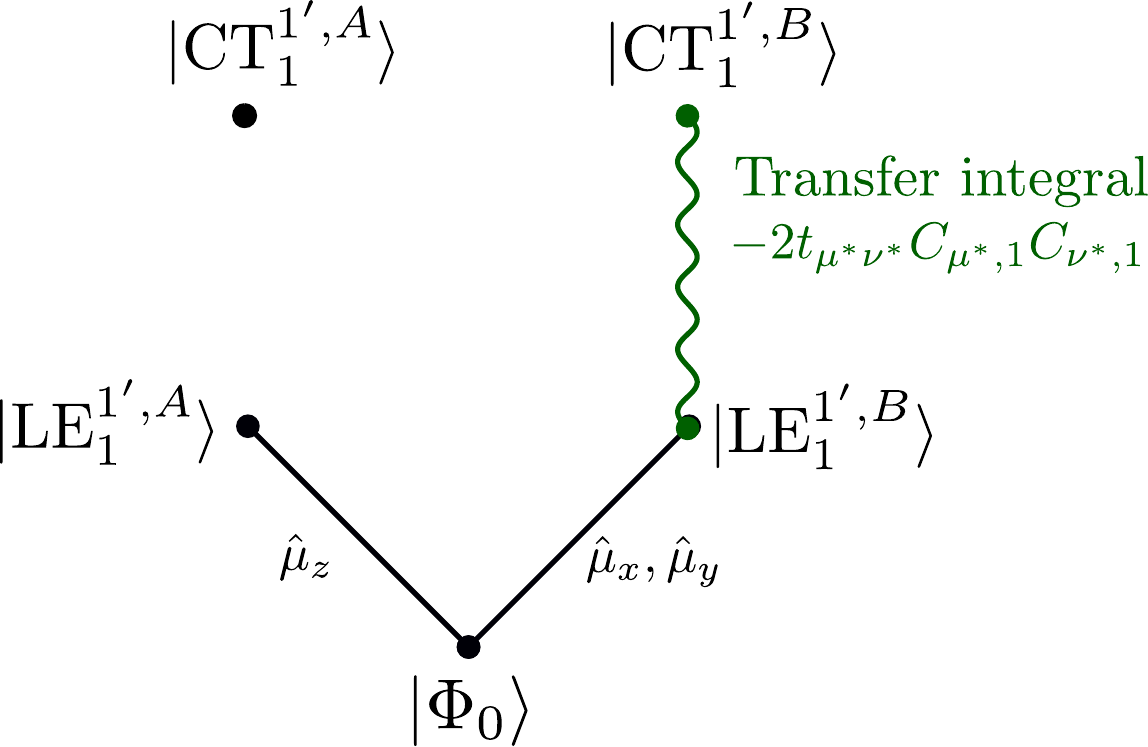}
 \caption{Intensity borrowing diagram illustrating Eq. 37 in the main text, showing how the antisymmetric HOMO to LUMO CT excitation $\ket{\rCT_1^{1',B}}$ can borrow intensity from antisymmetric HOMO to LUMO local excitation $\ket{\rLE_1^{1',B}}$, but there is no borrowing at first order between the corresponding symmetric ($A$) excitations.}
 \figl{hdim}
\end{figure}

\clearpage
\section{Supplementary information for `Trying it out on real molecules'}
\subsection{Aza-substitution}
\begin{figure}[h]
\includegraphics[width=.8\textwidth]{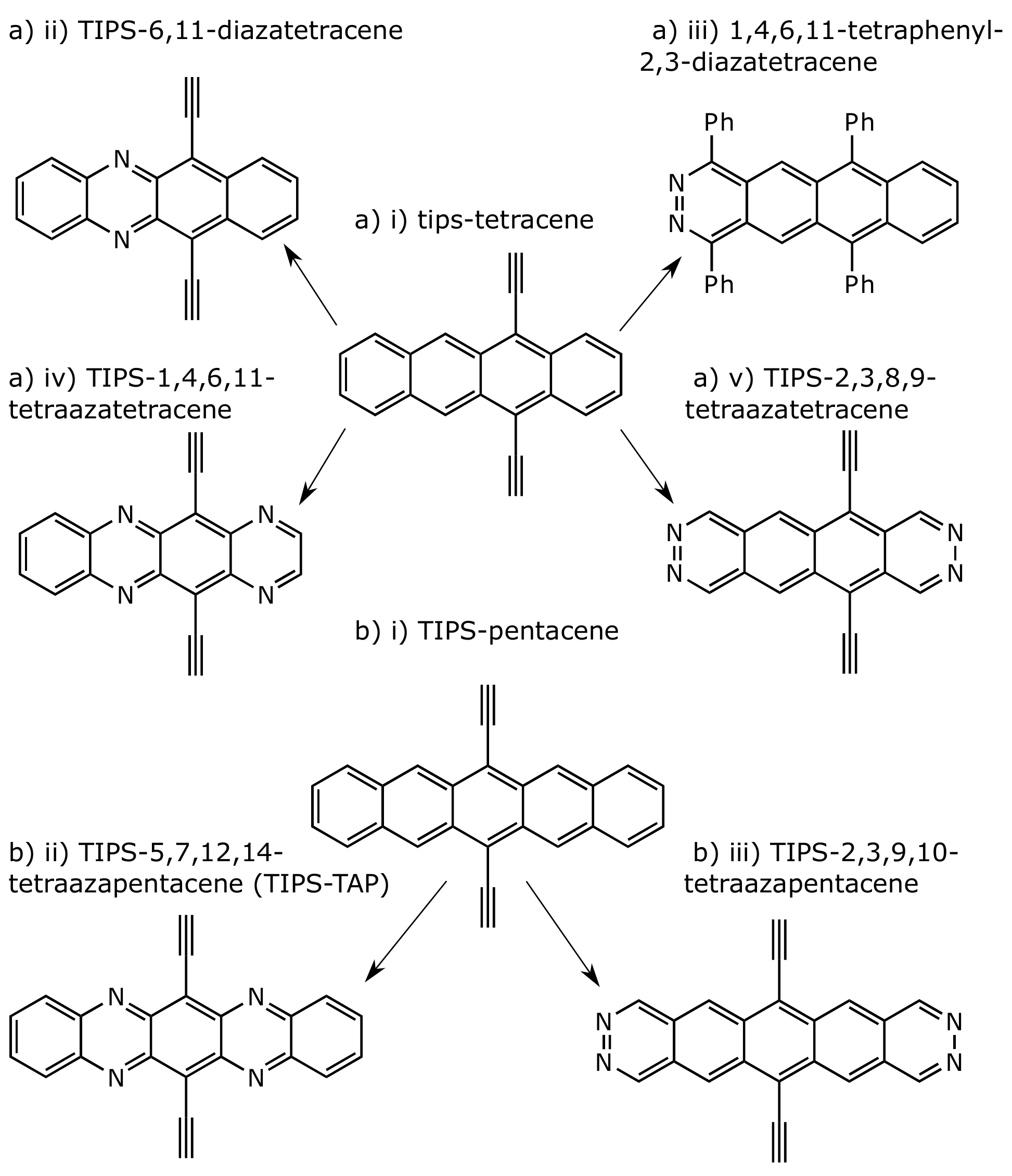}
\caption{Azaacenes discussed in this article. The spectra of tetraazatetracenes a i), ii), iii) and iv) are presented below and tetraazapentacenes b ii) and iii) are presented in the main text.}
\label{fig:aza_sub}
\end{figure}
\begin{figure}[h]
\includegraphics[width=.8\textwidth]{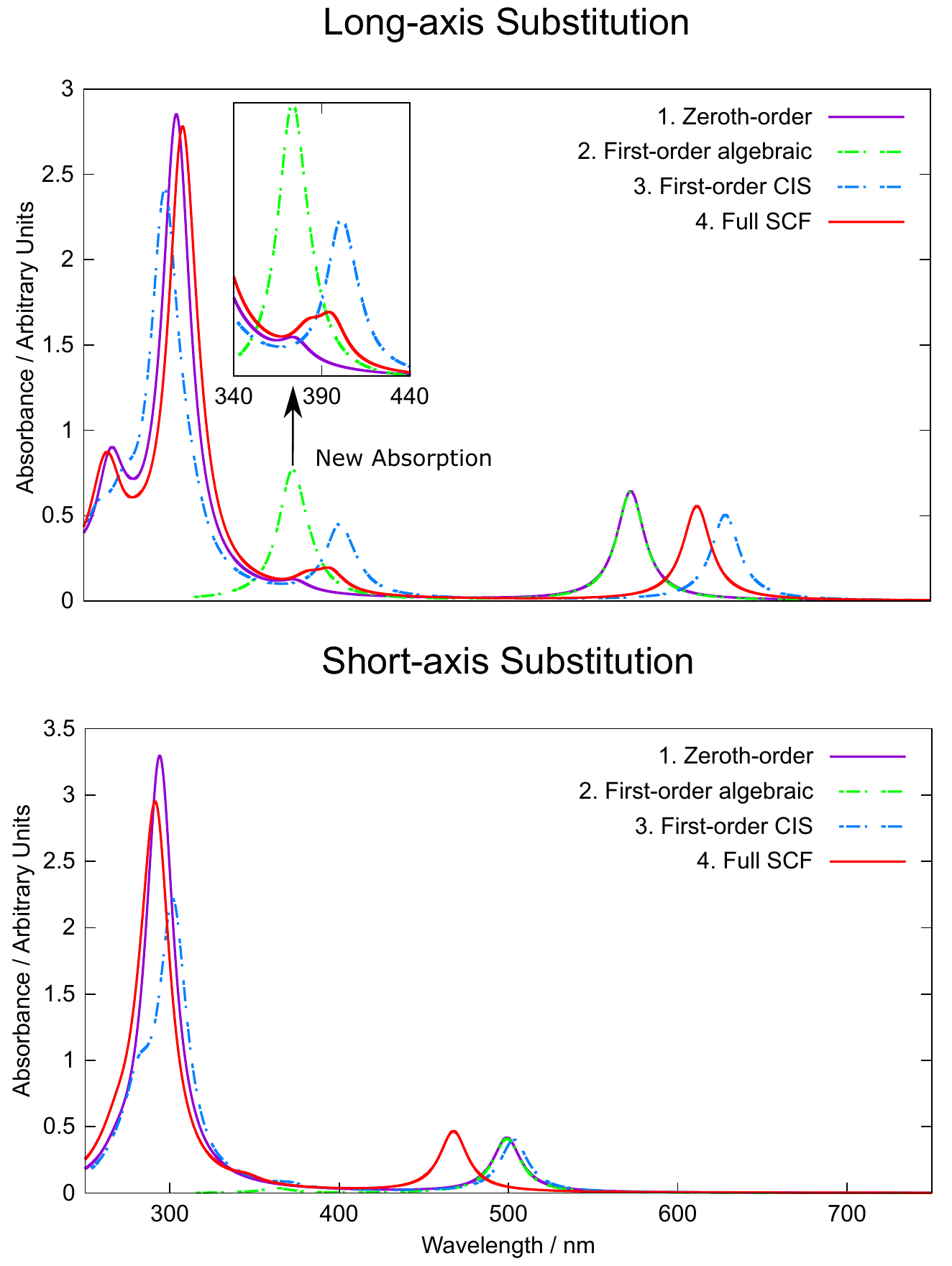}
\caption{Intensity borrowing upon diaza-substitution of TIPS-tetracene at long-axis positions (top) and 1,4,6,11-tetraphenyltetracene at short-axis positions (bottom). The unsubstituted TIPS-tetracene spectrum is shown as `Zeroth Order' along with predicted spectrum of TIPS-5,12-diazatetracene (top), and the unsubstituted 1,4,6,11-tetraphenyltetracene spectrum is shown as `Zeroth Order' along with predicted spectrum of 1,4,6,11-tetraphenyl-2,3-diazatetracene (bottom), at varying levels of theory. A new absorption can be seen around 400 nm after aza- substitution at the 5 and 12 positions as clearly shown by calculations at all levels of theory (top). Only a small new absorption can be seen after aza-substitution at the 2 and 3 positions (bottom).}
\label{fig:dat}
\end{figure}
\begin{figure}[h]
\includegraphics[width=.8\textwidth]{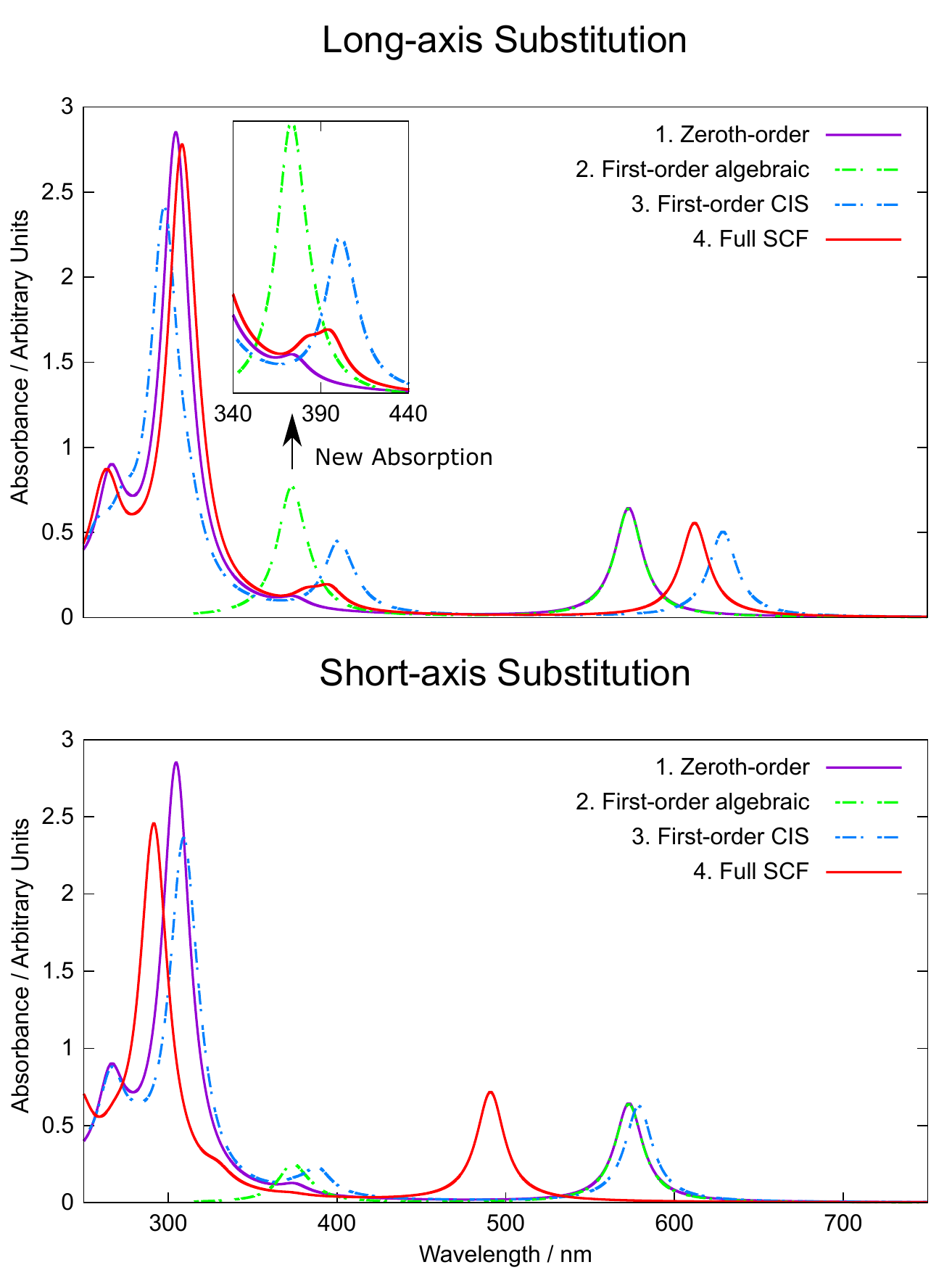}
\caption{Intensity borrowing upon tetra-aza-substitution of TIPS-tetracene at long-axis positions (top) and short-axis positions (bottom). The unsubstituted TIPS-tetracene spectrum is shown as `Zeroth Order' along with predicted spectra of i) TIPS-1,4,6,11-tetraazatetracene (top) and ii) TIPS-2,3,8,9-tetraazatetracene (bottom) at varying levels of theory. A new absorption can be seen around 400 nm after aza- substitution at the 1,4,6 and 11 positions as clearly shown by calculations at all levels of theory (top). Only a small new absorption can be seen after aza-substitution at the 2, 3, 8 and 9 positions (bottom).}
\label{fig:tat}%
\end{figure}
\clearpage
\subsection{Addition and dimerization}
\begin{figure}[h]
\includegraphics[width=.7\textwidth]{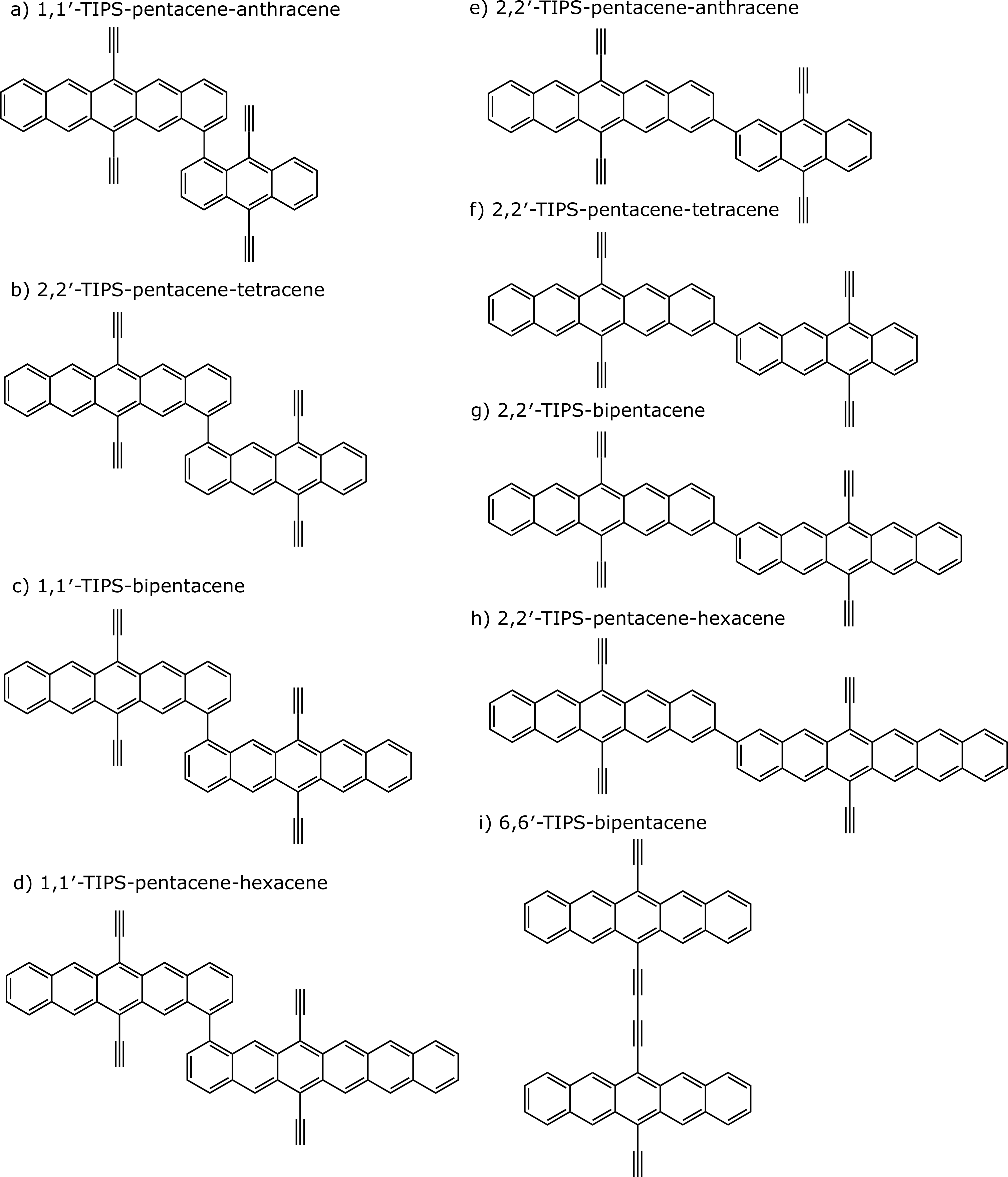}
\caption{Acene dimers discussed in this article. The spectra of TIPS-2,2$^{\prime}$-pentacene-tetracene and TIPS-2,2$^{\prime}$-bipentacene are presented in the main text, for all other dimers the spectra are presented below.}
\label{fig:aza_sub}
\end{figure}
\clearpage

\subsubsection{2,2$^{\prime}$ dimers}
\begin{figure}[!h]
\includegraphics[width=.7\textwidth]{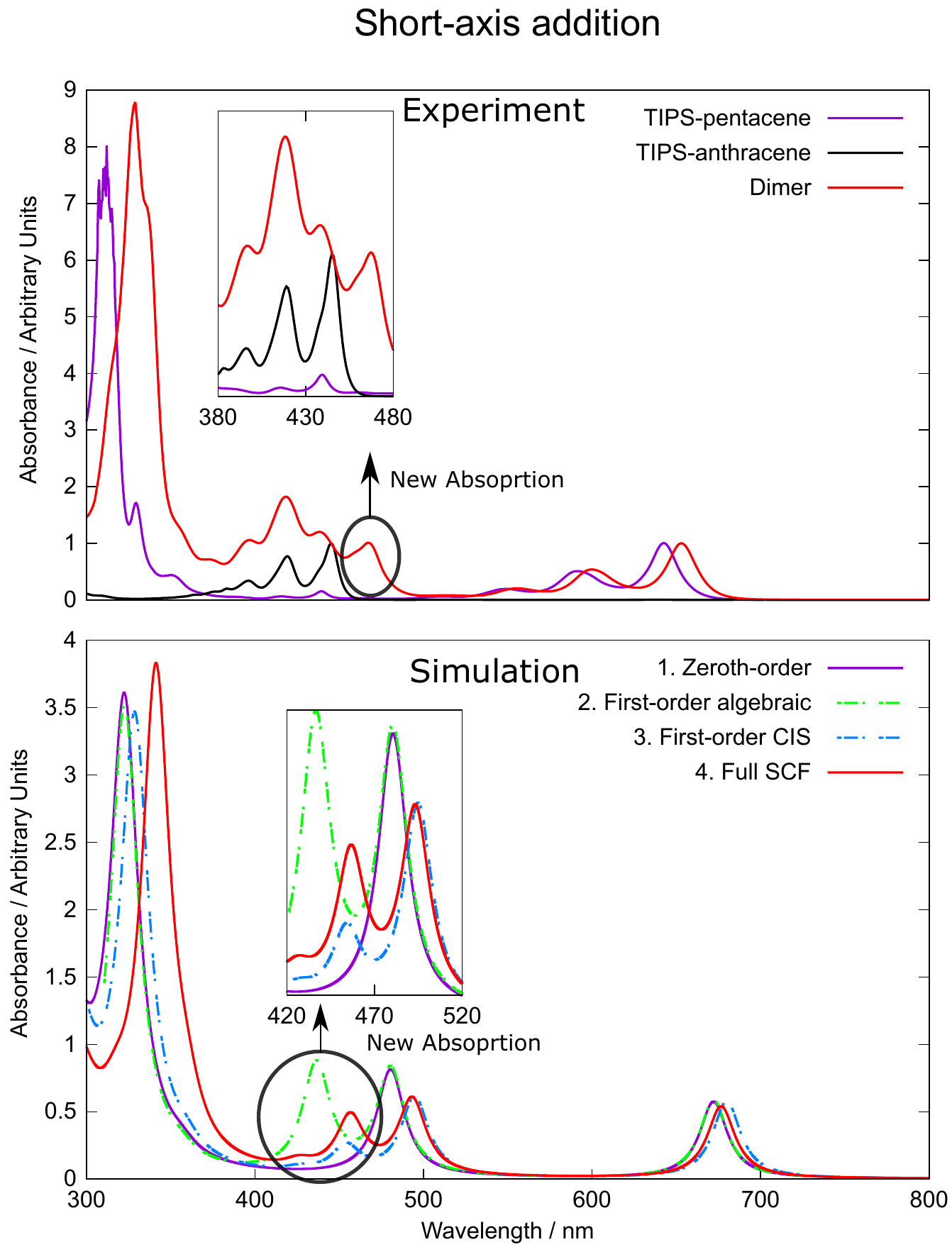}
\caption{Simulated UV/Visible spectrum of TIPS-2,2$^{\prime}$-pentacene-anthracene from PPP-SCF / CIS and intensity-borrowing perturbation theory calculations (bottom) along with the experimental spectra of the dimer the monomers (top). A new absorption is seen around 420 nm in the experimental spectrum, predicted by our calculations to be between 440 and 460nm and seen at all levels of theory above zeroth order.}
\label{fig:22pa}
\end{figure}
\begin{figure}[!h]
\includegraphics[width=.7\textwidth]{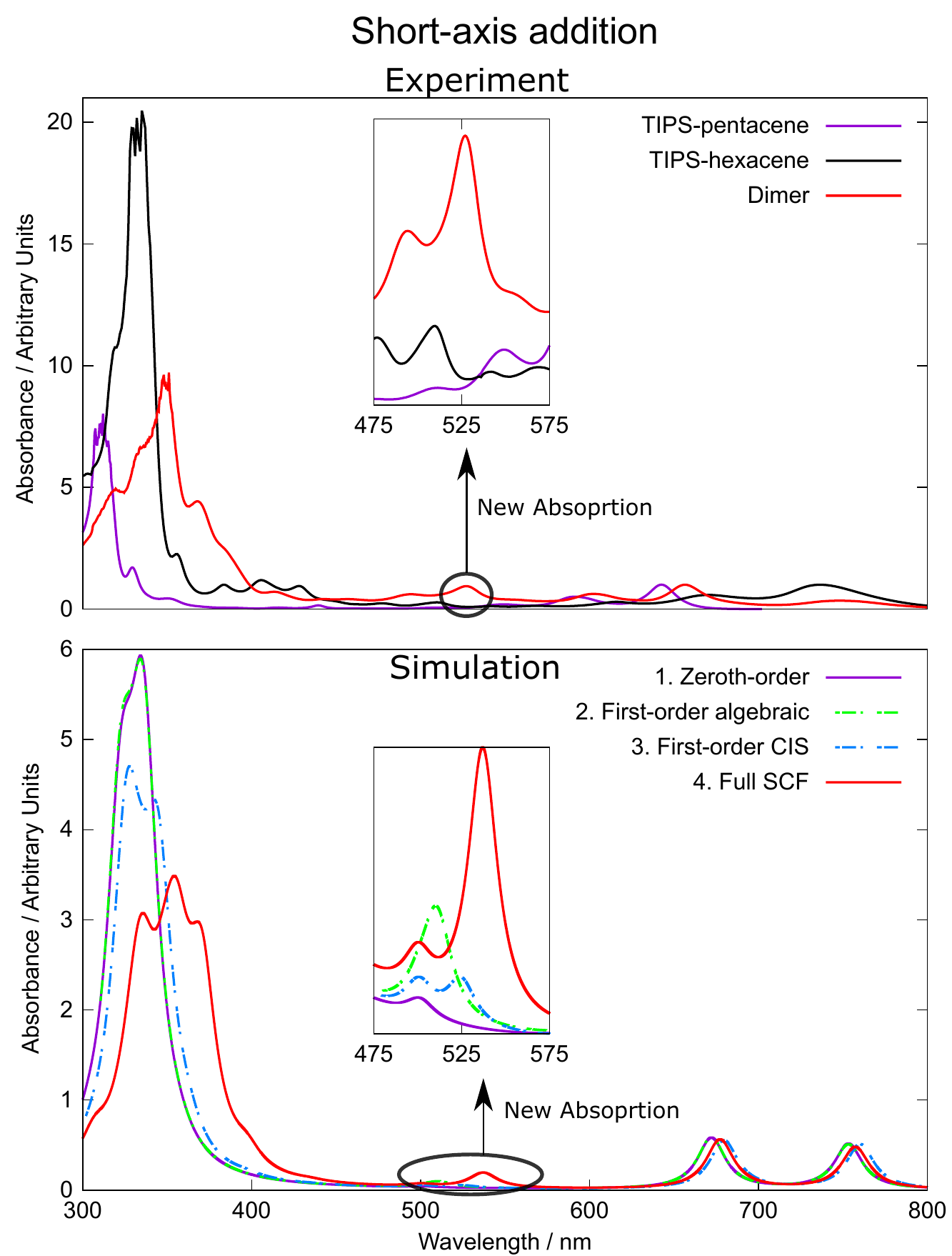}
\caption{Simulated UV/Visible spectrum of TIPS-2,2$^{\prime}$-pentacene-hexacene from PPP-SCF / CIS and intensity-borrowing perturbation theory calculations along with the experimental spectra of the dimer and monomers (top). A new absorption is seen around 530nm in the experimental spectrum, predicted by our calculations to be between 510 and 540nm and seen at all levels of theory, although predicted to be much weaker by the approximate methods.}
\label{fig:22ph}
\end{figure}
\clearpage

\subsubsection{1,1$^{\prime}$ dimers}
\begin{figure}[h]
\includegraphics[width=.8\textwidth]{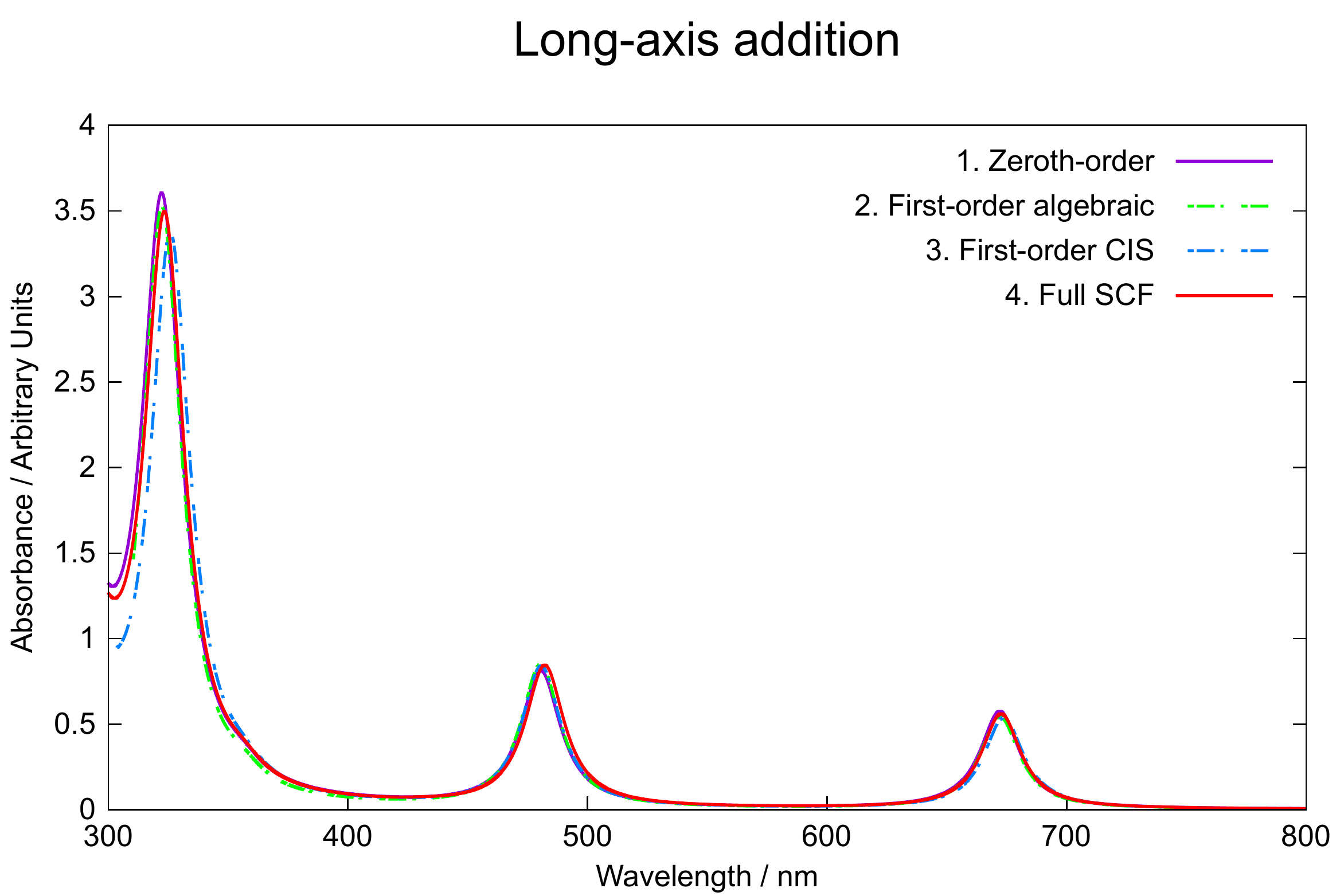}
\caption{Simulated UV/Visible spectrum of TIPS-1,1$^{\prime}$-pentacene-anthracene dimer from PPP-SCF / CIS and intensity-borrowing perturbation theory calculations. No new absorption can be seen after dimerisation at the 1,1$^{\prime}$ positions as shown by calculations at all levels of theory.}
\label{fig:11pa}
\end{figure}
\begin{figure}[h]
\includegraphics[width=.8\textwidth]{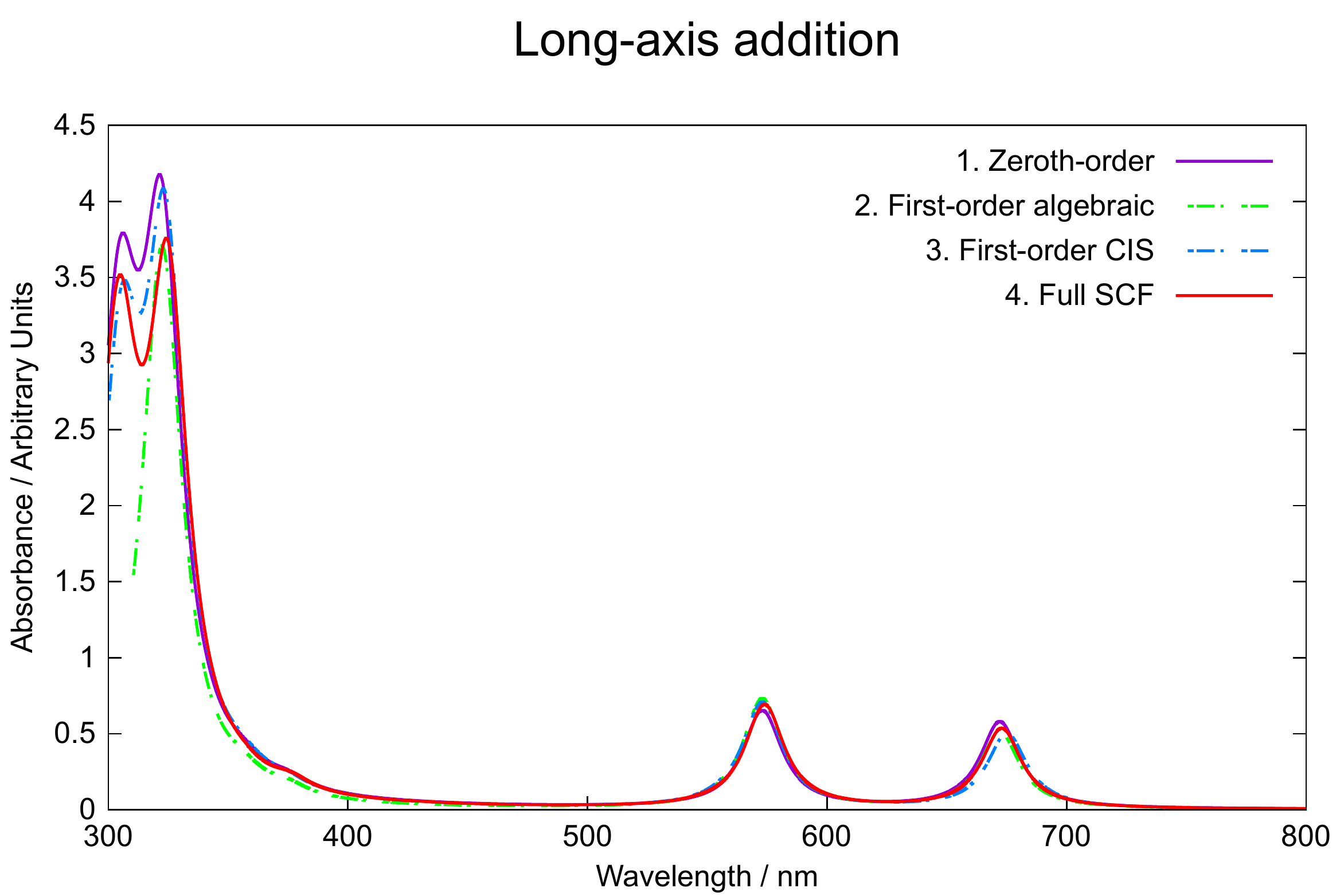}
\caption{Simulated UV/Visible spectrum of TIPS-1,1$^{\prime}$-pentacene-tetracene dimer from PPP-SCF / CIS and intensity-borrowing perturbation theory calculations. No new absorption can be seen after dimerisation at the 1,1$^{\prime}$ positions as shown by calculations at all levels of theory.}
\label{fig:11pt}
\end{figure}
\clearpage
\begin{figure}[h]
\includegraphics[width=.8\textwidth]{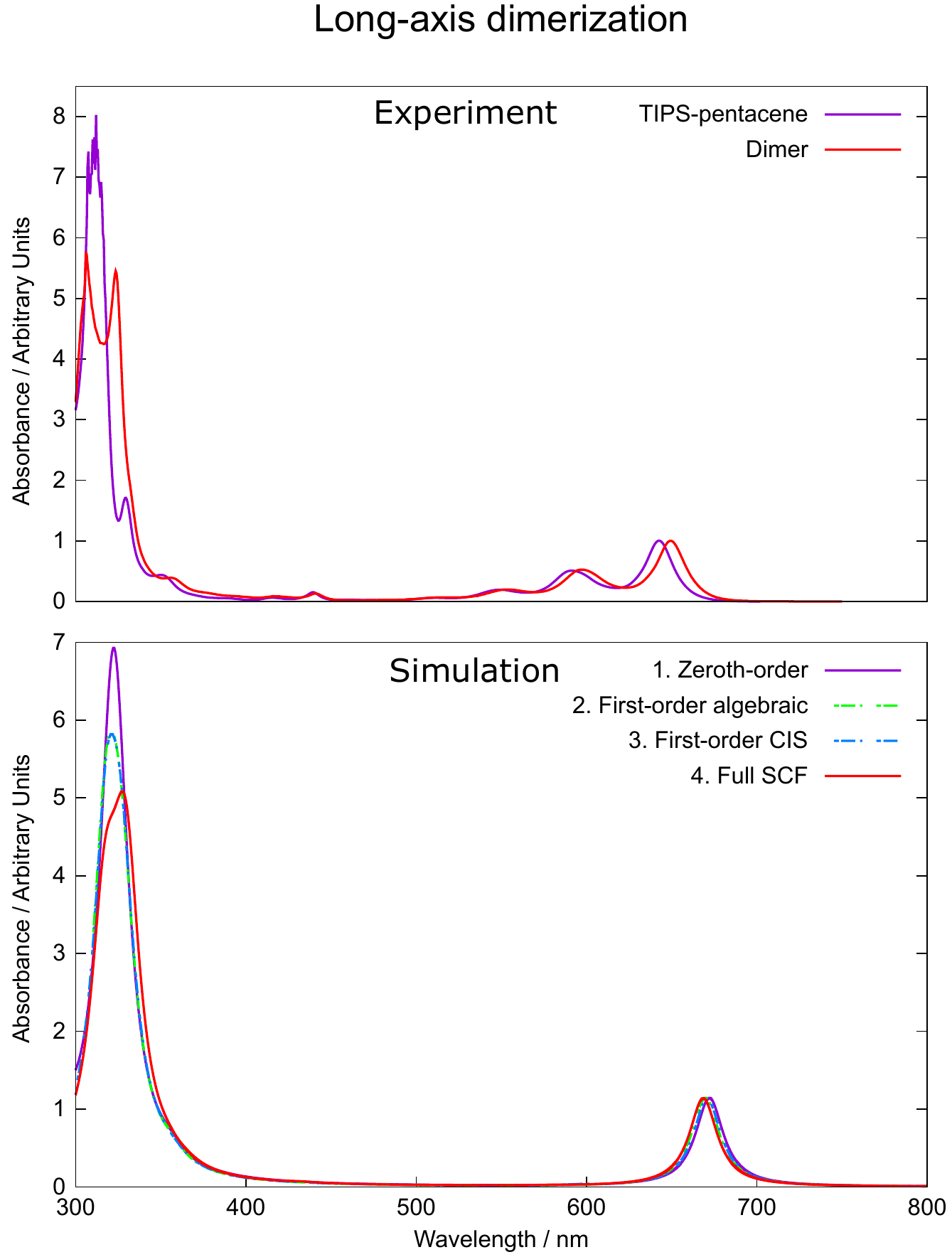}
\caption{Simulated UV/Visible spectrum of TIPS-1,1$^{\prime}$-bipentacene from PPP-SCF / CIS and intensity-borrowing perturbation theory calculations. No new absorption can be seen after dimerisation at the 1,1$^{\prime}$ positions as shown by calculations at all levels of theory.}
\label{fig:11bp}
\end{figure}
\begin{figure}[h]
\includegraphics[width=.8\textwidth]{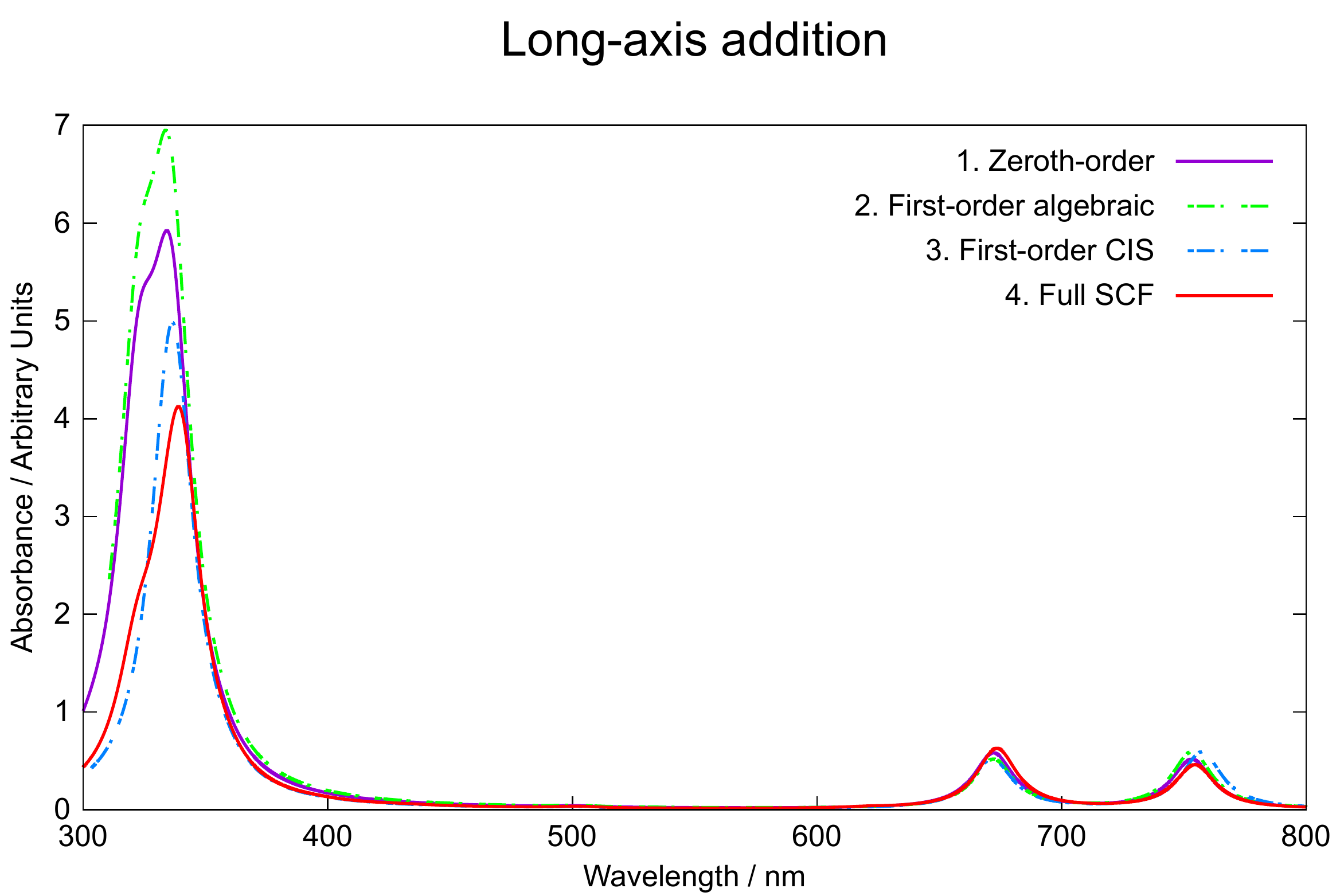}
\caption{Simulated UV/Visible spectrum of TIPS-1,1$^{\prime}$-pentacene-hexacene dimer from PPP-SCF / CIS and intensity-borrowing perturbation theory calculations. No new absorption can be seen after dimerisation at the 1,1$^{\prime}$ positions as shown by calculations at all levels of theory.}
\label{fig:11ph}
\end{figure}
\clearpage

\subsubsection{TIPS-6,6$^{\prime}$-bipentacene} \label{sec:66perturb}
\begin{figure}[h]
\includegraphics[width=.7\textwidth]{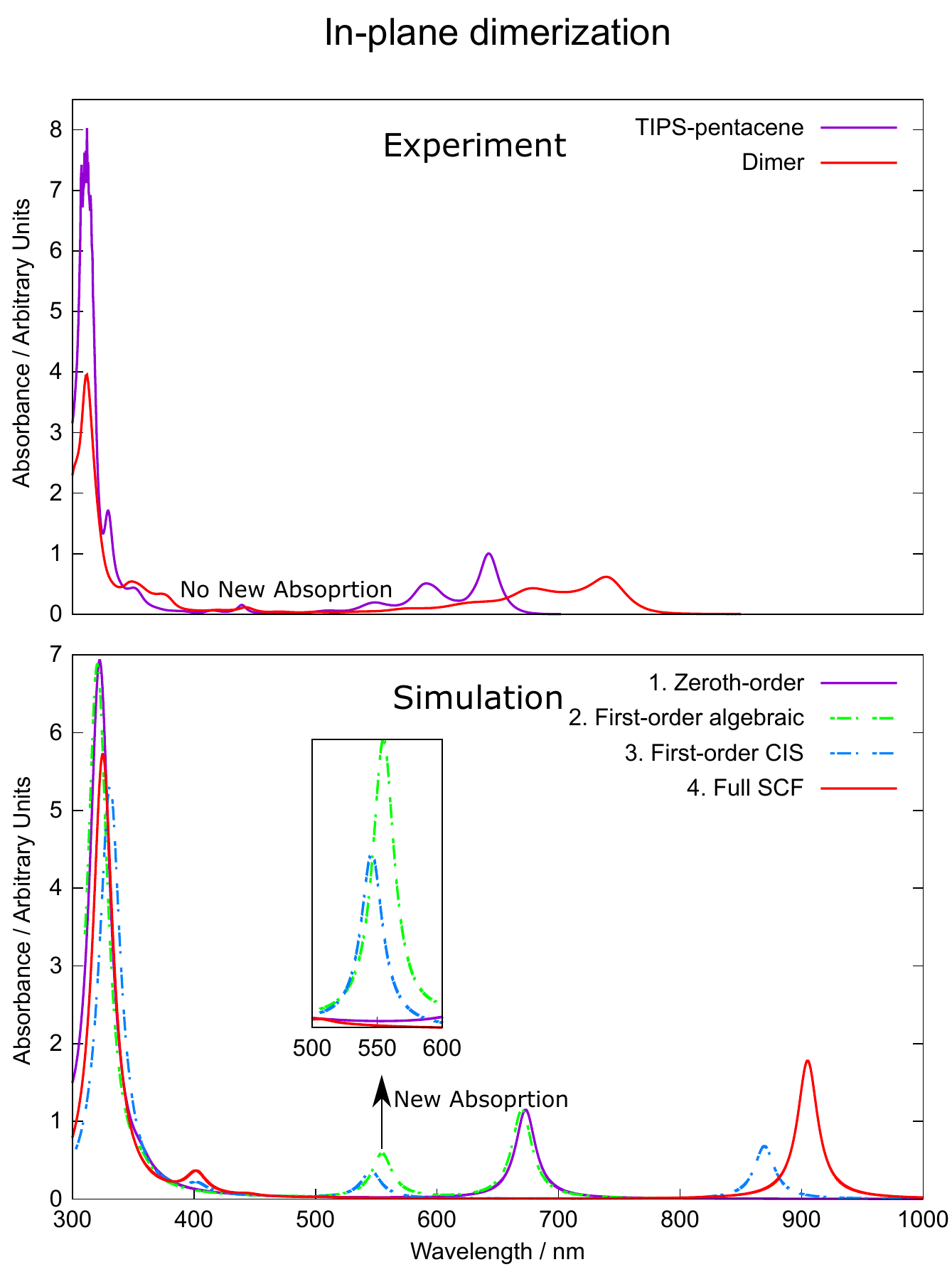}
\caption{Simulated UV/Visible spectrum of TIPS-6,6$^{\prime}$-bipentacene from PPP-SCF / CIS and intensity-borrowing perturbation theory calculations. Intensity borrowing perturbation theory incorrectly predicts an intense new absorption around 550 nm which is most clearly inaccurate for the algebraic perturbation calculation (see section \ref{sec:66perturb}). In the experimental spectrum of the dimer, the absoption is normalised to the TIPS-pentacene spectrum (per pentacene).}
\label{fig:66bp}
\end{figure}

Examination of the results of PPP/CIS and perturbation theory calculations on 6,6$^{\prime}$-bipentacene reveals an erroneus new absorption around 550nm predicted by perturbation theory at all levels of approximation, but not predicted by a full-SCF calculation, not seen in the experimental spectrum and not present in the zeroth-order spectrum. The origin of this new absorption is thought to be a dark HOMO-LUMO charge-transfer (CT) state $\ket{\rCT_{1}^{1^{\prime}, B}}$ borrowing intensity from the bright HOMO-LUMO local excitation (LE) state $\ket{\rLE_{1}^{1^{\prime}, B}}$ at first-order, where
\bse
\begin{align*}
\ket{\rCT_{1}^{1^{\prime}, B}} &= \frac{1}{\sqrt{2}}(\ket{\rCT_{n1}^{m1^{\prime}}} - \ket{\rCT_{m1}^{n1^{\prime}}})\\
\ket{\rLE_{1}^{1^{\prime}, B}} &= \frac{1}{\sqrt{2}}(\ket{\rLE_{n1}^{n1^{\prime}}} - \ket{\rLE_{m1}^{m1^{\prime}}}).
\end{align*}
\ese
The first-order Hamiltonian mixing element between these states is
\begin{align*}
\bra{\rCT_{1}^{1^{\prime}, B}} \hat V \ket{\rLE_{1}^{1^{\prime}, B}} &= \frac{1}{2}(\bra{\rCT_{n1}^{m1^{\prime}}} - \bra{\rCT_{m1}^{n1^{\prime}}}) \hat V (\ket{\rLE_{n1}^{n1^{\prime}}} - \ket{\rLE_{m1}^{m1^{\prime}}}) \\ 
& = \frac{1}{2}( \bra{\rCT_{n1}^{m1^{\prime}}} \hat V \ket{\rLE_{n1}^{n1^{\prime}}} - \bra{\rCT_{n1}^{m1^{\prime}}} \hat V \ket{\rLE_{m1}^{m1^{\prime}}} \\
& - \bra{\rCT_{m1}^{n1^{\prime}}} \hat V \ket{\rLE_{n1}^{n1^{\prime}}} + \bra{\rCT_{m1}^{n1^{\prime}}} \hat V\ket{\rLE_{m1}^{m1^{\prime}}} ) \\
& =\frac{1}{2}(F^{(1)}_{n1^{\prime},m1^{\prime}} + F^{(1)}_{m1,n1} + F^{(1)}_{n1,m1} + F^{(1)}_{m1^{\prime},n1^{\prime}}) \\
& = 2F^{(1)}_{m1,n1} \\
& = 2t_{\mu^*\nu^*}C_{\mu^*,m1}C_{\nu^*,n1},
\end{align*}
where $t_{\mu^*\nu^*}$ is the hopping term between the adjoining atoms of the two monomers, and $C_{\mu^*,m1}$ and $C_{\nu^*,n1}$ are the molecular orbital coefficients of the adjoining atoms in the HOMO (equal in magnitude to that of the LUMO). 
Note that 
\begin{align*}
F^{(1)}_{n1^{\prime},m1^{\prime}} &= F^{(1)}_{m1,n1} = F^{(1)}_{n1,m1} = F^{(1)}_{m1^{\prime},n1^{\prime}}
\end{align*}
due to the Coulson-Rushbrooke theorem\cite{cou40a} as TIPS-6,6$^{\prime}$-bipentacene is alternant. In order for $\bra{\rCT_{1}^{1^{\prime}, B}} \hat V \ket{\rLE_{1}^{1^{\prime}, B}}$ and hence the intensity borrowing from \eqr{alg_perturb} to be non-zero, both $C_{\mu^*,m1}$ and $C_{\nu^*,n1}$ must be non-zero. The HOMO-LUMO CT and LE states are separated by a large energy difference $\sim$1.4 eV at zeroth order, so one would expect mixing to be small, however the value of $t_{\mu^*\nu^*}C_{\mu^*,m1}C_{\nu^*,n1}$ will be relatively large given that $C_{\mu^*,m1}$ and $C_{\nu^*,n1}$ are non-zero in the monomer orbitals and $t_{\mu^*\nu^*}$ is at its maximum value as it depends on the cosine of the dihedral angle between monomers which is zero for TIPS-6,6$^{\prime}$-bipentacene (see section \ref{sec:param}). Therefore, intensity borrowing at first order would be expected. This clearly highlights a limitation in using intensity borrowing perturbation theory to describe the spectra of molecules whose excited states are poorly desribed by their zeroth-order states, such is the case for TIPS-6,6$^{\prime}$-bipentacene, as in such cases perturbation theory cannot be applied successfully. 

\clearpage
\section{Notation}
Fundamental constants
\begin{table}[h]
 \begin{tabular}{c|l|l}
 Symbol & Meaning & Refs \\ \hline
 $m_e$ & Electron mass & \\
 $c$ & Speed of light & \\
 $\ep_0$ & Permittivity of free space & \\
 $N_{\rm A}$ & Avogadro's constant & \\
 $\hbar$ & Reduced Planck's constant & \\
 $e$ & Fundamental electronic charge & \\
\end{tabular}
\caption{Fundamental constants}
\tabl{const}
\end{table}

\begin{table}[h]
 \begin{tabular}{c|l|l}
 Symbol & Meaning & Refs \\ \hline
 $\ket{\Psi_u}$ & Arbitrary eigenstate & \citenum{rob67a} \\
 $\hat \mu$ & Dipole moment operator & \citenum{sza89a} \\ 
 $f_u$ & Oscillator strength of $\ket{\Psi_0} \to \ket{\Psi_u}$ excitation & \citenum{zhe16a} \\
 $f_u^{(x,\pi)}$ & Oscillator strength of the $\pi$ electrons in the $x$ direction for $\ket{\Psi_0} \to \ket{\Psi_u}$ excitation & \\
 $N_e$ & Number of electrons in a molecule & \\
 $N_{\rm v}$ & Number of valence electrons in a molecule & \citenum{atk11a} \\
 $N_{\pi}$ & Number of $\pi$ electrons in a molecule & \\
 $E_u$ & Energy of $u$th state $\bra{\Psi_u} \hat H \ket{\Psi_u}$ for a specified $\hat H$ & \\ 
 $\hat \bmr$ & Vector position operator $(\hat x, \hat y, \hat z)\T$\\
 $x,y,z$ & Cartesian co-ordinates & \\
 $\eta_{\rm abs}$ & Absorption efficiency of a chromophore & Compare \citenum{for05a} \\
 $\nu$ & Frequency (in Hz unless otherwise stated) & \\
\end{tabular}
\caption{Oscillator strength algebra}
\end{table}

\begin{table}[h]
\begin{tabular}{c|l|l}
 Symbol & Meaning & Refs \\ \hline
 $u,v,w$ & Indices of arbitrary eigenstates & c.f.\citenum{kas65a} \\
 $\hat H$ & Total Hamiltonian & \\
 $\hat H_0$ & Zeroth-order Hamiltonian & \\
 $\hat V$ & Perturbation Hamiltonian & \\
 
 $\mu_{uv}$ & Dipole moment matrix element & \eqr{mudef}  \\
 $V_{uv}$ & Perturbation matrix element & \eqr{vdef} \\
 $E_{uv}$ & Energy gap & \eqr{edef}\\
 $c_{uv}$ & Degenerate-state PT expansion coefficient \\
 $\mathcal{D}$ & Degenerate subspace \\
\end{tabular}
\caption{Degenerate-State perturbation theory algebra}
\end{table}

\begin{table}[h]
 \begin{tabular}{c|l|l}
   Symbol & Meaning & Refs \\ \hline
   $\ket{\Phi_i^{j'}}$ & Singlet spin-adapted excitation & \\
   $S_{u,ij'}$ & Expansion coefficient & \citenum{par56a} \\
   $\mathbf{H}^{\rm cis}$ & CIS Hamiltonian matrix & \\
   $h_{ii}$ & One-electron part of electronic Hamiltonian & \citenum{sza89a}\\
   $F_{ij}$ & Fock matrix element $(i|\hat F|j)$ & \eql{fdef}\\
   $i,j,k,l$ & Molecular orbital indices (prime indicates antibonding) & \citenum{par56a} \\
   $(ij|kl)$ & Two-electron integral in chemists' notation & \citenum{sza89a}, \eqr{teintdef}  \\
   $J_{ii}$ & Coulomb integral & \citenum{sza89a} \\
   $K_{ii}$ & Exchange integral & \citenum{sza89a} \\
 \end{tabular}
\caption{Configuration interaction singles}
\tabl{cis}
\end{table} 

\begin{table}[h]
 \begin{tabular}{c|l|l}
 Symbol & Meaning & Refs \\ \hline
 $\ep_{\mu}$ & On-site energy on atom $\mu$ (H\"uckel $\alpha$) & \citenum{ary10a} \\
  $\Delta \ep_{\mu}$ & Change in on-site energy\\
 $t_{\mu\nu}$ & Hopping (transfer) term & \\
 $\sigma$ & spin & E.g.~\citenum{ary10a} \\
 $\mu,\nu,\lambda,\sigma$ & Atomic orbital indices & \citenum{pop53a} \\
 $\gamma_{\mu\nu}$ & PPP Coulomb repulsion term & \citenum{par56a} \\
 $\hat n_{\mu}$ & Number operator for atom $\mu$ & \\
 $\hat U_{\mu\mu}$ & Hubbard repulsion term on atom $\mu$ &  \\
 $\hat a^\dag_{\mu},\hat a_{\mu}$ & Creation and annihilation operators on atom $\mu$. & \citenum{sza89a} \\
 $\kappa$ & Effective dielectric constant & \citenum{ary10a,mot38a} \\
 $m, n$ & Monomers & c.f.~\citenum{ber13a} \\
 $\mathcal{M}$, $\mathcal{N}$ & Sets of atoms on $M$ and $N$ respectively \\
 $\mathcal{G}$ & Set of atoms which are substituted &  \\
 $\chi_{\mu}$ & $2p$ atomic orbital on atom $\mu$ & \citenum{par56a} \\
 $\phi_{ni}$ & Molecular orbital $i$ on monomer $n$ & c.f.\citenum{par56a} \\
 $p,q,r,s$ & Arbitrary monomer indices & \\
 $C_{\mu,ni}$ & Coefficient of orbital $ni$ on atom $\mu$ & c.f.\citenum{par56a,sza89a} \\
 $F_{ni,mj}^{(1)}$ & First-order perturbed Fock matrix element & \\
 $\rho$ & Charge density & \\
 $\theta$ & Dihedral angle between two $2p$ orbitals (or monomers) & \\
 $\ket{\rLE}$ & Frenkel (local) excitation & \\
 $\ket{\rCT}$ & Charge-transfer excitation & \\
 $A,B$ & Irreps of the $C_2$ point group & \citenum{atk11a} \\
\end{tabular}
\caption{PPP theory. See also \tabr{cis}}
\end{table}
\clearpage
Numbered superscripts such as $F^{(0)}$, $\ket{\Psi^{(1)}}$ refer to the zeroth and first-order quantities as in standard perturbation theory\cite{atk11a}. Note that $\Psi^{(1)}$ means the first-order correction to the wavefunction, not the wavefunction corrected to first order (which would also include the zeroth-order component). For clarity the zeroth-order superscript is generally omitted unless the level of perturbation is unclear from the context.

\section{Computational Details}
\subsection{General Methodology}
\label{sec:gen}
The optimised geometries of TIPS-pentacene (CASSCF), TIPS-1,1$^{\prime}$-bipentacene and TIPS-6,6$^{\prime}$-bipentacene (B3LYP/6-31G*) were taken from \citenum{hel19a}, and the optimised geometry of TIPS-2,2$^{\prime}$-bipentacene (B3LYP/6-31G*) was taken from \citenum{fue16a}. All other molecules were optimised in the ground $S_0$ state by DFT calculations with the B3LYP functional and 6-31G* basis set, using the free ORCA 4.2.1 quantum chemistry package.\cite{nee20a} Triisopropylsilyl (TIPS) groups were substituted with hydrogens for the input structures, as they have no significant effect on the electronic structure of the chromophore.\cite{fue16a} Then PPP/CIS calculations using an in-house code were carried out to calculate the excitation energies and oscillator strengths for the chromophores for the lowest 25 singly excited states, applying intensity borrowing perturbation thoery using four levels of approximation for aza-substitution and for addition/dimerisation, as described in the main text. For aza-substitution, the zeroth-order and intensity borrowing perturbation theory calculations were performed at the geometry of the unsubstituted acene. For addition and dimerisation, the monomer calculations were performed at the dimer geometry trunctaed at the joining atoms, as the monomer geometry in the optimised dimer is very similar to the optimised geometry of the isolated monomers and so this should not make any significant difference to the results. Peaks were broadened using a full width at half maximum value of 20 nm in the simulated spectra. In the algebraic perturbation calculations an energy cutoff of 4 eV ($\sim$310 nm) was employed for the excited states to be perturbed, above which the perturbation was not applied, in order to avoid issues with nearly degenerate high-energy states in the UV whose intensity borrowing is not relevant to the optoelectronic properties of these molecules.

\subsection{Parameters used for PPP calculations and simulation of spectra}
\label{sec:param}
The parameters used for PPP calculations are presented in Table \ref{tab:param} (below). The hopping parameter $t_{\mu\nu}$ for C-C bonds was set to 2.2 eV for single bonds, 2.4 eV for aromatic/double/conjugated bonds and 2.8 eV for triple bonds,\cite{ary10a,hel19a} and $t_{\mu\nu}$ was scaled by the cosine of the dihedral angle for intermonomer bonds as per previous work.\cite{ven06a,hel19a} For aromatic C-N and N-N bonds, hopping values of -2.576 eV and -2.75 eV were used respectively. The N-N hopping parameter is about 1.07 times larger than the C-N hopping parameter according to \citenum{van80a}, hence a value of 2.75 eV was used. Cutoffs of 1.6 {\AA}, 1.465 {\AA} and 1.3 {\AA} were used for single bonds, all aromatic/double/conjugated bonds and triple bonds respectively. These specific cutoffs were chosen as they reflect the ranges of single, aromatic/double/conjugated and triple C-C, C-N and N-N bond lengths in the DFT optimised structures of the oligoacenes, azaacenes, $\beta$-carotene and pentacene dimers that were studied. The two-electron integrals $\gamma_{\mu\nu}$ were parameterised as per the MN form: \cite{mat57a} 
\begin{align}
\gamma_{\mu\nu} = \frac{U}{ 1+\frac{r_{\mu\nu}} {r_0} },
\label{eqn:mn}
\end{align}
where the value of the Hubbard repulsion parameter $U$ for C was set to 8 eV\cite{ary10a}, $r_0$ was set to 1.328 {\AA} \cite{mat57a} for C-C interactions and $r_{\mu\nu}$ is the interatomic distance between centres $\mu$ and $\nu$. For azaacenes, the Hubbard parameter for N was set to 12.34 eV{\cite{hin71a}} and the average of the C and N hubbard parameters was used for C-N for two-centre interactions. The value of $r_0$ was set to 1.212 {\AA} for C-N interactions and 1.115 {\AA} for N-N interactions. \cite{mat57a} The value of $\epsilon$ (H{\"u}ckel $\alpha$ parameter) was set to 0 for C. A relative value of -2.96 eV was used for N in full PPP calculations where the two-electron integrals and hopping was also parameterised for N, in line with \citenum{mat57a} (see below). The value for N was calculated from the difference in the ionisation potentials of C and N according to the relation
\begin{align*}
\epsilon_N - \epsilon_C = I_C - I_N,
\end{align*}
where $I_C$ and $I_N$ are the ionisation potentials of C and N atoms, with values of 11.26 eV and 14.53 eV respectively. \cite{mat57a,hin71a} For perturbation theory calculations, where only the $\epsilon$ parameter is changed, a value of $\epsilon$ for N of -1.24 eV was employed, equal to $0.51t_{\mu\nu}$, according to \citenum{hin71a} (where $t_{\mu\nu}$ is the double bond hopping parameter for C discussed above). This parameterisation of PPP thoery was chosen as it has been shown to accurately predict the spectra of oligoacenes in previous studies, with particular accuracy for lower energy transitions, when compared to calculations at higher levels of theory such as CASSCF and experimental spectra. \cite{hel19a} 
\begin{table}[h]
\centering
\label{tab:param}
\caption{PPP parameter set used for simulation of spectra of $\pi$-systems.} \label{tab:ppp}
\vspace*{2mm}
 \begin{tabular}{c|c|c} 
  Parameter & Value & Unit \\ \hline
$\epsilon_C$ & 0 & eV \\
$\epsilon_N$(full calc.) & -2.96 & eV \\
$\epsilon_N$(perturbation thry.) & -1.24& eV \\
$t_{CC, double}$ & -2.4 & eV \\
$t_{CC, single}$ & -2.2 & eV \\
$t_{CC, triple}$ & -2.8 & eV \\
$t_{CN}$ & -2.576 & eV \\
$t_{NN}$ & -2.75 & eV \\
Single bond cutoff & 1.6 & {\AA} \\
Double bond cutoff & 1.465 & {\AA} \\
Triple bond cutoff & 1.3 &  {\AA} \\
$U_{CC}$ & 8 & eV\\
$U_{CN}$ & 10.17 & eV \\
$U_{NN}$ & 12.34 & eV \\
$r_{0, CC}$ & 1.328 & {\AA}\\
$r_{0, CN}$ & 1.212 & {\AA}\\
$r_{0, NN}$ & 1.115 & {\AA}\\
\end{tabular}
\end{table}

\clearpage
\subsection{Structures of simulated chromophores.}
Ethene\\
Cartesian coordinates / {\AA} (atom x y z)\\
 H      0.000000    0.923930   -1.238438\\
 C      0.000000    0.000000   -0.665298\\
 H     -0.000000   -0.923930   -1.238438\\
 C      0.000000   -0.000000    0.665298\\
 H     -0.000000    0.923930    1.238438\\
 H      0.000000   -0.923930    1.238438\\
\
Naphthalene\\
Cartesian coordinates / {\AA} (atom x y z)\\
C     -0.000000    0.717115    0.000000\\
C      1.245250    1.402900    0.000000\\
C      2.434294    0.708654   -0.000000\\
C      2.434294   -0.708654    0.000000\\
C      1.245250   -1.402900    0.000000\\
C      0.000000   -0.717115    0.000000\\
C     -1.245250   -1.402900   -0.000000\\
C     -2.434294   -0.708654   -0.000000\\
C     -2.434294    0.708654    0.000000\\
C     -1.245250    1.402900    0.000000\\
H     -3.378786   -1.245893   -0.000000\\
H     -1.242967   -2.490800   -0.000000\\
H     -3.378786    1.245893    0.000000\\
H     -1.242967    2.490800    0.000000\\
H      1.242967    2.490800   -0.000000\\
H      3.378786    1.245893   -0.000000\\
H      3.378786   -1.245893   -0.000000 \\
H      1.242967   -2.490800    0.000000\\
\
Anthracene\\
Cartesian coordinates / {\AA} (atom x y z)\\
  C     -0.031143    1.407508    0.000000\\
  C     -0.031143   -1.407508   -0.000000\\
  C     -1.212556    0.713363    0.000000\\
  C      1.224922    0.722779   -0.000000\\
  C     -1.212556   -0.713363   -0.000000\\
  C      1.224922   -0.722779   -0.000000\\
  C      2.449097    1.404027   -0.000000\\
  C      2.449097   -1.404027   -0.000000\\
  C      3.673477    0.722752   -0.000000\\
  C      3.673477   -0.722752   -0.000000\\
  C      6.110627    0.713249   -0.000000\\
  C      6.110627   -0.713249    0.000000\\
  C      4.929381    1.407490   -0.000000\\
  C      4.929381   -1.407490    0.000000\\
  H      2.448790    2.492659   -0.000000\\
  H      2.448790   -2.492659   -0.000000\\
  H      4.927003    2.495204   -0.000000\\
  H      4.927003   -2.495204    0.000000\\
  H      7.057264   -1.246551    0.000000\\
  H      7.057264    1.246551   -0.000000\\
  H     -0.029075   -2.495237   -0.000000\\
  H     -0.029075    2.495237    0.000000\\
  H     -2.158898   -1.247193    0.000000\\
  H     -2.158898    1.247193    0.000000\\
\
Tetracene\\
Cartesian coordinates / {\AA} (atom x y z)\\
  C     -0.012410    1.406852    0.000000\\
  C     -0.012410   -1.406852    0.000000\\
  C     -1.227971    0.726071   -0.000000\\
  C      1.223467    0.726342    0.000000\\
  C     -1.227971   -0.726071    0.000000\\
  C      1.223467   -0.726342    0.000000\\
  C     -2.489006    1.409709   -0.000000\\
  C      2.459177    1.406830    0.000000\\
  C     -2.489006   -1.409709    0.000000\\
  C      2.459177   -1.406830    0.000000\\
  C     -3.666978    0.715520   -0.000000\\
  C      3.674781    0.726067    0.000000\\
  C     -3.666978   -0.715520   -0.000000\\
  C      3.674781   -0.726067    0.000000\\
  C      6.113611    0.715498   -0.000000\\
  C      6.113611   -0.715498   -0.000000\\
  C      4.935765    1.409681    0.000000\\
  C      4.935765   -1.409681   -0.000000\\
  H     -2.487428    2.497371   -0.000000\\
  H      2.458861    2.495332    0.000000\\
  H     -2.487428   -2.497371    0.000000\\
  H      2.458861   -2.495332    0.000000\\
  H      4.933830    2.497364    0.000000\\
  H      4.933830   -2.497364   -0.000000\\
  H      7.061121   -1.247199   -0.000000\\
  H      7.061121    1.247199   -0.000000\\
  H     -0.012638   -2.495387    0.000000\\
  H     -0.012638    2.495387    0.000000\\
  H     -4.614245   -1.247686   -0.000000\\
  H     -4.614245    1.247686   -0.000000\\
\
Pentacene\\
Cartesian coordinates / {\AA} (atom x y z)\\
  C     -0.000000    1.409011   -0.000000\\
  C      0.000000   -1.409011   -0.000000\\
  C     -1.226910    0.728715   -0.000000\\
  C      1.226910    0.728715   -0.000000\\
  C     -1.226910   -0.728715   -0.000000\\
  C      1.226910   -0.728715   -0.000000\\
  C     -2.468420    1.408540   -0.000000\\
  C      2.468420    1.408540    0.000000\\
  C     -2.468420   -1.408540   -0.000000\\
  C      2.468420   -1.408540   -0.000000\\
  C     -3.679785    0.727851   -0.000000\\
  C      3.679785    0.727851    0.000000\\
  C     -3.679785   -0.727851   -0.000000\\
  C      3.679785   -0.727851    0.000000\\
  C     -6.119464    0.716687    0.000000\\
  C      6.119464    0.716687    0.000000\\
  C     -6.119464   -0.716687    0.000000\\
  C      6.119464   -0.716687    0.000000\\
  C     -4.943293    1.410943   -0.000000\\
  C      4.943293    1.410943    0.000000\\
  C     -4.943293   -1.410943    0.000000\\
  C      4.943293   -1.410943    0.000000\\
  H     -2.468389    2.497008   -0.000000\\
  H      2.468389    2.497008    0.000000\\
  H     -2.468389   -2.497008   -0.000000\\
  H      2.468389   -2.497008   -0.000000\\
  H     -4.941167    2.498617   -0.000000\\
  H      4.941167    2.498617    0.000000\\
  H     -4.941167   -2.498617    0.000000\\
  H      4.941167   -2.498617    0.000000\\
  H      7.067743   -1.247005    0.000000\\
  H     -7.067743   -1.247005    0.000000\\
  H      7.067743    1.247005    0.000000\\
  H     -7.067743    1.247005    0.000000\\
  H      0.000000   -2.497402   -0.000000\\
  H      0.000000    2.497402   -0.000000\\
$\beta$-carotene\\
\begin{figure}[h!]
\includegraphics[width=0.9\textwidth,trim=0cm 10cm 0cm 10cm,clip]{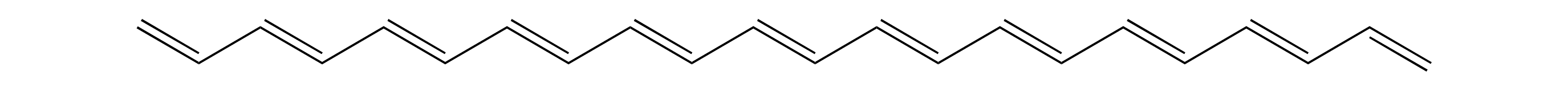}
\caption{Truncated structure of $\beta$-carotene used for PPP calculations in this work.}
\label{fig:struc_bcarot}
\end{figure}
Cartesian coordinates / {\AA} (atom x y z)\\
C       -0.246363000      0.048965000     -0.000004000\\
C        0.947670000      0.668421000     -0.000001000\\
C        2.224450000     -0.009911000     -0.000002000\\
C        3.426608000      0.624102000      0.000000000\\
C        4.698464000     -0.040498000      0.000000000\\
C        5.902039000      0.600655000      0.000002000\\
C        7.172402000     -0.058346000      0.000001000\\
C        8.376533000      0.586180000      0.000002000\\
C        9.646380000     -0.069921000      0.000002000\\
C       10.850659000      0.576366000      0.000002000\\
C       12.120349000     -0.078427000      0.000002000\\
C       13.324651000      0.568418000      0.000002000\\
C       14.594342000     -0.086374000      0.000002000\\
C       15.798620000      0.559914000      0.000001000\\
C       17.068467000     -0.096186000      0.000001000\\
C       18.272598000      0.548342000      0.000000000\\
C       19.542961000     -0.110657000      0.000001000\\
C       20.746535000      0.530499000     -0.000001000\\
C       22.018393000     -0.134099000      0.000000000\\
C       23.220550000      0.499917000     -0.000002000\\
C       24.497331000     -0.178412000     -0.000001000\\
C       25.691363000      0.441047000     -0.000003000\\
H       -1.177652000      0.605675000     -0.000004000\\
H       -0.324079000     -1.035883000     -0.000007000\\
H        0.978953000      1.758926000      0.000001000\\
H        2.203301000     -1.100889000     -0.000004000\\
H        3.440547000      1.715614000      0.000003000\\
H        4.687762000     -1.131704000     -0.000002000\\
H        5.910308000      1.692100000      0.000004000\\
H        7.165736000     -1.149662000      0.000000000\\
H        8.381907000      1.677604000      0.000004000\\
H        9.641838000     -1.161283000      0.000001000\\
H       10.854619000      1.667773000      0.000003000\\
H       12.116651000     -1.169815000      0.000002000\\
H       13.328349000      1.659807000      0.000002000\\
H       14.590381000     -1.177781000      0.000003000\\
H       15.803161000      1.651276000      0.000000000\\
H       17.063094000     -1.187609000      0.000003000\\
H       18.279262000      1.639658000     -0.000002000\\
H       19.534695000     -1.202102000      0.000003000\\
H       20.757235000      1.621704000     -0.000003000\\
H       22.004456000     -1.225611000      0.000002000\\
H       23.241696000      1.590894000     -0.000004000\\
H       24.466050000     -1.268917000      0.000002000\\
H       25.769076000      1.525894000     -0.000005000\\
H       26.622653000     -0.115662000     -0.000002000\\
TIPS-2,2$^{\prime}$-pentacene-anthracene\\ 
\begin{figure}[h!]
\includegraphics[width=0.9\textwidth,trim=4cm 6cm 4cm 6cm,clip]{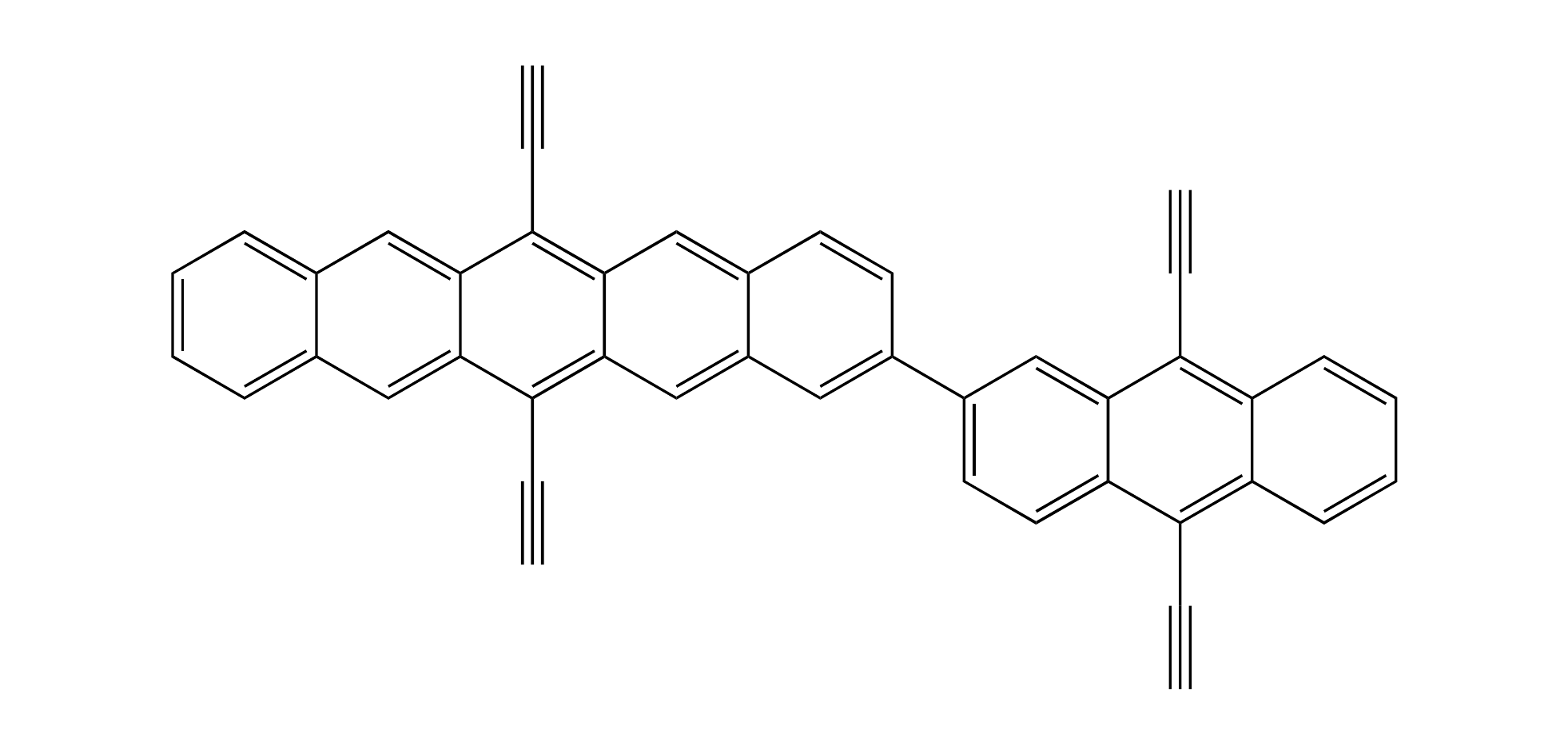}
\caption{Structure of TIPS-2,2$^{\prime}$-pentacene-anthracene.}
\label{fig:struc_22pa}
\end{figure}
Cartesian coordinates / {\AA} (atom x y z)\\
C       -6.596289000      1.179976000     -0.414733000\\
C       -7.071130000     -1.521979000      0.396220000\\
C       -6.359300000      2.524527000     -0.818581000\\
C       -7.307769000     -2.867332000      0.798504000\\
C       -6.156443000      3.669722000     -1.162549000\\
C       -7.510087000     -4.013005000      1.141016000\\
H       -5.977226000      4.676121000     -1.466363000\\
H       -7.688891000     -5.019950000      1.443239000\\
C       -7.925819000      0.671306000     -0.403514000\\
C       -5.501168000      0.358335000     -0.025373000\\
C       -8.167584000     -0.698279000      0.013064000\\
C       -5.742503000     -1.014732000      0.381153000\\
C       -9.027364000      1.467139000     -0.787503000\\
C       -4.175093000      0.843894000     -0.020712000\\
C       -9.495354000     -1.178170000      0.025542000\\
C       -4.638304000     -1.815001000      0.752862000\\
H       -8.842871000      2.489711000     -1.101968000\\
H       -3.998648000      1.871899000     -0.321374000\\
H       -9.673641000     -2.201343000      0.341279000\\
H       -4.820321000     -2.842650000      1.051910000\\
C      -10.331030000      0.982401000     -0.774267000\\
C       -3.097096000      0.048584000      0.356853000\\
C      -10.573172000     -0.383547000     -0.351304000\\
C       -3.338202000     -1.326219000      0.746702000\\
C      -11.451945000      1.789035000     -1.164533000\\
C       -1.754242000      0.543885000      0.370770000\\
C      -11.923636000     -0.868467000     -0.337778000\\
C       -2.209306000     -2.130346000      1.117689000\\
H      -11.266715000      2.812041000     -1.483101000\\
H       -1.588172000      1.571414000      0.057184000\\
H      -12.102746000     -1.892075000     -0.017717000\\
H       -2.384061000     -3.159595000      1.421816000\\
C      -12.722376000      1.284775000     -1.137280000\\
C       -0.688553000     -0.249277000      0.731216000\\
C      -12.961634000     -0.063573000     -0.716754000\\
C       -0.944556000     -1.619463000      1.106345000\\
H      -13.563181000      1.905272000     -1.435605000\\
H      -13.980370000     -0.442441000     -0.702214000\\
H       -0.108342000     -2.238695000      1.417035000\\
C        6.642058000     -1.155010000     -0.369556000\\
C        7.111931000      1.491217000      0.436307000\\
C        7.918946000     -0.655049000     -0.346312000\\
C        5.526611000     -0.349093000      0.013197000\\
C        8.157174000      0.685356000      0.063196000\\
C        5.768119000      1.009128000      0.424266000\\
C        4.199655000     -0.856802000     -0.001773000\\
C        4.675208000      1.834944000      0.803624000\\
C        3.108049000     -0.031964000      0.383185000\\
C        3.350877000      1.328667000      0.781473000\\
C        1.768526000     -0.518377000      0.382785000\\
C        2.228529000      2.131468000      1.150821000\\
H        1.603339000     -1.540634000      0.060469000\\
H        2.408777000      3.154311000      1.464319000\\
C        0.697091000      0.275077000      0.741304000\\
C        0.956190000      1.629782000      1.129348000\\
H        0.126337000      2.255729000      1.443872000\\
H        7.288824000      2.514532000      0.749670000\\
H        9.172975000      1.071757000      0.079427000\\
H        8.753656000     -1.284851000     -0.642015000\\
H        6.455353000     -2.177176000     -0.681103000\\
C        4.911821000      3.181901000      1.209286000\\
C        5.113581000      4.326270000      1.554283000\\
C        3.961250000     -2.203348000     -0.405728000\\
C        3.758010000     -3.347997000     -0.749491000\\
H        5.292060000      5.332497000      1.859249000\\
H        3.578225000     -4.354035000     -1.054358000\\
\
TIPS-2,2$^{\prime}$-pentacene-tetracene\\ 
\begin{figure}[h!]
\includegraphics[width=0.9\textwidth,trim=4cm 6cm 4cm 6cm,clip]{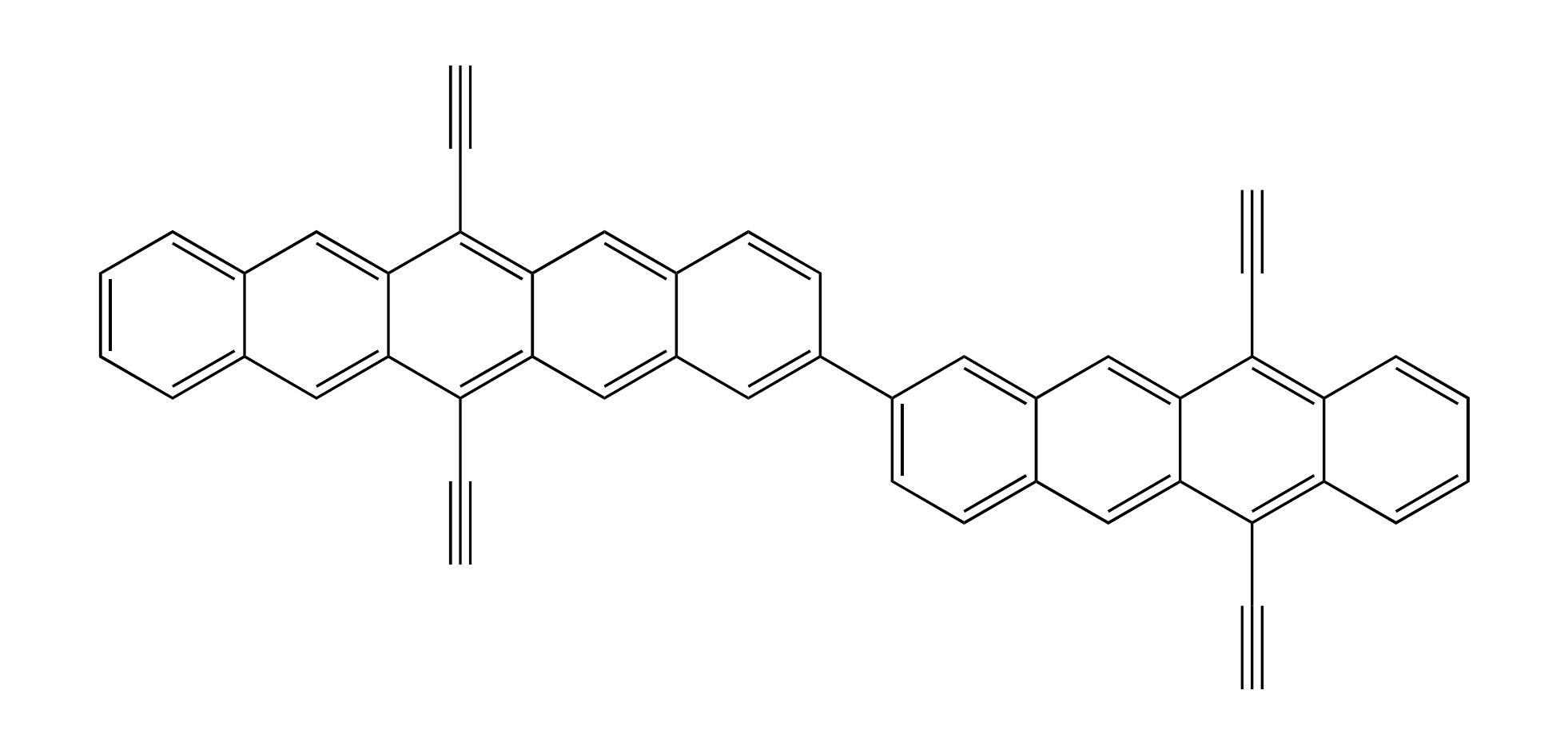}
\caption{Structure of TIPS-2,2$^{\prime}$-pentacene-tetracene.}
\label{fig:struc_22pt}
\end{figure}
\
Cartesian coordinates / {\AA} (atom x y z)\\
C       -6.595098000      1.175327000     -0.420144000\\
C       -7.077294000     -1.518107000      0.414475000\\
C       -6.354492000      2.515611000     -0.835797000\\
C       -7.317645000     -2.858847000      0.829715000\\
C       -6.148667000      3.657185000     -1.189898000\\
C       -7.523073000     -4.000581000      1.183341000\\
H       -5.966982000      4.660445000     -1.502461000\\
H       -7.704522000     -5.004100000      1.495208000\\
C       -7.925839000      0.669864000     -0.406020000\\
C       -5.502302000      0.354610000     -0.022245000\\
C       -8.171395000     -0.695215000      0.022886000\\
C       -5.747383000     -1.014425000      0.395489000\\
C       -9.025031000      1.464633000     -0.798880000\\
C       -4.175162000      0.837177000     -0.019978000\\
C       -9.500300000     -1.171842000      0.038230000\\
C       -4.645261000     -1.814670000      0.773389000\\
H       -8.837769000      2.483805000     -1.122549000\\
H       -3.996153000      1.862482000     -0.328196000\\
H       -9.681353000     -2.191644000      0.363114000\\
H       -4.829866000     -2.839587000      1.080073000\\
C      -10.329823000      0.982989000     -0.783079000\\
C       -3.099091000      0.042098000      0.363784000\\
C      -10.575748000     -0.378343000     -0.347668000\\
C       -3.343879000     -1.329430000      0.762900000\\
C      -11.448293000      1.788419000     -1.182725000\\
C       -1.755036000      0.533963000      0.374282000\\
C      -11.927303000     -0.860125000     -0.331809000\\
C       -2.216988000     -2.134293000      1.138235000\\
H      -11.260277000      2.807989000     -1.510491000\\
H       -1.586448000      1.559342000      0.055039000\\
H      -12.109196000     -1.880309000     -0.002574000\\
H       -2.394464000     -3.160746000      1.450132000\\
C      -12.719894000      1.287258000     -1.152870000\\
C       -0.690680000     -0.260041000      0.737399000\\
C      -12.962876000     -0.056544000     -0.720115000\\
C       -0.950855000     -1.626991000      1.121746000\\
H      -13.558812000      1.906799000     -1.458392000\\
H      -13.982476000     -0.433009000     -0.703784000\\
H       -0.116306000     -2.246203000      1.437138000\\
C        6.610452000     -1.170691000     -0.408139000\\
C        7.088558000      1.516505000      0.429114000\\
C        6.371059000     -2.511574000     -0.827457000\\
C        7.327596000      2.857691000      0.848489000\\
C        6.164808000     -3.651928000     -1.184069000\\
C        7.529830000      3.998563000      1.205595000\\
H        5.983001000     -4.654407000     -1.499131000\\
H        7.708891000      5.001525000      1.520659000\\
C        7.929387000     -0.662068000     -0.390031000\\
C        5.508308000     -0.351182000     -0.009373000\\
C        8.171927000      0.696879000      0.036844000\\
C        5.751228000      1.013116000      0.410730000\\
C        9.048505000     -1.462925000     -0.785791000\\
C        4.185940000     -0.836017000     -0.011937000\\
C        9.520248000      1.178112000      0.048334000\\
C        4.653328000      1.811622000      0.790096000\\
H        8.861821000     -2.481795000     -1.107697000\\
H        4.008538000     -1.860504000     -0.323939000\\
H        9.698433000      2.197869000      0.372007000\\
H        4.837662000      2.835543000      1.100389000\\
C       10.322274000     -0.962241000     -0.761188000\\
C        3.105746000     -0.041734000      0.372792000\\
C       10.561603000      0.377515000     -0.336922000\\
C        3.348939000      1.325837000      0.776638000\\
C        1.763337000     -0.534332000      0.377713000\\
C        2.223172000      2.129035000      1.153224000\\
H        1.595819000     -1.558498000      0.054011000\\
H        2.400183000      3.154079000      1.469973000\\
C        0.696825000      0.258657000      0.741179000\\
C        0.955912000      1.621859000      1.132203000\\
H        0.121051000      2.239961000      1.448949000\\
H       11.577577000      0.763436000     -0.320612000\\
H       11.157065000     -1.587377000     -1.066311000\\
\clearpage
TIPS-2,2$^{\prime}$-pentacene-hexacene\\ 
\begin{figure}[h!]
\includegraphics[width=0.9\textwidth,trim=3cm 6cm 3cm 6cm,clip]{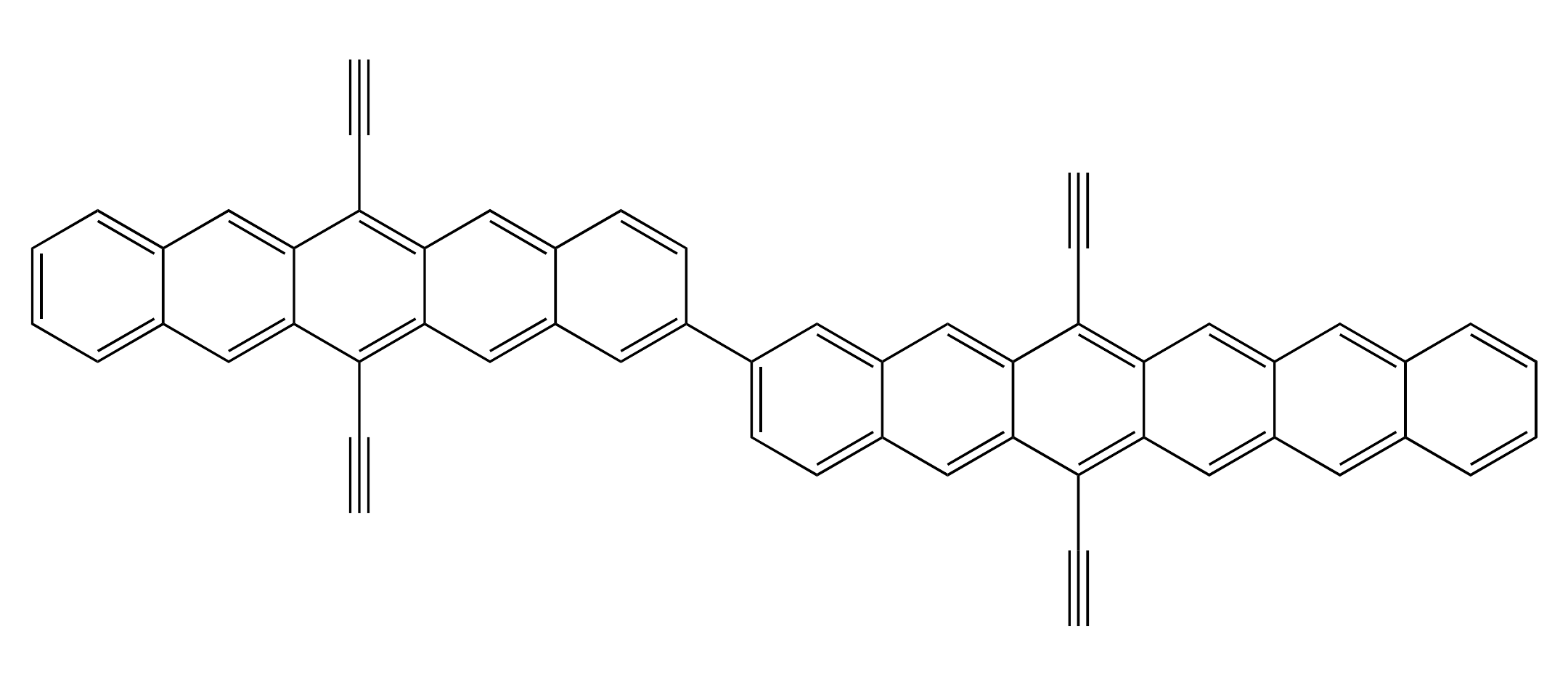}
\caption{Structure of TIPS-2,2$^{\prime}$-pentacene-hexacene.}
\label{fig:struc_22ph}
\end{figure}
\
Cartesian coordinates / {\AA} (atom x y z)\\
C       -6.594981000      1.172617000     -0.430728000\\
C       -7.078584000     -1.510487000      0.435529000\\
C       -7.925891000      0.667529000     -0.412260000\\
C       -5.502785000      0.356373000     -0.022378000\\
C       -8.172235000     -0.692105000      0.033092000\\
C       -5.748550000     -1.007850000      0.410360000\\
C       -9.024383000      1.457431000     -0.816608000\\
C       -4.175398000      0.838538000     -0.024359000\\
C       -9.501140000     -1.168359000      0.052534000\\
C       -4.646524000     -1.805004000      0.795503000\\
H       -8.836563000      2.472501000     -1.152544000\\
H       -3.995923000      1.860634000     -0.342738000\\
H       -9.682830000     -2.184082000      0.389541000\\
H       -4.831653000     -2.826763000      1.112199000\\
C      -10.329255000      0.976028000     -0.797035000\\
C       -3.099646000      0.046723000      0.366977000\\
C      -10.575954000     -0.379717000     -0.345068000\\
C       -3.344932000     -1.321058000      0.778511000\\
C      -11.446976000      1.776311000     -1.208818000\\
C       -1.755357000      0.538142000      0.372738000\\
C      -11.927462000     -0.861372000     -0.325747000\\
C       -2.217933000     -2.123765000      1.158442000\\
H      -11.258420000      2.791718000     -1.548885000\\
H       -1.587192000      1.561181000      0.045927000\\
H      -12.109942000     -1.877416000      0.015669000\\
H       -2.395581000     -3.147371000      1.479393000\\
C      -12.718627000      1.275456000     -1.175167000\\
C       -0.690705000     -0.254315000      0.739469000\\
C      -12.962363000     -0.062810000     -0.726125000\\
C       -0.951413000     -1.618092000      1.134670000\\
H      -13.556963000      1.891065000     -1.490062000\\
H      -13.981946000     -0.439203000     -0.707076000\\
H       -0.116786000     -2.235541000      1.453240000\\
C       -6.353606000      2.507503000     -0.862885000\\
C       -6.147070000      3.644421000     -1.231280000\\
H       -5.964424000      4.643198000     -1.557316000\\
C       -7.319786000     -2.845735000      0.867626000\\
C       -7.525999000     -3.982702000      1.235853000\\
H       -7.708537000     -4.981685000      1.561317000\\
C        6.598170000     -1.178801000     -0.414460000\\
C        7.083706000      1.513336000      0.432409000\\
C        7.935442000     -0.676785000     -0.398143000\\
C        5.510248000     -0.358164000     -0.015371000\\
C        8.183546000      0.689577000      0.040426000\\
C        5.757001000      1.011627000      0.406509000\\
C        9.025171000     -1.469285000     -0.793821000\\
C        4.180109000     -0.838352000     -0.014373000\\
C        9.504841000      1.163600000      0.064488000\\
C        4.652266000      1.813488000      0.781573000\\
H        8.836847000     -2.485768000     -1.124913000\\
H        4.000160000     -1.863234000     -0.323365000\\
H        9.687729000      2.180679000      0.396476000\\
H        4.837616000      2.838053000      1.088924000\\
C       10.341644000     -0.992156000     -0.774478000\\
C        3.106086000     -0.042570000      0.367928000\\
C       10.590568000      0.369219000     -0.325195000\\
C        3.351996000      1.330465000      0.766747000\\
C       11.440881000     -1.789837000     -1.175866000\\
C        1.760756000     -0.533042000      0.377962000\\
C       11.923843000      0.845671000     -0.298809000\\
C        2.223863000      2.137489000      1.137133000\\
H       11.252859000     -2.806569000     -1.514536000\\
H        1.592382000     -1.559234000      0.061308000\\
H       12.109421000      1.862671000      0.040383000\\
H        2.401541000      3.164324000      1.447642000\\
C       12.743806000     -1.309206000     -1.148068000\\
C        0.696826000      0.263057000      0.736378000\\
C       12.993585000      0.051730000     -0.692476000\\
C        0.957820000      1.631965000      1.117393000\\
H        0.122911000      2.252761000      1.428648000\\
C       14.349726000      0.525894000     -0.668326000\\
C       13.865283000     -2.110501000     -1.554252000\\
C       15.383070000     -0.275080000     -1.063398000\\
H       16.403738000      0.098052000     -1.039240000\\
C       15.137091000     -1.614069000     -1.514231000\\
H       15.975001000     -2.232376000     -1.824964000\\
H       14.535249000      1.541252000     -0.326036000\\
H       13.676592000     -3.125603000     -1.895616000\\
C        7.325127000      2.851745000      0.852683000\\
C        7.533377000      3.991798000      1.210529000\\
H        7.717392000      4.993494000      1.526692000\\
C        6.355865000     -2.516493000     -0.835499000\\
C        6.150090000     -3.656640000     -1.194688000\\
H        5.967820000     -4.658158000     -1.512418000\\
\
TIPS-1,1$^{\prime}$-pentacene-anthracene\\ 
\begin{figure}[h!]
\includegraphics[width=0.9\textwidth,trim=4cm 6cm 4cm 6cm,clip]{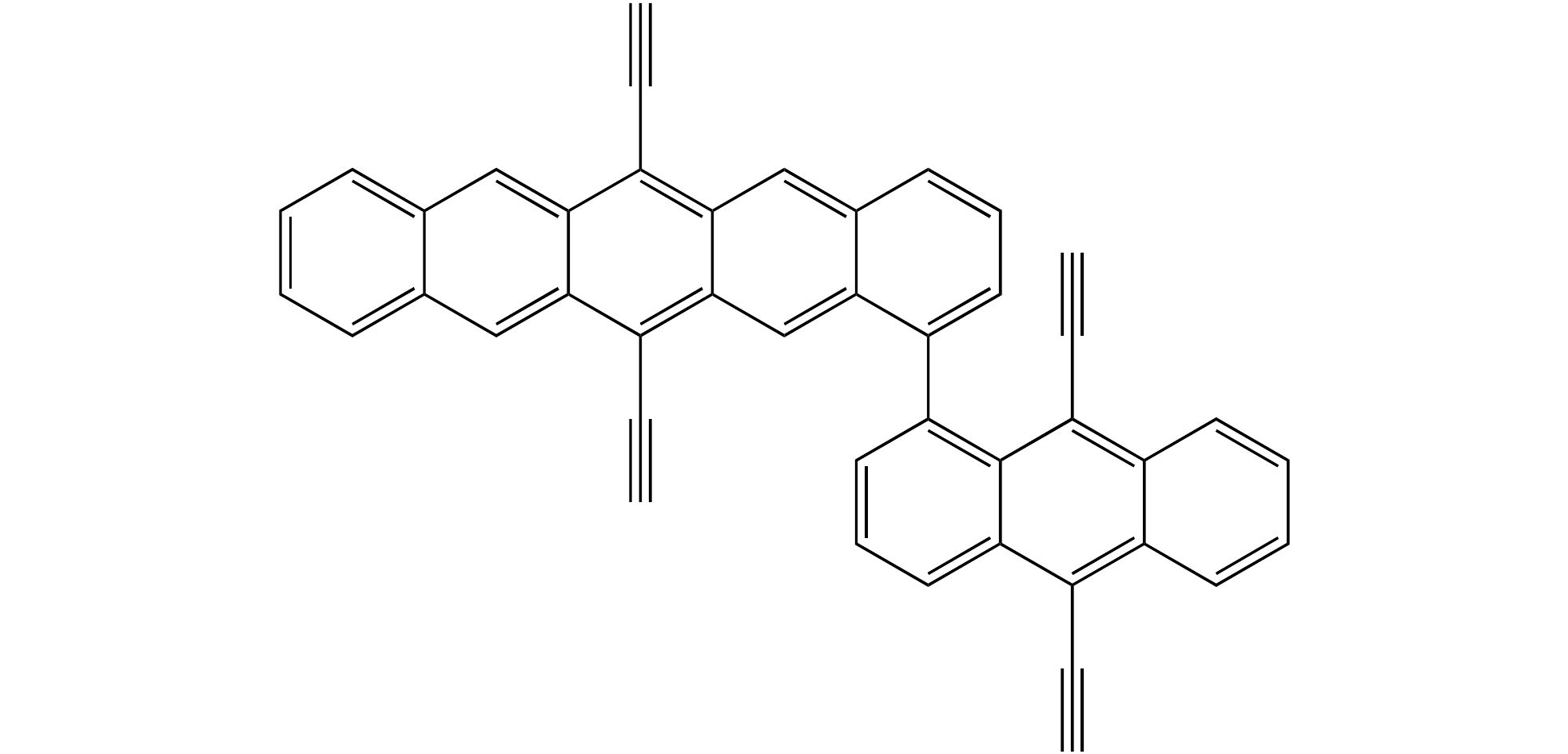}
\caption{Structure of TIPS-1,1$^{\prime}$-pentacene-anthracene.}
\label{fig:struc_11pa}
\end{figure}
\
Cartesian coordinates / {\AA} (atom x y z)\\
C        5.071164000     -1.653018000      0.155646000\\
C        3.653955000      0.822370000     -0.074033000\\
C        5.540972000     -0.653602000     -0.742489000\\
C        3.905193000     -1.430846000      0.940904000\\
C        4.821511000      0.602468000     -0.859177000\\
C        3.186095000     -0.176195000      0.825251000\\
C        6.697295000     -0.847615000     -1.529877000\\
C        3.419618000     -2.401491000      1.843236000\\
C        5.306165000      1.579957000     -1.755788000\\
C        2.031447000      0.023681000      1.616119000\\
H        7.233425000     -1.787412000     -1.440602000\\
H        3.958728000     -3.339501000      1.935995000\\
H        4.766011000      2.517922000     -1.841609000\\
H        1.503046000      0.965102000      1.521410000\\
C        7.166596000      0.123793000     -2.407922000\\
C        2.283237000     -2.193171000      2.617893000\\
C        6.447536000      1.378162000     -2.524942000\\
C        1.560716000     -0.939449000      2.501613000\\
C        8.344253000     -0.068912000     -3.205379000\\
C        1.810395000     -3.185764000      3.539079000\\
C        6.943988000      2.371687000     -3.433677000\\
C        0.376796000     -0.741977000      3.313209000\\
H        8.882024000     -1.009721000     -3.113964000\\
H        2.356010000     -4.123518000      3.614945000\\
H        6.402205000      3.310858000     -3.518361000\\
C        8.782022000      0.906696000     -4.057279000\\
C        0.712310000     -2.946676000      4.315905000\\
C        8.071026000      2.145321000     -4.173509000\\
C        0.001223000     -1.711936000      4.207802000\\
H        9.677145000      0.750580000     -4.653481000\\
H        0.366438000     -3.696100000      5.023343000\\
H        8.437910000      2.906858000     -4.856872000\\
H       -0.875143000     -1.550717000      4.829549000\\
C        5.774247000     -2.885873000      0.270857000\\
C        6.372341000     -3.936369000      0.369201000\\
C        2.948077000      2.053500000     -0.190801000\\
C        2.346977000      3.101806000     -0.290401000\\
H        6.897098000     -4.860608000      0.456201000\\
H        1.811047000      4.019824000     -0.375760000\\
C       -4.915468000      1.808883000     -0.043937000\\
C       -3.680122000     -0.690883000     -0.323965000\\
C       -5.398328000      0.900204000     -0.948768000\\
C       -3.782177000      1.505892000      0.771618000\\
C       -4.769588000     -0.366411000     -1.090360000\\
C       -3.145859000      0.225207000      0.635350000\\
C       -3.270983000      2.436273000      1.709425000\\
C       -2.015648000     -0.108343000      1.442241000\\
C       -2.148011000      2.111887000      2.516230000\\
C       -1.502416000      0.819927000      2.396953000\\
C       -1.655237000      3.069055000      3.453420000\\
C       -0.374209000      0.554936000      3.267158000\\
H       -2.157194000      4.028092000      3.517230000\\
C       -0.581118000      2.787295000      4.250218000\\
C        0.051672000      1.524128000      4.151981000\\
H       -0.211988000      3.522423000      4.960332000\\
H        0.900786000      1.302929000      4.792602000\\
H       -5.155927000     -1.082837000     -1.810781000\\
H       -6.263560000      1.145434000     -1.559067000\\
H       -3.203454000     -1.657487000     -0.434264000\\
H       -5.387778000      2.778512000      0.071681000\\
C       -1.440764000     -1.397696000      1.230730000\\
C       -1.042525000     -2.505859000      0.942625000\\
C       -3.896151000      3.712613000      1.841349000\\
C       -4.434782000      4.793559000      1.945738000\\
H       -0.624074000     -3.468946000      0.758279000\\
H       -4.906694000      5.745259000      2.039670000\\
\
TIPS-1,1$^{\prime}$-pentacene-tetracene\\ 
\begin{figure}[h!]
\includegraphics[width=0.9\textwidth,trim=4cm 6cm 4cm 6cm,clip]{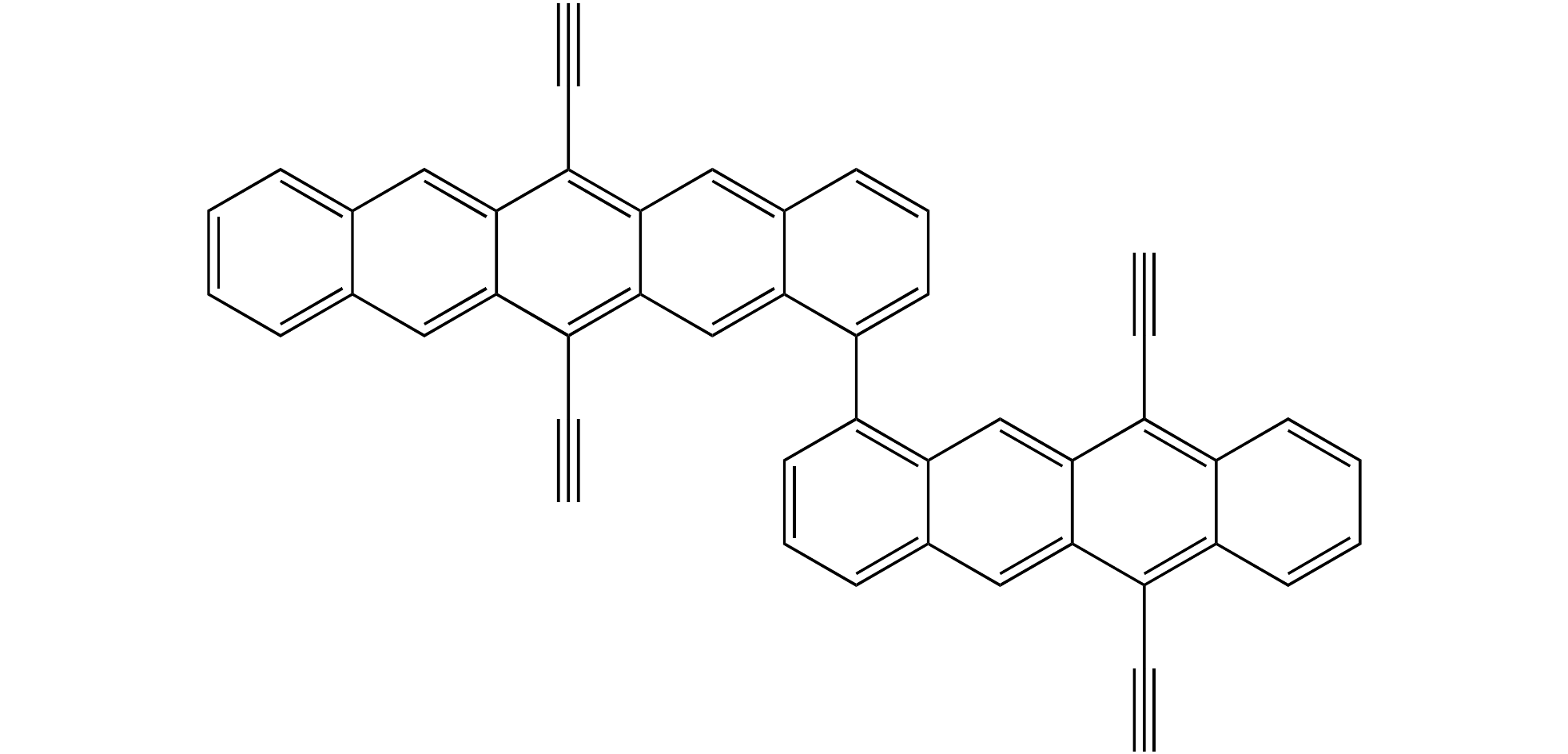}
\caption{Structure of TIPS-1,1$^{\prime}$-pentacene-tetracene.}
\label{fig:struc_11pt}
\end{figure}
\
Cartesian coordinates / {\AA} (atom x y z)\\
C        4.609088000     -1.911595000     -0.725946000\\
C        3.358961000      0.660038000     -0.840305000\\
C        5.062884000     -0.949560000     -1.671576000\\
C        3.542597000     -1.604212000      0.165073000\\
C        4.428002000      0.355274000     -1.729861000\\
C        2.908738000     -0.300615000      0.108398000\\
C        6.123140000     -1.227205000     -2.562447000\\
C        3.077018000     -2.535801000      1.117336000\\
C        4.895176000      1.294526000     -2.675307000\\
C        1.855343000     -0.014485000      1.006046000\\
H        6.596246000     -2.203251000     -2.516888000\\
H        3.551694000     -3.511111000      1.164130000\\
H        4.418154000      2.268856000     -2.717208000\\
H        1.392670000      0.964170000      0.957717000\\
C        6.576628000     -0.292677000     -3.487372000\\
C        2.039154000     -2.244321000      1.996477000\\
C        5.941910000      1.010338000     -3.545975000\\
C        1.402679000     -0.940046000      1.941061000\\
C        7.657593000     -0.570407000     -4.389773000\\
C        1.585067000     -3.200742000      2.963506000\\
C        6.421529000      1.965065000     -4.504013000\\
C        0.328398000     -0.648782000      2.869729000\\
H        8.132471000     -1.547531000     -4.342239000\\
H        2.067390000     -4.174818000      2.994609000\\
H        5.943560000      2.941009000     -4.544581000\\
C        8.082947000      0.370715000     -5.285496000\\
C        0.572079000     -2.890891000      3.825094000\\
C        7.455413000      1.657510000     -5.343754000\\
C       -0.055309000     -1.606395000      3.775701000\\
H        8.904722000      0.149831000     -5.961851000\\
H        0.230852000     -3.618067000      4.557342000\\
H        7.810880000      2.390170000     -6.063725000\\
H       -0.861212000     -1.385142000      4.470441000\\
C        2.734217000      1.937975000     -0.900781000\\
C        2.201050000      3.025728000     -0.952412000\\
C        5.228942000     -3.192123000     -0.668646000\\
C        5.756642000     -4.283031000     -0.619593000\\
H        1.723405000      3.978189000     -0.995818000\\
H        6.220935000     -5.242039000     -0.576220000\\
C       -4.773999000      1.855980000     -0.583493000\\
C       -3.523262000     -0.709334000     -0.691749000\\
C       -5.253928000      0.875138000     -1.480818000\\
C       -3.666214000      1.574280000      0.276043000\\
C       -4.620348000     -0.422826000     -1.536038000\\
C       -3.032584000      0.274831000      0.222556000\\
C       -6.364895000      1.128985000     -2.348105000\\
C       -3.169262000      2.527475000      1.184401000\\
C       -5.132179000     -1.393254000     -2.456157000\\
C       -1.946005000      0.008705000      1.079541000\\
H       -6.838154000      2.104098000     -2.301992000\\
H       -3.645017000      3.502472000      1.228653000\\
H       -4.652647000     -2.365665000     -2.493844000\\
H       -1.482846000     -0.969984000      1.034177000\\
C       -6.823499000      0.171050000     -3.211318000\\
C       -2.093519000      2.258476000      2.029504000\\
C       -6.197627000     -1.109028000     -3.266584000\\
C       -1.457361000      0.957076000      1.977565000\\
C       -1.604891000      3.236911000      2.954556000\\
C       -0.347798000      0.687216000      2.867656000\\
H       -2.088456000      4.210489000      2.983469000\\
C       -0.556226000      2.948667000      3.782123000\\
C        0.070875000      1.665979000      3.736645000\\
H       -0.188123000      3.693202000      4.483214000\\
H        0.904410000      1.460833000      4.403113000\\
H       -6.571595000     -1.861324000     -3.956116000\\
H       -7.670411000      0.383138000     -3.858673000\\
C       -5.396270000      3.137039000     -0.528635000\\
C       -5.922986000      4.228040000     -0.479185000\\
C       -2.896876000     -1.988213000     -0.748463000\\
C       -2.360583000     -3.074348000     -0.794008000\\
H       -1.881143000     -4.026072000     -0.834005000\\
H       -6.387693000      5.186898000     -0.436538000\\

\clearpage
TIPS-1,1$^{\prime}$-pentacene-hexacene\\ 
\begin{figure}[h!]
\includegraphics[width=0.9\textwidth,trim=4cm 6cm 4cm 6cm,clip]{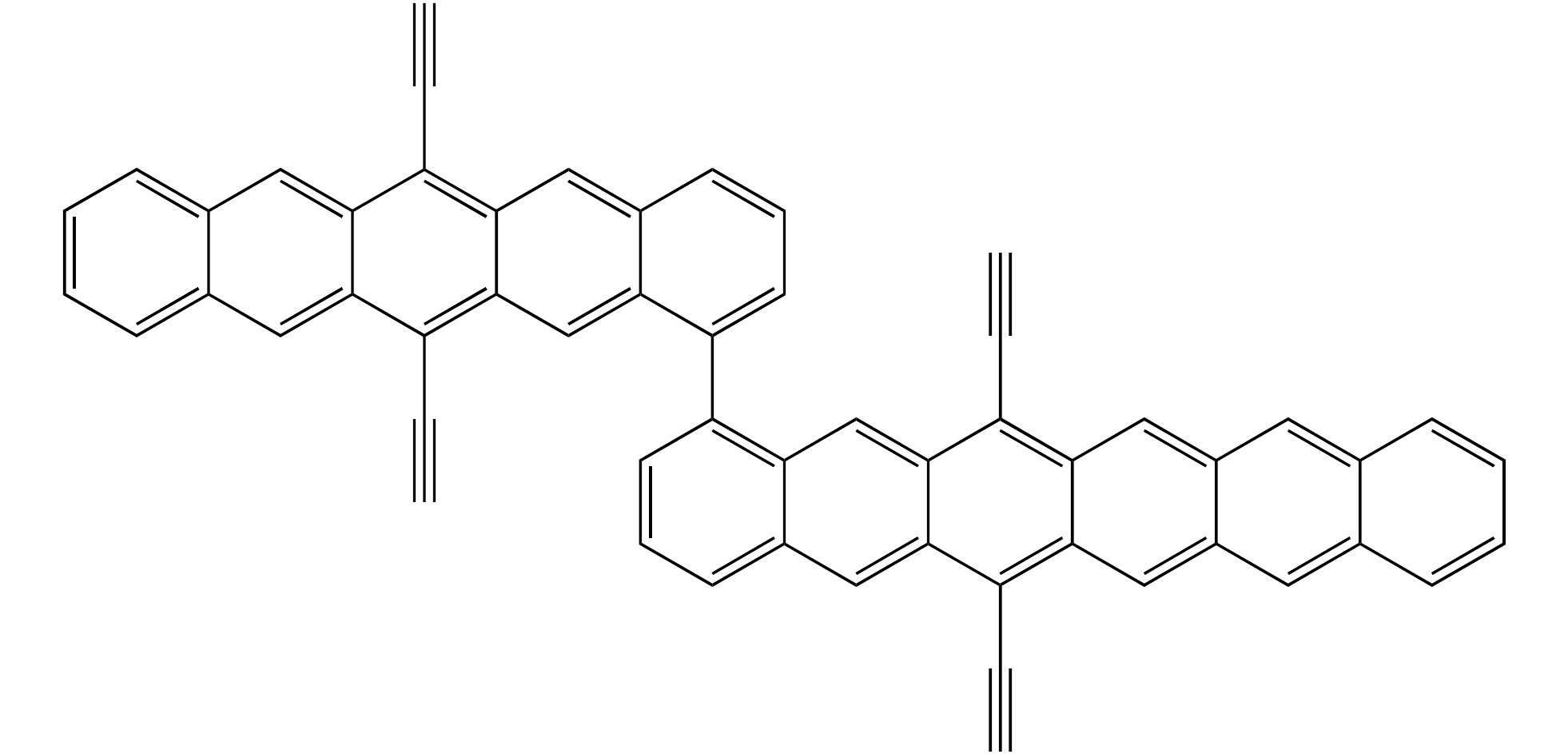}
\caption{Structure of TIPS-1,1$^{\prime}$-pentacene-hexacene.}
\label{fig:struc_11ph}
\end{figure}
\
Cartesian coordinates / {\AA} (atom x y z)\\
C        1.291225000      4.837265000      5.879928000\\
C        2.527139000      2.259963000      6.019886000\\
C        2.310292000      4.596505000      6.844011000\\
C        0.883956000      3.808667000      4.984621000\\
C        2.934487000      3.287359000      6.917463000\\
C        1.512140000      2.502984000      5.052012000\\
C        2.737502000      5.601761000      7.739315000\\
C       -0.120925000      4.020176000      4.016730000\\
C        3.939281000      3.069246000      7.885782000\\
C        1.098985000      1.499302000      4.146659000\\
H        2.271348000      6.580470000      7.680982000\\
H       -0.589771000      4.997795000      3.961301000\\
H        4.402107000      2.088742000      7.939789000\\
H        1.582039000      0.531035000      4.201204000\\
C        3.731611000      5.376366000      8.685664000\\
C       -0.523306000      3.024359000      3.132765000\\
C        4.350908000      4.067019000      8.763105000\\
C        0.104175000      1.716041000      3.197999000\\
C        4.171195000      6.397908000      9.592818000\\
C       -1.538374000      3.263759000      2.148515000\\
C        5.373555000      3.848471000      9.746046000\\
C       -0.317214000      0.694193000      2.259446000\\
H        3.706809000      7.379302000      9.530495000\\
H       -2.002502000      4.246351000      2.111757000\\
H        5.834949000      2.865395000      9.801669000\\
C        5.151376000      6.145784000     10.511887000\\
C       -1.907683000      2.279101000      1.277706000\\
C        5.760868000      4.851404000     10.590488000\\
C       -1.289019000      0.991022000      1.335397000\\
H        5.477277000      6.928749000     11.191715000\\
H       -2.676477000      2.463257000      0.531670000\\
H        6.537852000      4.673866000     11.329684000\\
H       -1.601337000      0.223110000      0.632735000\\
C        0.674544000      6.118657000      5.808994000\\
C        0.149047000      7.210031000      5.748062000\\
H       -0.310705000      8.170682000      5.694012000\\
C        3.139843000      0.976887000      6.092508000\\
C        3.660135000     -0.116566000      6.154334000\\
H        4.107397000     -1.083236000      6.207288000\\
C       -1.172056000     -4.786179000      5.946095000\\
C       -2.387593000     -2.198408000      6.126340000\\
C       -2.156775000     -4.540107000      6.950893000\\
C       -0.794918000     -3.762930000      5.036981000\\
C       -2.770488000     -3.222463000      7.046042000\\
C       -1.414423000     -2.451908000      5.123828000\\
C       -2.553689000     -5.538679000      7.855252000\\
C        0.173844000     -3.980906000      4.031570000\\
C       -3.731058000     -2.993362000      8.044856000\\
C       -1.030543000     -1.453349000      4.196940000\\
H       -2.097612000     -6.520879000      7.780622000\\
H        0.636398000     -4.960606000      3.962508000\\
H       -4.184226000     -2.009529000      8.115420000\\
H       -1.507426000     -0.483023000      4.265782000\\
C       -3.516647000     -5.309375000      8.845718000\\
C        0.545991000     -2.990898000      3.130528000\\
C       -4.123409000     -3.990449000      8.946470000\\
C       -0.075013000     -1.678785000      3.213099000\\
C       -3.922549000     -6.317996000      9.753515000\\
C        1.523383000     -3.237650000      2.109518000\\
C       -5.093875000     -3.760950000      9.952371000\\
C        0.313417000     -0.663090000      2.252351000\\
H       -3.469952000     -7.304463000      9.675691000\\
H        1.983098000     -4.221748000      2.060282000\\
H       -5.545293000     -2.773774000     10.027531000\\
C       -4.878782000     -6.080944000     10.732679000\\
C        1.862095000     -2.258662000      1.220991000\\
C       -5.482400000     -4.758932000     10.836974000\\
C        1.248512000     -0.967821000      1.294567000\\
H        2.601601000     -2.448636000      0.447301000\\
H        1.535500000     -0.205245000      0.575498000\\
C       -6.467769000     -4.536523000     11.859011000\\
C       -5.301108000     -7.099254000     11.654470000\\
C       -6.244171000     -6.839678000     12.607831000\\
H       -6.555060000     -7.621189000     13.296300000\\
H       -4.851403000     -8.086227000     11.573380000\\
H       -6.916541000     -3.548845000     11.935445000\\
C       -6.836028000     -5.537254000     12.712502000\\
H       -7.583870000     -5.353663000     13.479778000\\
C       -2.988301000     -0.911576000      6.218935000\\
C       -3.499764000      0.184975000      6.299359000\\
H       -3.938899000      1.154416000      6.367358000\\
C       -0.565596000     -6.070566000      5.855598000\\
C       -0.049652000     -7.165715000      5.779707000\\
H        0.402395000     -8.129142000      5.712003000\\
\
\clearpage
TIPS-tetracene\\ 
\begin{figure}[h!]
\includegraphics[width=0.9\textwidth,trim=4cm 7cm 4cm 6cm,clip]{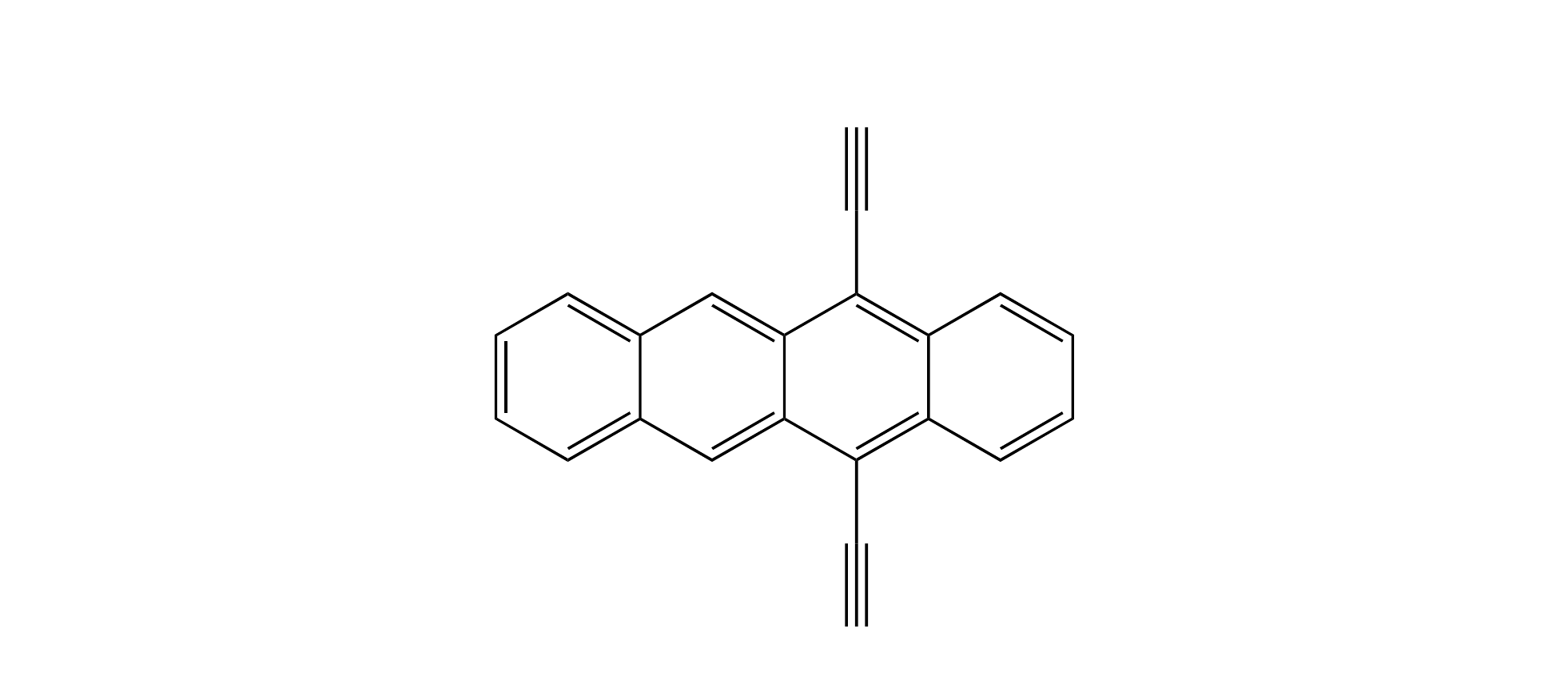}
\caption{Structure of TIPS-tetracene.}
\label{fig:struc_tt}
\end{figure}
\
Cartesian coordinates / {\AA} (atom x y z)\\
C        0.010068000      4.065376000      0.000000000\\
C        0.010068000     -4.065376000      0.000000000\\
C        0.012336000      2.853089000      0.000000000\\
C        0.012336000     -2.853089000      0.000000000\\
C        0.012602000      1.427747000      0.000000000\\
C        0.012602000     -1.427747000      0.000000000\\
C       -1.232486000      0.723983000      0.000000000\\
C        1.237453000      0.722575000      0.000000000\\
C       -1.232486000     -0.723983000      0.000000000\\
C        1.237453000     -0.722575000      0.000000000\\
C       -2.466704000      1.402165000      0.000000000\\
C        2.496616000      1.404465000      0.000000000\\
C       -2.466704000     -1.402165000      0.000000000\\
C        2.496616000     -1.404465000      0.000000000\\
C       -3.684422000      0.723903000      0.000000000\\
C        3.677424000      0.712866000      0.000000000\\
C       -3.684422000     -0.723903000      0.000000000\\
C        3.677424000     -0.712866000      0.000000000\\
C       -6.121039000      0.715286000      0.000000000\\
C       -6.121039000     -0.715286000      0.000000000\\
C       -4.942669000      1.410395000      0.000000000\\
C       -4.942669000     -1.410395000      0.000000000\\
H        0.008503000      5.132166000      0.000000000\\
H        0.008503000     -5.132166000      0.000000000\\
H       -2.463234000      2.488537000      0.000000000\\
H        2.491726000      2.489812000      0.000000000\\
H       -2.463234000     -2.488537000      0.000000000\\
H        2.491726000     -2.489812000      0.000000000\\
H       -4.938945000      2.498363000      0.000000000\\
H       -4.938945000     -2.498363000      0.000000000\\
H       -7.069130000     -1.246870000      0.000000000\\
H       -7.069130000      1.246870000      0.000000000\\
H        4.621965000     -1.250612000      0.000000000\\
H        4.621965000      1.250612000      0.000000000\\
\
1,4,6,11-tetraphenyltetracene\\ 
\begin{figure}[h!]
\includegraphics[width=0.9\textwidth,trim=4cm 6cm 4cm 5cm,clip]{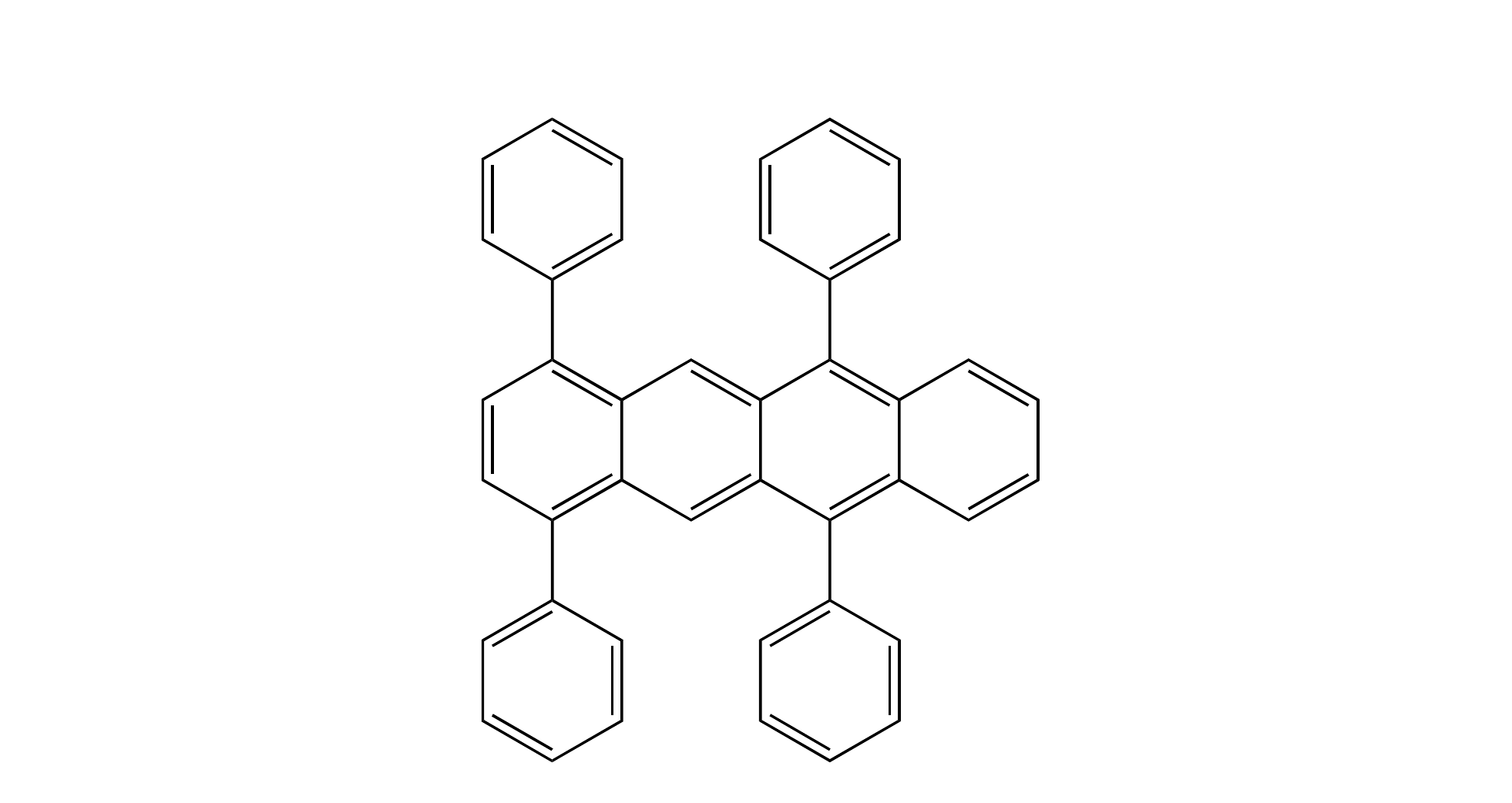}
\caption{Structure of 1,4,6,11-tetraphenyltetracene.}
\label{fig:struc_tphtet}
\end{figure}
\
Cartesian coordinates / {\AA} (atom x y z)\\
C       -1.353272000     -1.930091000     -0.000104000\\
C       -1.353272000     -4.795318000     -0.000027000\\
C       -2.587315000     -2.637193000     -0.000045000\\
C       -0.138859000     -2.635958000     -0.000120000\\
C       -2.587315000     -4.088215000     -0.000015000\\
C       -0.138859000     -4.089451000     -0.000075000\\
C       -3.830130000     -1.963856000      0.000006000\\
C        1.128954000     -1.958234000     -0.000185000\\
C       -3.830130000     -4.761554000      0.000031000\\
C        1.128953000     -4.767177000     -0.000086000\\
C       -5.050824000     -2.636584000      0.000066000\\
C        2.308557000     -2.649130000     -0.000196000\\
C       -5.050824000     -4.088825000      0.000049000\\
C        2.308557000     -4.076282000     -0.000144000\\
C       -7.480468000     -2.649391000      0.000231000\\
C       -7.480468000     -4.076020000      0.000195000\\
C       -6.313261000     -1.929297000      0.000161000\\
C       -6.313261000     -4.796113000      0.000089000\\
H       -3.831985000     -0.879961000      0.000010000\\
H        1.134230000     -0.873196000     -0.000226000\\
H       -3.831984000     -5.845448000      0.000060000\\
H        1.134228000     -5.852215000     -0.000049000\\
H       -8.432024000     -4.601894000      0.000222000\\
H       -8.432024000     -2.123517000      0.000303000\\
H        3.253613000     -2.111708000     -0.000246000\\
H        3.253613000     -4.613704000     -0.000152000\\
C       -6.360699000     -0.433860000      0.000190000\\
C       -6.398335000      0.280814000      1.207114000\\
C       -6.399092000      0.280837000     -1.206695000\\
C       -6.478409000      1.674701000      1.207531000\\
H       -6.368904000     -0.263687000      2.147778000\\
C       -6.479165000      1.674726000     -1.207037000\\
H       -6.370254000     -0.263642000     -2.147389000\\
C       -6.519901000      2.375352000      0.000266000\\
H       -6.511067000      2.212829000      2.151800000\\
H       -6.512412000      2.212864000     -2.151279000\\
H       -6.584943000      3.460795000      0.000301000\\
C       -6.360698000     -6.291550000      0.000035000\\
C       -6.397110000     -7.006254000     -1.206911000\\
C       -6.400315000     -7.006217000      1.206896000\\
C       -6.477188000     -8.400140000     -1.207374000\\
H       -6.366727000     -6.461778000     -2.147559000\\
C       -6.480383000     -8.400108000      1.207193000\\
H       -6.372430000     -6.461713000      2.147605000\\
C       -6.519899000     -9.100763000     -0.000132000\\
H       -6.508895000     -8.938291000     -2.151661000\\
H       -6.514580000     -8.938222000      2.151414000\\
H       -6.584941000    -10.186205000     -0.000208000\\
C       -1.352521000     -6.293130000      0.000012000\\
C       -1.348951000     -7.009703000      1.206912000\\
C       -1.348835000     -7.009767000     -1.206849000\\
C       -1.338822000     -8.405910000      1.207305000\\
H       -1.352496000     -6.464644000      2.147723000\\
C       -1.338701000     -8.405977000     -1.207166000\\
H       -1.352285000     -6.464764000     -2.147692000\\
C       -1.333045000     -9.107911000      0.000088000\\
H       -1.335225000     -8.944852000      2.151648000\\
H       -1.335007000     -8.944962000     -2.151485000\\
H       -1.324841000    -10.195287000      0.000122000\\
C       -1.352520000     -0.432279000     -0.000150000\\
C       -1.349062000      0.284287000     -1.207055000\\
C       -1.348722000      0.284365000      1.206706000\\
C       -1.338930000      1.680495000     -1.207456000\\
H       -1.352692000     -0.260777000     -2.147862000\\
C       -1.338591000      1.680574000      1.207015000\\
H       -1.352088000     -0.260633000      2.147553000\\
C       -1.333044000      2.382502000     -0.000243000\\
H       -1.335416000      2.219431000     -2.151802000\\
H       -1.334813000      2.219565000      2.151330000\\
H       -1.324839000      3.469878000     -0.000284000\\
\
TIPS-5,7,12,14-tetraazapentacene (TIPS-TAP)\\ 
\begin{figure}[h!]
\includegraphics[width=0.9\textwidth,trim=4cm 7cm 4cm 6cm,clip]{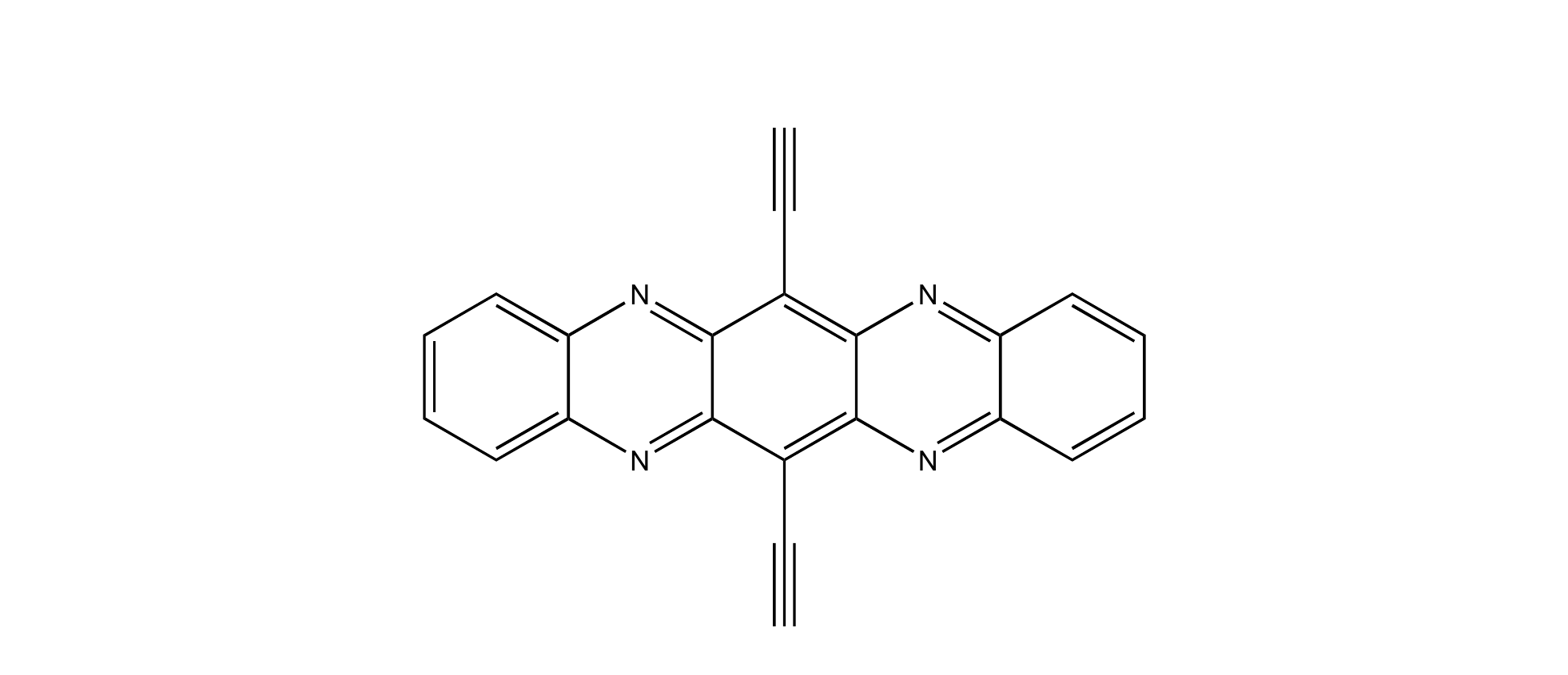}
\caption{Structure of TIPS-5,7,12,14-tetraaza-pentacene (TIPS-TAP).}
\label{fig:struc_ttp}
\end{figure}
\
Cartesian coordinates / {\AA} (atom x y z)\\
C        0.000000000      4.078700000      0.000000000\\
C        0.000000000     -4.078700000      0.000000000\\
C        0.000000000      2.867987000      0.000000000\\
C        0.000000000     -2.867986000      0.000000000\\
C        0.000000000      1.450024000      0.000000000\\
C        0.000000000     -1.450024000      0.000000000\\
C       -1.224668000      0.726742000      0.000000000\\
C        1.224668000      0.726742000      0.000000000\\
C       -1.224668000     -0.726742000      0.000000000\\
C        1.224668000     -0.726742000      0.000000000\\
N       -2.385370000      1.420393000      0.000000000\\
N        2.385370000      1.420392000      0.000000000\\
N       -2.385370000     -1.420392000      0.000000000\\
N        2.385370000     -1.420392000      0.000000000\\
C       -3.520163000      0.727556000      0.000000000\\
C        3.520163000      0.727556000      0.000000000\\
C       -3.520163000     -0.727556000      0.000000000\\
C        3.520163000     -0.727556000      0.000000000\\
C       -5.946360000      0.717869000      0.000000000\\
C        5.946360000      0.717869000      0.000000000\\
C       -5.946360000     -0.717869000      0.000000000\\
C        5.946360000     -0.717869000      0.000000000\\
C       -4.775241000      1.421921000      0.000000000\\
C        4.775241000      1.421921000      0.000000000\\
C       -4.775241000     -1.421921000      0.000000000\\
C        4.775241000     -1.421921000      0.000000000\\
H        0.000000000      5.145183000      0.000000000\\
H        0.000000000     -5.145183000      0.000000000\\
H       -4.748940000      2.507585000      0.000000000\\
H        4.748940000      2.507585000      0.000000000\\
H       -4.748940000     -2.507585000      0.000000000\\
H        4.748940000     -2.507585000      0.000000000\\
H        6.897584000     -1.243498000      0.000000000\\
H       -6.897584000     -1.243498000     -0.000001000\\
H        6.897584000      1.243498000      0.000000000\\
H       -6.897584000      1.243498000     -0.000001000\\
\
TIPS-2,3,9,10-tetraazapentacene\\ 
\begin{figure}[h!]
\includegraphics[width=0.9\textwidth,trim=4cm 7cm 4cm 6cm,clip]{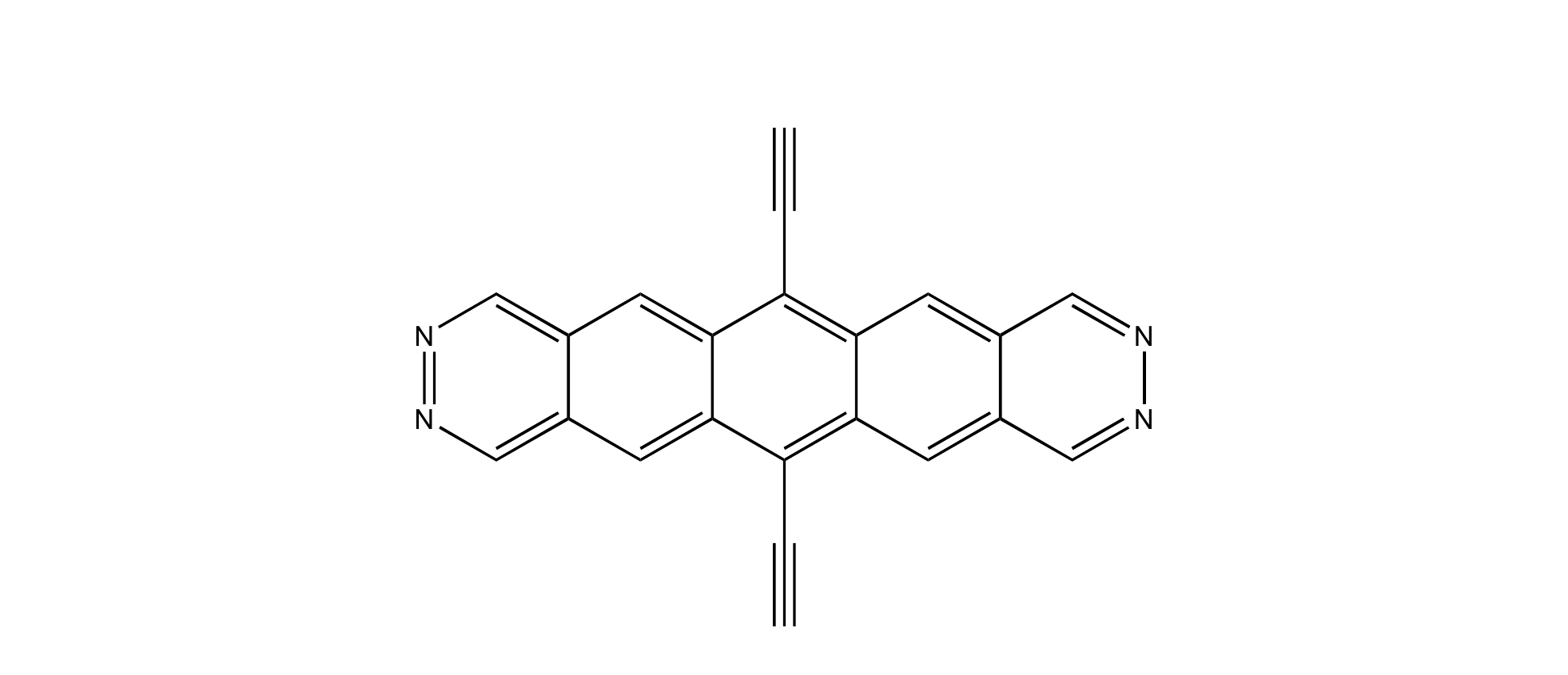}
\caption{Structure of TIPS-2,3,9,10-tetraaza-pentacene.}
\label{fig:struc_ttpb}
\end{figure}
\
Cartesian coordinates / {\AA} (atom x y z)\\
C        0.000000000      4.067631000      0.000000000 \\
C        0.000000000     -4.067631000      0.000000000\\
C        0.000000000      2.855557000      0.000000000\\
C        0.000000000     -2.855557000      0.000000000\\
C        0.000000000      1.432233000      0.000000000\\
C        0.000000000     -1.432233000      0.000000000\\
C       -1.235386000      0.726388000      0.000000000\\
C        1.235386000      0.726388000      0.000000000\\
C       -1.235386000     -0.726388000      0.000000000\\
C        1.235386000     -0.726388000      0.000000000\\
C       -2.474026000      1.412049000      0.000000000\\
C        2.474026000      1.412049000      0.000000000\\
C       -2.474026000     -1.412049000      0.000000000\\
C        2.474026000     -1.412049000      0.000000000\\
C       -3.671629000      0.716594000      0.000000000\\
C        3.671629000      0.716594000      0.000000000\\
C       -3.671629000     -0.716594000      0.000000000\\
C        3.671629000     -0.716594000      0.000000000\\
N       -6.104507000      0.695585000      0.000000000\\
N        6.104507000      0.695585000      0.000000000\\
N       -6.104507000     -0.695585000      0.000000000\\
N        6.104507000     -0.695585000      0.000000000\\
C       -4.969679000      1.335826000      0.000000000\\
C        4.969679000      1.335826000      0.000000000\\
C       -4.969679000     -1.335826000      0.000000000\\
C        4.969679000     -1.335826000      0.000000000\\
H        0.000000000      5.135057000      0.000000000\\
H        0.000000000     -5.135057000      0.000000000\\
H       -2.474674000      2.498112000      0.000000000\\
H        2.474674000      2.498112000      0.000000000\\
H       -2.474674000     -2.498112000      0.000000000\\
H        2.474674000     -2.498112000      0.000000000\\
H       -5.041347000      2.424448000      0.000000000\\
H        5.041347000      2.424448000      0.000000000\\
H       -5.041347000     -2.424448000      0.000000000\\
H        5.041347000     -2.424448000      0.000000000\\
\
1,4,6,11-tetraphenyl-2,3-diazatetracene\\ 
\begin{figure}[h!]
\includegraphics[width=0.9\textwidth,trim=4cm 6cm 4cm 5cm,clip]{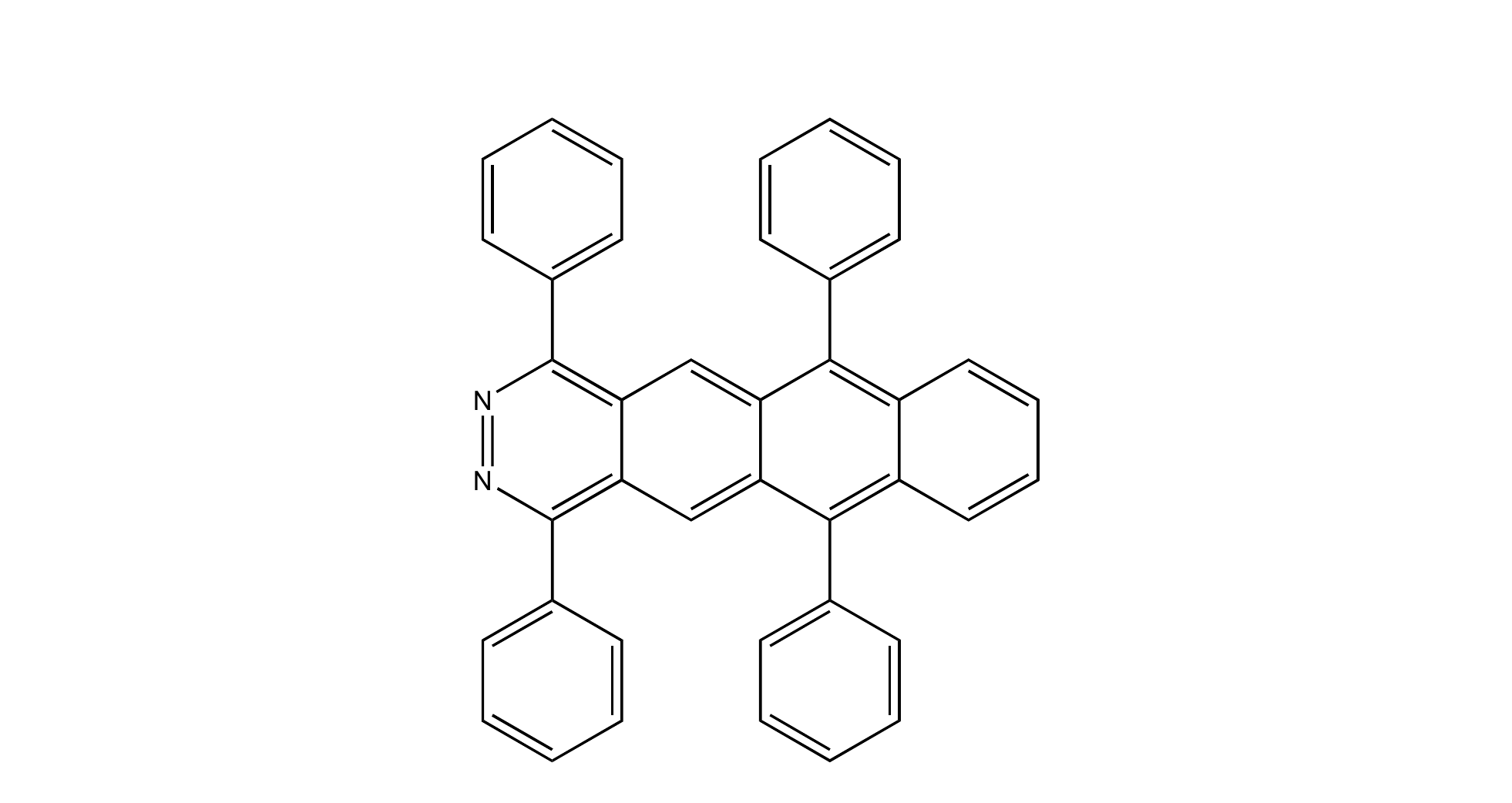}
\caption{Structure of 1,4,6,11-tetraphenyl-2,3-diazatetracene.}
\label{fig:struc_tdb}
\end{figure}
\
Cartesian coordinates / {\AA} (atom x y z)\\
C        0.174216000      1.793851000     -0.089003000\\
C        0.174351000     -1.074180000     -0.088847000\\
C       -1.055546000      1.084935000     -0.026161000\\
C        1.388116000      1.086368000     -0.147095000\\
C       -1.055495000     -0.365369000     -0.026080000\\
C        1.388183000     -0.366576000     -0.147025000\\
C       -2.296415000      1.761100000      0.067888000\\
C        2.652535000      1.764709000     -0.223398000\\
C       -2.296349000     -1.041605000      0.067969000\\
C        2.652667000     -1.044805000     -0.223254000\\
C       -3.506452000      1.080373000      0.126485000\\
C        3.830288000      1.073214000     -0.288330000\\
C       -3.506444000     -0.360930000      0.126466000\\
C        3.830355000     -0.353206000     -0.288257000\\
N       -5.890408000      1.043177000      0.553060000\\
N       -5.890426000     -0.323669000      0.552988000\\
C       -4.795348000      1.725063000      0.293950000\\
C       -4.795392000     -1.005599000      0.293827000\\
H        2.657301000      2.849481000     -0.230813000\\
H        2.657533000     -2.129579000     -0.230559000\\
H        4.773746000     -0.890083000     -0.346011000\\
H        4.773630000      1.610174000     -0.346134000\\
H       -2.290662000     -2.122241000      0.129605000\\
H       -2.290794000      2.841751000      0.129428000\\
C        0.172230000     -2.571741000     -0.102185000\\
C       -0.056159000     -3.275101000     -1.295603000\\
C        0.402802000     -3.300811000      1.075049000\\
C       -0.051208000     -4.671293000     -1.311987000\\
C        0.405272000     -4.697155000      1.059321000\\
C        0.179274000     -5.386028000     -0.134313000\\
H       -0.233335000     -2.720546000     -2.213798000\\
H        0.579602000     -2.766490000      2.005028000\\
H       -0.225354000     -5.199827000     -2.246046000\\
H        0.583069000     -5.245812000      1.980746000\\
H        0.182156000     -6.473178000     -0.146611000\\
C        0.171891000      3.291411000     -0.102547000\\
C       -0.056248000      3.994545000     -1.296145000\\
C        0.402172000      4.020699000      1.074608000\\
C       -0.051304000      5.390733000     -1.312796000\\
C        0.404582000      5.417042000      1.058625000\\
C        0.178859000      6.105691000     -0.135194000\\
H       -0.233160000      3.439814000     -2.214283000\\
H        0.578806000      3.486552000      2.004718000\\
H       -0.225174000      5.919086000     -2.247012000\\
H        0.582148000      5.965875000      1.979990000\\
H        0.181737000      7.192838000     -0.147696000\\
C       -4.984358000      3.201072000      0.225954000\\
C       -5.842829000      3.823251000      1.148625000\\
C       -4.390974000      3.985552000     -0.776907000\\
C       -6.081265000      5.194190000      1.083723000\\
C       -4.637230000      5.357868000     -0.844898000\\
C       -5.477942000      5.967630000      0.087673000\\
H       -6.321950000      3.212272000      1.906450000\\
H       -3.756902000      3.518031000     -1.524497000\\
H       -6.742353000      5.660824000      1.810071000\\
H       -4.176244000      5.947590000     -1.633008000\\
H       -5.668797000      7.036741000      0.034665000\\
C       -4.984619000     -2.481571000      0.225513000\\
C       -5.842784000     -3.103848000      1.148408000\\
C       -4.391920000     -3.265887000     -0.777888000\\
C       -6.081481000     -4.474729000      1.083271000\\
C       -4.638437000     -4.638148000     -0.846106000\\
C       -5.478770000     -5.248015000      0.086735000\\
H       -6.321432000     -2.492985000      1.906624000\\
H       -3.758222000     -2.798279000     -1.525730000\\
H       -6.742296000     -4.941438000      1.809818000\\
H       -4.177966000     -5.227738000     -1.634613000\\
H       -5.669832000     -6.317080000      0.033546000\\
\
TIPS-5,12-diazatetracene\\  
\begin{figure}[h!]
\includegraphics[width=0.9\textwidth,trim=4cm 7cm 4cm 6cm,clip]{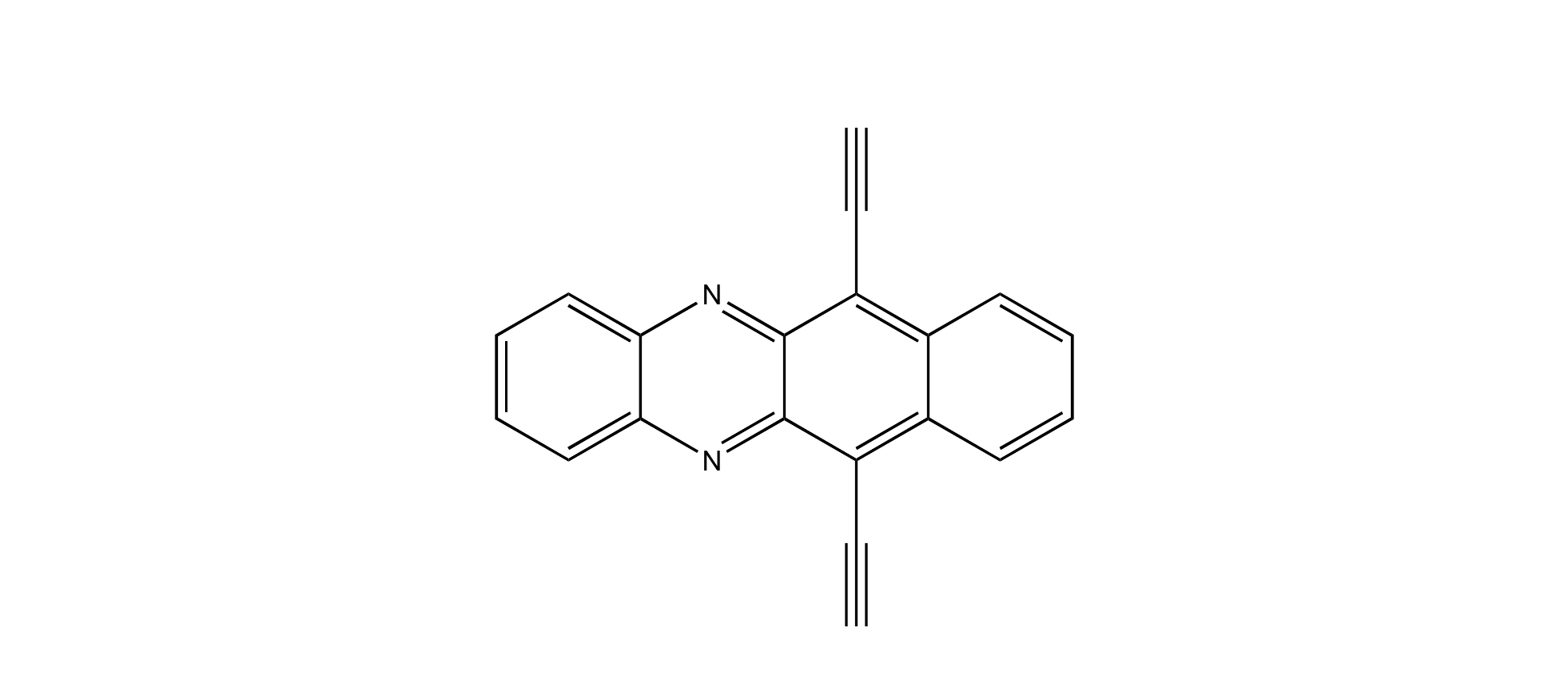}
\caption{Structure of TIPS-5,12-diaza-tetracene.}
\label{fig:struc_tda}
\end{figure}
\
Cartesian coordinates / {\AA} (atom x y z)\\
C       -0.045328000      4.071372000      0.000000000\\
C       -0.045328000     -4.071372000      0.000000000\\
C       -0.064413000      2.860187000      0.000000000\\
C       -0.064413000     -2.860186000      0.000000000\\
C       -0.045755000      1.437601000     -0.000001000\\
C       -0.045755000     -1.437601000     -0.000001000\\
C       -1.281732000      0.723702000     -0.000001000\\
C        1.175421000      0.723996000      0.000000000\\
C       -1.281732000     -0.723702000     -0.000001000\\
C        1.175421000     -0.723996000      0.000000000\\
N       -2.436591000      1.419651000     -0.000001000\\
C        2.434346000      1.405498000      0.000000000\\
N       -2.436591000     -1.419651000      0.000000000\\
C        2.434346000     -1.405498000      0.000000000\\
C       -3.575151000      0.725658000      0.000000000\\
C        3.614490000      0.712917000      0.000000000\\
C       -3.575151000     -0.725657000      0.000000000\\
C        3.614490000     -0.712917000      0.000000000\\
C       -6.001491000      0.716799000      0.000000000\\
C       -6.001491000     -0.716799000      0.000000000\\
C       -4.828470000      1.420116000      0.000000000\\
C       -4.828470000     -1.420116000      0.000000000\\
H       -0.058141000      5.137972000      0.000001000\\
H       -0.058141000     -5.137972000      0.000000000\\
H        2.426966000      2.490711000      0.000000000\\
H        2.426966000     -2.490711000      0.000000000\\
H       -4.801945000      2.505876000      0.000000000\\
H       -4.801944000     -2.505876000      0.000000000\\
H       -6.952125000     -1.243627000      0.000001000\\
H       -6.952126000      1.243626000      0.000000000\\
H        4.559567000     -1.249586000      0.000001000\\
H        4.559567000      1.249586000      0.000001000\\
\
TIPS-1,4,6,11-tetraazatetracene\\  
\begin{figure}[h!]
\includegraphics[width=0.9\textwidth,trim=4cm 7cm 4cm 6cm,clip]{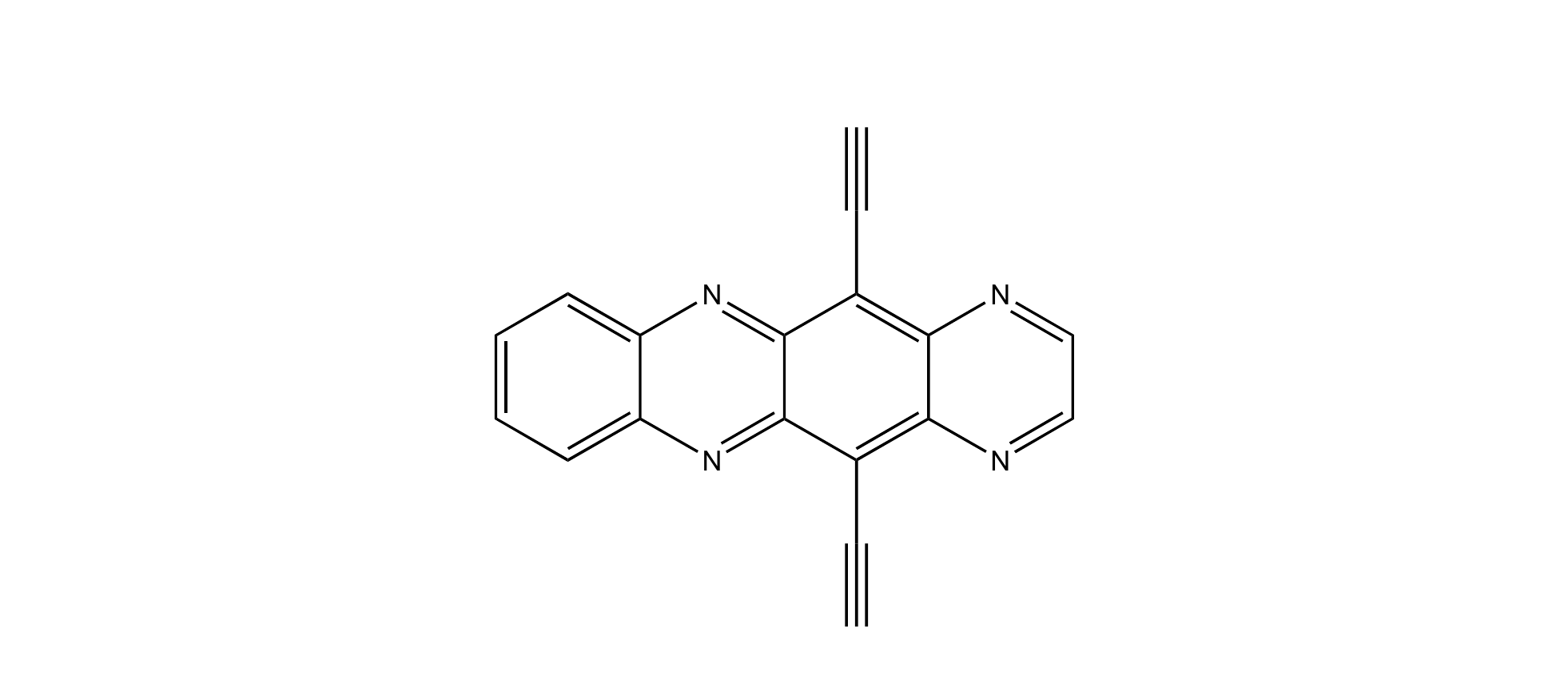}
\caption{Structure of TIPS-1,4,6,11-tetraaza-tetracene.}
\label{fig:struc_tta}
\end{figure}
\
Cartesian coordinates / {\AA} (atom x y z)\\
C       -0.022707000      4.077564000     -0.000001000\\
C       -0.022707000     -4.077564000     -0.000001000\\
C       -0.019156000      2.867313000      0.000000000\\
C       -0.019156000     -2.867313000      0.000000000\\
C       -0.019295000      1.448062000      0.000001000\\
C       -0.019295000     -1.448061000      0.000001000\\
C       -1.252889000      0.725035000      0.000001000\\
C        1.190060000      0.721590000      0.000001000\\
C       -1.252889000     -0.725035000      0.000001000\\
C        1.190060000     -0.721589000      0.000001000\\
N       -2.408233000      1.419230000      0.000001000\\
N        2.374612000      1.418217000      0.000000000\\
N       -2.408233000     -1.419230000      0.000001000\\
N        2.374612000     -1.418217000      0.000000000\\
C       -3.546687000      0.725792000      0.000000000\\
C        3.475883000      0.716955000      0.000000000\\
C       -3.546687000     -0.725792000      0.000000000\\
C        3.475883000     -0.716955000     -0.000001000\\
C       -5.971904000      0.716926000     -0.000001000\\
C       -5.971904000     -0.716926000      0.000000000\\
C       -4.799411000      1.420938000      0.000000000\\
C       -4.799411000     -1.420938000      0.000000000\\
H       -0.032380000      5.144077000     -0.000001000\\
H       -0.032380000     -5.144077000     -0.000001000\\
H       -4.772531000      2.506584000      0.000000000\\
H       -4.772530000     -2.506584000      0.000000000\\
H       -6.922706000     -1.243292000     -0.000001000\\
H       -6.922706000      1.243292000     -0.000001000\\
H        4.420024000     -1.261576000     -0.000001000\\
H        4.420024000      1.261576000      0.000000000\\
\
TIPS-2,3,8,9-tetraazatetracene\\  
\begin{figure}[h!]
\includegraphics[width=0.9\textwidth,trim=4cm 7cm 4cm 6cm,clip]{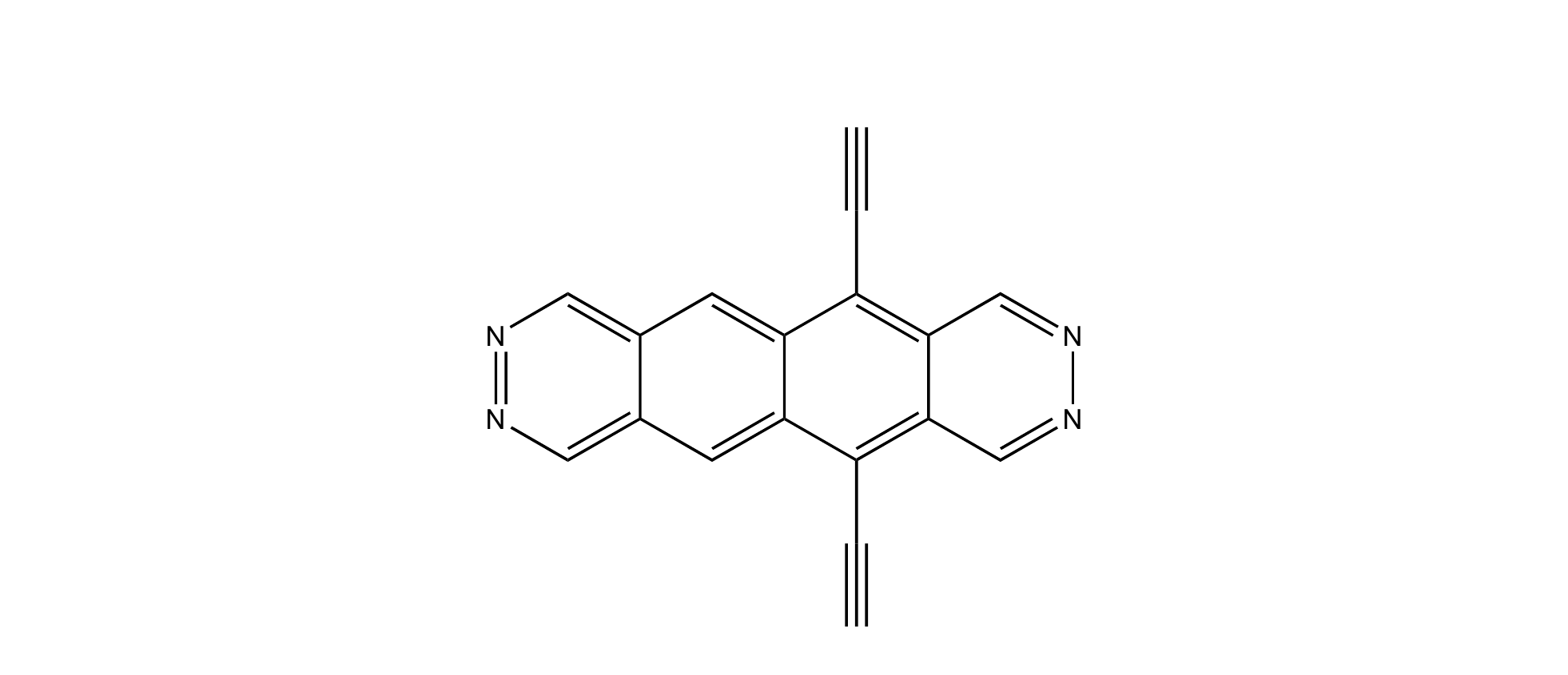}
\caption{Structure of TIPS-2,3,8,9-tetraaza-tetracene.}
\label{fig:struc_tta}
\end{figure}
\
Cartesian coordinates / {\AA} (atom x y z)\\
C        0.030988000      4.071530000      0.000000000\\
C        0.030988000     -4.071530000      0.000000000\\
C        0.016961000      2.860102000      0.000000000\\
C        0.016961000     -2.860102000      0.000000000\\
C        0.002366000      1.436550000      0.000000000\\
C        0.002366000     -1.436549000      0.000000000\\
C       -1.240974000      0.725709000      0.000000000\\
C        1.208977000      0.713152000      0.000000000\\
C       -1.240974000     -0.725709000      0.000000000\\
C        1.208977000     -0.713152000      0.000000000\\
C       -2.474128000      1.411322000      0.000000000\\
C        2.504878000      1.331563000      0.000000000\\
C       -2.474128000     -1.411322000      0.000000000\\
C        2.504878000     -1.331563000      0.000000000\\
C       -3.675481000      0.715024000      0.000000000\\
N        3.642268000      0.691032000      0.000000000\\
C       -3.675481000     -0.715024000      0.000000000\\
N        3.642268000     -0.691032000      0.000000000\\
N       -6.106700000      0.693785000      0.000000000\\
N       -6.106700000     -0.693785000      0.000000000\\
C       -4.971238000      1.335398000      0.000000000\\
C       -4.971238000     -1.335398000      0.000000000\\
H        0.046404000      5.138981000      0.000000000\\
H        0.046404000     -5.138981000      0.000000000\\
H        2.574546000      2.417673000      0.000000000\\
H        2.574546000     -2.417673000      0.000000000\\
H       -5.044018000      2.423740000      0.000000000\\
H       -5.044018000     -2.423740000      0.000000000\\
H       -2.473590000     -2.497571000      0.000000000\\
H       -2.473590000      2.497571000      0.000000000\\

TIPS-anthracene\\ 
\begin{figure}[h!]
\includegraphics[width=0.9\textwidth,trim=4cm 6cm 4cm 5cm,clip]{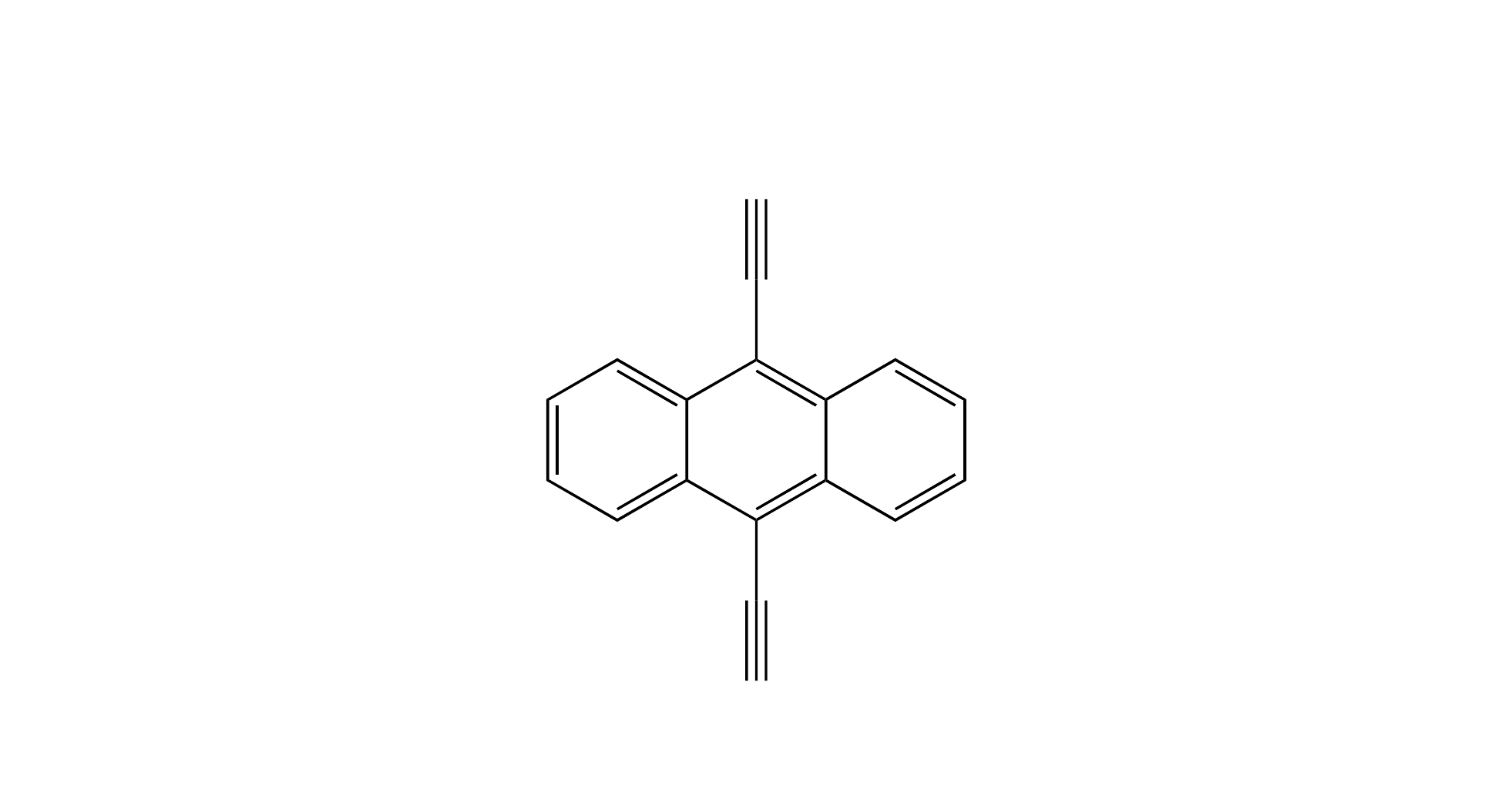}
\caption{Structure of TIPS-anthracene.}
\label{fig:struc_tipsant}
\end{figure}
Cartesian coordinates / {\AA} (atom x y z)\\
C       -0.038901000      1.402991000      0.000000000\\
C       -0.038901000     -1.402991000      0.000000000\\
C       -1.222412000      0.711022000      0.000000000\\
C        1.215630000      0.719838000      0.000000000\\
C       -1.222412000     -0.711022000      0.000000000\\
C        1.215630000     -0.719838000      0.000000000\\
C        2.448937000      1.424402000      0.000000000\\
C        2.448937000     -1.424402000      0.000000000\\
C        3.682244000      0.719837000      0.000000000\\
C        3.682244000     -0.719837000      0.000000000\\
C        6.120283000      0.711024000      0.000000000\\
C        6.120283000     -0.711024000      0.000000000\\
C        4.936776000      1.402995000      0.000000000\\
C        4.936776000     -1.402995000      0.000000000\\
H        4.931287000      2.488403000      0.000000000\\
H        4.931287000     -2.488403000      0.000000000\\
H        7.064470000     -1.249445000      0.000000000\\
H        7.064470000      1.249445000      0.000000000\\
H       -0.033424000     -2.488400000      0.000000000\\
H       -0.033424000      2.488400000      0.000000000\\
H       -2.166588000     -1.249463000      0.000000000\\
H       -2.166588000      1.249463000      0.000000000\\
C        2.448937000      2.850901000      0.000000000\\
C        2.448938000      4.062999000      0.000000000\\
H        2.448935000      5.129811000      0.000000000\\
C        2.448937000     -2.850901000      0.000000000\\
C        2.448938000     -4.062999000      0.000000000\\
H        2.448935000     -5.129811000      0.000000000\\

\begin{figure}[h!]
\includegraphics[width=0.9\textwidth,trim=4cm 6cm 4cm 5cm,clip]{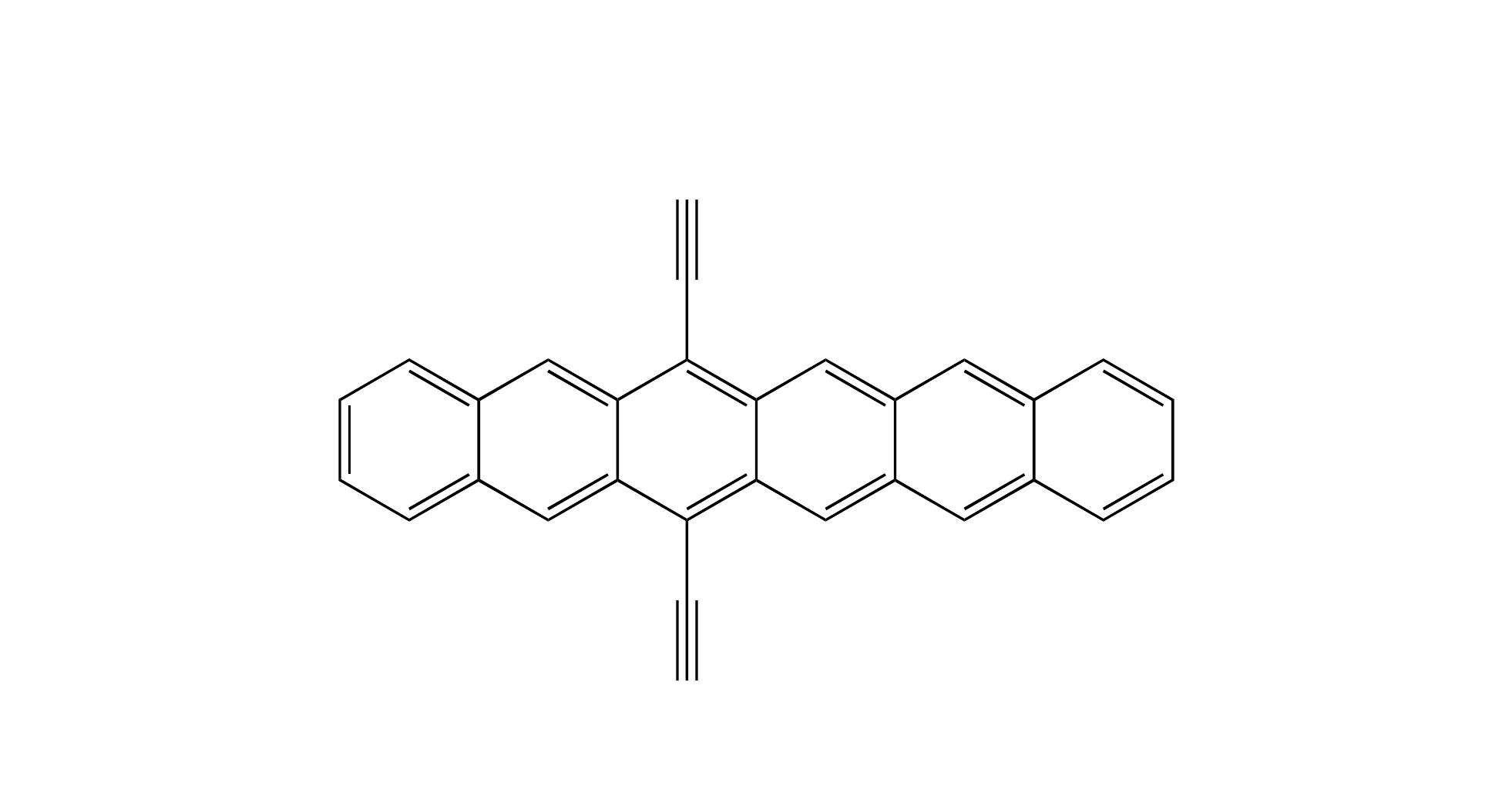}
\caption{Structure of TIPS-hexacene.}
\label{fig:struc_tipshex}
\end{figure}
Cartesian coordinates / {\AA} (atom x y z)\\
C        0.001225000      4.070632000      0.000000000\\
C        0.009138000     -4.065785000      0.000000000\\
C        0.001282000      2.857831000      0.000000000\\
C        0.006820000     -2.852986000      0.000000000\\
C        0.003031000      1.434462000      0.000000000\\
C        0.005800000     -1.429618000      0.000000000\\
C       -1.228833000      0.728291000      0.000000000\\
C        1.246441000      0.731855000      0.000000000\\
C       -1.227428000     -0.725841000      0.000000000\\
C        1.247849000     -0.724596000      0.000000000\\
C       -2.471363000      1.403759000      0.000000000\\
C        2.476361000      1.409934000      0.000000000\\
C       -2.468640000     -1.403735000      0.000000000\\
C        2.479092000     -1.400263000      0.000000000\\
C       -3.683226000      0.724787000      0.000000000\\
C        3.702308000      0.733669000      0.000000000\\
C       -3.681823000     -0.727133000      0.000000000\\
C        3.703713000     -0.721584000      0.000000000\\
C       -6.120848000      0.713284000      0.000000000\\
C        6.153898000      0.736928000      0.000000000\\
C       -6.119464000     -0.720377000      0.000000000\\
C        6.155290000     -0.720054000      0.000000000\\
C       -4.945275000      1.409506000      0.000000000\\
C        4.942941000      1.416829000      0.000000000\\
C       -4.942540000     -1.414312000      0.000000000\\
C        4.945666000     -1.402337000      0.000000000\\
H        0.000937000      5.137412000      0.000000000\\
H        0.010981000     -5.132564000      0.000000000\\
H       -2.469865000      2.490066000      0.000000000\\
H        2.472585000      2.496171000      0.000000000\\
H       -2.464969000     -2.490035000      0.000000000\\
H        2.477494000     -2.486506000      0.000000000\\
H       -4.943098000      2.497466000      0.000000000\\
H        4.940397000      2.505606000      0.000000000\\
H       -4.938217000     -2.502263000      0.000000000\\
H        4.945263000     -2.491114000      0.000000000\\
H       -7.067510000     -1.252023000      0.000000000\\
H       -7.069921000      1.243099000      0.000000000\\
C        7.420077000     -1.401878000      0.000000000\\
C        7.417429000      1.421117000      0.000000000\\
C        8.593983000      0.727773000      0.000000000\\
H        9.541927000      1.259601000      0.000001000\\
C        8.595377000     -0.706399000      0.000000000\\
H        9.544396000     -1.236286000      0.000000000\\
H        7.419349000     -2.490005000      0.000000000\\
H        7.414514000      2.509234000      0.000001000\\

\clearpage

\bibliography{refbig_persp2_short}